\documentclass[12pt,oneside,a4paper]{mr}
\usepackage{datetime}
\usepackage{xy}
\usepackage{ifpdf}
\usepackage{graphicx}
\usepackage{setspace}

\usepackage{grffile}
\usepackage{tabularx}
\usepackage{amsthm}
\usepackage{amsmath}
\usepackage{latexsym}
\usepackage{algorithm}
\usepackage{algorithmic}
\usepackage{amsfonts}
\usepackage{latexsym}
\usepackage{amsfonts}

\usepackage{myshortcuts}
\usepackage{amssymb}

\usepackage[titles]{tocloft}

\xyoption{all} 
\xyoption{frame}
\xyoption{line}
\xymatrixcolsep{3cm}
\newfloat{method}{h}{lom}
\floatname{method}{Method}

\def\CALL#1{\textsc{#1}}

\usepackage{a4wide}
\def\ifndef#1{\expandafter\ifx\csname#1\endcsname\relax} 

\newcommand\cobrx[2]
 {\begin{center}
  \ifndef{pdfpagewidth} 
    \includegraphics*[width=#1]{#2.eps}\else
    \includegraphics*[width=#1]{#2.pdf}
   \fi
  \end{center}}

\newcommand\coverpagestarts
{\pagenumbering{roman}\thispagestyle{empty}
\begin{center}
\large\sffamily
\University\\
\Faculty\\
\Department\\
\vglue 10mm
\cobrx{50mm}{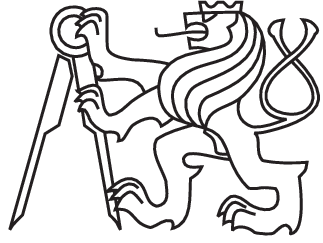}
\vglue 30mm
{\large\bf \DissTitle}\\
\bigskip
{\normalsize\rmfamily by}\\
\bigskip
{\large\it \FirstandFamilyName}\\
\vfill
{\normalsize\rmfamily A thesis submitted to\\the {\Faculty}, {\University},\\
in partial fulfilment of the requirements for the degree of Doctor.}\\
\vglue 20mm
{PhD programme: \PhDProgram}\\
{Specialization: \PhDSpecialization}\\
\bigskip
\Month\ \Year
\end{center}
\newpage
\rule{0pt}{0pt}
\vfill
\begin{description}
\item[Thesis Supervisor:]\ \\
\Supervisor\\
\SupervisorAffiliation
\end{description}
\ifndef{CoSupervisor}\relax\else\bigskip
\begin{description}
\item[Thesis Co-Supervisor:]\ \\
\CoSupervisor\\
\CoSupervisorAffiliation\\
\end{description}
\vglue 1cm
\fi
Copyright {\copyright} {\Year} by {\FirstandFamilyName}
}

\newcommand\mainbodystarts{
\pagestyle{headings}\pagenumbering{arabic}\setcounter{page}{1}
}

\newtheorem{lemma}{Lemma}
\newtheorem{note}[lemma]{Note}
\newtheorem{definition}[lemma]{Definition}
\newtheorem{corollary}[lemma]{Corollary}
\newtheorem{theorem}[lemma]{Theorem}
\newtheorem{proposition}[lemma]{Proposition}
\numberwithin{lemma}{chapter}

\newcommand{\bcen}{\begin{center}}
\newcommand{\ecen}{\end{center}}
\newcommand{\blem}{\begin{lemma}\sl}
\newcommand{\elem}{\end{lemma}\rm}
\newcommand{\bnote}{\begin{note}\rm}
\newcommand{\enote}{\end{note}}
\newcommand{\bcor}{\begin{corollary}\sl}
\newcommand{\ecor}{\end{corollary}\rm}
\newcommand{\bdefi}{\begin{definition}\rm}
\newcommand{\edefi}{\end{definition}}
\newcommand{\btheo}{\begin{theorem}\sl}
\newcommand{\etheo}{\end{theorem}\rm}
\newcommand{\bprop}{\begin{proposition}\sl}
\newcommand{\eprop}{\end{proposition}\rm}
\newcommand{\bexam}{\begin{example}\rm}
\newcommand{\eexam}{\end{example}}
\newcommand{\bfig}{\begin{figure}\begin{center}}
\newcommand{\efig}{\end{center}\end{figure}}
\newcommand{\btab}{\begin{table}\begin{center}}
\newcommand{\etab}{\end{center}\end{table}}
\newcommand{\benum}{\begin{enumerate}}
\newcommand{\eenum}{\end{enumerate}}
\newcommand{\bbibitem}{\begin{itemize}}
\newcommand{\ebibitem}{\end{itemize}}
\newcommand{\bbibflushr}{\begin{flushright}}
\newcommand{\ebibflushr}{\end{flushright}}

\settimeformat{ampmtime}
\ddmmyyyydate

\newcommand\DissTitle{Parallel algorithms for mining of frequent itemsets}
\newcommand\FirstandFamilyName{Robert Kessl}

\newcommand\Month{February}
\newcommand\Year{2011}
\newcommand\Supervisor{Pavel Tvrd\'{i}k}
\newcommand\SupervisorAffiliation{
Department of Computer Systems\\
Faculty of Information Technologies\\
Czech Technical University in Prague\\
Kolejn\'{i} 550/2\\
160 00 Praha 6\\
Czech Republic}

\newcommand\PhDProgram{Electrical Engineering and Information Technology}
\newcommand\PhDSpecialization{Computer Science and Engineering}
\newcommand\Department{Department of Computer Science and Engineering}
\newcommand\Faculty{Faculty of Electrical Engineering}
\newcommand\University{Czech Technical University in Prague}

\begin{document}

\coverpagestarts

\newpage

\chapter*{Abstract and contributions}


In the recent decade companies started collecting of large amount of
data. Without a proper analyse, the data are usually useless. The
field of analysing the data is called \emph{data
  mining}. Unfortunately, the amount of data is quite large: the data
do not fit into main memory and the processing time can become quite
huge. Therefore, we need parallel data mining algorithms.

One of the popular and important data mining algorithm is the
algorithm for generation of so called \emph{frequent itemsets}. The
problem of mining of \emph{frequent itemsets} can be explained on the
following example: customers goes in a store put into theirs baskets
some goods; the owner of the store collects the baskets and wants to
know the set of goods that are bought together in at least $p\%$ of
the baskets.

Currently, the sequential algorithms for mining of frequent itemsets
are quite good in the means of performance. However, the parallel
algorithms for mining of frequent itemsets still do not achieve good
speedup. 

In this thesis, we develop a parallel method for mining of frequent
itemsets that can be used for an arbitrary depth first search
sequential algorithms on a distributed memory parallel computer. Our
method achieves speedup of $\approx 6$ on 10 processors.  The method
is based on an approximate estimation of processor load from a
database sample -- however it always computes the set of frequent
itemsets from the whole database. In this thesis, we show a theory
underlying our method and show the performance of the estimation
process.

\bigskip

\noindent{\bf Keywords:}

~frequent itemset mining, parallel algorithms, approximate counting

\vfill



\newpage

\chapter*{Acknowledgements}

 
I would like to thanks to Petr Savick\'{y}. His advises contributed
substantially to the quality and completeness of the thesis.  I would
also like to thanks to Pascal Fleury and Pavel Kv\v{e}to\v{n} for
proofreading of this text.

This dissertation thesis was supported from the Czech Science
Foundation, grant number GA \v{C}R P202/10/1333.

Finally, my greatest thanks to my family for their support.

\newpage

\newpage
\vglue 6cm 

{\em To Lenka and my daddy}.



\newpage

\tableofcontents

\newpage

\listoffigures
\listoftables
\listofalgorithms



\newpage


\newpage

\mainbodystarts

\onehalfspacing


\chapter{Introduction}

Thanks to the automated data collection, companies collect huge amount
of data. It is impossible to manually analyse such amounts of
data. Therefore, automatic methods for analysis of the data are
developed in \emph{data mining}.

One of the important data mining tasks is the \emph{mining of
  association rules} or \emph{market basket analysis}
\cite{agrawal94fast}. The term market basket analysis comes from the
first historical application. The market basket analysis comes from
the need to analyse customer baskets of goods bought in a
supermarket. The supermarket stores the list of items of the basket,
called a \emph{transaction}, into a database. The owner of the
supermarket is interested in better shelf organization and wants to
analyse the behaviour of customers in the supermarket from the database
of the transactions. The result of the process are so called
\emph{association rules}, i.e. rules $X\Rightarrow Y$ such that $X,Y$
are sets of goods.

The association rules are mined in a two step process:

\begin{enumerate}
\item Mine all \emph{frequent itemsets} (FIs in short): find all sets
  of items that occur in a fraction of transactions at least of size
  $\rminsupp$. The $\rminsupp$, called the relative minimal support,
  is a parameter of the computation. An example of a frequent itemset
  is the set $\isetU=\{$\texttt{bread}, \texttt{milk},
  \texttt{butter}$\}$ with support $\supp(U)=0.3$, i.e., the set
  $\isetU$ occurs in $30\%$ of transactions.
\item Generate \emph{association rules}: from the FIs generate all
  association rules with minimal confidence $\minconf$. An example of
  an association rule is $\{$\texttt{bread}, \texttt{milk}$\}
  \Rightarrow\{$\texttt{butter}$\}$ with confidence $15\%$, i.e. the
  \texttt{butter} occurs in $15\%$ of transactions that also contains
  \texttt{bread} and \texttt{milk}.
\end{enumerate}

Because the mining of FIs is computationally expensive, we can only
mine some subsets of FIs, e.g. the mining of \emph{maximal frequent
  itemsets} (MFIs in short).

The problem of mining of FIs can be generalized to a
wide variety of problems, called frequent substructure mining:

\begin{enumerate}
\item mining of frequent subgraphs,
\item mining of frequent sequences,
\item mining of frequent episodes.
\end{enumerate}

An example of the frequent subgraph mining is the mining of the
structure of proteins. Proteins are complicated molecules that can be
viewed as a combinatorial graph. For a set of proteins, the task is to
find frequent subgraphs. The information computed from the database
can then help the chemists to search for proteins with similar effect,
for example.

\clearpage
\newpage

\chapter{Mathematical foundation}\label{chap:math-found}

\section{Basic notions}

Let $\baseset = \{ \bitem_i \}$ be a \emph{base set} of items (items
can be numbers, symbols, strings etc.). An arbitrary set of items
$\isetU \subseteq \baseset$ will be further called an \emph{itemset}.
Further, we need to view the baseset $\baseset$ as an ordered set. The
items are therefore ordered using an arbitrary order $<$: $\bitem_1 <
\bitem_2 < \ldots < \bitem_n, n = |\baseset|$. Hence, we can view an
itemset $\isetU = \{\bitem_{u_1}, \bitem_{u_2}, \ldots,
\bitem_{u_{|\isetU|}}\}$, $\bitem_{u_1} < \bitem_{u_2} < \ldots
\bitem_{u_{|\isetU|}}$, as an ordered set denoted by $\isetU =
(\bitem_{u_1}, \bitem_{u_2}, \ldots, \bitem_{u_{|\isetU|}})$. If it is
clear from context, we will not make difference between the set
$\{\bitem_{u_1}, \bitem_{u_2}, \ldots, \bitem_{u_{|\isetU|}}\}$ and
the ordered set $(\bitem_{u_1}, \bitem_{u_2}, \ldots,
\bitem_{u_{|\isetU|}})$. We denote the $i$th smallest item of $\isetU$
ordered by the arbitrary order $<$ by $\isetU[i] = \bitem_{u_i}$. We
denote the set of all itemsets, the powerset of $\baseset$, by
$\powerset(\baseset)$.

First, we define some necessary concepts:

\begin{definition}[Transaction] 
Let $\isetU\subseteq\baseset$ be an itemset and $\texttt{id}\in
\natnum$ a natural number, used as an identifier.  We call the pair
$(id, \isetU)$ a \emph{transaction}. The \texttt{id} is called the
\emph{transaction id}.
\end{definition}

A subset $\isetW$ of a transaction $t = (id, \isetU)$ will be further
denoted by $\isetW\trsubseteq t$, i.e., $\isetW$ is a subset of $t$ if
and only if $\isetW\subseteq\isetU$. A superset $\isetV$ of a
transaction will be denoted similarly, i.e., $t \trsubseteq
\isetV$. Because $\isetU$ can be viewed as an ordered set, we can also
view the transaction $t$ as an ordered set and denote $i$th item of
$t$ by $t[i]$.





\begin{definition}[Database]\label{def:database}
A \emph{database} $\db$ on $\baseset$ (or database $\db$ if $\baseset$
is clear from context) is a sequence of transactions $t \trsubseteq
\baseset$. Each transaction $t$ has an unique number in the database,
called \emph{the transaction id}.
\end{definition}

In our algorithms, we need to sample the database $\db$. A
\emph{database sample} is denoted by $\dbsmpl$.

\begin{definition}[Itemset cover and support]\label{cover}\label{def:cover-support} \cite{ndi}
Let $\isetU \subseteq \baseset$ be an itemset. Then the \emph{cover}
of $\isetU$, denoted by $\bcover(\isetU,\db)$ in a database $\db$, is
the subset of transactions $T = \{(\text{id}_i,\isetV_i) | \isetU
\subseteq \isetV_i\} \subseteq \db$. The number of transactions in
$\bcover(\isetU,\db)$ is called the \emph{support} of $\isetU$ in
$\db$, denoted by $\supp(\isetU,\db) = |\bcover(\isetU,\db)|$.
\end{definition}

We define the support as the number of transactions containing
$\isetU$, but in some literature, the relative support is defined by
$\supp^{*}(\isetU)=\supp(\isetU)/|\db|$.

\begin{definition}[Transaction id list]\label{def:tidlist}
Let $\isetU \subseteq \baseset$ be an itemset, $\db$ a database, and
$T = \bcover(\isetU, \db)$ the itemset cover of $\isetU$. The set
$\tidlist(\isetU, \db) = \{id | \text{exists } \isetV, (id, \isetV)
\in T \}$ is called the \emph{transaction id list} or \emph{tidlist}
in short.
\end{definition}

We omit $\db$ from $\tidlist(\isetV, \db)$, if clear from context.


Some algorithms use the concept of \emph{vertical representation} of a
database. The \emph{vertical representation} of a database $\db$ is
the set of pairs $\{(\{\bitem_i\}, \tidlist(\{\bitem_i\},\db)) |
\bitem_i \in \baseset\}$. The database described in Definition
\ref{def:database} is sometimes called the \emph{horizontal
  representation}. The vertical representation holds the same
information as the horizontal representation of the database $\db$.
The set of all transaction IDs can be denoted by
$\tidlist(\emptyset)$.


\begin{definition}[Transaction cover]\label{itemset-cover}
Let $\transaction\subseteq \db$ be a set of transactions from the
database $\db$. Then the \emph{cover} of $\transaction$, denoted by
$\tcover(\transaction,\db)$, is the greatest itemset $\isetU \subseteq
\baseset$ such that for all $t\in \transaction$ it holds that
$\isetU\trsubseteq t$.
\end{definition}


\begin{definition}[Frequent itemset]
Let $\db$ be a database on $\baseset$, $\isetU\subseteq \baseset$ an
itemset, and $\minsupp\in\natnum$ a natural number. We call $\isetU$
\emph{frequent} in database $\db$ if $\supp(\isetU,\db) \geq
\minsupp$.
\end{definition}

We can also define the frequent itemset using the relative support,
denoted by $\rminsupp,$ $0 \leq \rminsupp \leq 1$, i.e., an itemset is
frequent iff $\rsupp(\isetU,\db)\geq\rminsupp$.

We will denote the set of all frequent itemsets as $\allfi$.  In our
algorithms, we need to sample the set $\allfi$. A sample of frequent
itemsets is denoted by $\fismpl$. In the text, we use $\db$ and
$\minsupp$ ($\rminsupp$) generally, but may be omitted if they are
clear from the context.

\begin{definition}[Maximal Frequent Itemset (MFI in short)]
Let $\db$ be a database on $\baseset$, $\isetU\subseteq\baseset$ an
itemset, and $\minsupp \in Z$ a natural number. We call $\isetU$ a
\emph{maximal frequent itemset} if $\supp(\isetU,\db) \geq \minsupp$,
and $\supp(\isetV,\db) < \minsupp$ for any $\isetV$ such that
$\isetU\subsetneq \isetV$.
\end{definition}

\begin{definition}[Closure operator]
Let $\baseset$ be a base set of items. Let $\isetW \subseteq
\baseset$, we define an operator $\closure: \powerset(\baseset) \rightarrow
\powerset(\baseset)$ as $\closure(\isetW) = (\tcover \circ \bcover) (\isetW) =
\tcover(\bcover(\isetW))$.
\end{definition}

\begin{definition}[Closed itemset]\cite{zaki02charm}\label{def:closed-itemset}
An itemset $\isetU\subseteq\baseset$ is \emph{closed}, if and only if
$\isetU=\closure(\isetU)$.
\end{definition}

The concept of \emph{closed itemsets} (CIs in short) can be used to
reduce the size of the output of an FI algorithm. Additionally, the
compound projection $\tcover \circ \bcover$ can be used for
optimization of depth-first search algorithms (DFS in short) for
mining of FIs, see Appendix~\ref{appendix:seq-alg}.

\begin{definition}[Association rule]
Let $\db$ be a database on $\baseset$ and $\isetV, \isetW \subseteq
\baseset$ be itemsets such that $\isetV\cap \isetW = \emptyset$. Then
the ordered pair $(\isetV, \isetW)$, written $\isetV \Rightarrow
\isetW$, is called the \emph{association rule}. The itemset $\isetV$
is called the \emph{antecedent} and the itemset $\isetW$ is called the
\emph{consequent}.
\end{definition}

\begin{definition}[Confidence]
Let $\db$ be a database and $\isetV\Rightarrow \isetW$ an association rule.  The
\emph{confidence} of $\isetV \Rightarrow \isetW$ is defined as:
$$\conf(\isetV,\isetW,\db) = \frac{\supp(\isetV\cup
  \isetW,\db)}{\supp(\isetV,\db)}$$
\end{definition}

If it is clear from context, we omit the database $\db$ from the notation.

The association rules are mined in a two step process: 1) mine all FIs
$X=\isetV \cup \isetW, \isetV \cap \isetW = \emptyset$; 2) create
association rules $\isetV \Rightarrow \isetW$ from the FIs mined in
the first step, such that $\conf(\isetV, \isetW, \db) \geq
\minconf$. In our work, we consider only the first phase, i.e., we do
not consider the task of creation of association rules from all
frequent itemsets.

The values of $\minsupp$ (or $\rminsupp$) and $\minconf$ and a
database $\db$ are inputs to algorithms for the mining of association
rules. These algorithms first find all frequent itemsets, using the
$\minsupp$, and then generate association rules, using $\minconf$.

For the purpose of the description of the parallel algorithm, we
denote the number of processors by $\procnum$. The $i$th processor,
$1\leq i\leq\procnum$, is denoted by $\proc_i$.

At the start of the parallel algorithm, each processor $\proc_i$ has a
database partition $\dbpart_i$. Our parallel algorithms partitions the
database at the beginning into disjoint database partitions
$\dbpart_i, \dbpart_j$ such that $\bigcup_i \dbpart_i=\db$, $\dbpart_i
\cap \dbpart_j = \emptyset$, and $|\dbpart_i| \approx
|\db|/\procnum$. In our work, usually, processor $\proc_i$ loads
partition $\dbpart_i$ into main memory.

\section{The monotonicity of support}\label{sec:support-monotonicity}

The basic property of frequent itemsets is the so called
\emph{monotonicity of support}.  It is an important property for all
FIs mining algorithms and is defined as follows:

\begin{theorem}[Monotonicity of support] \label{monotonicity}
Let $\isetU, \isetV \subseteq \baseset$ be two itemsets such that
$\isetU \subsetneq \isetV$ and $\db$ be a database.  Then holds
$\supp(\isetU,\db) \geq \supp(\isetV, \db)$.
\end{theorem}

\begin{proof}
If a set $\isetU$ is contained in transactions $\tidlist(\isetU)$, then
a superset $\isetV\supseteq\isetU$ is contained in transactions
$\tidlist(\isetV)\subseteq \tidlist(\isetU)$.
\end{proof}

\begin{corollary}\label{monotonicity-subsets}
Let $\isetV$ be a frequent itemset, then all subsets $\isetU\subseteq
\isetV$ are also frequent.
\end{corollary}

\begin{proof}
Let $\isetU,\isetV\subseteq\baseset$ be two frequent itemsets such
that $\isetU \subsetneq \isetV$. Then by using the argument from
Theorem~\ref{monotonicity} it holds $\tidlist(\isetV) \subseteq
\tidlist(\isetU)$ and therefore $\supp(\isetV) \leq
\supp(\isetU)$. Because $\isetV$ is frequent, $\isetU$ must be frequent
as well.
\end{proof}


\section{The lattice of all itemsets}\label{sec:lattice-of-all-itemsets}

Zaki \cite{zaki00scalable} uses the set of all items,
$\powerset(\baseset)$, and the underlying lattice for description of
DFS algorithms. 

\begin{definition}
Let $P$ be finite ordered set, and let $S\subseteq P$. An element
$X\in P$ is an \emph{upper bound} (\emph{lower bound}) of $S$ if
$s\leq X$ ($s\geq X$) for all $s\in S$. A least upper bound is
called \emph{join} and is denoted by $\bigvee S$, and a greatest
lower bound, also called \emph{meet}, of $S$ is denoted $\bigwedge
S$. The greatest element of $P$, denoted by $\top$, is called the
\emph{top element}, and the least element of $P$, denoted by $\bot$,
is called the \emph{bottom element}.
\end{definition}

We denote the join (meet) of two elements $X,Y\in P$ by $X\vee Y$
($X\wedge Y$).

\begin{definition}
Let $\lattice$ be an ordered set, $\lattice$ is called a \emph{join}
(\emph{meet}) \emph{semilattice} if $X\vee Y$ ($X\wedge Y$) exists for
all $X,Y\in \lattice$. $\lattice$ is called a \emph{lattice} if it is
both a join and meet semilattice. $\lattice$ is \emph{complete
  lattice} if $\bigvee S$ and $\bigwedge S$ exist for all subsets
$S\subseteq \lattice$. An ordered set $M\subseteq \lattice$ is a
\emph{sublattice} of $\lattice$ if $X,Y\in M$ implies $X\vee Y\in M$
and $X\wedge Y\in M$.
\end{definition}

It is well known that for a set $S$ the powerset $\powerset(S)$ is a
complete lattice. The \emph{join} operation is the \emph{set union
  operation} and \emph{meet} the \emph{set intersection operation}.

For any ${\cal S}\subseteq \powerset(\baseset)$, ${\cal S}$ forms a
lattice of sets $({\cal S};\subseteq)$ if it is closed under finite
number of unions and intersections.

\begin{lemma}
The set of all \emph{frequent itemsets} forms a \emph{meet
  semilattice}.
\end{lemma}

\begin{proof}
The result follows from Corollary \ref{monotonicity-subsets} and
the fact that $\isetV\wedge\isetW=\isetV \cap \isetW$.
\end{proof}

\begin{corollary}
The set of maximal frequent itemsets delimits the set of
all frequent itemsets from above in the sense of set inclusion.
\end{corollary}

\begin{definition}
Let $P$ be an ordered set, and let $X,Y,Z\in P$. We say $X$ is covered
by $Y$, denoted $X\sqsubset Y$, if $X<Y$ and $X\leq Z < Y$ implies
$X=Z$, i.e., if there is no element $Z$ of $P$ with $X<Z<Y$.
\end{definition}

\begin{definition}\label{def:lattice-atom}
Let $\lattice$ be a lattice with bottom element $\bot$. Then
$\atom_i\in\lattice$ is called an atom if $\bot\sqsubset \atom_i$. The
set of atoms of $\lattice$ is denoted by $\atoms(\lattice)$.
\end{definition}

A set of all atoms of a lattice $\lattice = (\powerset(\baseset);
\subseteq)$ is $\atoms(\lattice)=\baseset$.

\section{The use of the lattice of frequent itemsets in algorithms} \label{sec:lattice-in-algorithms}

The lattice of frequent itemsets is the basic mathematical structure
for the description of sequential FIs mining algorithms. There are
many algorithms for mining of FIs, namely the Apriori algorithm, the
Eclat algorithm, and the FP-Growth algorithm.  All these algorithms
are based on the theory described in this section. Additionally, to
parallelize the sequential algorithms, we need to partition the set
$\allfi$ of all FIs into disjoint sets. The partitioning is also
described in this section.

To decompose $\powerset(\baseset)$ into disjoint sets, we need to
order the items in $\baseset$.  An equivalence relation partitions the
ordered set $\powerset(\baseset)$ into disjoint subsets called
\emph{prefix-based equivalence classes}:

\begin{definition}[prefix-based equivalence class (PBEC in short)]
\label{pbecdef}
Let $\isetU\subseteq\baseset, |\isetU|=n,$ be an itemset. We impose some
order on the set $\baseset$ and hence view $\isetU = (u_1, u_2,
\ldots, u_n), u_i\in\baseset$ as an ordered set. A \emph{prefix-based
  equivalence class} of $\isetU$, denoted by $[\isetU]_{\ell}$, is a set of
all itemsets that have the same prefix of length $\ell$,
i.e., $[\isetU]_{\ell}=\{\isetW=(w_1, w_2, \ldots, w_m) | u_i=w_i, i\leq \ell,
m = |\isetW| \geq \ell, \isetU,\isetW\subseteq\baseset\}$
\end{definition}

To simplify the notation, we use $[\isetW]$ for the prefix-based
equivalence class $[\isetW]_{\ell}$ iff $\ell = |\isetW|$. Each $[\isetW],
\isetW\subseteq\baseset$ is a sublattice of $(\powerset(\baseset),
\subseteq)$. 


\begin{definition}[Extensions]
\label{def:extension}
Let $\isetU\subseteq\baseset$ be an itemset. We impose some order $<$
on the set $\baseset=(\bitem_1,\bitem_2,\ldots,\bitem_n)$ and view
$\isetU = (u_1, u_2, \ldots, u_m), u_i\in\baseset,$ as an ordered
set. The \emph{extensions} of the prefix-based equivalence class
$[\isetU]$ is an ordered set $\prefixext\subseteq\baseset$ such that
$\isetU \cap \prefixext = \emptyset$ and for each $\isetW\in[\isetU]$
holds that $\isetW\setminus\isetU \subseteq \prefixext$. We denote the
prefix-based equivalence class together with the extensions
$\prefixext$ by $[\isetU | \prefixext]$.
\end{definition}

For example, let have $\baseset=\{1, 2, 3, 4, 5\}$, a prefix $\isetU =
\{1, 2\}$, and the extensions $\prefixext = \{ 3, 5 \}$. Then
$[\isetU|\prefixext] = [\{1, 2\} | \{ 3, 5 \} ] = \{ \{1,2,3\},
\{1,2,5\}, \{1,2,3, 5\}\}$.

\begin{proposition}\label{prop:disjoint-pbecs}
Let $\isetU_i = \{\bitem_i\}, \bitem_i\in\baseset$ for all $i, 1\leq i
\leq |\baseset|,$ and $\prefixext_i = \{ \bitem | \bitem > \bitem_i;
\bitem,\bitem_i\in\baseset\}$ then $[\isetU_i | \prefixext_i]$ are
disjoint.
\end{proposition}

\begin{proof}
The reason is obvious: each $\isetW \in [\isetU_i|\prefixext_i]$
contains $\bitem_i$ and \emph{does not contain} $\bitem < \bitem_i$.
\end{proof}

\begin{proposition}
Let $\prefixset = \{ (\isetU_i, \prefixext_i) \}$ be a set such that
$[\isetU_i | \prefixext_i]$ are disjoint and $q = (\isetV,
\prefixext_{\isetV}) \in \prefixset$. Let $\isetW_i = \isetV \cup \{ \bitem_i
\}, \bitem_i \in \prefixext_{\isetV}$ and $\prefixext_{\isetW_i} = \{
\bitem | \bitem_i < \bitem; \bitem_i,\bitem \in\prefixext_{\isetV} \}$ forms the
PBECs $[\isetW_i | \prefixext_{\isetW_i}]$. Let have a new set of
pairs $\prefixset' = (\prefixset \setminus \{q\}) \cup (\bigcup_i \{ (\isetW_i,
\prefixext_{\isetW_i}) \} )$. Then the pairs $(\isetU'_i, \prefixext'_i) \in
\prefixset'$ forms disjoint PBECs $[\isetU'_i | \prefixext'_i]$.
\end{proposition}

\begin{proof}

It suffices to show that the new PBECs $[\isetW_i | \prefixext_i]$ are
disjoint and that the union $(\bigcup_i[\isetW_i | \prefixext_i]) \cup
\{\isetV\} = [\isetV | \prefixext_{\isetV}]$.

The PBECs $[\isetW_i | \prefixext_i]$ are disjoint (using the same
argument as in Proposition~\ref{prop:disjoint-pbecs}), because each
$\isetW_i=\isetV\cup\{\bitem_i\}, \bitem_i \in \prefixext_{\isetV},$
contains one $\bitem_i \in \prefixext_{\isetV}$ and does not contain
any $\bitem\in\prefixext_{\isetV}$ such that $\bitem < \bitem_i$.

Each $X\in [\isetV | \prefixext_{\isetV}]$ has the form $X=\isetV\cup
Y, Y\subseteq \prefixext_{\isetV}, Y\neq\emptyset$. We can partition
the sets $Y$ on those having a prefix $\bitem_1 \in
\prefixext_{\isetV}$, those having a prefix $\bitem_2 \in
\prefixext_{\isetV}$, etc. But, this is exactly how the new PBECs
$[\isetW_i | \prefixext_{\isetW_i}]$ were created. Since we have used
all $\bitem\in\prefixext_{\isetV}$, it must be true that $(\bigcup_i
[\isetW_i | \prefixext_{\isetW_i}])\cup\{\isetV\} = [\isetV |
  \prefixext_{\isetV}]$.

\end{proof}


\label{pgref:extension-notation-simple} We simplify the notation and omit the extensions if clear from context.

\begin{lemma}
Let $\isetW\subseteq\baseset$ be an itemset. The equivalence class
$[\isetW]$ is a sublattice of the lattice $(\powerset(\baseset),
\subseteq)$.
\end{lemma}
\begin{proof}
Let $\isetU,\isetV$ be itemsets in class $[\isetW]$, i.e.,
$\isetU,\isetV$ share common prefix $\isetW$. $\isetW \subseteq \isetU
\cap \isetV$ implies that $\isetU\wedge \isetV \in [\isetW]$, and
$\isetU\vee \isetV \in [\isetW]$.  Therefore, $[\isetW]$ is a
sublattice of $(\powerset(\baseset),\subseteq)$.
\end{proof}


\begin{definition}
Let $\isetU,\isetW\subseteq\baseset$ and $[\isetU],[\isetW]$ be
prefix-based equivalence classes. Then $[\isetW]$ is a
\emph{prefix-based equivalence subclass} of $[\isetU]$ if and only if
$[\isetW]\subsetneq[\isetU]$.
\end{definition}

\begin{proposition}\label{prop:pbe-subclass}
Let $\isetW,\isetU\subseteq\baseset$. If $[\isetW]$ is a prefix-based
equivalence subclass of $[\isetU]$, then $\isetU\subsetneq\isetW$.
\end{proposition}

From Proposition \ref{prop:pbe-subclass}, it follows that the
prefix-based equivalence classes form a hierarchy.  The hierarchy of
classes is a tree, where each node corresponds to a prefix $\isetW$
and the children to the supersets $\isetU_i\supsetneq\isetW, 1\leq
i\leq n$ for some $n \geq 1$, such that $|\isetU_i| = |\isetW| +
1$. The items $\prefixext = \bigcup_i \isetU_i \setminus \isetW,
|\prefixext| = n,$ are the \emph{extensions} of $[\isetW]$.

Further, we need to partition $\allfi$ into $n$ disjoint sets, denoted
by $\fipart_1, \ldots, \fipart_n$, satisfying $\fipart_i\cap
\fipart_j=\emptyset, i\neq j$, and $\bigcup_i \fipart_i=\allfi$. This
partitioning can be done using the prefix-based equivalence
classes. The prefix-based equivalence classes can be collated to a
single partition: let have prefix-based equivalence classes
$[\isetU_l], (\bigcup_l[\isetU_l]) \cup (\bigcup_{l}
\powerset(\isetU_l)) = \allfi, 1\leq l\leq m$ and sets of indexes of
the prefix-based equivalence classes $\fipartidxset_i \subseteq \{k
| 1\leq k\leq m\}, 1\leq i\leq n$ such that $\fipartidxset_i \cap
\fipartidxset_j = \emptyset$ and $\sum_i|\fipartidxset_i| = m$ then
$\fipart_i = \bigcup_{l\in\fipartidxset_i}([\isetU_{l}] \cap
\allfi)$. To create prefix-based equivalence classes, we partition the
lattice into sublattices recursively. First, we partition the lattice
using prefixes of size $1$, i.e., $[(\bitem_i)],
\bitem_i\in\baseset$. Then, we can pick an arbitrary class and
partition it further on prefix-based equivalence classes with prefixes
of size $2$, etc. This recursive decomposition forms a depth-first
search (DFS in short) expansion tree, see
Example~\ref{ex:math-notion}.

\begin{definition}[Relative size of a PBEC]\label{def:relative-pbec-size}
Let $[\isetU|\prefixext]$ be a PBEC and a set of itemsets
$\mathcal{I}$. The itemsets in $\mathcal{I}$ are not necessarily
frequent. We define the \emph{relative size} of the PBEC as
$\frac{|[\isetU|\prefixext] \cap \mathcal{I}|}{|\mathcal{I}|}$.
\end{definition}

By the set $\mathcal{I}$ we \emph{usually} mean the set of all
frequent itemsets $\allfi$. But in Chapter~\ref{chap:approx-counting},
we also use other sets then $\allfi$.

The prefix-based equivalence classes decompose the lattice into
smaller parts that can be processed independently in main memory. That
is, for the computation of supports of itemsets in one prefix-based
equivalence class, we start with the tidlists of the atoms and
recursively construct the tidlists of itemsets belonging to that class
by intersecting them. Due to this, the computation of support in
different prefix-based equivalence classes is done independently.
This is important, because this independence makes parallelization
easier. Moreover, we can recursively decompose each equivalence class
into smaller prefix-based equivalence subclasses.

For the computation of the support of an itemset $\isetU \subseteq
\baseset$, we can use the tidlists of items:


\begin{lemma}
Let $\baseset$ be a baseset and $\isetU\subseteq\baseset, \isetU =
\bigcup_{u_i\in\isetU} \{u_i\}, u_i\in\baseset$. Then the support of
$\isetU$ can be computed by $\supp(\isetU) =
|\bigcap_{u_i\in\isetU}\tidlist(\{ u_i \})|$.
\end{lemma}
\begin{proof}
The support of $\isetU=\{u_i| 1 \leq i \leq n, u_i \in \baseset\}$ is
defined by $\supp(\isetU)=|\tidlist(U)|$, i.e., the number of
transactions containing all the items $u_i$. Hence, the set of all
transactions containing $\isetU$ is
$\tidlist(\isetU)=\bigcap_i\tidlist(u_i)$.
\end{proof}

\begin{corollary}
Let $\baseset$ be a baseset and $\isetU,\isetW_i\subseteq \baseset,
1\leq i\leq n$, for some $n \geq 1$ and $\isetU=\bigcup_i\isetW_i$
then $\supp(\isetU) = |\bigcap_i\tidlist(\isetW_i)|$.
\end{corollary}

It follows that for a prefix $\isetW$ and the extensions $\prefixext$
we can compute the support of $\isetW\cup\isetU, \isetU\subseteq
\prefixext$ using the tidlists of items in $\prefixext$ and the
tidlist $\tidlist(\isetW)$.

\begin{figure}[!ht]
\hrule
\medskip
\example[Illustration of the mathematical notion]{Illustration of the mathematical notion}{ 
\label{ex:math-notion}
\smallskip

\noindent\begin{tabular*}{15cm}{p{7.8cm}p{7.5cm}}
 Horizontal representation of the database $\db$ & Vertical representation of the database $\db$ \\
{
  \hfill
  \renewcommand{\arraystretch}{1}
{\scriptsize
\exampledatabase\hfill%
}
  \medskip
} &
{\scriptsize
  \renewcommand{\arraystretch}{1}
       \begin{tabular}{|c|c|c|c|c|c|c|} \hline
	 {\tt itemset} & $\{1\}$ & $\{2\}$ & $\{3\}$ & $\{4\}$ & $\{5\}$ & $\{6\}$ \\ \hline
	 {\tt TID}     & $1$  & $1$ & $1$  & $1$  & $2$    & $1$  \\ 
	 {\tt list}    & $3$  & $4$ & $2$  & $3$  & $5$    & $2$  \\ 
	               & $4$  & $6$ & $3$  & $5$  & $6$    & $4$  \\ 
	               & $5$  & $7$ & $6$  & $6$  & $7$    & $5$  \\ 
	               & $6$  & $8$ & $7$  & $7$  & $8$    & $9$  \\ 
	               & $11$ &$10$ & $8$  & $8$  & $9$    & $12$ \\ 
	               & $15$ &$11$ & $9$  & $9$  & $10$   & $13$ \\ 
	               &      &$12$ & $10$ & $10$ & $11$   & $14$ \\ 
	               &      &     & $11$ & $11$ & $12$   & $15$ \\ 
	               &      &     & $15$ & $12$ & $13$   &      \\ 
	               &      &     &      & $13$ & $14$   &      \\ 
	               &      &     &      & $14$ & $15$   &      \\ 
	               &      &     &      & $15$ &        &      \\ \hline
       \end{tabular}\hfill%
  \medskip
}
\if0\\
\multicolumn{2}{c}{%
  \renewcommand{\arraystretch}{1}
       \begin{tabular}{|c|c|c|c|c|c|c|c|c|c|} \hline
	 Support         & $3$  & $4$  & $4$  & $2$  & $2$  & $2$  & $4$  & $3$  & $2$   \\ \hline
	 {\tt itemset}    & $\{2,1\}$ & $\{1,3\}$ & $\{1,4\}$ & $\{1,5\}$ & $\{2,3\}$ & $\{2,4\}$ & $\{3,4\}$ & $\{3,5\}$ & $\{4,5\}$  \\ \hline
	 {\tt TID}        &    $1$    &    $1$    &    $1$    &    $5$    &    $1$    &    $1$    &    $1$    &    $2$    &    $5$     \\
	 {\tt list}       &    $4$    &    $3$    &    $3$    &    $6$    &    $6$    &    $6$    &    $3$    &    $5$    &    $6$     \\
	                  &    $6$    &    $5$    &    $5$    &           &           &           &    $5$    &    $6$    &            \\
	                  &           &    $6$    &    $6$    &           &           &           &    $6$    &           &            \\ \hline
       \end{tabular}
  \medskip%
}\fi
\end{tabular*}

\begin{center}
\includegraphics[type=mps,ext=.mps,read=.mps]{lattice_dfstree_pbec}
\end{center}

\hbox to 15cm{\hsize=15cm
\vtop{
\hsize=15cm
\noindent The picture shows the set $\allfi$ of the database $\db$
with $\minsupp=5$. The grey lines show the subset/superset
relationship. The arrows show the DFS expansion tree.}}

\centering

  \begin{itemize}
    \item Prefix-based equivalence class
      $[(2)]\cap\allfi=\{\{2\}, \{2,3\}, \{2,4\}, \{2,5\},\allowbreak \{2,3,4\},\allowbreak \{2,4,5\}\}$,
      marked in blue.
    \item Prefix-based equivalence class $[(2,3)]\cap\allfi=\{\{2,3\}, \{2,3,4\}\}$
      is a subclass of $[(2)]$, marked in red.
    \item The DFS expansion tree is highlighted using thicker lines with
      arrows. The extensions of the tree node $\{2\}$ is the set of
      nodes $\{3,4,5\}$, i.e., nodes $\{\{2,3\},\{2,4\},\{2,5\}\}$.
    \item The MFIs is the set $\mfi=\{\{1,3,4\}, \{2,3,4\}, \{2,4,5\}, \{3,4,5,6\}\}$
  \end{itemize}
}
\medskip
\hrule
\end{figure}

\clearpage

\section{Complexity of mining of frequent itemsets}\label{seq:fimi-mining-complexity}

The number of FIs is $2^{|\baseset|}$ in the worst case. In practice
the number of FIs is very small in comparison to $2^{|\baseset|}$.

Let $\mfi$ be the set of all maximal frequent itemsets. Let the size
of the longest MFI be $s=\max \{ |\isetU| : \isetU \in \mfi \}$. The
complexity of mining of all FIs is exponential in $s$. Let
$\smfi_i\in\mfi$ be an MFI then the computational complexity of mining
all FIs is $\Ocmplxty(\sum_i 2^{|\smfi_i|})=\Ocmplxty(2^{s}\cdot|\mfi|)$, where
$\Ocmplxty(\cdot)$ denotes the big O notation.

For a good introductory text on the computational complexity, see
\cite{AB09}. The notation used in this Section is based on \cite{AB09}.

\subsection{Maximal frequent itemsets}

We need to assess the computational complexity of the task of
enumeration of all MFIs. The NP-Completeness theory is mainly concerned
about existence of a solution. Fortunately, there are other complexity
classes, such as \#P and \#P-Complete \cite{AB09} that concerns about
counting the number of solutions. The \emph{counting Turing
  machine} \cite{AB09} is a standard non-deterministic Turing machine
that has an additional tape on which the number of accepting
computations is printed. The \#P is a problem that is solved by the
counting Turing machine in polynomial time.  The \#P-Complete problems
are those problems on which all other problems from \#P
reduce \cite{AB09}.

The counting problem $\Pi$ can be solved using an associated
enumeration problem $\Pi'$: we enumerate the solutions using $\Pi'$
and then count them. That is: if we know that a counting problem is
\#P-Complete then the associated enumeration problem must be
NP-Hard \cite{AB09}.

A bipartite graph $G_1=(U,V,E)$ is a subgraph of another bipartite
graph $G_2=(U',V',E')$ if $U\subseteq U'$, $V\subseteq V'$, and
$E\subseteq E'$. A bipartite graph $G_3=(U_3,V_3,E_3)$ is called
\emph{bipartite clique} if and only if $E_3=U_3\times V_3$, in
particular interest are bipartite cliques that appears as subgraphs in
another graph. We will omit $E_3$ from the notation of a bipartite
clique. A \emph{maximal bipartite clique} $G'=(U',V')$ in a given
graph $G=(U,V)$ is a clique such that there is no bipartite clique
$G''=(U'',V'')$, $U'\subseteq U''$, and $V'\subseteq V''$.

There is an intuitive correspondence between cliques and
transactions. Let have a bipartite graph $G=(\baseset,
\tidlist(\emptyset), E)$, such that $e=(\bitem_i,t)\in E$ if the
transaction $t$ contains the item $\bitem_i$, i.e., an edge of the
graph represent the fact that an item is contained in a transaction.

The problem of counting the number of maximal bipartite cliques can be
reduced to the problem of counting MFIs \cite{yang04complexity}.

\vbox{\begin{theorem}\cite{yang04complexity} 
The problem of counting the number of all bipartite cliques is
$\#$P-complete.
\end{theorem}
}

The previous discussion give us an evidence that mining of maximal
frequent itemsets is NP-hard.

\newpage

\chapter{Contribution of the thesis}

In this dissertation thesis, we present a novel method for
parallelization of an arbitrary algorithm for mining of all FIs. We
are able to parallelize depth-first search algorithms, which is a hard
task. Our method statically load-balance the computation using a
``double sampling process''. The ``double sampling process'' first
creates a database sample $\dbsmpl$ and using $\dbsmpl$ computes a
sample of FIs $\fismpl$. $\fismpl$ is then used for partitioning of
$\allfi$ into disjoint sets $\fipart_i$ such that $\bigcup_i \fipart_i
= \allfi$. The input of the whole process is the database $\db$, each
processor $\proc_i$ loads a database partition $\dbpart_i$ such that
$\db=\bigcup_i \dbpart_i$, the minimal support $\rminsupp$, and the
sampling parameters or the size of the database sample $\dbsmpl$ and
the size of the sample of FIs $\fismpl$.

The method consists of four phases: 1) creation of the database
sample $\dbsmpl$ and sample of FIs $\fismpl$; 2) creation of the
partitioning of $\allfi$; 3) exchanging of database partitions among
the processors; 4) the set $\allfi$ is computed in parallel.

We present three variants of our new method:

\begin{enumerate}
\item the \scparfimiseq{} method based on
  \textsc{Modified-Coverage-Algorithm}, see
  Section~\ref{sec:coverage-algorithm-sampling} and
  Chapter~\ref{chap:ppmfi-method}.
\item the \scparfimipar{} method based on parallel execution of
  \textsc{Modified-Coverage-Algorithm}, see
  Section~\ref{sec:coverage-algorithm-sampling},
  Chapter~\ref{chap:parallel-mfi}, and
  Chapter~\ref{chap:ppmfi-method}.
\item the \scparfimireserv{} method based on
  \textsc{Vitter-Reservoir-Sampling}, see
  Section~\ref{sec:reservoir-sampling} and
  Chapter~\ref{chap:ppmfi-method}.
\end{enumerate}

The three variants differ in the way the sample $\fismpl$ is
constructed in Phase~1. We present theoretical results regarding the
accuracy of the static load-balancing: see
Corollary~\ref{corollary:bounds-pbec-size-reservoir-probabilistic} of
Theorem~\ref{theorem:bounds-pbec-size-reservoir} and
Section~\ref{sec:estim-pbec-union}. We experimentally evaluate the
theoretical results in Chapter~\ref{chap:experimental-evaluation}. We
show that the speedup of our method, in the case of~\scparfimireserv,
is up to $13$ on $20$ processors. The results are valuable because we
apply our method to very fast sequential algorithm: this forces us to
make the process of statical load-balancing very efficient.

In order to make the execution of an arbitrary sequential algorithm
for mining of FIs efficient, we show how to execute the \textsc{Eclat}
algorithm in parallel in Section~\ref{chap:exec-eclat}. The execution
of other arbitrary algorithm for mining of FIs is very similar to the
algorithm shown in Section~\ref{chap:exec-eclat}.

\newpage

\chapter{Sequential algorithms for mining of FIs}

In this section, we show the taxonomy of the sequential algorithms for
mining of FIs. The existing algorithms for mining of FIs together with
their optimizations are described in Appendix \ref{appendix:seq-alg}.

\section{Taxonomy of sequential algorithms}

We can view the algorithms from many different point of views. The
basic division of the algorithms is by the way the lattice of FIs is
searched on two classes: 1) \emph{depth-first search}; 2)
\emph{breadth-first search}. The sequential algorithms can be designed
to mine:

\begin{enumerate}
\item all frequent itemsets;
\item maximal frequent itemsets;
\item concise representation of frequent itemsets, e.g., \emph{closed
  itemsets} (CIs in short), see~\cite{zaki02charm}.
\end{enumerate}

The algorithms can be also divided by the database
representation they use:

\begin{enumerate}
\item vertical representation;
\item horizontal representation.
\end{enumerate}

An incomplete list of the algorithms sorted by the expansion strategy
is the following:

\begin{enumerate}
\item Depth-first search: the Eclat algorithm \cite{zaki97newalgorithms}, the FPGrowth
  algorithm \cite{han00mining}, the H-mine \cite{hmine} algorithm, etc.
\item Breadth-first search: the Apriori algorithm \cite{agrawal94fast}, the DCI (Direct Count and
  Intersect) algorithm \cite{dci}, etc.
\end{enumerate}

Since, in this dissertation thesis, we are focused on parallel FI
mining algorithms, we skip the detailed description of the sequential
algorithms.  A reader is not familiar with the sequential
algorithms, can see Appendix~\ref{appendix:seq-alg} for the
description of the following algorithms:
\begin{enumerate}
\item The Apriori algorithm, Section~\ref{sec:apriori-algorithm},
\item The FPGrowth algorithm, Section~\ref{sec:fpgrowth-algorithm},
\item The Eclat Algorithm, Section~\ref{sec:eclat-algorithm}.
\end{enumerate}

\clearpage
\newpage

\chapter{Existing parallel algorithms}

We consider basically two categories of parallel computers:

\begin{enumerate}
\item shared memory (SM in short) machines;
\item distributed memory (DM in short) machines.
\end{enumerate}

Designing parallel algorithms for mining frequent itemsets on
\emph{shared memory machines} is relatively straightforward: the
machine hardware supports easy parallelization of the problem. All the
processors have access to the shared memory. If we store the database
in the shared memory and use a simple stack splitting algorithm with
arbitrary distributed termination detection and dynamic
load-balancing, the results must be very good. The reason is that
each processor has an access to the whole database and to the
datastructures created by other processors. To our best knowledge, the
parallel algorithms for shared memory machines use the datastructures
created by the other processors only for reading. Therefore the memory
pages containing the data structures are read by the processors and
there is no need for invalidation of the memory pages.


Parallel mining of FIs on DM machines is a hard task for couple reasons:
\begin{enumerate}
\item The databases are usually quite large and we want to have the
  database distributed among the processors so we utilize the main
  memory of all nodes. Re-distribution of the database due to dynamic
  load-balancing, i.e., regular exchange of large database parts
  during the exection, is out of question due to the size of the
  database. \label{enum:db-redistr}
\item The problem of parallel mining of FIs is highly irregular. For
  the same reasons as in \ref{enum:db-redistr} the dynamic
  load-balancing is out of question.
\end{enumerate}

In this chapter, we will briefly describe existing parallel algorithms
for mining of FIs. In \secref{sec:par-alg-shm-mlfpt}, we show an
example of a shared-memory parallel algorithm.
\secref{sec:apriori-based} describes Apriori-based DM algorithm,
\secref{sec:apriori-based:async} describes an asynchronous algorithm
that does not need a sequential FI mining algorithm,
\secref{sec:eclat-based} describes Eclat-based DM algorithms, and
\secref{sec:fpgrowth-based} describes FPGrowth-based DM parallel
algorithms.

During the whole chapter, we denote disjoint database partitions by
$\dbpart_i, 1\leq i\leq\procnum$. $\dbpart_i$ has always the size
$|\dbpart_i| \approx |\db| / \procnum$.

\section{Example of a shared memory algorithm}\label{sec:par-alg-shm-mlfpt}

An example of an algorithm that is designed for shared memory
multiprocessors is the Multiple Local Frequent Pattern Tree algorithm
(the MLFPT algorithm for short) \cite{zaiane01fast}. The MLFPT
algorithm is a parallelization of the \textsc{FPGrowth} algorithm. We
omit the details of the \textsc{FPGrowth} algorithm in this section.
The details of the \textsc{FPGrowth} algorithm can be found in
Appendix~\ref{appendix:seq-alg}. The algorithm works as follows:

\bigskip
\vbox{
\begin{algorithm}[H]
\caption{The Multiple Local Frequent Pattern Trees algorithm}
\label{MLFPT}
\hbox{\textsc{MLFPT}$($\inparam Database $\db$, \inparam Integer $min\_support$, \outparam Set $\allfi$)}
\begin{algorithmic}[1]
\FORPARALLEL{all processors $\proc_i$}
\item[] /* Parallel FPTree creation */ 
\STATE Load $i$-th partition $\dbpart_i$ of the database $\db$ into the main memory.
\STATE Count local support for each item $\bitem \in \baseset$.
\STATE Exchange local supports with other processors to compute global support for each $\bitem\in\baseset$ (hence an all-to-all broadcast takes place).
\STATE Prune not frequent items, i.e., remove from $\baseset$ all items $\bitem\in\baseset$ such that $\supp(\{\bitem\},\db)<\minsupp$.
\STATE Create FP-Tree $T_i$ from $\dbpart_i$
\STATE Barrier synchronization\footnote{Processor $\proc_i$ waits until all other processors has finished creation of the FP-Tree.}
\item[] /* Asynchronous FI mining phase */
  \STATE{} A modified \CALL{FPGrowth} algorithm is started: the
  modified algorithm is almost the same as the original
  \CALL{FPGrowth} algorithm but at the beginning it processes each
  FP-Tree $T_i$, creating a local FP-Tree that is used for further
  computations.
\STATE the computed FIs are put into the set $\allfi$
\ENDFOR
\end{algorithmic}
\end{algorithm}
}

The reported speedup of this algorithm is quite good, e.g., $53.35$ at
$64$ processors, $29.22$ at $32$ processors, and $7.53$ at $8$
processors with running time $\approx 25000$ seconds on single
processor. The experiments used databases of size $1$M, $5$M, $10$M,
$25$M, and $50$M transactions.

\section{Apriori-based parallel DM algorithms}\label{sec:apriori-based}

The first sequential FI mining algorithm was the Apriori algorithm. We
omit the details of the sequential Apriori algorithm in this
section. The details of the sequential Apriori algorithm can be found
in Appendix~\ref{appendix:seq-alg}.

There are many parallel algorithms based on the Apriori algorithm. The
first algorithm was described by Agrawal et
al.~\cite{agrawal96parallel}. Agrawal proposed three parallel
algorithms:

\begin{enumerate}
\item The Data Distribution algorithm.
\item The Count Distribution algorithm.
\item The Candidate Distribution algorithm.
\end{enumerate}

Because Agrawal evaluated \emph{the count distribution algorithm} as
the best of these three algorithms, we will describe this algorithm,
see Section~\ref{sec:apriori-based:count}. An improvement of the
Apriori algorithm, the Fast Parallel Mining algorithm (the FPM
algorithm in short) is described in
Section~\ref{sec:apriori-based:fpm}.

\subsection{The Count distribution algorithm}\label{sec:apriori-based:count}


To describe the algorithm, we need to define the candidate itemset:

\begin{definition}[candidate itemset on frequent itemset]\label{def:candidate-fi-itemset}
Let $k$ be an integer, $\isetU$ be an itemset of size $k$, $\db$ a
database, and $\fipart_{k-1}$ the set of all frequent itemsets of size
$k-1$. If each subset $\isetW\subseteq\isetU, |\isetW|=k-1$ is
frequent, $\isetW\in\fipart_k$, then $\isetU$ is called the
\emph{candidate itemset}. The set of all candidates of size $k$,
denoted by $C_k$, is:
$$C_{k} = \big\{\isetU | \isetU \subseteq \baseset, |\isetU|=k,\text{ and for each } \isetV\subsetneq\isetU, |\isetV|=k-1 \text{ follows that } \isetV \in \fipart_{k-1}\big\}.$$

\end{definition}

Since the computation of the support is the most computationally
expensive part, it computes the support for \emph{candidate itemsets}
in parallel. In the following text, we denote the set of all frequent
itemsets of size $k$ by $\fipart_k$ and the superset of all FIs,
called candidate itemsets, of size $k$ by $C_k$, i.e.,
$\fipart_k\subseteq C_k\subseteq\powerset(\baseset )$.

In the description of the Count Distribution algorithm, we use:

\begin{enumerate}
\item The \textsc{Compute-Support} procedure that computes the support
  of a set of itemsets from a database, see
  Algorithm~\ref{alg:compute-support-hashtrie} in
  Section~\ref{sec:apriori-algorithm}.
\item The \textsc{Generate-Candidates} function that generates
  candidates from a set of frequent itemsets, see
  Algorithm~\ref{alg:generate-candidates} in
  Section~\ref{sec:apriori-algorithm}.
\end{enumerate}

The understanding of the details of the \textsc{Compute-Support}
procedure and the \textsc{Generate-Candidates} function are not
important in order to understand the details of the Count Distribution
algorithm. Therefore, we omit the details in this Section and leave
them to Appendix~\ref{appendix:seq-alg}.

First, each processor $\proc_i$ loads its part of the database,
creates initial set of candidate itemsets $C_1=\big\{\{\bitem\} |
\bitem \in \baseset\big\}$, and computes its support in the database
part $\dbpart_i$. The support of candidates can be computed using the
\textsc{Compute-Support} procedure. Since each processor knows
$\baseset$, each processor has the same set of initial candidate
itemsets. Then, the local supports of the initial candidates are
broadcast, so each processor can compute the global support of the
initial candidates. $C_1$ is pruned and each processor gets frequent
itemsets of size $1$, i.e., $\fipart_1 = \big\{\isetU | \isetU\in
C_1\text{ and } \supp(\isetU,\db) \geq \minsupp\big\}$. Since each
processor has the same initial set of candidates and knows the global
supports, then each $\proc_i$ also has to have the same frequent
itemsets of size $1$. Thus, the first step is correct. All frequent
itemsets of size $k$ will be further denoted by $\fipart_k$.


In step $k$, processors create a set of candidates $C_k$ of size $k$
from the previous frequent itemsets $\fipart_{k-1}$ of size $k-1$.
The set $C_k$ can be computed using the \textsc{Generate-Candidates}
function. The candidates are generated by calling
$C_k=$\textsc{Generate-Candidates}$(\fipart_{k-1})$. Since each
processor $\proc_i$ has the same set of frequent itemsets of size
$k-1$, each processor generates the same set of candidates. Then each
processor $\proc_i$ computes the local support for these candidates
within its database part $\dbpart_i$ and broadcasts the local supports
to each other processor. Each processor updates local support,
computing global support for all these candidates, and creates
frequent itemsets of size $k$, i.e., $\fipart_k = \{\isetU | \isetU\in
C_k\text{ and } \supp(\isetU,\db) \geq \minsupp\}$. Since each
processor has correct frequent itemsets of size $k-1$ at the beginning
of step $k$, each processor has to have correct candidates
$C_k$. Thus, after exchanging and updating local supports and pruning
candidates, all processors have the correct frequent itemsets of size
$k$.  Note that only the support values of each $\isetU \in C_k$ must
be exchanged, because every processor has exactly the same set of
candidates.


The pseudocode for the \textsc{Apriori-Count-Distribution} algorithm
is shown in Algorithm~\ref{alg:apriori-count-distribution}:

\bigskip
\begin{algorithm}[H]
\caption{The \textsc{Apriori-Count-Distribution} algorithm}
\label{alg:apriori-count-distribution} 
\hbox{\textsc{Apriori-Count-Distribution}$($\inparam\ Database $\db$, \inparam Integer $min\_support$, \outparam Set $\allfi$)}
\begin{algorithmic}[1]
  \STATE $k\leftarrow 1$ 
  \FORPARALLEL{all processors $\proc_i$}
    \STATE Load the database part $\dbpart_i$. 
    \IF{$k=1$}
      \STATE Generate initial candidates $C_1 \leftarrow \big\{\{\bitem_\ell\} | \bitem_\ell \in \baseset\big\}$.
      \ELSE
      \STATE Generate candidates $C_k$ from frequent itemsets $\fipart_{k-1}$, by calling $C_k\leftarrow\textsc{Generate-Candidates}(F_{k-1})$.
    \ENDIF
    \STATE Count the support for candidates $C_k$ over local database partition using the \textsc{Compute-Support} procedure.
    \STATE Broadcast the local support of the itemsets in $C_k$ to each other processor (all-to-all broadcast).
    \STATE Prune candidates, creating $\fipart_k = \{\isetU | \isetU\in C_k, \supp(\isetU, \db) \geq \minsupp\}$.
    \IF{the set of frequent itemsets $F_k$ is empty}
      \STATE return all generated frequent itemsets, i.e., return $\allfi=\bigcup_k \fipart_k$ and terminate.
    \ENDIF 
    \STATE $k\leftarrow k + 1$
  \ENDFOR
\end{algorithmic}
\end{algorithm}




\subsection{The Fast Parallel Mining algorithm (FPM)}\label{sec:apriori-based:fpm}

Cheung \cite{cheung02effect, cheung99effect} proposed two pruning
techniques for the Count distribution algorithm. The pruning
techniques leverage two important relationships between a partitioned
database and frequent itemsets. Let $\db$ be a database partitioned
into $n$ disjoint parts $\dbpart_i$ of size $|\dbpart_i| \approx
|\db|/\procnum$, processor $\proc_i$ having database part
$\dbpart_i$. Cheung observed that if an itemset $\isetU$ is frequent
in a database $\db$, i.e., $\rsupp(\isetU, \db)\geq\rminsupp$, then
$\isetU$ must be frequent in at least one partition $\dbpart_i$,
i.e., there exists $i$ such that $\rsupp(\isetU,\dbpart_i) \geq
\rminsupp$. Note that we are using the \emph{relative supports},
instead of the \emph{absolute supports}. Cheung proposed two kind of
optimizations: 1) distributed pruning; 2) global pruning.

\emph{1) Distributed pruning:} uses an important relationship between
frequent itemsets and the partitioned database: \emph{every (globally)
  frequent itemset in the whole database $\db$ must also be (locally)
  frequent on some processors in the database part $\dbpart_i$.}

If an itemset $\isetU$ is globally frequent (i.e. $\rsupp(\isetU,
\db)\geq \rminsupp$) and locally frequent on some processor $\proc_i$
(i.e. $\rsupp(\isetU, \dbpart_i)\geq \rminsupp$), then $\isetU$ is
called \emph{gl-frequent}. We will use $GL_{k(i)}$ to denote the
gl-frequent itemsets of size $k$ at $\proc_i$. As in the Apriori
Count-Distribution algorithm, we denote the set of all FIs of size $k$
computed in step $k$ by $\fipart_k$. Note that $\forall i, 1\leq i\leq
\procnum, GL_{k(i)} \subseteq \fipart_k$.

\begin{lemma}\label{lemma:gl-frequent}\cite{cheung99effect}
If an itemset $\isetU$ is globally frequent, then there exists a processor
$\proc_i$ such that $\isetU$ and all its subsets are \emph{gl-frequent}
at processor $\proc_i$.
\end{lemma}

For the next theorem, we need a function that creates the set of
candidates:
$$CG_{k(i)} = \big\{\isetU | \isetU \subseteq \baseset, |\isetU|=k,\text{
  and for each } \isetV\subsetneq\isetU, |\isetV|=k-1 \text{ follows
  that } \isetV \in GL_{k-1(i)}\big\}.$$ $CG_{k(i)}$ can be computed from
$LG_{k(i)}$ using the algorithm \textsc{Generate-Candidates} by
calling $CG_{k(i)}=$ \textsc{Generate-Candidates}$(GL_{k-1(i)})$,
see Appendix~\ref{appendix:seq-alg} for
Algorithm~\ref{alg:generate-candidates}.

It follows from Lemma~\ref{lemma:gl-frequent} that if $\isetU\in \fipart_k$,
then there exists a processor $p_i$, such that all its subsets of size
$k-1$ are gl-frequent at processors $p_i$, i.e., they belong to
$GL_{k-1(i)}$.

\begin{theorem}\cite{cheung99effect}
For every $k>1$, the set of all frequent itemsets of size $k$,
$\fipart_k$, is a subset of $\fipart_k\subseteq
CG_{(k)}=\bigcup_{i=1}^n CG_{k(i)}$, where $CG_{k(i)} = \{\isetU |
\isetU \subseteq \baseset, |\isetU|=k,\text{ and for each }
\isetV\subsetneq\isetU, |\isetV|=k-1 \text{ follows that } \isetV \in
GL_{k-1(i)}\}$.

\end{theorem}

In \cite{cheung99effect} it is shown that $CG_{k}$, which is a subset of
the Apriori candidates, could be much smaller then the number of the
Apriori candidates.

\emph{2) Global pruning:} after the supports of all itemsets are
exchanged among the processors, the local support counts
$\supp(\isetU, \dbpart_i)$ are also available for all processors. Let
$|\isetU|=k$. At each partition $\dbpart_i$, the monotonicity
principle holds for all itemsets, i.e., $\supp(\isetU, \dbpart_i)\leq
\supp(\isetV, \dbpart_i)$ iff $\isetV\subsetneq \isetU$. Therefore the
local support $\supp(\isetU, \dbpart_i)$ is bounded
by $$maxsupp(\isetU,\dbpart_i)=\min_{\isetV} \big\{\supp(\isetV,
\dbpart_i)|\isetV\subsetneq\isetU, \text{ and } |\isetV| = |\isetU| -
1 \big\}$$ from above, i.e., $\supp(\isetU, \dbpart_i) \leq
maxsupp(\isetU, \dbpart_i)$.  Because the global support
$\supp(\isetU, \db)=\sum_{1\leq i\leq \procnum} \supp(\isetU,
\dbpart_i)$ is the sum of its local support counts at all the
processors, the value:

$$ \sum_{1\leq i\leq \procnum} maxsupp(\isetU, \dbpart_i) $$



\noindent is an upper bound of $\supp(U, \dbpart_i)$. If $\sum_{1\leq
  i\leq \procnum} maxsupp(\isetU, \dbpart_i)<\rminsupp \times |\db| =
\minsupp$, then $\isetU$ can be pruned away. The pseudocode of the
\textsc{FPM} algorithm is shown in Algorithm~\ref{alg:fpm}:

\bigskip
\begin{algorithm}[H]
\caption{The \textsc{FPM} algorithm (Fast Parallel Mining algorithm)}\label{alg:fpm}
\hbox{\textsc{FPM}$($\inparam Database $\db$, \inparam Set $\baseset$, \inparam Integer $\minsupp$, \outparam Set $\allfi$)}
\label{FPM}
\begin{algorithmic}[1]
  \FORPARALLEL{all processors $\proc_i$}
  \STATE Compute the candidate sets $CG_{(k)}=\bigcup_{i=1}^{\procnum} \textsc{Generate-Candidates}(GL_{k-1(i)})$. (distributed pruning)
  \STATE Apply global pruning to prune the candidates in $CG_k$.
  \STATE Scan partition $D_i$ to find out
      the local support counts $\supp(\isetU, \dbpart_i)$ for all remaining
      candidates $\isetU\in CG_k$.
  \STATE Exchange $\{\supp(\isetU, \dbpart_i)\}$
      with all other processors to find out the global support counts
      $\supp(\isetU, \db)$.
  \STATE Compute $GL_{k(i)}=\{\isetU\in CG_k | \rsupp(\isetU,\dbpart)\geq\rminsupp\times |\dbpart|$ and 
      $\rsupp(U, \dbpart_i)\geq\rminsupp\times |\dbpart_i|\}$ and exchange the
      result with other processors.
  \ENDFOR
  \STATE \textbf{return} $\allfi \leftarrow \bigcup_{i=1}^{\procnum} GL_{k(i)}$
\end{algorithmic}
\end{algorithm}

\section{The asynchronous parallel FI mining algorithm}\label{sec:apriori-based:async}

Veloso \cite{veloso03new} proposed another parallelization of the
frequent itemset mining process. This algorithm is based on the fact
that if we know MFIs, we are able to mine all frequent itemsets that
are subsets of MFIs asynchronously.

Each processor $\proc_i$ reads its partition of the database $\dbpart_i$
and computes the local support for all items in $\dbpart_i$. By
exchanging the local supports the processors gets the support of all
items in $\db$.

The algorithm uses the fact that if an itemset is frequent, it must be
frequent in at least one partition $\dbpart_i$. Every processor
$\proc_i$ then finds all MFIs in its local database partition $D_i$
and broadcasts them, together with the support, to other
processors. Because the MFIs are MFIs computed using $\dbpart_i$, the
processors makes global MFIs. Now the processors know the boundaries
of $\allfi$ (in the whole database) and can proceed in a top-down
fashion to compute the support of all itemsets. At the end, the
processors exchange counts of the itemsets and prunes infrequent
itemsets.

The pseudocode of the algorithm is shown in
Algorithm~\ref{alg:asynchronous-apriori}:

\bigskip
\begin{algorithm}[H]
\caption{The \textsc{Asynchronous-FI-Mining} algorithm}
\label{alg:asynchronous-apriori}
\hbox{\textsc{Asynchronous-FI-Mining}$($\inparam Database $\db$, \inparam Integer $min\_support$, \outparam Set $\allfi$)}
\begin{algorithmic}[1]
  \FORPARALLEL{all processors $\proc_i$}
  \item[] /* Phase 1: computation of MFIs */
  \STATE Read its local database partition $\dbpart_i$.
  \STATE Compute all local MFIs, denoted by $M_i$.
  \item[] /* Phase 2 */
  \STATE Broadcast $M_i$ (hence an all-to-all broadcast takes place). 
  \STATE Compute $\bigcup_{1\leq i\leq\procnum} M_i$.
  \item[] /* Phase 3 (every node has $\bigcup_{1\leq i\leq\procnum} M_i$). */
  \STATE Enumerate itemsets $\isetU\subseteq\smfi, \smfi\in M_i$ in a top-down fashion.
  \item[] /* Phase 4 (reduction of results) */
  \STATE Perform sum-reduction operation and removes itemsets $\isetU,
  \supp(\isetU)\leq \minsupp$, i.e.  processor $\proc_i$ sends its
  frequent itemsets to $\proc_{i+1}$ and the last processor removes
  all infrequent itemsets.
  \ENDFOR
\end{algorithmic}
\end{algorithm}

The authors in \cite{veloso03new} reports that the speedup range from
$5$ to $10$ on $16$ processors. Unfortunately, the paper
\cite{veloso03new} is missing a table of speedups, therefore we have
estimated the speedup from graphs of the running time. Additionally,
the problem is that in \cite{veloso03new} there is no mention to the
algorithm used as a base for the computation of speedup, i.e., a
sequential algorithm that is used for computation of the speedup of
the method. If the used sequential algorithm is the Apriori algorithm,
then we have to argue that the Apriori algorithm itself is slow and
the speedup could be much worse if the execution time of the parallel
algorithm is compared with some other, quicker, sequential algorithm.

\section{Eclat-based parallel algorithms}\label{sec:eclat-based}

\subsection{The bitonic scheduling}


Zaki et. al. \cite{zaki97localized} proposed a parallelization of the
Eclat algorithm \cite{zaki97newalgorithms}. The algorithm is similar
to our method in the sense that it partitions $\allfi$ into
prefix-based equivalence classes. However, it uses the \emph{bitonic
  scheduling} \cite{zaki96parallel}, a heuristic for scheduling the
prefix-based classes on the processors that is not able to capture the
real size of each prefix-based equivalence class.

The bitonic scheduling works this way: each PBEC with $n$ atoms, see
Definition~\ref{def:lattice-atom}, is assigned a weight ${n \choose
  2}$, and the equivalence classes are assigned to processors
$\proc_i$ using a best-fit algorithm. The best-fit algorithm is in
fact the same algorithm, we use for assigning of the prefix-based
equivalence classes, see Algorithm~\ref{alg:lpt-schedule} in
Section~\ref{sec:phase-2-detailed} and Graham~\cite{Graham69} for
reference. The problem with this heuristic is that it does not capture
the real size of the equivalence classes. This algorithm achieves
speedups of $\approx 2.5$--$10.5$ on $24$ processors, $\approx
2$--$10$ on $16$ processors, $\approx 1.4$--$8$ on $8$ processors, and
$\approx 3$--$3.5$ on $4$ processors. The experiments were performed
on databases generated by the IBM generator with average transaction
size $10$ and database sizes $800$k, $1.6$M, $3.2$M, and $6.4M$
transactions. Our hypothesis is that in many real-world applications,
the average size of maximal potentially frequent item is much bigger
than $10$.

\bigskip
\begin{algorithm}[H]
\caption{The \textsc{Parallel-Eclat} algorithm}
\label{Parallel-Eclat}
\hbox{\textsc{Parallel-Eclat}$($\inparam Database $\db$,  \inparam Integer $\minsupp$, \outparam Set $\allfi$)}
\begin{algorithmic}[1]
\FORPARALLEL{all processors $\proc_i$}
\item[] /* Initialization phase */
\STATE Scan local database partition $\dbpart_i$.
\STATE Compute local support for all itemsets of size $2$,
\par\noindent denoted by $C_2=\{\isetU | \isetU \subseteq \baseset,
|\isetU| = 2\}$.
\STATE Broadcast the local support of itemsets in $C_2$, \par\noindent
creating global support of itemsets in $C_2$.
\item[] /* Transformation Phase */
\STATE Partition $C_2$ into equivalence classes
\STATE Schedule the equivalence classes on all processors $\proc_i$
\STATE Transform local database into vertical form
\STATE Send to each other processor the tidlists, needed by other
process for computation of its assigned portion of the equivalence
classes.
\item[] /* FI computation phase */ \STATE All processors computes
  frequent itemsets from the assigned equivalence classes.
\item[] /* Final Reduction Phase */
\STATE Aggregate results and output FIs into $\allfi$
\ENDFOR
\end{algorithmic}
\end{algorithm}

\section{FPGrowth-based parallel algorithms}\label{sec:fpgrowth-based}

The FPGrowth algorithm is an important sequential FI mining
algorithm. In this section, we show two parallel algorithms based on
the FPGrowth algorithm. The details of the FPGrowth algorithm are
described in Section~\ref{sec:fpgrowth-algorithm}.

\subsection{A trivial parallelization}

A trivial distributed-memory parallelization of the FP-Growth
algorithm is proposed in \cite{pramudiono03p}. The parallelization
uses dynamic load-balancing. The idea is that each processor creates
its local FP-Tree, broadcast the local FP-Tree to other processors
(resulting in global FP-Tree on every processor) and assign
prefix-based equivalence classes to processors using a hash
function. The problem is that the amount of assigned work is
unpredictable and the resulting computational load could be highly
unbalanced. The solution to the unbalanced computation is the use of
dynamic load-balancing.

The dynamic load-balancing uses \emph{minimal
  path-depth}~\cite{pramudiono03p} threshold to estimate the
granularity of a subtree. We define the \emph{path-depth} as the
maximal length of a path from the root to a list in an FP-Tree. Since
the path-depth of the FP-Tree is non-increasing during the
computation, the dynamic load-balancing works as follows: if a
processor finishes its assigned work, it starts requesting work from
other, busy, processors. The busy processors sends part of their
assigned work to the requesting processor if and only if the
path-depth is bigger than the \emph{minimal path-depth} threshold.

The result of this approach is that the aggregate memory is not used
efficiently. \cite{pramudiono03p} reports speedup of $\approx 4$--$20$
on $32$ processors on a \emph{single} database with $100$K and maximal
potentially frequent itemset size were set to $25$, and $20$.
transactions. However, the speedup of $20$ is achieved in only two
experiments from five. In the rest of the experiments, the maximum
speedup is $\approx 8$ at $30$ processors. The maximum execution time
of the sequential algorithm was $\approx 900$ seconds.

\subsection{The Parallel-FPTree algorithm}

The {\sc Parallel-FPtree} is proposed by Javed and Khokhar in
\cite{javed04}:

\bigskip
\begin{algorithm}[H]
\caption{The \textsc{Parallel-FPTree} algorithm}
\label{Parallel-FPTree}
\hbox{\textsc{Parallel-FPTree}$($\inparam Database $D$, \inparam Itemset $\baseset$, \inparam Integer $\rminsupp)$}
\begin{algorithmic}[1]
\FORPARALLEL{all processors $\proc_i$}
\STATE Scan its assigned partition and computes the
       support for single items sets based on items in the local database.
\STATE Exchange the local supports and compute the global
       support for each itemset with each other processor.
\STATE Sort the global support for the single itemsets
       and discards all the non-frequent items.
\STATE Scan the assigned partition again and
       constructs a local FP-Tree.
\STATE The header table is partitioned into $\procnum$ disjoint sets
       and each processor is assigned to mine frequent patters for distinct
       set of item.
\STATE Identify the information from its local tree needed by other
       processors. The prefix paths of the single itemsets assigned to a
       processor in step $4$ constitute the complete information needed for
       the mining step. This is identified using a bottom up scan of the
       local FP-Tree.
\STATE The information in step $6$ is communicated in $\log\procnum$
       rounds employing a recursive merge of the tree structure over
       processors. For example, processor $\proc_i$ communicates with
       processor $\proc^r_{\procnum/2+1}\%\procnum$ in round $r$ where $1\leq
       i\leq\procnum$ and $0\leq r\leq\log\procnum$. At the end of each
       round, a processor simply unpacks the received information into its
       local FP-tree and prepares a new message for the next round of the
       merge. 
\STATE Mine FIs in its PBECs with prefix of size 1 constructed from the assigned itemsets.
\ENDFOR
\end{algorithmic}
\end{algorithm}

The problem with this approach is obvious: the computation must be
unbalanced. However, in \cite{javed04} present different results: an
almost linear speedup. The reason for such results could be the very
small running time of the algorithm (up to couple of seconds) and very
small database ($10000$ transactions).

\subsection{Summary and conclusion}

We have described parallel algorithms based on the Apriori, the
FPGrowth and the Eclat algorithm. The biggest problem of the Apriori
algorithm is its slowness and memory consumption. Therefore,
parallelization of the Apriori algorithm is not practical. The biggest
advantage of the parallel Apriori algorithms is that they use the
aggregate memory of the cluster efficiently. That is: every processor
has a database partition of size $|\db|/\procnum$. The parallel
Apriori algorithms usually works in iterations that correspond to the
sequential Apriori iterations, except that they are done in
parallel. The authors claim that static load-balancing is used. We
must argue that the load is not statically balanced at all: parallel
execution of the sequential iterations should not be considered as
static load-balancing .

The parallelizations of the Eclat and the FPGrowth algorithms use an
estimate of the sizes of the prefix-based classes. However, the estimates are
very simple and do not capture the real amount of work assigned to the
processors. Dynamic load-balancing on distributed-memory parallel
computers also does not work. The reason is that the computation is
quite fast and exchanging large portions of the database among
processors can be quite time-consuming.

Parallelizations of other algorithms than the Apriori algorithm do not
achieve good speedups. But, the Apriori itself is quite slow.

The best solution should:
\begin{enumerate}
\item distribute the computation: computation time of each
  processors should be approximately the same.
\item distribute the database: the database should be distributed among
  the processors so that processor $\proc_i$ has database partition of
  size $|\db|/\procnum$.
\end{enumerate}

All parallel algorithms based on the Apriori algorithm have the
previous two properties. However, the sequential Apriori algorithm is
very slow and very memory consuming. Therefore, we would like to
parallelize faster and less memory consuming algorithm with the
described properties.

It seems that the major difference in the sequential algorithms is in
the used datastructures. Therefore, we would like have a universal
parallelization method for an arbitrary sequential algorithm.




\clearpage
\newpage

\chapter{Approximate counting by sampling}\label{chap:approx-counting}

Our method for parallel mining of FIs is based on efficient estimation
of the number of FIs in a given prefix-based equivalence class (PBEC
in short). Unfortunately, as discussed in
Section~\ref{seq:fimi-mining-complexity}, counting the number of FIs
is \#P-Complete problem, i.e., computing the number of FIs in a given
PBEC is also \#P-Complete. Fortunately, to estimate the (relative)
number of FIs in a PBEC, we do not need to count the relative number
of FIs exactly. We can estimate the relative sizes of FIs in PBECs
with a sampling algorithm that approximately counts the relative
number of FIs in a PBEC.  Further, when talking about the relative
(absolute) size of a PBEC, we always mean the relative (absolute)
number of FIs in the PBEC.



In this chapter, we show two sampling algorithms for estimating the
relative size of a given PBEC, or a set of PBECs. Both sampling
algorithms need the support of an itemset $\isetU$ to decide whether
an itemset is frequent or not. This decision can be made with an
\emph{estimate} of the support of $\isetU$. The support of $\isetU$ is
estimated using a \emph{database sample}. Therefore, in this chapter,
we also derive minimum sample size needed to achieve a small error of
the support estimate with high probability. Finally, we bound the
error of the size of a PBEC estimated using $\fismpl$.



To describe the sampling methods and our method for parallel mining of
FIs, we need an additional notation. We extend the notation introduced
in \secref{chap:math-found}. A database is denoted by $\db$ and a
database sample is denoted by $\dbsmpl$.  The set of all FIs
\emph{computed from} $\db$ is denoted by $\allfi$.  The set of all FIs
\emph{computed from }$\dbsmpl$ is denoted by $\fiapprox$. The set of
all MFIs \emph{computed from} $\db$ is denoted by $\mfi$. The set of
all MFIs \emph{computed from} $\dbsmpl$ is denoted by $\mfiapprox$.
$\mfiapprox$ is the upper bound on $\fiapprox$ in the sense of the set
inclusion, i.e., for all $\isetU\in\fiapprox$ there exists an $m \in
\mfiapprox$ such that $\isetU\subseteq m$.  The sample of $\fiapprox$,
which is computed using $\fiapprox$ or $\mfiapprox$, is denoted by
$\fismpl$. The additional notation is summarized in
\figref{fig:sampling-notation}.

\begin{table}[!ht]
\centering
\begin{tabular}{|l|p{10cm}|} \hline
Symbol       & Description \\ \hline
$\dbsmpl$    & A database sample computed from $\db$.  \\ \hline
$\fiapprox$  & The set of all FIs computed from $\dbsmpl$. \\ \hline
$\mfiapprox$ & The set of all MFIs computed from $\dbsmpl$.
               $\mfiapprox$ also bounds the set $\fiapprox$, i.e., for
               each $\isetU\in\fiapprox$ exists $m\in\mfiapprox$ such that
               $\isetU\subseteq m$.\\ \hline
$\fismpl$    &  A sample of $\fiapprox$, computed using $\fiapprox$ or $\mfiapprox$. \\ \hline
\end{tabular}
\caption{The new notation used to describe the sampling algorithms.}\label{fig:sampling-notation}
\end{table}

This chapter is organized as follows: first, in
Section~\ref{sec:db-sample-support-estimtate} we show how to estimate
support of an itemset from a database sample. In
Section~\ref{sec:pbec-relative-size-estimate} we show the two methods
for estimating the size of a PBEC.

\section{Estimating the support of an itemset from a database sample}\label{sec:db-sample-support-estimtate}

The time complexity of the decision whether an itemset $\isetU$ is
frequent or not is in fact the complexity of computing the
\emph{relative support} $\rsupp(\isetU, \db)$ in the input database
$\db$. If we know the approximate relative support of $\isetU$, we can
decide whether $\isetU$ is frequent or not with certain
probability. We can estimate the relative support $\rsupp(\isetU,\db)$
from a database sample $\dbsmpl$, i.e., we can use $\rsupp(\isetU,
\dbsmpl)$ instead of $\rsupp(\isetU, \db)$.

An approach of estimating the relative support of $\isetU$ was
described by Toivonen \cite{toivonen96sampling}.  Toivonen uses a
\emph{database sample} $\dbsmpl$ for the \emph{sequential} mining of
frequent itemsets and for the efficient estimation of theirs
supports. Toivonen's algorithm works as follows: 1) create a database
sample $\dbsmpl$ of $\db$; 2) compute all frequent itemsets,
$\fiapprox$, in $\dbsmpl$; 3) check that all these FIs computed using
$\dbsmpl$ are also FIs in $\db$ and correct the output. If an itemset
is frequent in $\db$ and not in $\dbsmpl$, correct the output using
$\db$, see \cite{toivonen96sampling} for details. Toivonen's algorithm
is based on an efficient probabilistic estimate of the support of an
itemset $\isetU$.

We reuse this idea of estimating the support of $\isetU$ in our method
for parallel mining of FIs, i.e., we use only the first two steps. We
define the error of the estimate of $\rsupp(\isetU,\db)$ from a
database sample $\dbsmpl$ by:

$$\supperr(\isetU,\dbsmpl)=|\rsupp(\isetU,\db)-\rsupp(\isetU,\dbsmpl)|$$

The database sample $\dbsmpl$ is sampled with replacement. The
estimation error can be analyzed using the Chernoff bound without
making other assumptions about the database. The error analysis then
holds for a database of arbitrary size and properties.

\begin{theorem}\label{theorem:chernoff-supp-estimate}\cite{toivonen96sampling}
Given an itemset $U\subseteq \baseset$, two real numbers
$\epsilon_{\dbsmpl}, \delta_{\dbsmpl}, 0\leq \epsilon_{\dbsmpl},
\delta_{\dbsmpl} \leq 1$, and a sample $\dbsmpl$ drawn from database
$\db$ of size

$$ |\dbsmpl| \geq \frac{1}{2\epsilon_{\dbsmpl}^2}\ln\frac{2}{\delta_{\dbsmpl}}, $$

then the probability that $\supperr(\isetU,\dbsmpl)>\epsilon_{\dbsmpl}$ is at
most $\delta_{\dbsmpl}$.
\end{theorem}
\begin{proof}
We denote the probability of an event by $\prob[\cdot]$. Toivonen used
in the original paper the Chernoff bounds described in
\cite{book:alon-probabilistic}. We use the same equation, see
(\ref{eqn:noga-chernoff}) or Appendix A in
\cite{book:alon-probabilistic}. The Chernoff bounds gives:





$$\begin{array}[t]{rl}
\prob\left[\supperr(\isetU,\dbsmpl)>\errsize\right]=&\prob\left[|\rsupp(\isetU,\db)-\rsupp(\isetU,\dbsmpl)|>\epsilon_{\dbsmpl}\right] \\[3mm]
                                          =&\prob\left[|\rsupp(\isetU,\db)-\rsupp(\isetU,\dbsmpl)|\cdot|\dbsmpl|>\epsilon_{\dbsmpl}\cdot|\dbsmpl|\right]\leq2e^{-2(\epsilon_{\dbsmpl}\cdot|\dbsmpl|)^2/|\dbsmpl|} \\
\end{array}$$


give an upper bound for the probability:
$$2e^{-2(\epsilon_{\dbsmpl}\cdot|\dbsmpl|)^2/|\dbsmpl|} \leq \delta_{\dbsmpl}$$

$$ |\dbsmpl| \geq \frac{1}{2\epsilon_{\dbsmpl}^2}\ln\frac{2}{\delta_{\dbsmpl}} $$
\end{proof}

Using a database sample $\dbsmpl$ with size given by the previous
theorem, we can estimate $\rsupp(\isetU,\db)$ with error
$\epsilon_{\dbsmpl}$ that occurs with probability at most
$\delta_{\dbsmpl}$: it follows from
Theorem~\ref{theorem:chernoff-supp-estimate} that if we compute the
approximation $\fiapprox$ of $\allfi$ from the database sample
$\dbsmpl$ of size $|\dbsmpl| \geq \frac{1}{2\epsilon_{\dbsmpl}^2} \ln
\frac{2}{\delta_{\dbsmpl}}$, we should get an estimate of the supports
of itemsets $\isetU\in\fiapprox$, i.e., potentially, $\fiapprox$
closely approximates $\allfi$.

\section{Estimating the relative size of a PBEC}\label{sec:pbec-relative-size-estimate}

In our parallel method for mining FIs, we need to estimate the
relative size of a PBEC $[\isetU]$: $|[\isetU] \cap
\fiapprox|/|\fiapprox|$, see
Definition~\ref{def:relative-pbec-size}. This can be estimated using
$\fismpl$. There are two ways for constructing $\fismpl$:

\begin{enumerate}
\item Compute $\mfiapprox$ and get $\fismpl$ using the \emph{modified
  coverage algorithm};

\item Compute $\fiapprox$ and get $\fismpl$ using the \emph{reservoir
  sampling}.

\end{enumerate}

These two algorithms are presented in the next sections.


\subsection{The coverage algorithm and its modification}\label{sec:coverage-algorithm-sampling}

Let us have the MFIs $\mfiapprox$, computed from $\dbsmpl$. The set of
all MFIs $\mfiapprox$ is the upper bound on the set $\fiapprox$, i.e.,
$\fiapprox = \bigcup_{\smfi\in\mfiapprox} \powerset(m)$. To construct
a sample $\fismpl\subseteq\fiapprox$ of independent and identically
distributed (i.i.d.) elements chosen from the uniform distribution, we
can use the \emph{coverage algorithm}~\cite{motwani} that uses
$\mfiapprox$ for construction of the sample $\fismpl$. To make the
sampling in our parallel method for mining of FIs faster, we have
modified the algorithm, so it does not constructs sample from
\emph{uniform} distribution, but it creates only \emph{independently}
distributed sample.  The coverage algorithm (or its modification)
produces only the sample $\fismpl$.

The coverage algorithm estimates the relative size of a set
$F\subseteq\fiapprox$. The coverage algorithm takes as input the set
$\mfiapprox$ of the MFIs computed from the database sample $\dbsmpl$,
a number $0\leq\rho\leq 1$ representing the fraction $\rho =
\frac{|F|}{|\fiapprox|}$ where in our case $F = [\isetW]\cap\fiapprox$
is the smallest PBEC we want to estimate, and two real numbers $0 \leq
\epsilon_{\fismpl}, \delta_{\fismpl}\leq 1$ where $\epsilon_{\fismpl}$
is the error of the approximated size and $\delta_{\fismpl}$ its
probability \textbf{or} the number of samples $N = |\fismpl|$ instead
of the sampling parameters. The output of the coverage algorithm is
the sample $\fismpl$ that is used for the estimation of the relative
sizes of PBECs.

In our parallel method, the set $F$ represents the set of FIs in some
PBECs: let have a set of prefixes $\prefixset = \{\isetU_i | \isetU_i
\neq \isetU_j, \isetU_i\subseteq\baseset, \text{ and } [\isetU_i] \cap
     [\isetU_j] = \emptyset \text{ for all } i\neq j\}$ be a set of
     itemsets (prefixes) then in our method $F = (\bigcup_{\isetU\in
       \prefixset} [ \isetU ]) \cap \fiapprox$.  However, the theory
     we show holds for an arbitrary set of itemsets $F$.

The coverage algorithm follows:

\vbox{
\begin{algorithm}[H]
\caption{The \textsc{Coverage-Algorithm} algorithm}
\label{alg:coverage-algorithm}
\vbox{\textsc{Coverage-Algorithm}(\vtop{\inparam Set $\mfiapprox=\{\smfi_i\}$,
  \par\noindent \inparam Integer $N$, 
  \par\noindent\outparam Set $\fismpl$)}}
\begin{algorithmic}[1]
  \STATE $\fismpl\leftarrow\emptyset$
  \WHILE{$|\fismpl|\neq N$}
     \STATE pick index $i\in\lcint1,|\mfiapprox|\rcint$ with probability
            $\prob[i]=\frac{|\powerset(\smfi_i)|}{\sum_j|\powerset(\smfi_j)|}$ \label{alg:coverage-algorithm:pick-mfi}
     \STATE pick a random set $\isetU\in\powerset(\smfi_i)$ with probability
            $\frac{1}{|\powerset(\smfi_i)|}$ \label{alg:coverage-algorithm:pick-itemset}
     \STATE \texttt{found}$\leftarrow$ \texttt{false}
     \FOR{$l\leftarrow i - 1$ to 1}\label{alg:coverage-algorithm:forloop}
        \IF{$\isetU\subseteq\smfi_l$}\label{alg:coverage-algorithm:test}
             \STATE \texttt{found}$\leftarrow$ \texttt{true}
        \ENDIF
     \ENDFOR
     \IF{\texttt{found}$=$\texttt{false}}
          \STATE $\fismpl\leftarrow \fismpl\cup\{\isetU\}$
     \ENDIF
  \ENDWHILE
\end{algorithmic}
\end{algorithm}
}

Let's have a set of itemsets $\fipart\subseteq\fiapprox$. Now, we
analyze the dependency of the error $\epsilon_{\fismpl}$ of the
estimated relative size $|\fipart|/|\fiapprox|$ estimated using the
sample $\fismpl$ by $|\fipart \cap \fismpl|/|\fismpl|$ on the size of
$|\fismpl|$.

First, we define the \emph{multiset} $\mathcal{S} =
\biguplus_{\smfi\in\mfiapprox} \powerset(\smfi)$ that contains as many
copies of $\isetW\in \fiapprox$ as the number of $\powerset(\smfi)$
containing $\isetW$. The sample obtained by uniform sampling of
$\mathcal{S}$ is therefore a non-uniform sample of $\fiapprox$.

We can represent every itemset in $s\in \mathcal{S}$ as a pair
$s=(\isetW,i)$ that correspond to $\isetW\in\powerset(\smfi_i)$, that
is $\mathcal{S}=\{(\isetW,i)|\isetW\in\powerset(\smfi_i)\}$.  We
define a function $g: \mathcal{S} \rightarrow \{\text{true},
\text{false} \}$ as follows:

$$ g((\isetW,i)) = \left\{ \begin{array}{l} \text{true},\quad \text{if }i=\min\{j|\isetW\in\powerset(\smfi_j)\}\\
                               \text{false}, \quad \text{otherwise}\end{array}\right. $$

To make the sample of $\fiapprox$ uniform, we must sample the set
$\mathcal{S'} = \{s | s\in \mathcal{S}\text{ and } g(s) =
\text{true}\}$. Each element of $\mathcal{S'}$ corresponds to one
element of $\fiapprox$. Therefore, by sampling $\mathcal{S'}$, we
sample $\fiapprox$. It is clear that $|\mathcal{S}|\geq |\mathcal{S'}|
= |\fiapprox|$.

The coverage algorithm, described in
Algorithm~\ref{alg:coverage-algorithm} in fact samples
$\mathcal{S'}$. To sample $\mathcal{S'}$,
Algorithm~\ref{alg:coverage-algorithm} picks $i$ with probability
proportional to $|\powerset(\smfi_i)|$ (line
\ref{alg:coverage-algorithm:pick-mfi}) and then it picks
$\isetW\in\powerset(\smfi_i)$ uniformly at random (line
\ref{alg:coverage-algorithm:pick-itemset}). In order to sample the set
$\mathcal{S'}$ instead of $\mathcal{S}$, the algorithm must assure
that we choose only those itemsets $\isetW, g((\isetW,i)) =
\text{true}$ for some integer $i$. That is: we must check that there
does not exists $\smfi_j, j<i$, such that $\isetW \in
\powerset(\smfi_j)$. This is performed at line
\ref{alg:coverage-algorithm:forloop}.


\begin{theorem}[estimation error of the size of a subset $\fipart\subseteq\fiapprox$]
\label{theorem:monte-carlo-error}\cite{motwani}
Let $\mfiapprox$ be the set of MFIs such that $\fiapprox =
\bigcup_{\smfi_i\in\mfiapprox} \powerset(\smfi_i)$, $\fipart \subseteq
\fiapprox$, $\rho = |\fipart|/|\fiapprox|$, two real numbers
$\epsilon_{\fismpl}, \delta_{\fismpl}$ such that $0 \leq
\epsilon_{\fismpl}, \delta_{\fismpl} \leq 1$, and $\fismpl$ is the
\emph{independent and identically distributed} sample of $\fiapprox$
obtained by the coverage algorithm by calling
\textsc{Coverage-Algorithm} $(\mfiapprox, N_{\fismpl}, \fismpl)$. Then the estimate:

$$\frac{|\fipart\cap\fismpl|}{|\fismpl|}$$

is an estimation of $|\fipart|/|\fiapprox|$ with error at most
$\epsilon_{\fismpl}$ with probability at least $1 - \delta_{\fismpl}$
provided

$$ N_{\fismpl} = |\fismpl| \geq \frac{4}{\epsilon_{\fismpl}^2\rho}\ln\frac{2}{\delta_{\fismpl}}.$$
\end{theorem}

\begin{proof}
The proof of the theorem is again based on the Chernoff bounds. We
know that:

$$ \prob\left[|\fipart\cap\fismpl|\geq (1 + \epsilon_{\fismpl})\rho|\fismpl|\right] \leq e^{-|\fismpl|\rho\epsilon_{\fismpl}^2/4} $$

and similarly for the lower bound:

$$\prob\left[|\fipart\cap\fismpl|\leq (1 - \epsilon_{\fismpl})\rho|\fismpl|\right] \leq e^{-|\fismpl|\rho\epsilon_{\fismpl}^2/4}$$



Therefore:
$$ \prob\left[(1-\epsilon_{\fismpl})|\fismpl|\rho\leq |\fipart\cap\fismpl| \leq (1+\epsilon_{\fismpl})|\fismpl|\rho \right] \geq 1-2 e^{-|\fismpl|\rho\epsilon_{\fismpl}^2/4}\geq 1-\delta_{\fismpl}$$

\end{proof}


An important part of the coverage algorithm is the for-loop at the
line \ref{alg:coverage-algorithm:forloop}. It guarantees that each
$\isetU\in\fiapprox$ is selected with probability
$\frac{1}{|\fiapprox|}$. Unfortunately, this loop will prevent many
selected samples $\isetU\subseteq\smfi_i$ to not make it into
$\fismpl$ because the $\isetU$ is contained in an MFI $\smfi_j \in
\mfiapprox, j < i$ with lower index. Additionally, the set $\mfiapprox$
can be quite large and the loop has the complexity
$\Ocmplxty(\mfiapprox)$.




The coverage algorithm runs in  polynomial time given 1) the number of
samples   $N_{\fismpl}$    which   is   a    function   of   $1/\rho$,
$1/\epsilon_{\fismpl}$,  $1/\delta_{\fismpl}$; 2)  $|\mfiapprox|$; and
3) $|\baseset|$ if the following properties holds:

\begin{enumerate}
\item For all $i$, $|\powerset(\smfi_i)|, \smfi_i \in \mfiapprox$ is
  computable in polynomial time in $|\smfi_i|$.
\item It is possible to sample uniformly from any $\powerset(\smfi_i),
  \smfi_i\in\mfiapprox$.
\item For all $\isetU\in \powerset(\baseset)$, it can be determined in
  polynomial time in $|\smfi_i|$ whether $\isetU\in
  \powerset(\smfi_i)$.
\item We are using $\fismpl$ for estimating the size of a set
  $\fipart\subseteq\fiapprox$. Therefore, $\rho=\frac{\left|
  F\right|}{|\fiapprox|}$ must be polynomial in $|\fiapprox|$ in order
  to make the \textsc{Coverage-Algorithm} polynomial.
\end{enumerate}


\textbf{The modification of the coverage algorithm:} in our parallel
method for mining FIs, we need to compute the sample even faster. To
do so, we resign on the uniform sampling of $\fiapprox$ (i.e. on
sampling $\mathcal{S'}$) by omitting the checks against MFIs with
lower index, i.e., we omit the for-loop at line
\ref{alg:coverage-algorithm:forloop} and therefore, we sample the set
$\mathcal{S} = \biguplus_{\smfi\in\mfiapprox}\powerset(\smfi)$.  This
makes the sampling non-uniform because it prefers samples $\isetU$
that are contained in many $\powerset(\smfi_i)$, i.e., the sampling
prefers sets $\powerset(\smfi_i)\cap\powerset(\smfi_j), i\neq
j$. Therefore, the estimate of the relative size of a set
$\fipart\subseteq\fiapprox$ (computed using the modified algorithm)
\emph{is just a heuristic!}. However, this heuristic is much faster
then the \textsc{Coverage-Algorithm} and the estimates of the relative
sizes made using the sample obtained by the modified coverage
algorithm are sufficient for our purposes.


\vbox{
\begin{algorithm}[H]
\caption{The \textsc{Modified-Coverage-Algorithm} algorithm}
\label{alg:modified-coverage-algorithm}
\textsc{Modified-Coverage-Algorithm}(\vtop{\inparam Set $\mfiapprox=\{\smfi_i\}$,
  \par\noindent \inparam Integer $N$,
  \par\noindent\outparam Set $\fismpl$)}
\begin{algorithmic}[1]
  \STATE $\fismpl\leftarrow\emptyset$
  \WHILE{$|\fismpl| \neq N$}
      \STATE pick $\smfi\in\mfiapprox$ with probability $\frac{|\powerset(\smfi)|}{\sum_{\smfi'\in\mfiapprox}|\powerset(\smfi')|}$
      \STATE pick subset $S\subseteq\smfi$ with probability $\frac{1}{|\powerset(\smfi)|}$
      \STATE $\fismpl\leftarrow\fismpl\cup\{S\}$
  \ENDWHILE\label{alg:mfi-sampling-seq:send}
\end{algorithmic}
\end{algorithm}
}

\subsection{The reservoir sampling}\label{sec:reservoir-sampling}


In this section, we show the \emph{reservoir sampling algorithm} that
constructs an uniformly but not independently distributed sample
$\fismpl$ of $\fiapprox$ on the contrary of the previous section.

Vitter \cite{vitter85reservoir} formulates the problem of
\emph{reservoir sampling} as follows: given a stream of records, the
task is to construct a sample of size $n$ \emph{without replacement} from the
stream of records without any prior knowledge of the length of the
stream.

We can reformulate the original problem in the terms of $\fiapprox$
and $\fismpl$: let's consider a sequential algorithm that outputs all
frequent itemsets $\fiapprox$ from a database $\dbsmpl$. We can view
$\fiapprox$ as a stream of FIs. We do not know $|\fiapprox|$ in
advance and we need to take $|\fismpl|$ samples of $\fiapprox$. We
take the samples $\fismpl$ using the reservoir sampling algorithm.
This solves our problem of making a uniform sample $\fismpl \subseteq
\fiapprox$. The sampling is done using an array of FIs (a buffer, or
in the terminology of \cite{vitter85reservoir} a reservoir) that holds
$\fismpl$.


The reservoir sampling uses the following two procedures:

\begin{enumerate}
\item \textsc{ReadNextFI($L$)}: reads next FI from an output of an arbitrary
  sequential algorithm for mining of FIs and stores the itemset at the location
  $L$ in memory.
\item \textsc{SkipFIs($k$)}: skips $k$ FIs from the output of an arbitrary
  algorithm for mining of FIs.
\end{enumerate}

\noindent and the following function:

\begin{enumerate}
\item \textsc{Random()} which returns an uniformly distributed real
  number from the interval $\lcint 0,1\rcint$
\end{enumerate}

The simplest reservoir sampling algorithm is summarized in
Algorithm~\ref{alg:simple-reservoir-algorithm}. It takes as an input
an array $R$ (reservoir/buffer) of size $n=|\fismpl|$, the function
\textsc{ReadNextFI($L$)} that reads an FI from the output of an FI
mining algorithm and stores it in memory at location $L$, and finally
the function \textsc{SkipFIs($k$)} that skips $k$ FIs. The algorithm
samples $|\fismpl|$ FIs and stores them in memory into the buffer $R$. 

The \textsc{Simple-Reservoir-Sampling}
follows:

\vbox{
\begin{algorithm}[H]
\caption{The \textsc{Simple-Reservoir-Sampling} algorithm}
\label{alg:simple-reservoir-algorithm}
\textsc{Simple-Reservoir-Sampling}(\vtop{\inoutparam Array $R$ of size $n$, 
  \par\noindent \inparam Integer $n$, 
  \par\noindent \inparam Procedure ReadNextFI, 
  \par\noindent \inparam Procedure SkipFIs)}
\begin{algorithmic}[1]
  \FOR{$j\leftarrow 0$ to $n-1$}
     \STATE ReadNextFI($R[j]$)
  \ENDFOR
  \STATE $t\leftarrow n$
  \WHILE{not eof}
     \STATE $t\leftarrow t+1$
     \STATE $m\leftarrow \lfloor t\times\textsc{Random}()\rfloor$ \COMMENT{pick uniformly a number from the set $\{0, \ldots, t-1\}$}
     \IF{$m < n$}
        \STATE ReadNextFI$(R[m])$
     \ELSE
        \STATE SkipFIs$(1)$
     \ENDIF
  \ENDWHILE
\end{algorithmic}
\end{algorithm}
}

The \textsc{Simple-Reservoir-Sampling} is quite slow, it is linear in
the number of input records read by \textsc{ReadNextFI(R)}, i.e., it
is linear in $|\fiapprox|$. Vitter \cite{vitter85reservoir} created a
\emph{faster} algorithm that has the same parameters as the
\textsc{Simple-Reservoir-Sampling} algorithm. We denote the Vitter's
variant of the algorithm by
\textsc{Vitter-Reservoir-Sampling}(\inoutparam Array $R$ of size $n$,
\inparam Integer $n$, \inparam Procedure ReadNextFI, \inparam
Procedure SkipFIs).

The \textsc{Vitter-Reservoir-Sampling} runs with the average running
time $\Ocmplxty(|\fismpl|(1+\log \frac{|\fiapprox|}{|\fismpl|}))$,
where $|\fismpl|$ is the size of the array $R$ used by
\textsc{Vitter-Reservoir-Sampling}. Vitter in his analyse does not
consider the time needed to read the record using the
\textsc{ReadNextFI} and to skip the records using the
\textsc{SkipFIs}. That is: the formula represents only the time needed
by the execution of the \textsc{Vitter-Reservoir-Sampling} algorithm,
see~\cite{vitter85reservoir} for details.




Now, we analyse the relative size of a PBEC using the samples taken by
the reservoir algorithm.  The reservoir sampling samples the set
$\fiapprox$ \textbf{without replacement}, resulting in $\fismpl$.  In
Theorem~\ref{theorem:monte-carlo-error} we analysed the error of the
approximation of the relative size of an arbitrary set using an
i.i.d. sample using the Chernoff bounds. In the case of the reservoir
sampling, we cannot use the Chernoff bounds because the elements of
the sample $\fismpl$ are \emph{identically} but unfortunately
\textbf{not independently} distributed due to the use of the
reservoir. The reservoir sampling process can be modeled using the
\emph{hypergeometric distribution}, see
Appendix~\ref{appendix:discrete-distributions} or
\cite{book:johnson-probability}. In the rest of this chapter, we
analyze the bounds on the relative size of a set of itemsets using the
sample made by the reservoir sampling using a \emph{hypergeometric
  distribution}.

Using the bounds from Appendix~\ref{appendix:sec:hypergeometric}, we
can state a theorem similar to Theorem~\ref{theorem:monte-carlo-error}
(using the Chernoff bounds and an i.i.d sample) but now for the
\emph{hypergeometric distribution}, i.e., estimation of the relative
size of a PBEC but using a uniformly but not \emph{independently}
distributed sample:

\begin{theorem}[Estimation error of the size of a subset $\fipart\subseteq\fiapprox$]
\label{theorem:reservoir-size-estim}
Let $\fipart\subseteq\fiapprox$ be a set of itemsets. The relative
size of $\fipart$, $\frac{|\fipart|}{|\fiapprox|}$, is estimated with
error $\epsilon_{\fismpl}, 0\leq \epsilon_{\fismpl}\leq 1,$ with
probability $\delta_{\fismpl}, 0\leq \delta_{\fismpl}\leq 1,$ from a
hypergeometrically distributed sample $\fismpl\subseteq\fiapprox$ with
parameters $N=|\fiapprox|, M=|\fipart|$ (see
Appendix~\ref{appendix:discrete-distributions}) of size

$$|\fismpl| \geq -\frac{\log(\delta_{\fismpl}/2)}{D(\rho+\epsilon_{\fismpl}||\rho)} $$

Where $D(x||y)$ is the Kullback-Leibler divergence of two
hypergeometrically distributed variables with parameters $x, y$ and
$\rho=|F|/|\fiapprox|$.

The expected value of the size $|\fipart \cap\fismpl|$ is
$\expect[|\fipart\cap\fismpl|] = |\fismpl| \cdot
\frac{|\fipart|}{|\fiapprox|}$.


\end{theorem}
\begin{proof}
The proof is based on bounds provided in \cite{hypergeometric-skala}
which is a summarization of \cite{chvatal79hypergeometric}, see
(\ref{eqn:hypergeometric-kullback-leibler}) where $p=\rho,
\epsilon=\epsilon_{\fismpl}, n=|\fismpl|$, and the fact that
$D(p+\epsilon||p) > D(p-\epsilon||p)$:

$$ 1-(e^{-|\fismpl|\cdot D(p-\epsilon||p)} + e^{-|\fismpl|\cdot D(p+\epsilon||p)}) \leq 1-\delta_{\fismpl}$$

$$ 1-2e^{-|\fismpl|\cdot D(p+\epsilon||p)} \leq 1-\delta_{\fismpl} $$

\end{proof}

Since $\fipart$ is a union of PBECs and the reservoir sampling
algorithm give us identically distributed sample, we are able to bound
the error of relative size of a union of PBECs made by the ``double
sampling process'', i.e., estimating the size of a union of PBECs
using a database sample:


\begin{theorem}[bounds on the size of a set of FIs from a given PBEC]\label{theorem:bounds-pbec-size-reservoir}
Let $\isetV_i\subseteq\baseset, 1\leq i\leq n, [\isetV_i]\cap[\isetV_j] =
\emptyset, i\neq j$. We use two sets of itemsets:

\begin{enumerate}
\item $A = \{ \isetU| \rsupp(\isetU,\db)<\rminsupp \text{ and }
  \rsupp(\isetU,\dbsmpl)\geq\rminsupp \}$, i.e., the collection of
  itemsets $\isetU$ infrequent in $\db$ and frequent in $\dbsmpl$ --
  wrongly added FIs to $\fiapprox$.
\item $B=\{ \isetU | \rsupp(\isetU,\db) \geq \rminsupp \text{ and }
  \rsupp(\isetU,\dbsmpl) < \rminsupp \}$, i.e., the collection of
  itemsets $\isetU$ frequent in $\db$ and infrequent in $\dbsmpl$ --
  wrongly removed FIs from $\fiapprox$.
\end{enumerate}

The relative size of $A$ is denoted by $a=\frac{|A|}{|\allfi|}$ and the relative
size of $B$ is denoted by $b=\frac{|B|}{|\allfi|}$. Then for two sets of itemsets
$C = \bigcup_i [\isetV_i] \cap \allfi$ and $\widetilde{C} = \bigcup_i [\isetV_i] \cap
\fiapprox$, we have:

$$\frac{|\widetilde{C}|}{|\fiapprox|}(1+a-b) - a \leq \frac{|C|}{|\allfi|} \leq \frac{|\widetilde{C}|}{|\fiapprox|}\cdot (1+a-b) + b$$

\end{theorem}

\begin{proof}
From the assumptions follows: $|\fiapprox|=|\allfi|(1+a-b)$. Therefore:
$\frac{|\fiapprox|}{(1+a-b)}=|\allfi|$.


We know that the fraction $a$ of FIs is not frequent in $\db$ but is
frequent in $\dbsmpl$ are present in $\fiapprox$.  Therefore, we can
compute the lower bound of the relative size of $C$:

\begin{equation}\label{eqn:reservoir-below-1}
|\widetilde{C}|\leq |C| + a\cdot|\allfi|
\end{equation}

\begin{equation}\label{eqn:reservoir-below-2}
\frac{|\widetilde{C}|}{|\allfi|} \leq \frac{|C|}{|\allfi|} + a
\end{equation}

(\ref{eqn:reservoir-below-3}) follows from (\ref{eqn:reservoir-below-2}) using
the fact that $|\allfi|=\frac{|\fiapprox|}{(1+a-b)}$.

\begin{equation}\label{eqn:reservoir-below-3}
\frac{|\widetilde{C}|}{|\fiapprox|}(1+a-b) \leq \frac{|\widetilde{C}|}{|\allfi|} \leq \frac{|C|}{|\allfi|} + a
\end{equation}

\begin{equation}\label{eqn:reservoir-below-4}
\frac{|\widetilde{C}|}{|\fiapprox|}(1+a-b) - a  \leq \frac{|C|}{|\allfi|} 
\end{equation}

We compute the upper bound of $\frac{|C|}{|\allfi|}$ using similar
computations as for the lower bound. The fraction $b$ of FIs $\allfi$
was not frequent in $\dbsmpl$ and frequent in $\db$ and therefore the
lower bound of the size $|\widetilde{C}|$ is:

\begin{equation}\label{eqn:reservoir-above-1}
|C| - b\cdot|\allfi|\leq |\widetilde{C}|
\end{equation}
\begin{equation}\label{eqn:reservoir-above-2}
\frac{|C|}{|\allfi|} - b \leq \frac{|\widetilde{C}|}{|\allfi|}
\end{equation}
\begin{equation}\label{eqn:reservoir-above-3}
\frac{|C|}{|\allfi|} \leq \frac{|\widetilde{C}|}{|\fiapprox|}\cdot (1+a-b) + b
\end{equation}



\end{proof}

\begin{corollary}\label{corollary:bounds-pbec-size-reservoir-probabilistic}
If the size of $\frac{|\widetilde{C}|}{|\fiapprox|}$ is estimated with error
$\epsilon_{\fismpl}, 0\leq \epsilon_{\fismpl}\leq 1$, with probability
$0\leq\delta_{\fismpl}\leq 1$ then:

$$\frac{|\widetilde{C}|}{|\fiapprox|}(1-\epsilon_{\fismpl})(1+a-b) - a \leq
\frac{|C|}{|\allfi|} \leq \frac{|\widetilde{C}|}{|\fiapprox|} (1-\epsilon_{\fismpl})
(1+a-b) + b$$ with probability $\delta_{\fismpl}$.
\end{corollary}

Set $C$ can be viewed as a partition processed by a single
processor. We estimate the relative size of $|C|/|\allfi|$ from
$\fismpl$ and we are able to bound the error made while estimating the
size of a partition. Unfortunately, the bounds are not very tight and
making tighter bounds is hard.

Because our modification of the coverage algorithm does not give an
identically distributed sample,
Corollary~\ref{corollary:bounds-pbec-size-reservoir-probabilistic}
cannot be used to bound the size of $\fipart\subseteq\fiapprox$
using the sample taken by \textsc{Modified-Coverage-Algorithm}.


\section{Estimating the size of a union of PBECs}\label{sec:estim-pbec-union}

Let $\isetU_i\subseteq\baseset, 1\leq i\leq n,$ be prefixes and
$[\isetU_i]$ corresponding PBECs. We are constructing the PBECs by
recursive splitting and estimating the size using the sample, i.e.,
$|[\isetU_i] \cap \fiapprox| / |\fiapprox| \approx |[\isetU_i] \cap
\fismpl| / |\fismpl|$.
Let $\fipartidxset\subseteq \lcint 1, n\rcint$ be the set of indexes
of the PBECs. The set of indexes is chosen in such a way that
$|\bigcup_{i\in\fipartidxset}[\isetU_i] \cap \fismpl| / |\fismpl|
\approx 1/\procnum$. That is: the set $\fipartidxset$ is dependent on
the constructed sample $\fismpl$. Therefore, we are not able to use
the Chernoff bounds (or the estimates using the Kullback-Leibler
divergence) with the same sample $\fismpl$ (used for construction of
PBECs) for estimation of the relative size of $\fipart = \bigcup_{j
  \in \fipartidxset} [\isetU_j] \cap \fiapprox$ because the sets
$[\isetU_j]$, the set $\allfi$ and the sample $\fismpl$ are not
independent.  Instead, we must choose $\epsilon_{\fismpl}$ such that
$\epsilon_{\fismpl} \cdot |\fipartidxset|$ is small enough. In
Chapter~\ref{chap:experimental-evaluation}, we experimentally show the
error and its probability made by a particular choice of the number of
samples.

\clearpage
\newpage

\chapter{Approximate parallel mining of MFIs}\label{chap:parallel-mfi}

In our method, we need to compute an approximation of the maximal
frequent itemsets (MFIs), $\mfiapprox$ (see
Chapter~\ref{chap:approx-counting} for notation). Because we have
$\procnum$ processors at our disposal, we could execute an arbitrary
algorithm for mining of MFIs in parallel. Unfortunately, parallel
mining of MFIs using a DFS algorithm, is a hard task. We can relax the
requirement of computing $\mfiapprox$ to a requirement of computing
the set $M$ such that $\mfiapprox \subseteq
M\subseteq\fiapprox$. Recall that we denote a prefix-based equivalence
class with prefix $\isetU$ and extensions $\isetW$ by
$[\isetU|\isetW]$, to emphasize that the items in the extensions are
important as described in Definition~\ref{def:extension}.


We define a \emph{candidate} on an MFI as follows:

\begin{definition}[candidate itemset on MFI]
Let $\isetU\subseteq\baseset$ be a frequent itemset and $\prefixext$
the extensions used by a DFS MFI algorithm for extending $\isetU$.  We
call $\isetU$ a candidate itemset (or candidate in short) on an MFI if
for each $\bitem \in \prefixext$ the itemset $\isetU\cup \{\bitem\}$
is not frequent, i.e., $\supp(\isetU \cup \{\bitem\}) < \minsupp$.
\end{definition}

A ``template'' of a DFS algorithm for mining of MFIs is shown in
Algorithm~\ref{alg:dfs-mfi-schema}. The difference between algorithms
for mining of MFIs is in the way they implement the depth-first search
of the PBECs. The DFS MFI algorithms optimize the search so they visit
as small number of FIs as possible. Other difference is in the used
datastructures, and the way the algorithms implement the test at
line~\ref{alg:dfs-mfi-schema:prune}.

The candidates  on the  MFIs are  the leafs of  the DFS  algorithm for
mining of MFIs. An example of  the candidate on the MFI is the itemset
${5,6}$ in Example~\ref{example:mfi-candidate}.

\begin{definition}[longest subset of a MFI in a PBEC]
Let $\isetW$ be a maximal frequent itemset, $\bitem\in\baseset$ an
item, the set $\prefixext=\{\bitem'\in\baseset : \bitem<\bitem'\}$,
and $[\{\bitem\} | \prefixext]$ the PBEC. We call the set $\isetU =
\isetW \cap (\prefixext \cup \{\bitem\})$ \emph{the longest subset of
$\isetW$ in the PBEC $[\{\bitem\} | \prefixext]$}.

\end{definition}

For example, let $\isetU=\{1\}$ be a prefix and $\prefixext=\{2, 3,
5\}$ its extensions. For the MFI $\smfi=\{1,3,4,5\}$ the longest
subset of $\smfi$ in $[\isetU|\prefixext]$ is the set $\{1, 3, 5\}$.

The longest subset of a MFI in a PBEC can be a candidate set, but
there exists longest subsets that are not candidates. We say that
$\isetW$ is a candidate on the MFI $\isetU, \isetW \subsetneq \isetU$
in a PBEC, if it is a \emph{candidate} and a longest subset of
$\isetU$ in the PBEC, i.e., it is a leaf of a DFS tree and it is a
longest subset.

Recall, that we omit the extensions in a PBEC $[\bitem | \{\bitem' |
  \bitem < \bitem'; \bitem',\bitem\in\baseset\}]$, see
page~\pageref{pgref:extension-notation-simple}. We use the extensions
in the notation if we want to emphasise them.

A sequential schema for mining of MFIs is shown in
Algorithm~\ref{alg:dfs-mfi-schema}:


\vbox{
\begin{algorithm}[H]
\caption{The schema of a DFS algorithm for mining of MFIs.}
\label{alg:dfs-mfi-schema}
\vbox{\textsc{DFS-MFI-Schema}(\inparam Database $\dbsmpl$, \inparam $\rminsupp$, \inparam Set $B$, \outparam Set $\mfiapprox$)}
\begin{algorithmic}[1]
  \REQUIRE $\baseset=\{\bitem_i\}$ to be an ordered
    set $\bitem_1\leq\ldots\leq\bitem_{|\baseset|}$ and
    $\rsupp(\{\bitem_i\}, \dbsmpl)\geq\rminsupp$.
  \STATE $\mfiapprox\leftarrow\emptyset$ 
  \FOR{each $\bitem_{i}\in B$ in ascending order}
  \STATE perform depth-first search of $[(\bitem_i) | \{\bitem :
    \bitem\in\baseset, \bitem > \bitem_i\}]\cap\fiapprox$
  \par\noindent(visiting/discovering candidates $\isetU\in [(\bitem_i)
    | \{\bitem : \bitem\in\baseset, \bitem > \bitem_i\}] \cap \fiapprox$
  on an MFI) \label{alg:dfs-mfi-schema:dfs-search}
  \FOR{each candidate to maximal itemset $\isetU \in [(\bitem_i) |
      \{\bitem_{i+1},\ldots,\bitem_{|\baseset|}\}]\cap\fiapprox$} \label{alg:dfs-mfi-schema:prune}
        \IF{exists no $\isetW\in\mfiapprox$ such that $\isetU\subseteq\isetW$}
           \STATE $\mfiapprox\leftarrow\mfiapprox\cup\{\isetU\}$\label{alg:dfs-mfi-schema:add-to-mfiapprox}
        \ENDIF
     \ENDFOR
  \ENDFOR
\end{algorithmic}
\end{algorithm}
}

A maximal frequent itemset $\isetW = (\bitem_{w_1}, \ldots,
\bitem_{w_{|\isetW|}}), \bitem_{w_1} < \ldots < \bitem_{w_{|\isetW|}}$
is visited(discovered) by a DFS MFI algorithm by expanding first
$[(\bitem_{w_1})]$, then $[(\bitem_{w_1}, \bitem_{w_2})]$, etc.  To
our best knowledge, all MFIs DFS mining algorithms follow the schema
in Algorithm~\ref{alg:dfs-mfi-schema}: the algorithm initializes
$\mfiapprox\leftarrow\emptyset$ and starts a depth-first search on the
lattice of all FIs, skipping some FIs. The PBECs are expanded in the
order of $\bitem_i$, i.e.,
$[(\bitem_1)|(\bitem_2,\bitem_3,\ldots,\bitem_{|\baseset|})]$ is
processed first, then $[(\bitem_2) | (\bitem_3, \ldots,
  \bitem_{|\baseset|})]$ is processed, etc. Therefore, if $\isetU =
(\bitem_{u_1}, \ldots, \bitem_{u_{|\isetU|}})$ is an MFI and
$\isetW\subseteq\isetU$ be a candidate itemset. $\isetW$ are visited
\emph{after} visiting $\isetU$. If the algorithm finds a candidate
itemset $\isetW$, it looks into $\mfiapprox$ and if $\mfiapprox$
\emph{contains a superset} of $\isetW$, the algorithm skips $\isetW$
(\emph{not} storing $\isetW$ in $\mfiapprox$). If $\mfiapprox$
\emph{does not contains} a superset of $\isetW$, it is an MFI and is
stored into $\mfiapprox$ (see
line~\ref{alg:dfs-mfi-schema:add-to-mfiapprox}).

\begin{proposition}\label{prop:candidate-discovery-order}
Let $\mfiapprox$ be a set of all MFIs mined with some value of
$\rminsupp$ in a database $\dbsmpl$, $\isetU \in \mfiapprox$ be an
MFI, and $\isetW$ be a candidate on the MFI such that $\isetW
\subsetneq \isetU$. Then $\isetW$ is visited by
Algorithm~\ref{alg:dfs-mfi-schema} after visiting the MFI $\isetU$.
\end{proposition}

\begin{proof}
The proposition follows from the fact that the items in the baseset
$\baseset$ are ordered and Algorithm~\ref{alg:dfs-mfi-schema}
processes $[(\bitem_i)]$ and its extensions in the order of the items
in $\baseset$.
\end{proof}



We can execute Algorithm~\ref{alg:dfs-mfi-schema} in parallel with
dynamic load-balancing as shown in
Algorithm~\ref{alg:parallel-dfs-mfi-schema}.


\vbox{
\begin{algorithm}[H]
\caption{The parallel schema of a DFS algorithm for mining of MFIs.}
\label{alg:parallel-dfs-mfi-schema}
\vbox{\textsc{Parallel-DFS-MFI-Schema}(\vtop{\inparam Database $\dbsmpl$, \par\noindent\inparam $\rminsupp$, \par\noindent\inparam Set $\baseset$, \par\noindent\outparam Set $M$)}}
\begin{algorithmic}[1]
  \REQUIRE $\baseset=\{\bitem_i\}$ to be an ordered
    set $\bitem_1\leq\ldots\leq\bitem_{|\baseset|}$ and
    $\rsupp(\{\bitem_i\}, \dbsmpl)\geq\rminsupp$.
  \FORPARALLEL{each $\proc_i$}
     \STATE $M_i\leftarrow\emptyset$

     \STATE $S_i\leftarrow \{\bitem_j | \bitem_j\in \baseset, i =
     \lceil  j\cdot \procnum /|\baseset| \rceil\}$.

     \FOR{each $\bitem_k\in S_i$ in ascending order} 

        \STATE perform depth-first search of $[(\bitem_k) | \{\bitem_{k+1},
        \ldots, \bitem_{|\baseset|}\}]\cap\fiapprox$, visiting/discovering
        candidates $\isetU\in [(\bitem_k) | \{\bitem_{k+1}, \ldots,
        \bitem_{|\baseset|}\}] \cap \fiapprox$ on an MFI by calling \textsc{DFS-MFI-Schema}$(\dbsmpl, \rminsupp, \{\bitem_k\}, M'_i)$
        \STATE \textbf{Dynamic load-balancing:} during the depth-first
        search we have to perform dynamic-load balancing. Each
        $\proc_i$ has to check if it has work and if not it asks other
        processors for a PBEC. The processors can send to other
        processors only a PBEC with prefix of size $1$. Therefore, at
        this point the set $S_i$ can be modified, removing $\bitem\in
        S_i$ if it has been processed or scheduled to other processor
        and adding $\bitem\in\baseset$ to $S_i$ if it was send by
        another processor.  We omit other details from the
        description of the algorithm.
        \FOR{each maximal itemset in $\isetU\in M'_i$}
           \IF{there is no $\isetW\in M_i$ such that $\isetU\subseteq\isetW$}\label{alg:parallel-dfs-mfi-schema:if}
              \STATE $M_i \leftarrow M_i \cup \{\isetU\}$
           \ENDIF
        \ENDFOR
     \ENDFOR
  \ENDFOR
\end{algorithmic}
\end{algorithm}
}


Algorithm~\ref{alg:parallel-dfs-mfi-schema} works in the following
way: because $\dbsmpl$ is much smaller than the whole database $\db$,
the processors replicates $\dbsmpl$, i.e., every processor has a copy
of the database sample $\dbsmpl$ and knows the items that are frequent
in the database $\db$ (note that $\db$ is distributed among the
processors).  All processors partition the base set $\baseset$ to
$\procnum$ blocks of size $\approx |\baseset| / \procnum$. Processor
$\proc_i$ runs a sequential DFS MFI algorithm in the $i$-th part of
$\baseset$, where the items $\bitem_i$ are interpreted as
$1$-prefixes, i.e., prefix-based equivalence classes
$[(\bitem_i)]$. When a processor finishes its assigned items, it asks
other processors for work. The computation is terminated using the
Dijkstra's token termination detection algorithm. The output of the
algorithm is a superset of all MFIs.

The approach described in the
Algorithm~\ref{alg:parallel-dfs-mfi-schema} computes the set
$M=\bigcup_i M_i$ such that $\mfiapprox\subseteq M$. The reason is the following:

\begin{enumerate}
\item every processor has its copy of the database sample $\dbsmpl$;
\item an arbitrary algorithm for mining of MFIs always correctly
  computes the support of an arbitrary itemset.
\end{enumerate}


We demonstrate the parallel execution (the parallel processing of
assigned PBECs) of a sequential DFS algorithm for mining of MFIs on
the following example (for simplicity without dynamic load balancing):
because the computation is distributed, the algorithm is unable to
check the candidate against all already computed MFIs which results in
a superset of all MFIs.  Let $B=\{1,2,3,4,5,6\}$ and $\procnum=3$ and
assume that the prefix-based equivalence classes $[(1)|(2,3,4,5,6)],
[(2)|(3,4,5,6)]$ were assigned to $\proc_1$; the prefix-based
equivalence classes $[(3)|(4,5,6)],[(4)|(5,6)]$ were assigned to
$\proc_2$; and the classes $[(5)|(6)],[(6)|\emptyset]$ to
$\proc_3$. The MFIs $\{\{1,3,4\}, \{2,3,4\}, \{2,4,5\}\}$ are
correctly computed by $\proc_1$. The processor $\proc_2$ correctly
computes the MFI $\{3,4,5,6\}$, but processor $\proc_3$ computes also
the itemset $\{5,6\}$ as an MFI. The reason is that $\proc_3$ does not
know that the MFI $\{3,4,5,6\}$ was already computed by processor
$\proc_2$. In Figure \ref{fig:mfi-parallel} the FIs, MFIs, and the
additional itemset computed as MFI are shown.

\begin{figure}[!ht]
\hrule
\medskip
\example[MFIs computed in parallel by a trivial parallelization of a
  DFS algorithm for mining of MFIs.]{\label{fig:mfi-parallel} MFIs
  computed in parallel by a trivial parallelization of a DFS algorithm
  for mining of MFIs.}{



The FIs are computed from the database from Example
\ref{fig:ppmfi-start} are marked by a dot, except the set $\emptyset$
which is not an FI.  The MFIs are marked in blue, the additionally
computed itemset $\{5,6\}$, which is a candidate on the MFI
$\{3,4,5,6\}$ in the PBEC $[(5)]$, is marked in orange. In this case
the itemset $\{5,6\}$ is also the longest subset of the MFI
$\{3,4,5,6\}$ in the PBEC $[(5)]$ The MFIs are computed with dynamic
load-balancing on $\procnum=3$ processors. The processor $\proc_1$ is
scheduled with $[(1)],[(2)]$; $\proc_2$ with $[(3)],[(4)]$; and
$\proc_3$ with $[(5)],[(6)]$. The following picture shows the lattice
of all FIs.

\begin{center}
\includegraphics[type=mps,ext=.mps,read=.mps]{lattice_mfiparallel}
\end{center}
\medskip
\hrule
}\label{example:mfi-candidate}
\end{figure}

\newpage

\begin{lemma}\label{lemma:num-of-cand-itemsets}
Let $\isetW = (\bitem_{w_1}, \ldots, \bitem_{w_{|\isetW|}})$ be an
MFI, $\bitem \in \isetW$ any of its element. There exists \emph{at
  most one candidate} on the MFI $\isetW$ in the PBEC $[(\bitem)]$. If
such candidate exists then it is the longest subset $S_{\isetW} =
\{\bitem' | \bitem' \in \isetW, \bitem \leq \bitem' \}$ of the MFI
$\isetW$ in the PBEC $[(\bitem)]$.
\end{lemma}

\begin{proof}
In each PBEC $[(\bitem)]$ all frequent sets $X\in [(\bitem)]$ such
that $X\subseteq\isetW$ are always subsets of $S_{\isetW}$. Consider
sets $X\subsetneq S_{\isetW}$: $X$ cannot be a candidate because there
exists an item $\bitem\in S_{\isetW}$ such that $X \cup \{ \bitem \}$
is frequent, due to the monotonicity of the support, see
Theorem~\ref{monotonicity}. 
\end{proof}


Note that $S_{\isetW}$ is a candidate if and only if there is no
frequent itemset in $[(\bitem)]$, which is a proper superset of
$S_{\isetW}$.

As stated in the proof of the lemma, in some cases the longest subset
$S_{\isetW}$ is not a candidate on the MFI $\isetW$. Let have an
arbitrary other MFI $\isetU = (\bitem_{u_1}, \ldots,
\bitem_{u_{|\isetU|}})$, the item $\bitem \in \isetU, \isetW$ and
$S_{\isetU} = \{ \bitem' | \bitem,\bitem' \in \isetU ; \bitem \leq
\bitem' \}$ be the longest subset of the MFI $\isetU$ in the PBEC
$[(\bitem)]$. We discuss the cases of the MFI $\isetW$ and its longest
subset $S_{\isetW}$ in the PBEC $[(\bitem)]$. If for $S_{\isetW}$
holds $S_{\isetW}\subsetneq S_{\isetU}$ then the candidate on the MFI
$\isetW$ does not exists in the PBEC $[(\bitem)]$ because there exists
$\bitem'\in S_{\isetU}$ such that $S_{\isetW}\cup\{\bitem'\}$ is
frequent. If for $S_{\isetW}$ holds $S_{\isetW} = S_{\isetU} = S$ then
there is a candidate $S$ on both MFIs $\isetW, \isetU$. Therefore, the
number of candidates of the MFI $\isetW$ depends on all other mined
MFIs and subset/superset relations of the longest subsets of all MFIs.

The following theorem is a corollary of
Lemma~\ref{lemma:num-of-cand-itemsets}:

\begin{theorem}\label{thm:parallel-mfi-count}
Let have a baseset $\baseset$ and $1 < \procnum < |\baseset|$
processors $\proc_1,\ldots,\proc_\procnum$, a database $\dbsmpl$,
$M_i$ be a set of itemsets computed by $\proc_i$ in
Algorithm~\ref{alg:parallel-dfs-mfi-schema}, and $M = \bigcup_{1\leq
  i\leq\procnum} M_i$. Let $\isetW$ be the \emph{longest} MFI, i.e.,
for all $\isetU, \isetW \in \mfiapprox$ holds that $|\isetU| \leq
|\isetW|$. An arbitrary DFS algorithm for mining MFIs that is executed
in parallel, e.g., in Algorithm~\ref{alg:parallel-dfs-mfi-schema},
computes a set of itemsets $M$, such that $\mfiapprox\subseteq M$, of
size:

$$|\mfiapprox| < |M| = \left|\bigcup_{1\leq i\leq \procnum} M_i\right|
\leq |\isetW| \cdot|\mfiapprox|.$$
\end{theorem}

\begin{proof}
The proof of this theorem follows from the
Lemma~\ref{lemma:num-of-cand-itemsets} and the fact that for each MFI
$\isetU$ there are at most $|\isetU|$ PBECs that contains some subsets
of $\isetU$, i.e., in the worst case the dynamic load-balancing causes
that Algorithm~\ref{alg:parallel-dfs-mfi-schema} discovers all
candidates on a single MFI.
\end{proof}

If we do not use dynamic load-balancing and assign the items
statically (each processor processing $|\baseset|/\procnum$ PBECs), for
an MFI $\isetU = (\bitem_{u_1}, \ldots, \bitem_{u_{|\isetU|}})$ each
$\proc_i$ computes the candidate on the MFI $\isetU$, if it exists, in
each of its assigned PBECs with prefix of size $1$. The \textbf{if}
condition at line \ref{alg:parallel-dfs-mfi-schema:if} of
Algorithm~\ref{alg:parallel-dfs-mfi-schema} assures that from these
candidates will be picked the longest candidate on the MFI $\isetU$
(in the sense of the cardinality of the candidates). Denote the
longest MFI by $\isetW$, as in the previous theorem. If we statically
assign the PBECs to each processor and do not use dynamic
load-balancing, the upper bound on $|M|$ is $|M| < \procnum \cdot
|\mfiapprox|$. The two bounds can be combined: $|M| < \min(\procnum,
|\isetW|) \cdot |\mfiapprox|$.

\clearpage
\newpage

\chapter{Proposal of a new DM parallel method}\label{chap:ppmfi-method}


In this section, we present our \emph{new method}, called \scparfimi{} that has
three variants: \scparfimiseq{} \cite{kessl06balancing}, \scparfimipar{}
\cite{kessl07improved}, and \scparfimireserv{}. The method provides
parallelizations of the DFS (or BFS) sequential frequent itemsets mining
algorithm. The method has the following advantages over current existing
algorithms:



\begin{enumerate}
\item \emph{It is universal}: with our method it is possible to parallelize any
  DFS algorithms for mining of frequent itemsets.  It is even possible to
  parallelize BFS algorithms, though the performance of the Apriori algorithm
  could suffer in the candidate pruning phase.


\item \emph{The computation is balanced statically}: if the database is very
  large, the dynamic load-balancing is out of question as the overhead of
  exchanging large partitions of a database and/or large data structures during
  dynamic-load balancing is too expensive.

  Static load-balancing of the computation is not easy, as the amount
  of work for each prefix-based equivalence class is unknown.

  In our approach, the static load-balancing is based on a heuristic and a
  sampling algorithm for estimating the size of the PBECs. The PBECs are then
  assigned to the processors, so that the processors perform approximately the
  same amount of work.
\end{enumerate}

Our method also has the following property:

\bigskip
\hbox to\hsize{\hfill
\vbox{\hsize=15cm
 \emph{Result distribution}: at the end of the execution of our parallel
  method, the frequent itemsets are distributed among the processors. This
  is an advantage, if we need to query for particular frequent itemsets. For
  example, we need to find all frequent itemsets containing the set $\{5, 8\}$ as
  a subset. Each processor gets the set $\{5, 8\}$, finds the FIs and sends them
  to the querying processor. In some cases, we need to send the FIs to a
  particular processor for further processing. The FIs distributed among
  processors could help for parallel computation of association rules. However,
  parallel computation of association rules goes beyond the scope of our work.
}\hfill}
\medskip


In this chapter, if we talk about \emph{size of a PBEC or relative size of a
  PBEC}, we mean the \emph{relative number of FIs} in the particular PBEC. If we
talk about a partition $\fipart \subseteq \allfi$ or $\fipart \subseteq
\fiapprox$ then the relative size of $\fipart$ is $|\fipart| / |\allfi|$ or
$|\fipart| / |\fiapprox|$.

Our new method is called \emph{Parallel Frequent Itemset MIning} (Parallel-FIMI
in short). This method works for any number of processors $\procnum \ll
|\baseset|$. The basic idea is to partition all FIs into $\procnum$ disjoint
sets $\fipart_i$, using PBECs, of relative size $\frac{|\fipart_i|}{|\allfi|}
\approx \frac{1}{\procnum}$. Each processor $\proc_i$ then processes partition
$\fipart_i$.

The input and the parameters of the whole method are the following:
\begin{enumerate}
\item Minimal support: the real number $\rminsupp$, see
  Definition~\ref{def:cover-support}.
\item The sampling parameters: real numbers $0 \leq \epsilon_{\dbsmpl}, \delta_{\dbsmpl},
  \epsilon_{\fismpl}, \delta_{\fismpl} \leq 1$, see Section~\ref{chap:approx-counting}.
\item The relative size of a smallest PBEC: the parameter $\rho, 0\leq\rho\leq 1$, see
  Sections~\ref{chap:approx-counting}.
\item Partition parameter: real number $\alpha, 0\leq\alpha\leq 1$, see Section~\ref{sec:phase-2-detailed}.

\item Database parts $\dbpart_i, 1\leq i\leq\procnum$: processor $\proc_i$ loads
  its database partition $\dbpart_i$ to a local memory. The database partitions
  $\dbpart_i$ has the following properties: $\dbpart_i \cap \dbpart_j =
  \emptyset, i\neq j$, and $|\dbpart_i| \approx
  \frac{|\db|}{\procnum}$. 
\end{enumerate}

Additionally, without loss of generality, we expect that each
$\bitem_i\in\baseset$ is frequent.  Otherwise, each processor $\proc_i$ computes
local support of all items $\bitem_j\in\baseset$ in its database part
$\dbpart_i$. The support is then broadcast and each $\proc_i$ removes all
$\bitem_j$ that are not globally frequent.

The whole method consists of four phases. The first three phases are designed in
such a way that they statically balance the load of the computation of all
FIs. Phases~1--2 prepare the PBECs and its assignment to the processors for
Phase~4, i.e., the static load-balancing is precompute in Phases~1--2. In the
Phase~3, we redistribute the database partitions so each processor can proceeds
independently with the assigned PBECs. In the Phase~4, we execute an arbitrary
algorithm for mining of FIs and the processors computes the FIs in it assigned
PBECs. To speed-up Phases~1--2, we can execute each of Phase~1--2 in
parallel. The four phases are summarized below:


\noindent\textbf{Phase~1} (sampling of FIs): the input of Phase~1 is the minimal
support $\rminsupp$, a partitioning of the database $\db$ into $\procnum$
disjoint partitions $\dbpart_i$, and the real numbers $0 \leq
\epsilon_{\dbsmpl}, \delta_{\dbsmpl}, \epsilon_{\fismpl}, \delta_{\fismpl} \leq
1$. Output of Phase~1 is a sample of frequent itemsets $\fismpl$. Generally, the
purpose of the first phase is to compute a sample $\fismpl$ and create the
database sample $\dbsmpl$. First, each processor samples $\dbpart_i$ (in
parallel) and creates part $\db'_i$ and broadcasts them to other processors
(all-to-all broadcast\if0\footnote{all-to-all scatter is a well known communication
  operation: each processor $\proc_i$ sends a message $m_{ij}$ to processor
  $\proc_j$ such that $m_{ij}\neq m_{ik}, i\neq k$}\fi). Each processor $\proc_i$
then creates $\dbsmpl=\bigcup_i\db'_i$. Then from $\dbsmpl$ is computed
$\fismpl$. We propose three methods for creation of $\fismpl$.

\noindent\textbf{Phase~2} (lattice partitioning): the input of this phase is the
sample $\fismpl$, the database sample $\dbsmpl$ (both computed in Phase~1) and
the parameter $\alpha$. In Phase~2, the algorithm creates prefixes $\isetU_i
\subseteq \baseset$ and the extensions $\prefixext_i$ of disjoint PBECs
$[\isetU_i | \prefixext_i]$, and estimates the size of $[\isetU_i | \prefixext]
\cap \allfi$ using $\fismpl$. $\proc_1$ assigns the PBECs $[\isetU_i |
  \prefixext_i]$ to all processors and the PBECs together with the assignment
are broadcast to all processors.

\noindent\textbf{Phase~3} (data distribution): the input of this phase is the
assignment of the prefixes $\isetU_i$ and the extensions $\prefixext_i$ to the
processors $\proc_i$ and the database partitioning $\dbpart_i,
i=1,\ldots,\procnum$. Now, the processors exchange database partitions:
processor $\proc_i$ sends $S_{ij} \subseteq \dbpart_i$ to processor $\proc_j$ such
that $S_{ij}$ contains transactions needed by $\proc_j$ for computing support of
the itemsets of its assigned PBECs.

\noindent\textbf{Phase~4} (computation of FIs): as the input to each processor
are the prefixes $\isetU_i \subseteq \baseset$, the extensions $\prefixext_i$,
and the database parts needed for computation of supports of itemsets
$\isetV\in[\isetU_i]\cap\allfi$ and the original $\dbpart_i$. Each processor
computes the FIs in $[\isetU_i]\cap\allfi$ by executing an arbitrary sequential
algorithm for mining of FIs. Additionally, each processor computes support of
$\isetW\subseteq\isetU_i$ in $\dbpart_i$, i.e., $\supp(\isetW,\dbpart_i)$. The
supports are then send to $\proc_1$ and $\proc_1$ computes $\supp(\isetW,\db) =
\sum_{1\leq i\leq\procnum} \supp(\isetW,\dbpart_i)$

In this chapter, we use the database in Example~\ref{fig:ppmfi-start} to
demonstrate Phases~1--4.

\begin{figure}[!ht]
\hrule
\medskip
\example[Start of the running example.]{\label{fig:ppmfi-start}\textbf{(start of the running example)}}{

The four phases of our method will be further demonstrated on the following
database $\db$ with $\minsupp=5$ and $\baseset=\{1,2,3,4,5,6\}$ (or equivalently $\rminsupp=0.3$):

\begin{center}
{
  \scriptsize
  \renewcommand{\arraystretch}{1}
  \exampledatabase
}
\end{center}
}
\medskip
\hrule
\end{figure}

\section{Detailed description of Phase 1}\label{sec:phase-1-detailed}

In Phase~1, we create a sample $\fismpl$ of all frequent itemsets.  The input of
this phase, for processor $\proc_i$, are the database partitions $\dbpart_i$
such that $\dbpart_i \cap \dbpart_j = \emptyset, i\neq j$, $|\dbpart_i| \approx
|\db|/\procnum$, the relative minimal support $\rminsupp$, and the real numbers
$0 \leq \epsilon_{\dbsmpl}, \epsilon_{\fismpl}, \delta_{\dbsmpl},
\delta_{\fismpl} \leq 1$. The output of this phase is the sample of FIs
$\fismpl$ and the database sample $\dbsmpl$. We propose three methods for
creation of $\fismpl$. The input and the output is the same for all of the three
proposed variants of Phase~1.  For the details on sampling, see the
Sections~\ref{sec:db-sample-support-estimtate}
and~\ref{sec:pbec-relative-size-estimate}.

Without the knowledge of the process that creates the sample $\fismpl$, we can
demonstrate the purpose of $\fismpl$ and the idea behind Phase~1 and the
consequences of Phase~1 on the whole process of mining of FIs. The idea of
Phase~1 is to create the sample $\fismpl$ so that we can estimate the relative
size of PBECs using $\fismpl$ and making a set of PBECs that can be processed
by a single processor, see Example \ref{fig:mfi-sampling}.

This section is organized as follows: first, in Section
\ref{sec:phase1-mfi-sampling} we propose two variants based on our modification
of the \emph{coverage algorithm}. Then, in Section
\ref{sec:phase1-reservoir-sampling}, we propose a variant based on the reservoir
sampling algorithm, i.e., we propose three variants of the first phase:


\begin{enumerate}
\item Compute the boundary $M$ of $\fiapprox$, in the sense of set inclusion: 
  \begin{enumerate}
    \item Sequentially: the boundary in this case is the set $M=\mfiapprox$, see
      \cite{kessl06balancing}. This variant of Phase~1 is denoted by
      \textsc{Phase-1-Coverage-Sampling-Sequential}, resulting in the
      \scparfimiseq{} method.
    \item In parallel: the boundary in this case is a set $M$, such that
      $\mfiapprox \subsetneq M\subsetneq\fiapprox$, see
      \cite{kessl07improved}. This variant of Phase~1 is denoted by
      \textsc{Phase-1-Coverage-Sampling-Parallel}, resulting in the
      \scparfimipar{} method.
  \end{enumerate}
  Using the boundary $M$, we create a sample $\fismpl$ using the \emph{modified
    coverage algorithm}, see Section~\ref{sec:phase1-mfi-sampling}. The details
  of parallel mining of MFIs and the boundary $M$ are in
  Chapter~\ref{chap:parallel-mfi}.
\item Create the sample $\fismpl$ by putting together an arbitrary sequential
  algorithm for mining of FIs and the so called \emph{reservoir sampling}, see
  Section \ref{sec:phase1-reservoir-sampling}. For the details on the reservoir
  sampling algorithm, see Section~\ref{sec:reservoir-sampling}. This variant of
  Phase~1 is denoted by \textsc{Phase-1-Reservoir-Sampling}, resulting in the
  \scparfimireserv{} method.
\end{enumerate}

\begin{figure}[!ht]
\hrule
\smallskip
\small
\example[Example of Phase~1.]{\label{fig:mfi-sampling}\textbf{(the running example)}
  Example of Phase~1.}{ In this part of the example, we will show the sample
  obtained in Phase~1.  The pictures show only the FIs $\allfi$ of the database
  $\db$ and the FIs $\fiapprox$ of $\dbsmpl$ with $\rminsupp=0.3$. The red
  circles mark the sampled frequent itemsets $\fismpl$ and the blue
  circles mark the MFIs $m\in\mfi$ or $m\in\mfiapprox$.

\begin{center}
\noindent\begin{tabular*}{15cm}{p{5cm}p{7.5cm}}
 Horizontal representation of the database $\db$: & Horizontal representation of the database sample $\dbsmpl$ \\
{
  \hfill\scriptsize
  \renewcommand{\arraystretch}{1}
\exampledatabase
\hfill%
  \medskip
} & {
\scriptsize
\databasesample
}
\\
 \multicolumn{2}{c}{Prefix based equivalence classes with its relative sizes:} \\
 \multicolumn{2}{c}{
  \renewcommand{\arraystretch}{1}\small
       \begin{tabular}{|c|p{4cm}|p{4cm}|p{4cm}|} \hline
	 Prefix      & Real relative size computed from $\db$& Real relative size computed from $\dbsmpl$ & Estimated relative size   \\ \hline
	 $\{1\}$     &$4/25=0.1600$ & $2/25=0.08$  & $1/10=0.1$  \\ \hline
	 $\{2\}$     &$6/25=0.2399$ & $8/25=0.32$  & $3/10=0.3$   \\ \hline
	 $\{3\}$     &$8/25=0.3200$ & $8/25=0.32$  & $4/10=0.4$  \\ \hline
	 $\{4\}$     &$4/25=0.1600$ & $4/25=0.16$  & $1/10=0.1$  \\ \hline
	 $\{5\}$     &$2/25=0.0800$ & $2/25=0.08$  & $1/10=0.1$ \\ \hline
	 $\{6\}$     &$1/25=0.0400$ & $1/25=0.04$  & $0$         \\ \hline
       \end{tabular}\hfill%
  \medskip
} 
\end{tabular*}
\end{center}

\noindent The lattice representing the FIs $\allfi$ and the MFIs $\mfi$ in
the database $\db$:
\begin{center}
\includegraphics[type=mps,ext=.mps,read=.mps]{lattice_mfidb}
\end{center}

\noindent The lattice representing the FIs $\fiapprox$ and $\mfiapprox$ in the database sample $\dbsmpl$:
\begin{center}
\includegraphics[type=mps,ext=.mps,read=.mps]{sample_mfis}
\end{center}

}
\smallskip
\hrule
\end{figure}


\clearpage

\subsection{The modified coverage algorithm based sampling}\label{sec:phase1-mfi-sampling}

In this section, we propose two variants of Phase~1 based on our modification of
the \emph{coverage algorithm}, see
Algorithm~\ref{alg:modified-coverage-algorithm}. Additionally, we put together
the fragments of the algorithms shown in previous chapters.

The workflow of the Phase~1 is summarized in
Figure~\ref{fig:workflow-mfi-based-sampling}:

\begin{figure}[hbt!]
\centering
\smallskip
\begin{tabular}{c}
      \xymatrix{
	\db\ar[r]^{1.~sample} & \dbsmpl\ar[r]^{2.~\text{compute}}  &  M\ar[r]^{3.~\text{sample}}    & \fismpl  \\
      }
\end{tabular}

\begin{enumerate}
\item The sampling produces a database sample $\dbsmpl$ of size $|\dbsmpl| \geq
  \frac{1}{2\epsilon_{\dbsmpl}^2} \ln \frac{2}{\delta_{\dbsmpl}}$. For details
  see Section~\ref{sec:db-sample-support-estimtate}.

\item Computation of the boundary $M$ of the set $\fiapprox$ using $\dbsmpl$ is
  described in Chapter~\ref{chap:parallel-mfi}. The boundary $M$ is then used
  for creation of the sample $\fismpl \subseteq \fiapprox =
  \bigcup_{\smfi_i\in M} \powerset(\smfi_i)$. The boundary is created: 
  \begin{enumerate}
  \item sequentially, producing $M=\mfiapprox$;
  \item in parallel, producing $\mfiapprox\subseteq M\subsetneq\fiapprox$.
  \end{enumerate}
\item Creation of the sample $\fismpl$ using $M$ is performed using the
  \textsc{Modified-Coverage-Algorithm}. For details see
  Section~\ref{sec:coverage-algorithm-sampling}. The sample $\fismpl$ is an
  independently but \textbf{not identically} distributed sample. Therefore, the
  estimates of a size of a PBEC using this sample is a \emph{heuristic} for
  estimating the size of a prefix-based equivalence class.
\end{enumerate}
\caption{The workflow of the coverage algorithm based sampling}\label{fig:workflow-mfi-based-sampling}
\smallskip
\end{figure}

\newpage

\textbf{(a) $\mfiapprox$ is computed sequentially \cite{kessl06balancing}:} the
$\mfiapprox$ is computed on processor $\proc_1$ using an arbitrary algorithm for
mining of MFIs. The sampling is performed sequentially by processor $\proc_1$
using the \textsc{Modified-Coverage-Algorithm}. Phase~1 based on the sequential
computation of MFIs. The pseudocode of Phase~1 is given in
Algorithm~\ref{alg:mfi-sampling-seq}:

\vbox{
\begin{algorithm}[H]
\caption{The \textsc{Phase-1-Coverage-Sampling-Sequential} algorithm}
\label{alg:mfi-sampling-seq}
\vbox{\textsc{Phase-1-Coverage-Sampling-Sequential}(\vtop{\hsize=5cm\inparam Database $\dbpart_i$, 
                   \par\noindent\inparam Double $\rminsupp$, 
                   \par\noindent\inparam Double $\epsilon_{\dbsmpl}$, 
                   \par\noindent\inparam Double $\delta_{\dbsmpl}$, 
                   \par\noindent\inparam Double $\epsilon_{\fismpl}$,
                   \par\noindent\inparam Double $\delta_{\fismpl}$,
                   \par\noindent\inparam Double $\rho$,
                   \par\noindent\outparam Set $\fismpl$,
                   \par\noindent\outparam Database $\dbsmpl$)}}
\begin{algorithmic}[1]
  \FORPARALLEL{all $\proc_i$} \label{alg:mfi-sampling-seq:dbsmpl_beg}
  \STATE $N_{\dbsmpl} \leftarrow \frac{1}{2\epsilon_{\dbsmpl}^2}\ln\frac{2}{\delta_{\dbsmpl}}$

  \STATE $\dbpart'_i\leftarrow$ an i.i.d. sample of $\dbpart_i$ of size  $N_{\dbsmpl}/\procnum$
  \STATE send $\dbpart'_i$ to $\proc_1$ (an all-to-one gather)
  \ENDFOR \label{alg:mfi-sampling-seq:dbsmpl_end}
  \BLOCK{\textbf{processor} $\proc_1$ \textbf{executes}:}\label{alg:mfi-sampling-seq:sbeg}
  \STATE $\dbsmpl\leftarrow\bigcup_{1\leq j\leq\procnum} \dbpart'_j$ \label{alg:mfi-sampling-seq:dbsmpl_collect}
  \STATE compute the approximation of MFIs $\mfiapprox$ from $\dbsmpl$ using an
  arbitrary algorithm for mining of
  MFIs \label{alg:mfi-sampling-seq:compute_mfis}.
  \item[] // \emph{The modified coverage algorithm}
  \STATE $N_{\fismpl} \leftarrow \frac{4}{\epsilon_{\fismpl}^2\rho}\ln\frac{2}{\delta_{\fismpl}}$
  \STATE call \textsc{Modified-Coverage-Algorithm}$(\mfiapprox, N_{\fismpl}, \fismpl)$

  \ENDBLOCK{\textbf{end of} $\proc_1$ \textbf{execution}}\label{alg:mfi-sampling-seq:send}
\end{algorithmic}
\end{algorithm}
}

At the lines
\ref{alg:mfi-sampling-seq:dbsmpl_beg}--\ref{alg:mfi-sampling-seq:dbsmpl_end}, the
\textsc{Phase-1-Coverage-Sampling-Sequential} algorithm creates the database
sample $\dbsmpl$ from the database partitions $\dbpart_i$ that are collected by
processor $\proc_1$ at the line \ref{alg:mfi-sampling-seq:dbsmpl_collect}. Then at line
\ref{alg:mfi-sampling-seq:compute_mfis}, $\proc_1$ executes an arbitrary algorithm for
mining of MFIs.  The creation of the sample $\fismpl$ is performed at lines
\ref{alg:mfi-sampling-seq:sbeg}--\ref{alg:mfi-sampling-seq:send}. The sampling
is a heuristic based on the \emph{modified coverage algorithm}, see
Section~\ref{sec:coverage-algorithm-sampling}.



\newpage

\textbf{(b) The set $M, \mfiapprox\subseteq M\subsetneq\fiapprox$ ($\mfiapprox$
  plus some additional frequent itemsets) is computed in parallel
  \cite{kessl07improved}:} the parallel variant of the MFI based sampling is
summarized in Algorithm~\ref{alg:mfi-sampling-par}.

\vbox{
\begin{algorithm}[H]
\caption{The \textsc{Phase-1-Coverage-Sampling-Parallel} algorithm}
\label{alg:mfi-sampling-par}
\vbox{\textsc{Phase-1-Coverage-Sampling-Parallel}(\vtop{\hsize=5cm\inparam Database $\dbpart_i$,
                   \par\noindent\inparam Double $\rminsupp$,
                   \par\noindent\inparam Double $\epsilon_{\dbsmpl}$, 
                   \par\noindent\inparam Double $\delta_{\dbsmpl}$, 
                   \par\noindent\inparam Double $\epsilon_{\fismpl}$,
                   \par\noindent\inparam Double $\delta_{\fismpl}$,
                   \par\noindent\inparam Double $\rho$,
                   \par\noindent\outparam Set $\fismpl$,
                   \par\noindent\outparam Database $\dbsmpl$)}}
\begin{algorithmic}[1]
  \FORPARALLEL{all $\proc_i$}
     \STATE $N_{\dbsmpl} \leftarrow \frac{1}{2\epsilon_{\dbsmpl}^2}\ln\frac{2}{\delta_{\dbsmpl}}$
     \STATE $N_{\fismpl} \leftarrow \frac{4}{\epsilon_{\fismpl}^2\rho}\ln\frac{2}{\delta_{\fismpl}}$\label{alg:mfi-sampling-par:sbeg}
     \STATE $\dbpart'_i\leftarrow$ an i.i.d. sample of $\dbpart_i$ of size  $N_{\dbsmpl}/\procnum$
     \STATE broadcast $\dbpart'_i$ (an all-to-all broadcast).
     \STATE $\dbsmpl \leftarrow \bigcup_{1\leq j\leq\procnum} \dbpart'_j$.

  \STATE Execute an arbitrary algorithm for mining of MFIs in parallel,
  $\proc_i$ computing $M_i$ from $\dbsmpl$, e.g., call
  \textsc{Parallel-DFS-MFI-Schema}($\dbsmpl$, $\rminsupp$, $M_i$).

     \item[] // \emph{Create the sample using the modified coverage algorithm}
     \STATE broadcast $s_i=\sum_{\smfi\in M_i}|\powerset(\smfi)|$ (hence an all-to-all scatter takes place).\label{alg:mfi-sampling-par:par-sample-beg}
     \STATE $s\leftarrow \sum_{1\leq i\leq\procnum} s_i$
     \STATE $\fipart_i\leftarrow\emptyset$.
     \STATE call \textsc{Modified-Coverage-Algorithm}$(M_i, N_{\fismpl}\cdot\frac{s_i}{s}, \fipart_i)$
     \STATE send $\fipart_i$ to $\proc_1$.\label{alg:mfi-sampling-par:par-sample-end}
  \ENDFOR
  \STATE processor $\proc_1$ computes $\fismpl=\bigcup_{1\leq i\leq\procnum}\fipart_i$.
\end{algorithmic}
\end{algorithm}
}

In Algorithm~\ref{alg:mfi-sampling-par}, an arbitrary modified DFS sequential
algorithm for mining of MFIs is executed in parallel, see
Chapter~\ref{chap:parallel-mfi}. The modified algorithm for mining of MFIs does
not compute $\mfiapprox$, but instead it computes a set $M = \bigcup_{1\leq i
  \leq \procnum} M_i$ such that $\mfiapprox\subseteq M\subsetneq\fiapprox$.  The
computed sets are distributed among the processors and the number of these sets
can be large. Therefore, we perform the sampling in parallel. For details of
parallel mining of MFIs, see Chapter~\ref{chap:parallel-mfi}.

\emph{The parallel sampling of $\fiapprox$ using $M$}, steps
\ref{alg:mfi-sampling-par:par-sample-beg}--\ref{alg:mfi-sampling-par:par-sample-end},
is performed in the following way: every processor $\proc_i$ broadcasts the sum
$s_i = \sum_{\smfi\in M_i} |\powerset(\smfi)|$ of sizes of powersets of its
local MFIs (hence, an all-to-all broadcast takes place), creates a fraction of
sample $\fismpl$ of size $|\fismpl| \cdot \frac{s_i}{\sum_{1\leq j\leq\procnum}
  s_j}$, and finally sends them to $\proc_1$. We should pick the number of
samples chosen by each processor from a multivariate binomial distribution with
parameters $p_i = \frac{s_i}{\sum_{1\leq j\leq\procnum} s_j}$ and $n =
\sum_{1\leq j\leq\procnum} s_j$, see Appendix~\ref{sec:multivariate-binomial} in
order to be able to give guarantees on the error of the estimate. However, using
the modified coverage algorithm makes from the sample just a heuristic.
Therefore, we do not have any guarantees and $\proc_i$ takes the number of
samples $|\fismpl| \cdot \frac{s_i}{\sum_{1\leq j\leq\procnum} s_j}$.

\newpage

\subsection{The sampling based on the reservoir algorithm}\label{sec:phase1-reservoir-sampling}

In the previous section, we have proposed a variant of Phase~1, based
on the \emph{modified coverage algorithm}, that samples $\allfi$
non-uniformly. In this Section, we propose another variant of Phase~1:
a sampling process based on the \emph{reservoir sampling}
\cite{vitter85reservoir} that samples $\fiapprox$ uniformly, i.e., it
creates an identically distributed sample of $\fiapprox$. The workflow
of the reservoir sampling algorithm is shown in
Figure~\ref{fig:workflow-reservoir-sampling}.

\begin{figure}[hbt!]
\centering
\begin{tabular}{c}
      \xymatrix{
	\db\ar[r]^{1.~\text{compute}} & \dbsmpl\ar[r]^{2.~\text{compute in parallel}}  &  \fiapprox\ar[r]^{3.~\text{sample in parallel}}    & \fismpl  \\
      }
\end{tabular}
\begin{enumerate}
\item As in the previous section, we first need to produce the database sample
  $\dbsmpl$ of size $|\dbsmpl| = \frac{1}{2\epsilon_{\dbsmpl}^2} \ln
  \frac{2}{\delta_{\dbsmpl}}$. For details see
  Section~\ref{sec:db-sample-support-estimtate}.
\item From the database sample $\dbsmpl$, we compute all FIs $\fiapprox$,
  using an arbitrary sequential algorithm for mining of FIs.
\item The output of the sequential algorithm for mining of FIs is sampled using
  the \emph{reservoir sampling}, we produce $\fismpl$ of size $|\fismpl| =
  -\frac{\log(\delta_{\fismpl}/2)}{D(\rho+\epsilon_{\fismpl}||\rho)}$. For
  details see Section \ref{sec:reservoir-sampling}.
\end{enumerate}
\caption{The workflow of the reservoir based sampling}\label{fig:workflow-reservoir-sampling}
\end{figure}

In our parallel method, we are using the \textsc{Vitter-Reservoir-Sampling}
Algorithm, the faster reservoir sampling algorithm. To speedup the sampling
phase of our parallel method, we execute the reservoir sampling in parallel. The
database sample $\dbsmpl$ is distributed among the processors -- each processor
having a copy of the database sample $\dbsmpl$. The baseset $\baseset$ is
partitioned into $\procnum$ parts $B_i\subseteq\baseset$ of size $|B_i| \approx
|\baseset|/\procnum$ such that $B_i \cap B_j = \emptyset, i\neq j$. Processor
$\proc_i$ then takes part $B_i$ and executes an arbitrary sequential DFS
algorithm for mining of FIs, enumerating $[(\bitem_j)] \cap \fiapprox,
\bitem_j\in B_i$. The output, the itemsets $[(\bitem_j)] \cap \fiapprox$, of the
sequential DFS algorithm are read by the reservoir sampling algorithm. If a
processor finished its part $B_i$, it asks other processors for work, hence
performing dynamic load-balancing. For terminating the parallel execution, we
use the Dijkstra's token termination algorithm.

The task of the Phase~1 is to take $|\fismpl| =
-\frac{\log(\delta_{\fismpl}/2)}{D(\rho+\epsilon_{\fismpl}||\rho)}$ samples, see
Theorem \ref{theorem:reservoir-size-estim}.  Because the reservoir algorithm and
the sequential algorithm is executed in parallel, it is not known how many FIs
is computed by each processor. Denote the unknown number of FIs computed on
$\proc_i$ by $f_i$, the total number of FIs is denoted by $f=\sum_{1\leq
  i\leq\procnum} f_i$.  Because, we do not know $f_i$ in advance, each processor
samples $|\fismpl|$ frequent itemsets using the reservoir sampling algorithm,
producing $\fismpl$, and counts the number of FIs computed by the sequential
algorithm. When the reservoir sampling finishes, processor $\proc_i$ sends $f_i$
to $\proc_1$. $\proc_1$ picks $\procnum$ random variables $X_i, 1\leq
i\leq\procnum$ from multivariate hypergeometrical distribution, see
Appendix~\ref{appendix:discrete-distributions}, with parameters $M_i=f_i$. The
value of $X_i$ is send to $\proc_i$.  $\proc_i$ then choose $X_i$ itemsets
$\isetU\in\fismpl$ at random out of the $|\fismpl|$ sampled frequent itemsets
computed by $\proc_i$. The samples are then send to processor $\proc_1$.
$\proc_1$ stores the received samples in $\fismpl$. This process is summarized
in Algorithm~\ref{alg:phase-1-simple-reservoir-algorithm}.

\vbox{
\begin{algorithm}[H]
\caption{The \textsc{Phase-1-Reservoir-Sampling} algorithm}
\label{alg:phase-1-simple-reservoir-algorithm}
\vbox{\textsc{Phase-1-Reservoir-Sampling}(\vtop{\inparam Database $\dbpart_i$,
    \par\noindent\inparam Double $\rminsupp$,
    \par\noindent\inparam Double $\epsilon_{\dbsmpl}$,
    \par\noindent\inparam Double  $\delta_{\dbsmpl}$,
    \par\noindent\inparam Double $\epsilon_{\fismpl}$,
    \par\noindent\inparam Double $\delta_{\fismpl}$,
    \par\noindent\inparam Double $\rho$,
    \par\noindent\outparam Set $\fismpl$,
    \par\noindent\outparam Database $\dbsmpl$)}}
\begin{algorithmic}[1]
  \FORPARALLEL{all processors $\proc_i$}
  \STATE $N_{\fismpl}\leftarrow -\frac{\log(\delta_{\fismpl}/2)}{D(\rho+\epsilon_{\fismpl}||\rho)}$
  \STATE $N_{\dbsmpl} \leftarrow \frac{1}{2\epsilon_{\dbsmpl}^2}\ln\frac{2}{\delta_{\dbsmpl}}$
  \STATE $\dbpart'_i\leftarrow$ an i.i.d. sample of $\dbpart_i$ of size  $N_{\dbsmpl}/\procnum$
  \STATE broadcast $\dbpart'_i$ to $\proc_1$ (an all-to-all scatter).
  \STATE $\dbsmpl\leftarrow\bigcup_{1\leq j\leq\procnum} \dbpart_j$.
  \STATE $R\leftarrow$ array of size $N_{\fismpl}$
  \STATE Partition $\baseset$ on $\procnum$ parts $\baseset_i$, such that
         $\baseset_i\cap\baseset_j=\emptyset, i\neq j$. 
  \STATE Execute \textsc{Vitter-Reservoir-Sampling}(R,
  $N_{\fismpl}$, \textsc{ReadNextFI}, \textsc{SkipFIs}) and the
  \textsc{ReadNextFI($R$)} reads the output of an arbitrary sequential algorithm
  $A_{FI}$ for mining of FIs with minimal support $\rminsupp$ and
  \textsc{SkipFIs($n$)} skips $n$ FIs from the output of the algorithm. $A_{FI}$
  at processor $\proc_i$ processes $[(\bitem_k)], \bitem_k\in\baseset_i$. If
  $A_{FI}$ finishes its $\baseset_i$ it asks other processors for work,
  performing dynamic load-balancing. The algorithm terminates using the
  Dijkstra's token termination algorithm.
  \item[] // \emph{The number of \textbf{all} FIs computed by $\proc_i$ is denoted by $f_i$.}
  \STATE $f_i$ is broadcast to other processors (all-to-all-broadcast)
  \STATE $\proc_1$ picks the random numbers $X_i$ from the multivariate hypergeometric distribution with parameters $M_i=f_i$.
  \STATE $\proc_1$ broadcasts $X_i$ to other processors and each processor creates a sample $S_i\subseteq R, |S_i|=X_i$.
  \STATE $S_i$ is send to processor $\proc_1$.
  \STATE $\proc_1$ creates $\fismpl\leftarrow\bigcup_{1\leq j \leq \procnum}S_j$.
  \ENDFOR 
\end{algorithmic}
\end{algorithm}
}



\clearpage

\section{Detailed description of Phase 2}\label{sec:phase-2-detailed}

In Phase~2 the method partitions $\allfi$ sequentially on processor
$\proc_1$. As an input of the partitioning, we use the samples $\fismpl$, the
database $\dbsmpl$ (computed in Phase~1), and a real number $\alpha,
0<\alpha\leq 1$. Recall that, we denote the prefixes by $\isetU_k$, the
extensions of $\isetU_k$ by $\prefixext_k$, i.e., $\isetU_k$ and $\prefixext_k$
forms a PBEC $[\isetU_k|\prefixext_k]$. In the following text, we omit
$\prefixext_k$ from the notation, i.e., a PBEC $[\isetU_k | \prefixext_k]$ is
denoted by $[\isetU_k]$ if clear from context or if $\prefixext_k$ is
unnecessary. The set of the indexes of the PBECs assigned to processor
$\proc_i$ is denoted by $\fipartidxset_i$, and the set of all FIs assigned to
processor $\proc_i$ is denoted by $\fipart_i$. Each $\fipart_i$ is the union of
FIs in one or more PBECs $[\isetU_k|\prefixext_k]$, i.e., $\fipart_i =
\bigcup_{k\in\fipartidxset_i} ([\isetU_k | \prefixext_k]) \cap \allfi$. Each
processor $\proc_i$ then in Phase~4 processes the FIs contained in
$\fipart_i$. \emph{The output of Phase~2} are the index sets $\fipartidxset_i$
of PBECs, computed on $\proc_1$, and the PBECs $[\isetU_k|\prefixext_k]$.

\emph{The partitioning of $\allfi$ is a two step process:}
\begin{enumerate}
\item[(1)] $\proc_1$ creates a list of prefixes $\isetU_{k}$ such that the
  estimated relative size of the PBEC $[\isetU_k] \cap \allfi$ satisfies
  $\frac{|[\isetU_k]\cap\fismpl|}{|\fismpl|} \leq \alpha \cdot
  \frac{1}{\procnum}$, where $0 < \alpha < 1$ is a parameter of the computation
  set by the user. The reason for making the PBECs of relative size $\leq \alpha
  \cdot \frac{1}{\procnum}$ is to make the PBECs small enough so that they can
  be scheduled and the schedule is balanced, i.e., each processor having a
  fraction $\approx 1/\procnum$ of FIs. Smaller number of large PBECs could make
  the scheduling unbalanced.


\item[(2)] $\proc_1$ creates set of indexes $\fipartidxset_i$ such that $|\fipart_i|
  / |\allfi|\approx 1 / \procnum$.
\end{enumerate}

\textbf{(1) The creation of the prefixes $\isetU_k$ proceeds as follows:}
processor $\proc_1$ initially set $\isetU_k=\{\bitem_k\}, \bitem_k\in\baseset$
and estimate the size of $[\isetU_k]\cap\allfi$ using $\fismpl$. The extensions
of the initial $\isetU_k$ are the sets $\prefixext_k=\{b_i | \bitem_k, \bitem_i
\in \baseset, b_k\in\isetU_k\text{ and } b_k < b_i \}$. After the construction
of $\isetU_k$ and $\prefixext_k$ is finished, we estimate the relative size of
$[\isetU_k|\prefixext_k] \cap \allfi$ by $\frac{|[\isetU_k|\prefixext_k] \cap
  \fismpl|}{|\fismpl|}$. If some of the PBEC $[\isetU_k|\prefixext_k]$ is too
big, i.e., $\frac{|[\isetU_k|\prefixext_k] \cap \fismpl|}{|\fismpl|} > \alpha
\cdot \frac{1}{\procnum}$, the algorithm recursively partitions
$[\isetU_k|\prefixext_k]$ into smaller disjoint prefix-based equivalence
subclasses with prefix $\isetU_k\cup\{\bitem_i\}, \bitem_i \in \prefixext_k$
with extensions $\prefixext'_i=\{\bitem_j | \bitem_j \in \prefixext_k \text{ and
} \bitem_i < \bitem_j\}$, i.e., the PBEC $[\isetU_k\cup\{\bitem_i\} |
  \prefixext'_i]$. The size of the parameter $\alpha$ influence the granularity
of the partitioning.

The result of this process are the PBECs $[\isetU_i|\prefixext_i]$ that are
assigned to the processors and used in Phase~3 and 4.  The pseudocode of the
partitioning of the PBEC $[\isetU | \prefixext]$, i.e., partitioning a prefix
$\isetU$ and its extensions $\prefixext$, is summarized in
Algorithm~\ref{alg:partition}.

\vbox{
\begin{algorithm}[H]
\caption{The \textsc{Partition} algorithm}
\label{alg:partition}
\vbox{\textsc{Partition}(\vtop{\inparam Prefix $\isetU$, \par\noindent\inparam Extensions $\prefixext$, \par\noindent\inparam Database $\dbsmpl$, \par\noindent\inparam Sample $\fismpl$, \par\noindent\outparam Set $\prefixset$)}}
\begin{algorithmic}[1]
  \STATE sort $\bitem_i\in\prefixext_k$ by the support in ascending order, i.e.,
  \par\hskip 1 cm $\supp(\isetU\cup\{\bitem_1\}, \dbsmpl) < \supp(\isetU\cup\{\bitem_2\}, \dbsmpl) < \ldots
  < \supp(\isetU\cup\{\bitem_{|\prefixext_k|}\}, \dbsmpl)$\label{alg:partition:estim-order}
  \FOR{$\bitem\in \prefixext_k$} 
     \STATE $\isetU' \leftarrow \isetU\cup\{\bitem\}$
     \STATE $\prefixext' \leftarrow \{\bitem_i|\bitem_i\in\prefixext_k, \bitem < \bitem_i\}$
     -- use the ordering created at line~\ref{alg:partition:estim-order}.
     \STATE $s\leftarrow |[\isetU'|\prefixext']\cap\fismpl|$, i.e., estimate the number of FIs in $[\isetU'|\prefixext']$
     \STATE $\prefixset\leftarrow \prefixset\cup\{(\isetU', \prefixext', s)\}$. 
  \ENDFOR
\end{algorithmic}
\end{algorithm}
}

\begin{proposition}
Let $\isetW\subseteq\baseset$ be a prefix and
$\prefixext\subseteq\baseset$ its extensions and $\dbsmpl$ a database
sample and $\fismpl$ a sample of FIs. Let $\prefixset =
\{(\isetU_k,\prefixext_k, s_k)\}$ be the PBECs created by the
\textsc{Partition} algorithm by calling $\textsc{Partition}(\isetW,
\prefixext, \dbsmpl, \fismpl, \prefixset)$. The PBECs $[\isetU_k|\prefixext_k],
(\isetU_k,\prefixext_k, s_k)\in \prefixset,$ are disjoint.


\end{proposition}

\begin{proof}
Trivially follows from the fact that the prefixes $\isetU_k$ are distinct, i.e.,
$\isetU_i \cap \isetU_j = \emptyset$, and the process of creation of the
extensions.
\end{proof}

In Chapter \ref{chap:math-found}, we defined without loss of generality a single
order of $\bitem_i\in\baseset$: $\bitem_1 < \bitem_2 < \ldots <
\bitem_{|\baseset|}$. But: a sequential DFS algorithms (like Eclat and FPGrowth)
expands every prefix $\isetW_k$ using the extensions $\prefixext_k$ sorted by
the support in ascending order by the support of $\bitem, \bitem' \in
\prefixext_k$ and $\bitem < \bitem'$ if and only if $\supp(\isetW \cup
\{\bitem\}, \db) < \supp(\isetW \cup \{\bitem'\}, \db)$, i.e., each prefix
$\isetW_k$ can have different order of the extensions $\prefixext_k$. The
dynamic re-ordering of items can significantly reduce the execution time of the
sequential algorithm executed in Phase~4. To make the parallel algorithm fast,
we have to use the same order as the sequential algorithm for mining of FIs, see
Section~\ref{sec:dynamic-item-ordering}. To make the order the same as the
sequential algorithm, we estimate the order of extensions $\prefixext_k$ for
prefix $\isetW_k$ using the supports from $\dbsmpl$, i.e., $\supp(\isetW \cup
\{\bitem\}, \dbsmpl), \supp(\isetW \cup \{\bitem'\}, \dbsmpl)$. The different
order of items for different prefix does not influence the output of a
sequential algorithm for mining of FIs. The details of the influence of the
order of the items of $\baseset$ in sequential algorithms are discussed in the
\secref{sec:dynamic-item-ordering}.

\textbf{(2) The creation of the assignment, i.e., the index sets
  $\fipartidxset_i$ of the prefix-based classes $[\isetU_k]$ proceeds as
  follows:} we need to create index sets $\fipartidxset_i$, such that $\fipart_i
= \bigcup_{k\in\fipartidxset_i}([\isetU_k] \cap \allfi)$ and $\max_i|\fipart_i|
/ |\allfi|$ is minimized, i.e., we want to schedule $\sum_i|\fipartidxset_i|$
tasks on $\procnum$ equivalent processors. The scheduling task is known
NP-complete problem with known approximation algorithms. We use the
\textsc{LPT-Schedule} algorithm (LPT stands for least processing time). The
\textsc{LPT-Schedule} algorithm (see \cite{Graham69} for the proofs) is a
best-fit algorithm, see Algorithm~\ref{alg:lpt-schedule}:

\medskip
\vbox{
\begin{algorithm}[H]
\caption{The \textsc{LPT-Schedule} algorithm}\label{alg:lpt-schedule}
\vbox{\textsc{LPT-Schedule}(\inparam Set $S=\{(\isetU_i, \prefixext_i, s_i)\}$, \outparam Sets $\fipartidxset_i$)}
\begin{algorithmic}[1]
  \STATE Sort the set $S$ such that $s_i<s_j, i\neq j$.
  \STATE Assign each $(\isetU_i, \prefixext_i, s_i)$ (in decreasing order by
  $s_i$) to the least loaded processor $\proc_k$. The indexes assigned to
  $\proc_k$, are stored in $\fipartidxset_k$.
\end{algorithmic}
\end{algorithm}
}

\begin{lemma}\cite{Graham69} \textsc{LPT-Schedule} is $4/3$-approximation algorithm.
\end{lemma}

Let OPT be the time of the optimum schedule. The lemma says that the
\textsc{LPT-Schedule} algorithm finds a schedule with the time at most $4/3
\cdot \text{OPT}$.

The index sets $\fipartidxset_i$ together with $\isetU_k$ and $\prefixext_k$ are
then broadcast to the remaining processors. 

The pseudocode of Phase~2 is summarized in Algorithm~\ref{alg:phase-2}. An
example of the partitioning process is in Figure \ref{fig:pbec-assignment}.

\vbox{
\begin{algorithm}[H]
\caption{The \textsc{Phase-2-FI-Partitioning} algorithm}
\label{alg:phase-2}
\vbox{\textsc{Phase-2-FI-Partitioning}(\vtop{\inparam Set $\fismpl$,
  \par\noindent\inparam Database $\dbsmpl$,
  \par\noindent\inparam Double $\alpha$,
  \par\noindent\outparam Set $\prefixset$,
  \par\noindent\outparam Sets $\fipartidxset_i$)}}
\begin{algorithmic}[1]
  \STATE create initial prefixes and extensions\par\noindent $\prefixset \leftarrow \left\{(\isetU_k,\prefixext_k, s) |
  \isetU_k=\{\bitem_k\}, \bitem_k\in\baseset\text{ and } \prefixext_k=\{\bitem_i
  | \bitem_k < \bitem_i\}, s=\frac{|[\isetU_k]\cap\fismpl|}{|\fismpl|}\right\}$
  \WHILE{exists $q = (\isetU,\prefixext,s)\in \prefixset$ such that $s > \alpha\cdot\frac{1}{\procnum} \cdot |\fismpl|$}
     \STATE select $q = (\isetU,\prefixext,s)\in \prefixset$ such that $s > \alpha\cdot\frac{1}{\procnum} \cdot |\fismpl|$
     \STATE \textsc{Partition}($\isetU$, $\prefixext$, $\dbsmpl$, $\prefixset'$)
     \STATE $\prefixset \leftarrow (\prefixset \setminus\{q\})\cup \prefixset'$
  \ENDWHILE
  \STATE $\fipartidxset_i\leftarrow\emptyset, i=1,\ldots,\procnum$
  \STATE call \textsc{LPT-Schedule}($\prefixset$, $\fipartidxset_i$)
\end{algorithmic}
\end{algorithm}
}

\begin{figure}[!ht]
\medskip
\example[Example of Phase~2.]{\label{fig:pbec-assignment}\textbf{(the running example)} Example of Phase~2.}{

This example shows the prefix-based classes, samples created using the modified
coverage algorithm or the reservoir sampling, and the final assignment of the
PBECs to the processors. The samples are marked by a red color.

\begin{center}
\hbox{The lattice partitioning computed from the database sample $\dbsmpl$:}
\includegraphics[type=mps,ext=.mps,read=.mps]{sample_partitioning}

\bigskip

\begin{tabular}{|c|c|p{3cm}|p{3cm}|} \hline
Processor & Assigned prefix-based classes & The estimated amount of work & Real amount of work \\ \hline
$\proc_1$  & $[(3)]$                     &       $0.4$                 &   $0.3076$           \\ \hline
$\proc_2$  & $[(2)]$                     &       $0.3$                 &   $0.2307$           \\ \hline
$\proc_3$  & $[(1)], [(4)], [(5)], [(6)]$ &      $0.3$                 &   $0.4228$           \\ \hline
\end{tabular}
\end{center}

}
\medskip
\end{figure}

\clearpage

\section{Detailed description of Phase 3}\label{sec:phase-3-detailed}

The input of Phase~3 for a processor $\proc_i$ is the set of indexes of the
assigned PBECs $\fipartidxset_i$ together with the prefixes $\isetU_k$ and its
extensions $\prefixext_k$. The processor $\proc_i$ needs for the computation of
$\fipart_i = \bigcup_{k\in\fipartidxset_i}([\isetU_k] \cap \allfi)$ a database
partition $\dbpart'_i$ that contain all the information needed for computation
of $\fipart_i$. At the beginning of this phase, the processors has disjoint
database partitions $\dbpart_i$ such that $|\dbpart_i| \approx
\frac{|\db|}{\procnum}$. For the description of the algorithm of Phase~3, we
expect that we have a distributed memory machine whose nodes are interconnected
using a network such as Myrinet \cite{myrinet} or Infiniband \cite{infiniband},
i.e., a network that is not congested while an arbitrary permutation of two
nodes communicates with each other. The problem is the congestion of the network
in Phase~3.

To construct $\dbpart'_i$ on processor $\proc_i$, every processor $\proc_j,
i\neq j$, has to send a part of its database partition $\dbpart_j$ needed by the
other processors to all other processors (an all-to-all scatter takes
place\footnote{all-to-all scatter is a well known communication operation: each
  processor $\proc_i$ sends a message $m_{ij}$ to processor $\proc_j$ such that
  $m_{ij}\neq m_{ik}, i\neq k$}). That is: processor $\proc_i$ send to processor
$\proc_j$ the set of transactions $\{t | t\in\dbpart_i, k\in\fipartidxset_j,
\text{ and } \isetU_k\trsubseteq t\}$, i.e., all transactions that contain at
least one $\isetU_k, k\in\fipartidxset_j$ as a subset. Each processor then has
the database part $\dbpart'_j=\bigcup_i \{t | t\in\dbpart_i, k\in\fipartidxset_j,
\text{ and } \isetU_k\trsubseteq t\} = \{t | t\in\db, \text{exists }
k\in\fipartidxset_j, \isetU_k\trsubseteq t \}$.


Each round of the all-to-all scatter is done in $\lfloor \frac{\procnum}{2}
\rfloor$ parallel communication steps.  We can consider the scatter as a
round-robin tournament of $\procnum$ players \cite{round-robin}. Creating the
schedule for the tournament is the following procedure: if $\procnum$ is odd, a
dummy processor(player) can be added, whose scheduled opponent waits for the
next round and the processors(player) performs $\procnum$ communication
rounds(games).  If $\procnum$ is even, then we perform $\procnum-1$ rounds of
the parallel communication steps(games). For example let have $14$ processors,
in the first round the following processors exchange their database partitions:

\begin{center}
\begin{tabular}{|c|c|c|c|c|c|c|} \hline
1  & 2  & 3  & 4  & 5  & 6 & 7 \\ \hline
14 & 13 & 12 & 11 & 10 & 9 & 8 \\ \hline
\end{tabular}
\end{center}

The processors are paired by the numbers in the columns. That is, database parts
are exchanged between processors $\proc_{1}$ and $\proc_{14}$, $\proc_2$ and
$\proc_{13}$, etc.

In the second round one processor is fixed (number one in this case) and the
other are rotated clockwise:

\begin{center}
\begin{tabular}{|c|c|c|c|c|c|c|} \hline
\textbf{1}  & 14  & 2 & 3  & 4  & 5  & 6 \\ \hline
        13  & 12 & 11 & 10 & 9  & 8  & 7 \\ \hline
\end{tabular}
\end{center}

This process is iterated until the processors are almost in the
initial position:

\begin{center}
\begin{tabular}{|c|c|c|c|c|c|c|} \hline
\textbf{1}  & 3  & 4  & 5  & 6 & 7 & 8 \\ \hline
        2   & 14 & 13  & 12 & 11 & 10 & 9  \\ \hline
\end{tabular}
\end{center}

In the description of the \textsc{DB-Partition-Exchange},
Algorithm~\ref{alg:db-exchange}, we borrow the \emph{ternary operator}
\texttt{?:} from C. The ternary operator has the form ``\texttt{test} ?
\texttt{value1} : \texttt{value2}'' that means: if \texttt{test} is true then
return \texttt{value1} else return \texttt{value2}. The shift operation works
like a bit shift operator, e.g., given an array $A=(1,2,3,4,5)$, the result of
$A$ shifted to the right is the array $(\text{undef},1,2,3,4)$.


\medskip
\vbox{
\begin{algorithm}[H]
\caption{The \textsc{Phase-3-DB-Partition-Exchange} algorithm}\label{alg:db-exchange}
\vbox{\textsc{Phase-3-DB-Partition-Exchange}(\vtop{\inparam Integer $P$, 
    \par\noindent\inparam Prefixes $\{U_k\}$, 
    \par\noindent\inparam Indexsets $\fipartidxset_i$,
    \par\noindent\outparam Database parts $\dbpart'_i$)}}
\begin{algorithmic}[1]
  \FORPARALLEL{all processors $\proc_i$}
  \IF{$\procnum$ is odd}
      \STATE $A_s\leftarrow (P-1)/2$
    \ELSE
      \STATE $A_s\leftarrow P/2$
  \ENDIF
  \STATE $A_1\leftarrow$ \textbf{new} array of size $A_s$;  $A_2\leftarrow$ \textbf{new} array of size $A_s$
  \STATE $\dbpart'_i\leftarrow$ empty database
  \STATE Rounds $\leftarrow$ $\procnum$ is odd ? $\procnum$ : $\procnum-1$
  \FOR{$q\leftarrow 1$ to $A_s$}
       \STATE $A_1[q]\leftarrow q$; $A_2[A_s-q+1]\leftarrow A_s+q$
  \ENDFOR
     \FOR{$m\leftarrow 1$ to Rounds}
        \STATE $\ell\leftarrow$ index $\ell$ such that $A_1[\ell]=i$ or $A_2[\ell]=i$
        \STATE opponent $\leftarrow$ $A_1[\ell]=i$ ? $A_2[\ell]$ : $A_1[\ell]$
        \STATE $T\leftarrow$ all transactions $t\in\dbpart_i$ such that $\isetU_k\trsubseteq t$ and $k\in\fipartidxset_{\text{opponent}}$
        \IF{($P$ is odd and $i\neq P+1$) or $\procnum$ is even}
           \IF{$i < $opponent}
              \STATE send transactions $T$ to $\proc_{\text{opponent}}$
               \STATE receive transactions $T$ from $\proc_{\text{opponent}}$ and store them in $\dbpart'_i$
           \ELSE
              \STATE receive transactions $T$ from $\proc_{\text{opponent}}$ and store them in $\dbpart'_i$
              \STATE send transactions $T$ to $\proc_{\text{opponent}}$
           \ENDIF
        \ENDIF
        \STATE $\text{tmp}_1\leftarrow A_1[A_s]$;  $\text{tmp}_2\leftarrow A_2[1]$
        \STATE shift $A_1[2..A_s]$ to the right
        \STATE shift $A_2$ to the left
        \STATE $A_1[2]\leftarrow \text{tmp}_2$; $A_2[A_s]\leftarrow \text{tmp}_1$
     \ENDFOR
  \ENDFOR
\end{algorithmic}
\end{algorithm}
}

\section{Detailed description of Phase 4}\label{sec:phase-4-detailed}

The input to this phase, for processor $\proc_q, 1\leq q\leq \procnum,$ is the
database partition $\dbpart_q$ (the database partition that is the input of the
whole method, the database partition), the set $\prefixset = \{(\isetU_k,
\prefixext_k) | \isetU_k\subseteq\baseset, \prefixext_k\subseteq\baseset,
\isetU_k\cap\prefixext_k = \emptyset \}$ of prefixes $\isetU_k$ and the
extensions $\prefixext_k$, and the sets of indexes $\fipartidxset_q$ of prefixes
$\isetU_k$ and extensions $\prefixext_k$ assigned to processor $\proc_q$, and
$\dbpart'_q = \bigcup_{1\leq i\leq \procnum} \{t | t\in\db, \text{such that for
  each } k\in\fipartidxset_i \text{ holds } \isetU_k\trsubseteq t\}$ (the
database received in Phase~3 from other processors).


In Phase~4, we execute an arbitrary algorithm for mining of FIs. The sequential
algorithm is run on processor $\proc_q$ for every prefix and extensions
$(\isetU_k,\prefixext_k) \in \prefixset, k \in \fipartidxset_q$ assigned to the
processor, i.e., $\proc_q$ enumerates all itemsets $\isetW\in[\isetU_k |
  \prefixext_k], k\in\fipartidxset_q$. Therefore, the datastructures used by a
sequential algorithm, must be prepared in order to execute the sequential
algorithm for mining of FIs with particular prefix and extensions. To make the
parallel execution of a DFS algorithm fast, we prepare the datastructures by
simulation of the execution of the sequential DFS algorithm, e.g., to enumerate
all FIs in a PBEC $[\isetU_k|\prefixext_k]$ Phase~4 simulates the sequential
branch of a DFS algorithm for mining of FIs up to the point the sequential
algorithm can compute the FIs in $[\isetU_k | \prefixext_k]$. An example of such
a simulation is in Chapter~\ref{chap:exec-eclat}. We describe Phase~4 in
Algorithm~\ref{alg:phase-4-compute-fi} ($\dbpart_q$ is the database partition
loaded in Phase~2, $\dbpart'_q$ is the database partition received in Phase~3):


\begin{algorithm}[H]
\caption{The \textsc{Phase-4-Compute-FI} algorithm}\label{alg:phase-4-compute-fi}
\vbox{\textsc{Phase-4-Compute-FI}(\vtop{\inparam Set of prefixes $\prefixset = \{(\isetU_k, \prefixext_k)\}$, 
  \par\noindent\inparam Indexsets $\fipartidxset_q$,
  \par\noindent\inparam Database $\dbpart_q$,
  \par\noindent\inparam Database $\dbpart'_q$,
  \par\noindent\inparam Integer $\minsupp$,
  \par\noindent\outparam Set $\allfi_q$)}}
\begin{algorithmic}[1]
  \FORPARALLEL{all processors $\proc_i$}
  \STATE compute support of itemsets $\isetW\subseteq\isetU_k$ in $\dbpart_q$, i.e., $\supp(\isetW, \dbpart_q)$
  \STATE send $\supp(\isetW, \dbpart_q)$ to $\proc_1$
  \ENDFOR
  \STATE $\proc_1$ outputs $\isetW$ such that $\sum_{1\leq i\leq\procnum}\supp(\isetW, \dbpart_i) \geq \minsupp$.

  \STATE all $\proc_q$ execute an arbitrary algorithm for mining of FIs in
  parallel that computes supports of $\supp(\isetW,\dbpart'_q),
  \isetW\in\bigcup_{k\in\fipartidxset_q}[\isetU_k|\prefixext_k]$,
  $(\isetU_k,\prefixext_k) \in \prefixset$ and adds them to $\allfi_q$.

\end{algorithmic}
\end{algorithm}

\section{The summary of the new parallel FIMI methods}
From the previous discussion it follows that we can create three
parallel FIMI methods. Two of the methods are based on the
\emph{modified coverage algorithm}.  The third method leverages the
\emph{reservoir sampling}. The methods are described in such a way
that they can be parametrized using an arbitrary algorithm for mining
of MFIs and/or an arbitrary algorithm for mining of FIs. In this section,
we show pseudocodes for the three methods.

\subsection{The \scparfimiseq{} method}
The \scparfimiseq{} method is based on the \emph{coverage algorithm} and
computes the MFIs sequentially. It samples $\fiapprox$ non-uniformly, see
Method~\ref{method:parallel-fimi-seq}.

\bigskip
\vbox{
\begin{method}[H]
\caption{The \textsc{\parfimiseq} method}\label{method:parallel-fimi-seq}
\vbox{\textsc{Parallel-FIMI-Seq}(\vtop{\noindent\inparam Double $\rminsupp$, 
  \par\noindent\inparam Double $\epsilon_{\dbsmpl}$,
  \par\noindent\inparam Double  $\delta_{\dbsmpl}$,
  \par\noindent\inparam Double $\epsilon_{\fismpl}$,
  \par\noindent\inparam Double $\delta_{\fismpl}$,
  \par\noindent\inparam Double $\rho$,
  \par\noindent\inparam Double $\alpha$,
  \par\noindent\outparam Sets $\allfi_i$)}}
\begin{algorithmic}[1]
     \STATE // \emph{Phase 1: sampling.}
     \STATE Each processor $\proc_i$ reads $\dbpart_i$.
     \STATE $\proc_1$ calls \textsc{Phase-1-Coverage-Sampling-sequential}($\dbpart_i$, $\rminsupp$, $\epsilon_{\dbsmpl}$,
            $\delta_{\dbsmpl}$, $\epsilon_{\fismpl}$, $\delta_{\fismpl}$, $\rho$, $\fismpl$, $\dbsmpl$).
     \STATE // \emph{Phase 2: partitioning.}
     \STATE $\proc_1$ does: $\fipartidxset_i\leftarrow\emptyset, 1\leq i\leq \procnum$.
     \STATE $\proc_1$ calls \textsc{Phase-2-FI-Partitioning}($\fismpl$, $\dbsmpl$, $\alpha$, $\prefixset$, sets $\fipartidxset_i$).
     \STATE $\proc_1$ broadcasts $\prefixset$ and the sets $\fipartidxset_i$ to each other processor.
  \FORPARALLEL{all processors $\proc_i$}
     \STATE // \emph{Phase 3: data distribution.}
     \STATE \textsc{Phase-3-DB-Partition-Exchange}($\procnum$, $\{\isetU_k|u=(\isetU_k, \prefixext_k, s)\in \prefixset\}$, $\fipartidxset_i$, $\dbpart'_i$).
     \STATE // \emph{Phase 4: execution of arbitrary sequential algorithm for computation of FIs.}
     \STATE $\prefixset'\leftarrow \{(\isetU_k,\prefixext_k) | u=(\isetU_k, \prefixext_k, s)\in \prefixset\}$.
     \STATE \textsc{Phase-4-Compute-FI}($\prefixset'$, $\fipartidxset_i$, $\dbpart_i$, $\dbpart'_i$, $\rminsupp \cdot \sum_i|\dbpart_i|$, $\allfi_i$).
  \ENDFOR
\end{algorithmic}
\end{method}
}

\newpage

\subsection{The \scparfimipar{} method}
The \scparfimipar{} method is based on the coverage algorithm and computes
the MFIs in parallel. It samples $\fiapprox$ non-uniformly, see
Method~\ref{method:parallel-fimi-par}.

\bigskip
\vbox{
\begin{method}[H]
\caption{The \textsc{\parfimipar} method}\label{method:parallel-fimi-par}
\vbox{\textsc{\parfimipar}(\vtop{\noindent\inparam Double $\rminsupp$, 
  \par\noindent\inparam Double $\epsilon_{\dbsmpl}$,
  \par\noindent\inparam Double  $\delta_{\dbsmpl}$,
  \par\noindent\inparam Double $\epsilon_{\fismpl}$,
  \par\noindent\inparam Double $\delta_{\fismpl}$,
  \par\noindent\inparam Double $\rho$,
  \par\noindent\inparam Double $\alpha$,
  \par\noindent\outparam Sets $\allfi_i$)}}
\begin{algorithmic}[1]
  \FORPARALLEL{all processors $\proc_i$}
     \STATE // \emph{Phase 1: sampling.}
     \STATE Read $\dbpart_i$.
     \STATE \textsc{Phase-1-Coverage-Sampling-Parallel}($\dbpart_i$, $\rminsupp$, $\epsilon_{\dbsmpl}$,
            $\delta_{\dbsmpl}$, $\epsilon_{\fismpl}$, $\delta_{\fismpl}$, $\fismpl$, $\rho$, $\dbsmpl$).
     \ENDFOR
     \STATE // \emph{Phase 2: partitioning.}
     \STATE $\proc_1$ does: $\fipartidxset_i\leftarrow\emptyset, 1\leq i\leq \procnum$.
     \STATE $\proc_1$ calls \textsc{Phase-2-FI-Partitioning}($\fismpl$, $\dbsmpl$, $\alpha$, $\prefixset$, sets $\fipartidxset_i$).
     \STATE $\proc_1$ broadcasts $\prefixset$ and the sets $\fipartidxset_i$ to each other processor.
     \FORPARALLEL{all $\proc_i$}
     \STATE // \emph{Phase 3: data distribution.}
     \STATE \textsc{Phase-3-DB-Partition-Exchange}($\procnum$, $\{\isetU_k | u=(\isetU_k, \prefixext_k, s)\in \prefixset\}$, $\fipartidxset_i$, $\dbpart'_i$).
     \STATE // \emph{Phase 4: execution of arbitrary sequential algorithm for computation of FIs.}
     \STATE $\prefixset'\leftarrow \left\{(\isetU_k,\prefixext_k) | u=(\isetU_k, \prefixext_k, s)\in \prefixset\right\}$, in parallel.
     \STATE \textsc{Phase-4-Compute-FI}($\prefixset'$, $\fipartidxset_i$, $\dbpart_i$, $\dbpart'_i$, $\rminsupp  \cdot \sum_i|\dbpart_i|$, $\allfi_i$).
  \ENDFOR
\end{algorithmic}
\end{method}}

\newpage

\subsection{The \scparfimireserv{} method}
This method samples $\fiapprox$ using the \emph{reservoir sampling}.
The reservoir sampling samples $\fiapprox$ uniformly. To make the
reservoir sampling algorithm faster, the reservoir sampling is
executed in parallel.  The \textsc{Parallel-FIMI-Reservoir} method
follows:

\bigskip
\vbox{
\begin{method}[H]
\caption{The \textsc{Parallel-FIMI-Reservoir} method}
\label{method:parallel-fimi-reserv}
\vbox{\textsc{Parallel-FIMI-Reservoir}(\vtop{\noindent\inparam Double $\rminsupp$, 
  \par\noindent\inparam Double $\epsilon_{\dbsmpl}$,
  \par\noindent\inparam Double $\delta_{\dbsmpl}$,
  \par\noindent\inparam Double $\epsilon_{\fismpl}$,
  \par\noindent\inparam Double $\delta_{\fismpl}$,
  \par\noindent\inparam Double $\rho$,
  \par\noindent\inparam Double $\alpha$,
  \par\noindent\outparam Sets $\allfi_i$)}}
\begin{algorithmic}[1]
  \FORPARALLEL{all processors $\proc_i$}
  \STATE // \emph{Phase 1: sampling.}
  \STATE Read $\dbpart_i$.
  \STATE \textsc{Phase-1-Reservoir-Sampling}($\dbpart_i$, $\rminsupp$, $\epsilon_{\dbsmpl}$,
         $\delta_{\dbsmpl}$, $\epsilon_{\fismpl}$, $\delta_{\fismpl}$, $\rho$, $\fismpl$, $\dbsmpl$)
  \ENDFOR
  \STATE // \emph{Phase 2: partitioning.}
  \STATE $\proc_1$ does: $\fipartidxset_i\leftarrow\emptyset, 1\leq i\leq \procnum$.
  \STATE $\proc_1$ calls \textsc{Phase-2-FI-Partitioning}($\fismpl$, $\dbsmpl$, $\alpha$, $\prefixset$, sets $\fipartidxset_i$).
  \STATE $\proc_1$ broadcasts $\prefixset$ and the sets $\fipartidxset_i$ to each other processor.
  \FORPARALLEL{all processors $\proc_i$}
  \STATE // \emph{Phase 3: data distribution.}
  \STATE \textsc{Phase-3-DB-Partition-Exchange}($\procnum$, $\{\isetU_k | u=(\isetU_k, \prefixext_k, s)\in \prefixset\}$, $\fipartidxset_i$, $\dbpart'_i$).
  \STATE // \emph{Phase 4: execution of arbitrary sequential algorithm for computation of FIs.}
  \STATE $\prefixset'\leftarrow \{(\isetU_k,\prefixext_k) | u=(\isetU_k, \prefixext_k, s)\in \prefixset\}$.
  \STATE \textsc{Phase-4-Compute-FI}($\prefixset'$, $\fipartidxset_i$, $\dbpart_i$, $\dbpart'_i$, $\rminsupp  \cdot \sum_i |\dbpart_i|$, $\allfi_i$).
  \ENDFOR
\end{algorithmic}
\end{method}
}

\clearpage
\newpage

\chapter{Execution of the Eclat algorithm in Phase 4}\label{chap:exec-eclat}

In Chapter \ref{chap:experimental-evaluation}, we are evaluating the
method with the sequential Eclat algorithm used in Phase 4. Therefore,
in this Chapter, we show how to \emph{efficiently execute the Eclat
  algorithm} on a processor $\proc_q$, $1\leq q\leq \procnum$ in
Phase~4, so it efficiently process the assigned PBECs. The Eclat
algorithm is a DFS algorithm that works with the tidlists, see
Definition~\ref{def:tidlist} and
Section~\ref{sec:lattice-in-algorithms} for details.  We omit the
details of the Eclat algorithm, they are described in
Section~\ref{sec:eclat-algorithm}. To execute the Eclat algorithm for
one assigned PBEC, we have to prepare the tidlists for every assigned
prefix and its extensions, by simulating one branch of the Eclat
algorithm.

We denote the set of indexes of PBECs assigned to processor $\proc_q$
by $\fipartidxset_q$.  The task is to efficiently prepare the tidlists
used by the Eclat algorithm in order to enumerate FIs $\isetW\in
\bigcup_{i \in \fipartidxset_q} [\isetU_i | \prefixext_i] \cap \allfi$
(and its supports). We denote the database partition that was received
by processor $\proc_q$ in Phase 3 by $\dbpart'_q$, i.e., database
$\dbpart'_q$ contains all the necessary information needed to compute
the support of itemsets $\isetW\in \bigcup_{i \in \fipartidxset_q}
[\isetU_i|\prefixext_i]$.

In order to explain the execution of the Eclat algorithm in Phase~4,
we need to define the lexicographical order:

\begin{definition}[lexicographical order of two itemsets]
Let $\isetU=(\bitem_{u_1},\ldots, \bitem_{u_{|\isetU|}}),
\isetW=(\bitem_{w_1},\ldots, \bitem_{w_{|\isetW|}})$ be two itemsets. We say
that $\isetU<\isetW$ ($\isetU$ is lexicographically smaller then
$\isetW$) if and only if: 

\begin{enumerate}
\item $\bitem_{u_i} = \bitem_{w_i}$ for each $1 \leq i < k$ and
  $\bitem_{u_k} < \bitem_{w_k}$ for some $k \leq
  \min(|\isetU|,|\isetW|)$.
\item $|\isetU| < |\isetW|$ and $\bitem_{u_i} = \bitem_{w_i}$ for all
  $1 \leq i \leq |\isetU|$.
\end{enumerate}
\end{definition}

At the start of Phase 4, processor $\proc_q$, $1\leq q\leq \procnum$,
creates tidlists $\tidlist (\bitem_i, \dbpart'_q), \bitem_i \in
\baseset$, i.e., $\proc_q$ creates tidlists for each item $\bitem_i$
in its partition of the database $\dbpart'_q$.  


Processor $\proc_q$ has been assigned a set of prefixes. Let one PBEC,
assigned to $\proc_q$, be $[\isetU_i | \prefixext_i]$. The Eclat
algorithm uses the tidlists for computation of supports. To prepare
the execution of the Eclat algorithm for processing of $[\isetU_i |
  \prefixext_i]$, we have to compute the tidlists of each itemset
$\isetU_i\cup\{\bitem\}, \bitem\in\prefixext_i$, i.e.,
$\tidlist(\isetU_i\cup\{\bitem\}, \dbpart'_q)$. Each processor
$\proc_q$ has been assigned with a set of such prefixes. We denote the
prefix an itemset $X$ of size $k < |X|$ by $X^{k-1}$. To make the
preparation of $\tidlist(\isetU_i \cup \{ \bitem \}, \dbpart'_q)$
efficient, we sort the PBECs $[\isetU_i | \prefixext_i],
i\in\fipartidxset_q$ lexicographically in ascending order by
$\isetU_i$ and prepare the tidlist of each prefix of $\isetU_i$ of
length $\leq|\isetU_i|$, i.e., $\isetU^j_i, j\leq|\isetU_i|$. We reuse
the tidlists for preparation of processing of subsequent PBECs
$[\isetU_k | \prefixext_k], k>i$. The sorting of prefixes allows
reuse of the already prepared tidlists. This is preformed using an
array $C$ that serves as a cache of tidlists. The array $C$ contains
at position $j$ a pair $C[j] = (\bitem, T)$, where $\bitem\in\baseset$
is the $j$-th item in the prefix $\isetU_k$, and $T = \{(\bitem',
\tidlist(\{\bitem'\} \cup \isetU^j_k))\}, \bitem'\in\prefixext_k,$ is
the set of pairs of the extensions of the prefix $\prefixext_k$ and
its tidlist in the database $\dbpart'_q$.  We omit the details of the
preparation of the tidlists, the details can be found in
Section~\ref{sec:lattice-in-algorithms} and in
Appendix~\ref{appendix:seq-alg}.

The PBECs are then processed sequentially one by one: when preparing
the tidlists for the next prefix, $\isetU_{i+1}$, we reuse the
elements in the cache $C$ that represents the longest common prefix of
$\isetU_{i}$ and $\isetU_{i+1}$.








Further, we denote the $i$th item of an itemset
$\isetU=(\bitem_{u_1},\ldots,\bitem_{u_{|\isetU|}})$ by $\isetU[i] =
\bitem_{u_i}$.  The algorithm \textsc{Prepare-Tidlists}, see Algorithm
\ref{alg:prepare-tidlists}, summarizes the preparation of the tidlists
for the sequential run of the Eclat algorithm.  The algorithm
\textsc{Exec-Eclat}, see Algorithm \ref{alg:exec-eclat}, summarizes
the execution of the Eclat algorithm needed to processes the assigned
PBECs. The \textsc{Exec-Eclat} is executed in parallel on each
processor $\proc_q$. The Phase~4 parametrized with the Eclat algorithm
is summarized in the \textsc{Phase-4-Eclat} algorithm.

\begin{algorithm}[H]
\caption{The \textsc{Prepare-Tidlists} algorithm}
\label{alg:prepare-tidlists}
\vbox{\textsc{Prepare-Tidlists}(\inoutparam Array $C$ of size $|\baseset|$, \inparam Pair $(\isetU,\prefixext)$)}
\begin{algorithmic}[1]
  \item[\textbf{Notation:}] $\isetU=(\bitem_{u_1},\ldots,\bitem_{u_{|\isetU|}})$, $\isetU[i]=\bitem_{u_i}$. \par\noindent
    $C[i]=(\bitem_i, T_i)$, $C[i].\texttt{item}=\bitem_i$, $C[i].\texttt{tidlists}=T_i$.
    
  \STATE $n\leftarrow -1$
  \FOR{$i\leftarrow 1,\ldots,|\isetU|$}
    \IF{$C[i].$\texttt{item}$ \not= \isetU[i]$}
      \STATE $n\leftarrow i$
      \STATE \textbf{break}
    \ENDIF
    \STATE $C[i].\texttt{tidlist}\leftarrow$ prepare tidlists using $\prefixext$ and $C[i-1].\texttt{tidlist}$
  \ENDFOR
  \FOR{$i\leftarrow n,\ldots,|\isetU|-1$}
      \STATE $C[i]\leftarrow$ create new array element from $C[i-1]$  using $\prefixext$ and $C[i-1].\texttt{tidlists}$
  \ENDFOR
  \FOR{$i\leftarrow |C|,\ldots,|\baseset|-1$}
    \STATE $C[i]\leftarrow$ \emph{null}
  \ENDFOR
\end{algorithmic}
\end{algorithm}

\vbox{
\begin{algorithm}[H]
\caption{the \textsc{Exec-Eclat} algorithm}
\label{alg:exec-eclat}
\vbox{\textsc{Exec-Eclat}(\vtop{\inparam Prefixes and extensions $Q=\{(\isetU_k, \prefixext_k)\}$, 
    \par\noindent\inparam Integer $\minsupp$,
    \par\noindent\inparam Database $\db$,
    \par\noindent\outparam Set $F$)}}
\begin{algorithmic}[1]
  \STATE sort $Q$ lexicographically by $\isetU_k$,
  i.e., $(\isetU_i,\prefixext_i), (\isetU_j,\prefixext_j)\in Q$ and
  $\isetU_i < \isetU_j, i < j$
  \STATE $Q_{tidlists}\leftarrow$ array of size $|\baseset|$ with $Q_{tidlists}[i]\leftarrow$\emph{null}
  \STATE $Q_{tidlists}[0]\leftarrow(\emptyset,\{(\bitem_i,\tidlist(\bitem_i, \db))|\bitem_i\in\baseset\})$
  \FORALL{$q=(\isetU_i,\prefixext_i)\in Q$ such that $i=1,\ldots,|Q|$}
    \STATE\CALL{Prepare-Tidlists}$(Q_{tidlists}, q)$
    \STATE run the Eclat algorithm with prepared tidlists and
    extensions that are stored in $Q_{tidlists}[|\isetU_i|]$ with
    support value $\minsupp$. Output FIs into $F$.
  \ENDFOR
\end{algorithmic}
\end{algorithm}
}

\begin{algorithm}[H]
\caption{The \textsc{Phase-4-Eclat} algorithm}
\label{alg:phase-4-eclat}
\vbox{\textsc{Phase-4-Eclat}(\vtop{\inparam Set of prefixes $S=\{(\isetU_k, \prefixext_k)\}$, 
  \par\noindent\inparam Indexsets $\fipartidxset_q$,
  \par\noindent\inparam Database $\dbpart_q$,
  \par\noindent\inparam Database $\dbpart'_q$,
  \par\noindent\inparam Integer $\minsupp$,
  \par\noindent\outparam Set $F$)}}
\begin{algorithmic}[1]
  \FORPARALLEL{all $\proc_i$}
  \STATE computes support of itemsets $\isetW\subset\isetU_k$ in $\dbpart_q$, i.e., $\supp(\isetW, \dbpart_q)$
  \STATE send $\supp(\isetW, \dbpart_q)$ to $\proc_1$
  \ENDFOR
  \STATE $\proc_1$ puts all $\isetW$ into $F'$ such that $\sum_i\supp(\isetW,\dbpart_i) > \minsupp$
  \STATE each $\proc_q$ executes \textsc{Exec-Eclat}($\{u=(\isetU_k,
  \fipartidxset_k)|u\in S\text{ and } k\in\fipartidxset_q\}$, $\minsupp$, $\dbpart'_q$, $F_q$) in parallel.
  \STATE $F\leftarrow(\cup_{1\leq i\leq\procnum} F_q)\cup F'$
\end{algorithmic}
\end{algorithm}

\newpage

\clearpage
\newpage

\chapter{The database replication factor}\label{chap:db-minimalization}


In Phase 3 of the \textsc{Parallel-FIMI} methods, the processors must
exchange database partitions in order to start Phase 4, i.e., Phase 3
re-distributes the database in such a way that the sequential
algorithm for mining of FIs used in Phase 4 can compute the FIs.
After Phase 3, the database must not be distributed evenly among the
processors. To measure how the database is distributed among the
processors, we use the \emph{database replication factor}.

We define the database replication factor as a real number that
determines the number of copies of a database that is spread among the
processors.  Let $\dbpart'_i$ be the database partition received by
$\proc_i$ in Phase~3. The database replication factor is defined as:

$$\frac{\sum_{i=1}^{P}|\dbpart'_i|}{|\db|}$$

The database replication factor measures memory efficiency of our
method by measuring the number of replications of $\db$ among
processors.  We can handle the database replication factor in two
ways:

\begin{enumerate}
\item hope that the database replication factor will be small;
\item reduce the database replication factor.
\end{enumerate}

In this section, we will describe how to reduce the database replication
factor. The measurements of the database replication factor of our method is
given in Chapter \ref{chap:experimental-evaluation}.


\section{Reduction of the replication factor}


The \textsc{LPT-Schedule} algorithm assigns the prefix-based
equivalence classes to the processors based solely on their sizes. If we
want to reduce the database replication, we have to consider the
mutual sharing of the database partitions among the prefix-based
classes. In this section, we will show how the problem of scheduling
of the prefix-based classes with respect to the mutual share of
transactions is related to the Quadratic Knapsack Problem (QKP in
short). For a good source of information on knapsack problems,
see~\cite{qkp-book}. 

The QKP can be defined as follows: let have $n$ items and the $j$-th item having
a positive integer weight $w_j$, and a limit on the total weight of the chosen
items is given by a positive integer knapsack capacity $c$. In addition, we have
a $n\times n$ \emph{profit matrix} $\sharem=(\sharem_{ij})$, where
$\sharem_{ij}$ is the profit of having item $i$ together with item $j$ in the
knapsack. Additionally, we have indicator variables $x_i \in \{0, 1\}$ where
$x_i=1$ if the item $i$ was selected to the knapsack and $0$ otherwise.  The QKP
selects subset of items that fit in the knapsack and have maximal profit. The
problem can be stated in the following way:

\begin{center}
\begin{tabular}{rl}
maximize    & $\displaystyle\sum_{i}\sum_{j}\sharem_{ij} x_i x_j$ \\
subject to  & $\sum_jw_jx_j\leq c$ \\
\end{tabular}
\end{center}

We can reformulate the QKP in the terms of our problem: let have a
list of prefixes $P=\{\isetU_i | \isetU_i\subseteq\baseset\}$. The
\emph{profit matrix} $S$, contains the number of shared transactions
for every two PBECs, i.e., $\sharem_{ij} = |\tidlist(\isetU_i \cup
\isetU_j)|, i\neq j$ and $\sharem_{ii}=0$. The weight $w_i$ is
defined as the size of the prefix-based class
$[\isetU_i]\cap\allfi$. The size $|[\isetU_i]\cap\allfi|$ is
determined by the relative number of samples $\fismpl$ belonging to
$[\isetU_i]$, i.e., $|[\isetU_i] \cap \fismpl| / |\fismpl|$. The task
is to put prefix-based equivalence classes into the knapsack, such
that the size of the knapsack $c=\sum_i s_i/\procnum$ while maximizing
the share of transactions. This task is the same as solving the
QKP. When we have a set of prefixes, we assign them to a processor,
remove them from the set $\prefixset$, update the matrix and the
weight vector, and repeat the process until we assign all the
prefix-based classes.

For the purpose of the database replication reduction algorithm,
we denote the prefix-based equivalence class by a tuple $(\isetU_i,
\proc^S_{j})$, where $\isetU_i$ is a prefix and $\proc^S_j$ is the
scheduled processor. The algorithm is summarized in Algorithm
\ref{alg:db-repl-min}.


\begin{algorithm}[H]
\caption{The \textsc{DB-Repl-Min} algorithm}\label{alg:db-repl-min}
\vbox{\textsc{DB-Repl-Min}\vbox{(\inoutparam Prefixes $\prefixset = \{(\isetU_i, \proc^S_i)\}$, \inparam Profit matrix $S$)}}
\begin{algorithmic}[1]
\STATE $p\leftarrow 1$
\FORALL{$i$}
\STATE $\proc^S_i\leftarrow 0$
\ENDFOR
\FOR{$p\leftarrow 1,\ldots,\procnum$}
\STATE $\sharem'\leftarrow$ a submatrix of $\sharem$ such that for all columns $j$
       and rows $i$ $\proc^S_i=0$ and $\proc^S_j=0$
\STATE Using a QKP: find a subset of prefix-based classes, $x_i=1$, with: $\sum
       x_i\cdot w_i\leq c=\sum_i s_i/\procnum$.
\FORALL{$i$, $x_i=1$}
\STATE $\proc^S_i=p$
\ENDFOR
\ENDFOR
\end{algorithmic}
\end{algorithm}

This algorithm will not give the optimal solution, however, it should
have better results, from the replication point of view, than the
\textsc{LPT-Schedule} algorithm because \textsc{LPT-Schedule} does not
consider the sharing of transactions.

\clearpage
\newpage



\chapter{Experimental evaluation}\label{chap:experimental-evaluation}

Our proposed method is a two step sampling process: 1) sampling of the database,
creation of $\dbsmpl$; 2) creation of a sample of FIs, $\fismpl$, from the sampled
database. Since the whole process is quite complicated and, as shown in the
previous sections, theoretically we can make big error of the estimate of the
size of a PBEC, we must experimentally show the performance of our method.

This chapter is organized as follows: in
Section~\ref{sec:impl-exp-setup} we describe our implementation and
the experimental setup, in Section~\ref{sec:datasets} we describe the
databases, in Section~\ref{sec:exp-eval-pbec}, we experimentally show
the error of the estimate of the size of a set of PBECs, in
Section~\ref{sec:exp-eval-speedup} we experimentally evaluate the
speedup of the proposed method and in
Section~\ref{sec:exp-eval-db-repl} we evaluate the database
replication factor.

\section{Implementation and experimental setup}\label{sec:impl-exp-setup}

We have implemented our methods using the \texttt{C++} language and
the \texttt{g++} compiler version 4.4.3 with the \texttt{-O4} option
(highest optimizations on speed of the resulting code). As the
sequential algorithm, we have used the Eclat algorithm
\cite{zaki00scalable}. As the algorithm for mining of MFIs, we have
chosen the \emph{fpmax*} \cite{grahne03fpmax} algorithm. The choice of
the two algorithms is not accidental: we choose very fast
algorithms. This makes our result more valuable because it is harder
to achieve good speedup results: a very fast algorithm for mining of
FIs and MFIs forces us to make the process of statical load-balancing
more efficient. If we have used the Apriori algorithm for computation
of FIs, we could have better speedup.

We have to modify the Eclat algorithm so it can be executed in
parallel and the output could be read by the reservoir algorithm. We
had also modified the \emph{fpmax*} algorithm so it runs in
parallel. Both algorithms utilize the dynamic load-balancing with the
Dijkstra's token termination algorithm. The dynamic load-balancing is
limited to PBECs with prefix of size $1$, see
Section~\ref{chap:parallel-mfi}. As the implementation of the
\emph{fpmax*} algorithm, we have used the implementation from the FIMI
workshop \cite{fimi03}. The implementation of the Eclat algorithm was
downloaded from \cite{goethalssoft}.

We have preformed all the experiments with our methods on a cluster of
workstations interconnected with the Infiniband network. Every node in
the cluster has two dual-core 2.6GHz AMD Opteron processors and 8GB of
main memory.

\section{Databases}\label{sec:datasets}

The experiments were performed on databases generated using the IBM database
generator -- which is a standard way for assessing the algorithms for mining of
all FIs. We would like to use real datasets, however the standart datasets used
as benchmarks are too small. We have used databases with 500k transactions and
supports for each database such that the sequential run of the Eclat algorithm
is between $100$ and $12000$ seconds ($\approx 3.3$ hours) and two cases with
running time $33764$ seconds ($9.37$ hours) and $132186$ seconds ($36.71$
hours). The IBM generator is parametrized by the average transaction length
\texttt{TL} (in thousands), the number of items \texttt{I} (in thousands), by
the number of patterns \texttt{P} used for creation of the parameters, and by
the average length of the patterns \texttt{PL}. To clearly differentiate the
parameters of a database we are using the string {\tt T[number in
    thousands]I[items count in 1000]P[number]PL[number]TL[number]}, e.g. the
string \texttt{T500I0.4P150PL40TL80} labels a database with $500$K transactions
$400$ items, $150$ patterns of average length $40$ and with average transaction
length $80$. All speedup experiments were performed with various values of the
support parameter on $2$, $4$, $6$, $10$, $16$, and $20$ processors. The
databases and supports used for evaluation of our methods is summarized in
Table~\ref{fig:datasets}. \emph{We have chosen the parameters of the IBM
  generator so that the distribution of the lengths of FI, the lengths of
  intersections of MFI, and of length of MFIs are similar to the same
  characteristics of some of the real databases.}  However, mimicking the real
dataset using the IBM generator is a hard task. The database characteristics are
the following:




\begin{enumerate}
\item The distribution of intersections of MFIs: let have a set of MFIs
  $\mfi$. We have measured $|\smfi_i\cap\smfi_j|, \smfi_i,\smfi_j \in \mfi$ for
  particular choice of $\rminsupp$ and compared the histograms of real databases
  and databases generated by the IBM generator.
\item The distribution of FIs of certain length: let have a set of FIs
  $\allfi$. We have measured $|\isetU|, \isetU\in\allfi$. We have measured the
  lengths for various values of $\rminsupp$: we have split the interval $\lcint
  0,1\rcint$ on $n=1000$ values $i\cdot\frac{1}{n}$ for $i=0,\ldots ,n-1$ and
  compared histograms of real datbases and databases generated by the IBM
  generator for each value of $i\cdot\frac{1}{n}$.
\item The distribution of lengths of MFI: let have a set of MFIs $\mfi$. We have
  drawn the histograms of $|\smfi|, \smfi\in\mfi$ for various values of
  $\rminsupp$ and compared the histograms of the databases generated by the IBM
  generator to the histograms of real databases.

\end{enumerate}

We have chosen the datasets so these characteristics are close to the
characteristics of real datasets, e.g., \texttt{connect}, \texttt{pumsb},
etc. The only exception to this choice is the \texttt{T500I1P100PL20TL50}
dataset. We omit details of the measurements because they are out of the scope
of this thesis.

\begin{table}
\centering
\begin{tabular}{|l|l|} \hline
Database               & Supports \\ \hline
T500I0.1P100PL20TL50  & $0.11, 0.12, 0.13, 0.14, 0.15, 0.16, 0.17, 0.18$ \\ \hline
T500I0.1P250PL10TL40  & $0.05, 0.07, 0.09, 0.1$ \\ \hline
T500I0.1P50PL10TL40   & $0.09, 0.1, 0.13, 0.15, 0.18$ \\ \hline
T500I0.1P50PL20TL40   & $0.05, 0.07, 0.09, 0.1$ \\ \hline
T500I0.4P250PL10TL120 & $0.2, 0.25, 0.26, 0.27, 0.3$ \\ \hline
T500I0.4P250PL20TL80  & $0.02, 0.03, 0.05, 0.07, 0.09$ \\ \hline
T500I0.4P50PL10TL40   & $0.02, 0.05, 0.07, 0.09$ \\ \hline
T500I1P100PL20TL50    & $0.02, 0.03, 0.05, 0.07, 0.09$ \\ \hline
\end{tabular}
\caption{Databases used for measuring of the speedup and used supports
  values for each database.}\label{fig:datasets}
\end{table}

\section{Evaluation of the estimate of the size of PBECs}\label{sec:exp-eval-pbec}

In the previous chapters, we have shown that the parallel mining of FIs is a two
stage sampling process. Some of the shown theorems suggest that the results of
the double sampling process can be very bad, e.g.,
Theorem~\ref{theorem:bounds-pbec-size-reservoir},
Corollary~\ref{corollary:bounds-pbec-size-reservoir-probabilistic}, and
Section~\ref{sec:estim-pbec-union}. In this section, we show that the results
are not that pessimistic, as shown in Section~\ref{sec:estim-pbec-union}. The
estimates are always made only using the samples taken by the
\textsc{Vitter-Reservoir-Sampling} algorithm. The reason why we do not consider
the sample taken by the \textsc{Modified-Coverage-Algorithm} algorithm is that
the estimates using the sample are just heuristics and we consider the
\scparfimireserv{} as the \emph{major} result of this thesis.


We use the notation from our previous chapters: by $\isetU_j$, we denote the
prefixes of PBECs, by $\fipartidxset_i$ we denote a set of indexes of prefixes
assigned to processor $\proc_i$. The indexsets $\fipartidxset_i$ are chosen as
described in Section~\ref{sec:phase-2-detailed}, i.e.,
$\frac{|\bigcup_{j\in\fipartidxset_i}[\isetU_j] \cap \fismpl|}{|\fismpl|}
\approx 1/\procnum$.

\begin{figure*}[!p]
\centering
\vbox{\hbox{
\scalebox{0.45}{\includegraphics{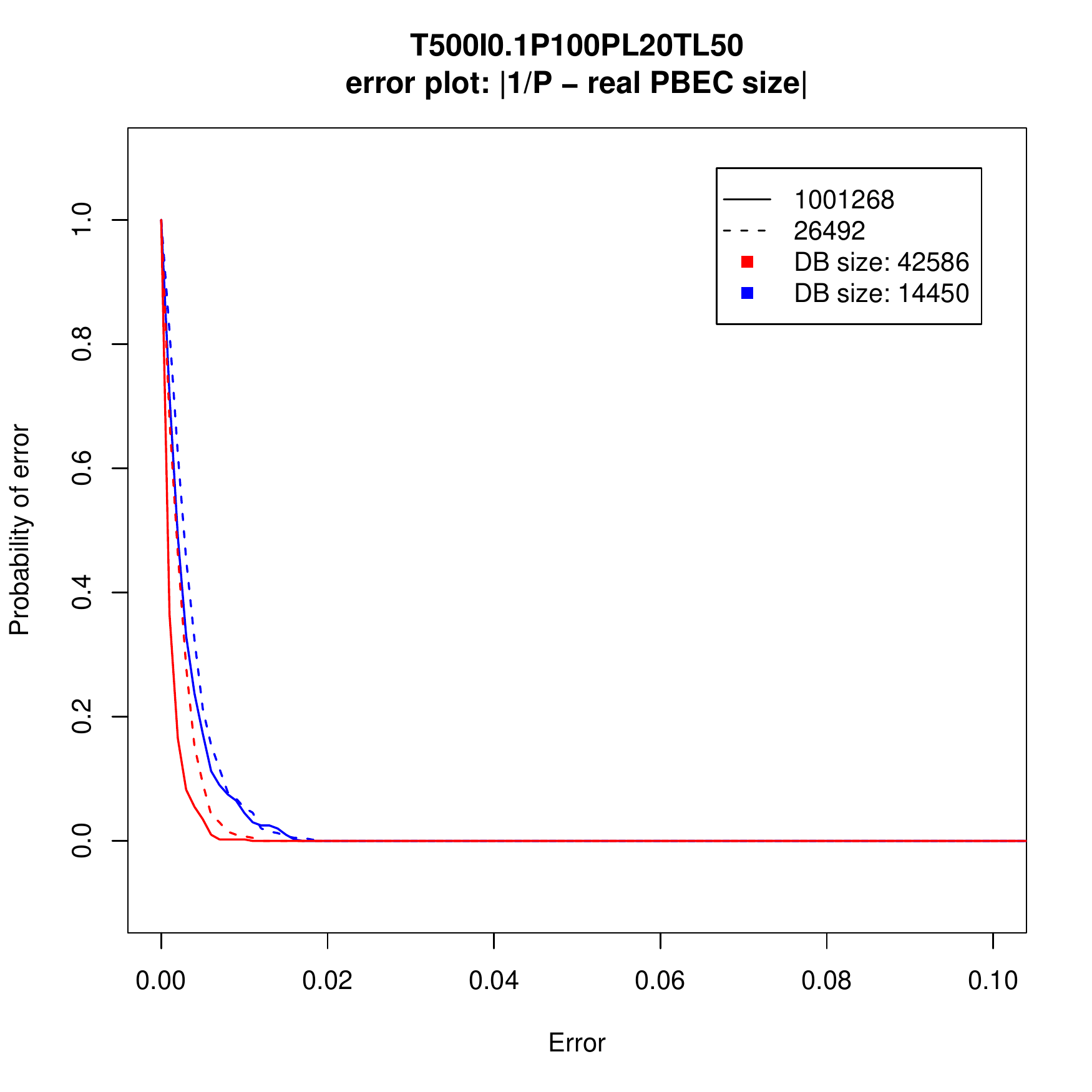}}
\scalebox{0.45}{\includegraphics{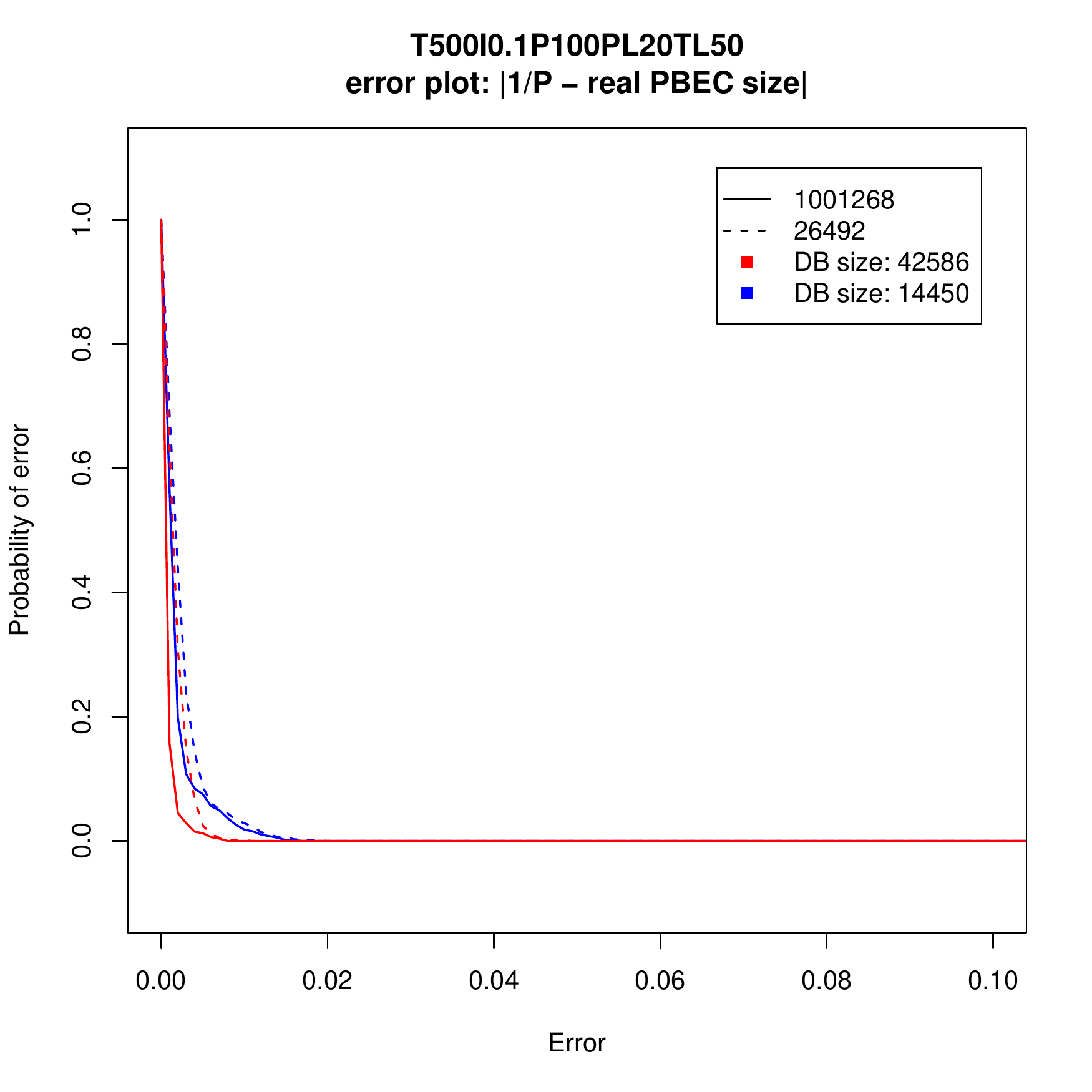}}
}}
\caption[Probability of errors made in Phases 1 and 2]{Probability of
  error of the estimation of the union of PBECs using a database
  sample created in Phase~1 and 2. Experiments made using $\procnum=5$
  processors (left) and $\procnum=10$ processors (right). The
  \texttt{T500I0.1P100PL20TL50} database.}
\label{fig:scheduler-error-1}
\end{figure*}

\begin{figure*}[!p]
\centering
\vbox{\hbox{
\scalebox{0.45}{\includegraphics{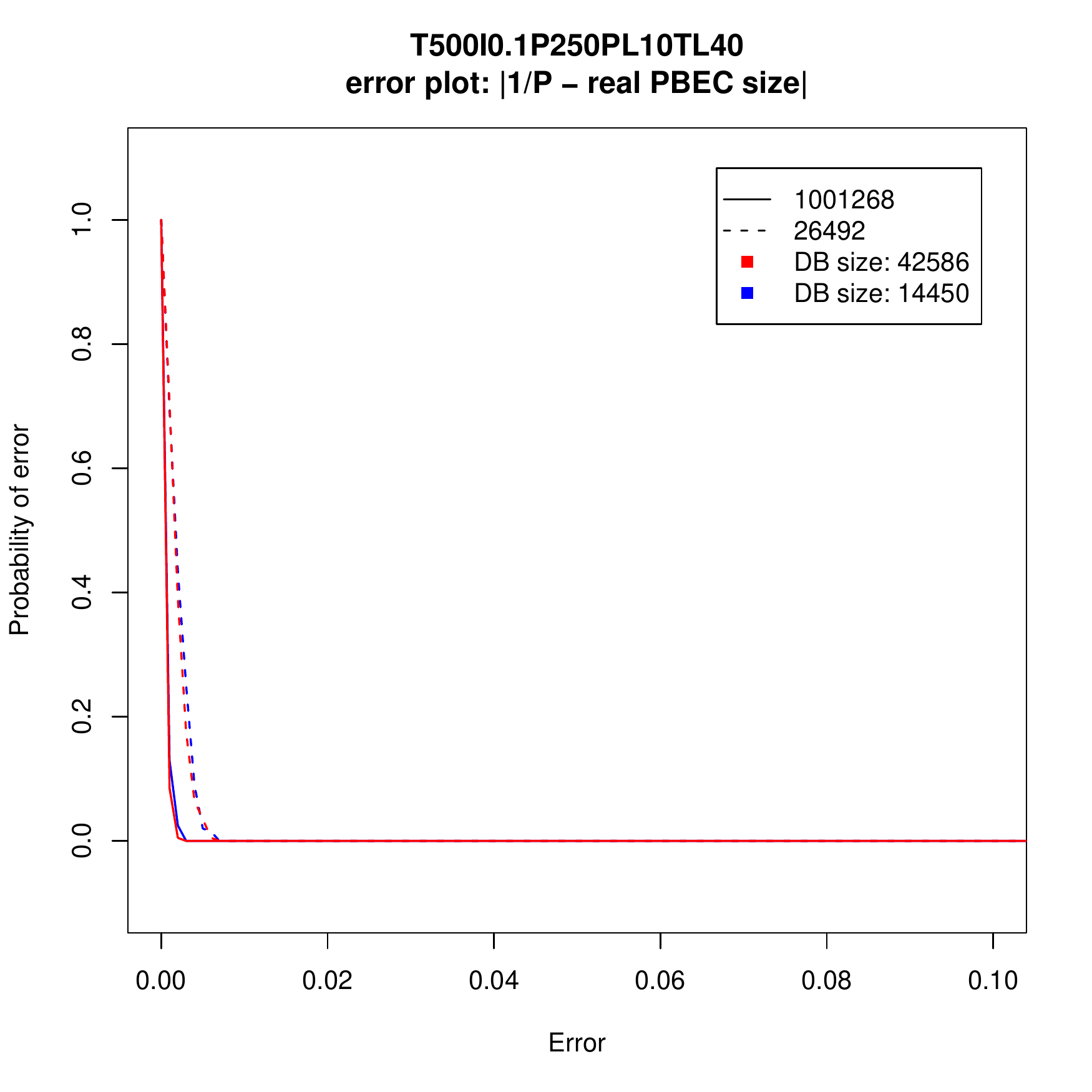}}
\scalebox{0.45}{\includegraphics{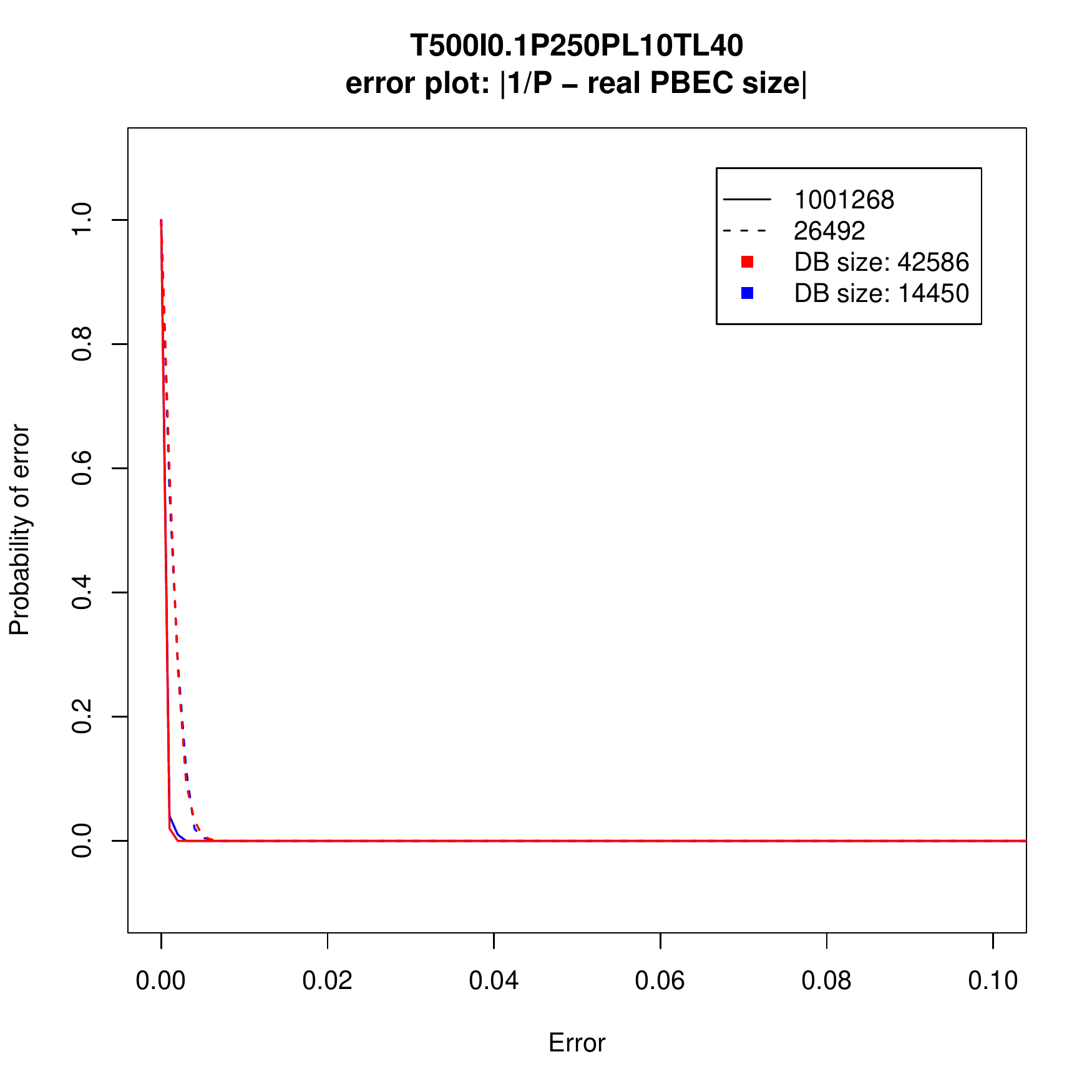}}
}}
\caption[Probability of errors made in Phases 1 and 2]{Probability of
  error of the estimation of the union of PBECs using a database
  sample created in Phase~1 and 2. Experiments made using $\procnum=5$
  processors (left) and $\procnum=10$ processors (right). The
  \texttt{T500I0.1P250PL10TL40} database.}
\label{fig:scheduler-error-2}
\end{figure*}

\begin{figure*}[!p]
\centering
\vbox{\hbox{
\scalebox{0.45}{\includegraphics{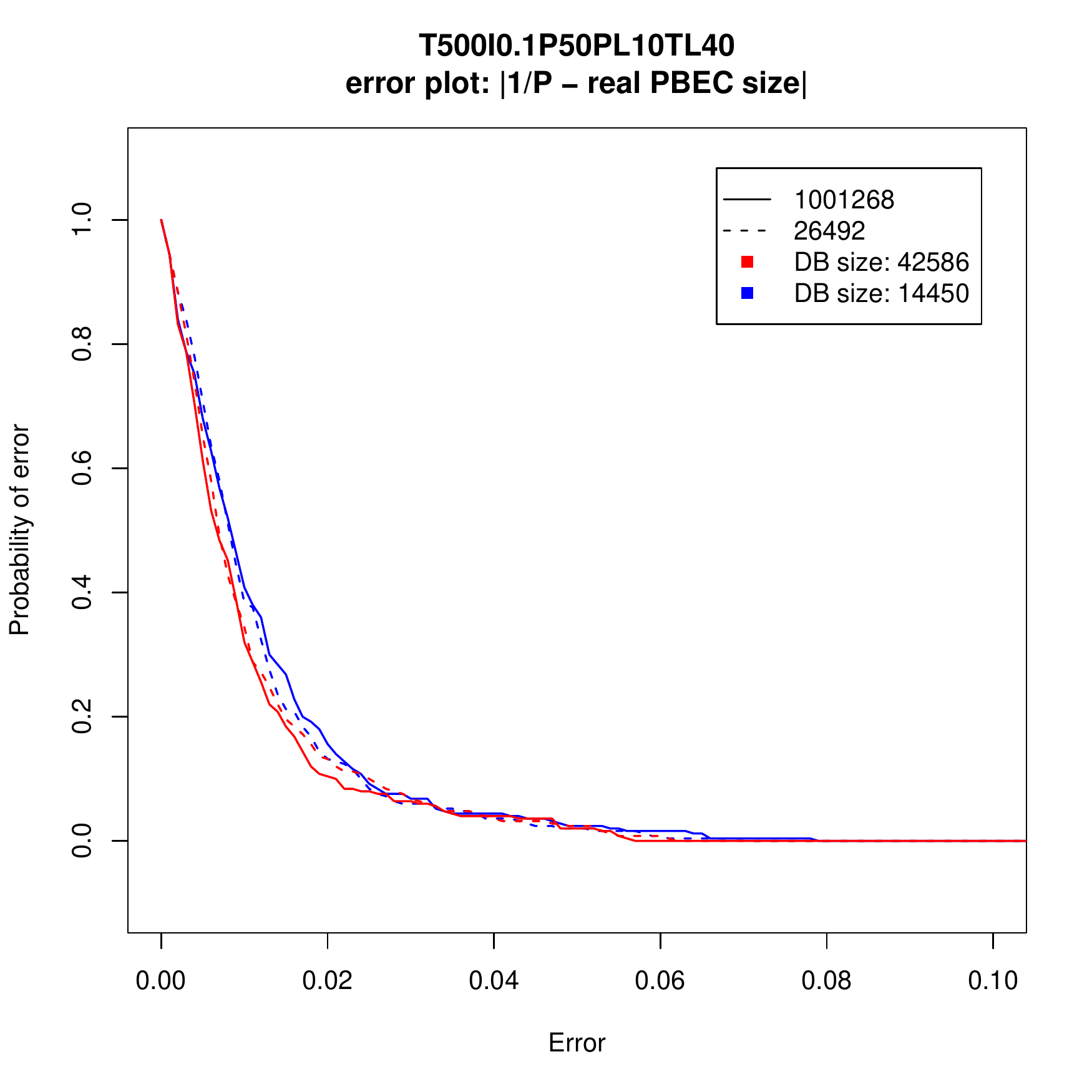}}
\scalebox{0.45}{\includegraphics{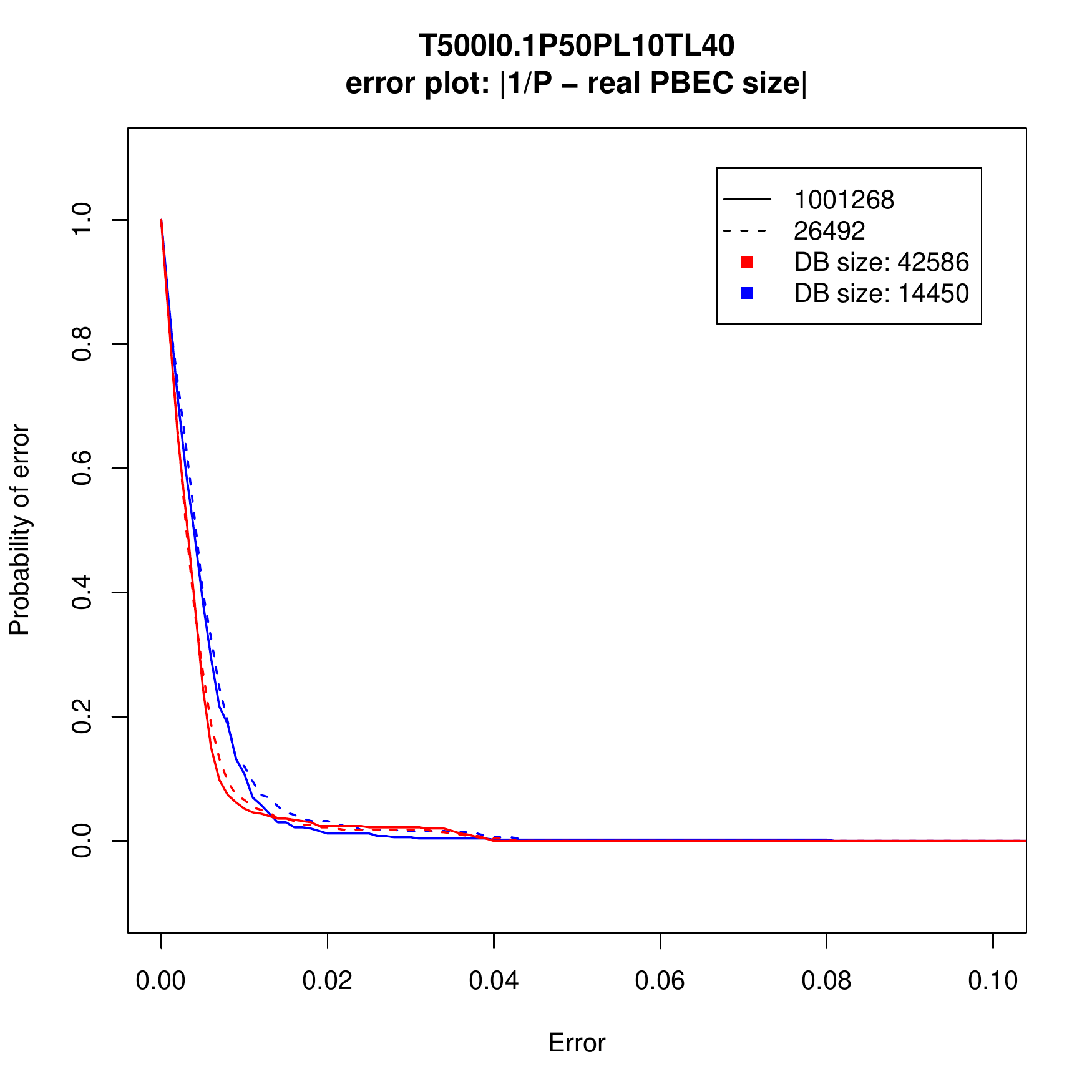}}
}}
\caption[Probability of errors made in Phases 1 and 2]{Probability of
  error of the estimation of the union of PBECs using a database
  sample created in Phase~1 and 2. Experiments made using $5$
  processors (left) and $10$ processors (right). The
  \texttt{T500I0.1P50PL10TL40} database.}
\label{fig:scheduler-error-3}
\end{figure*}

\begin{figure*}[!p]
\centering
\vbox{\hbox{
\scalebox{0.45}{\includegraphics{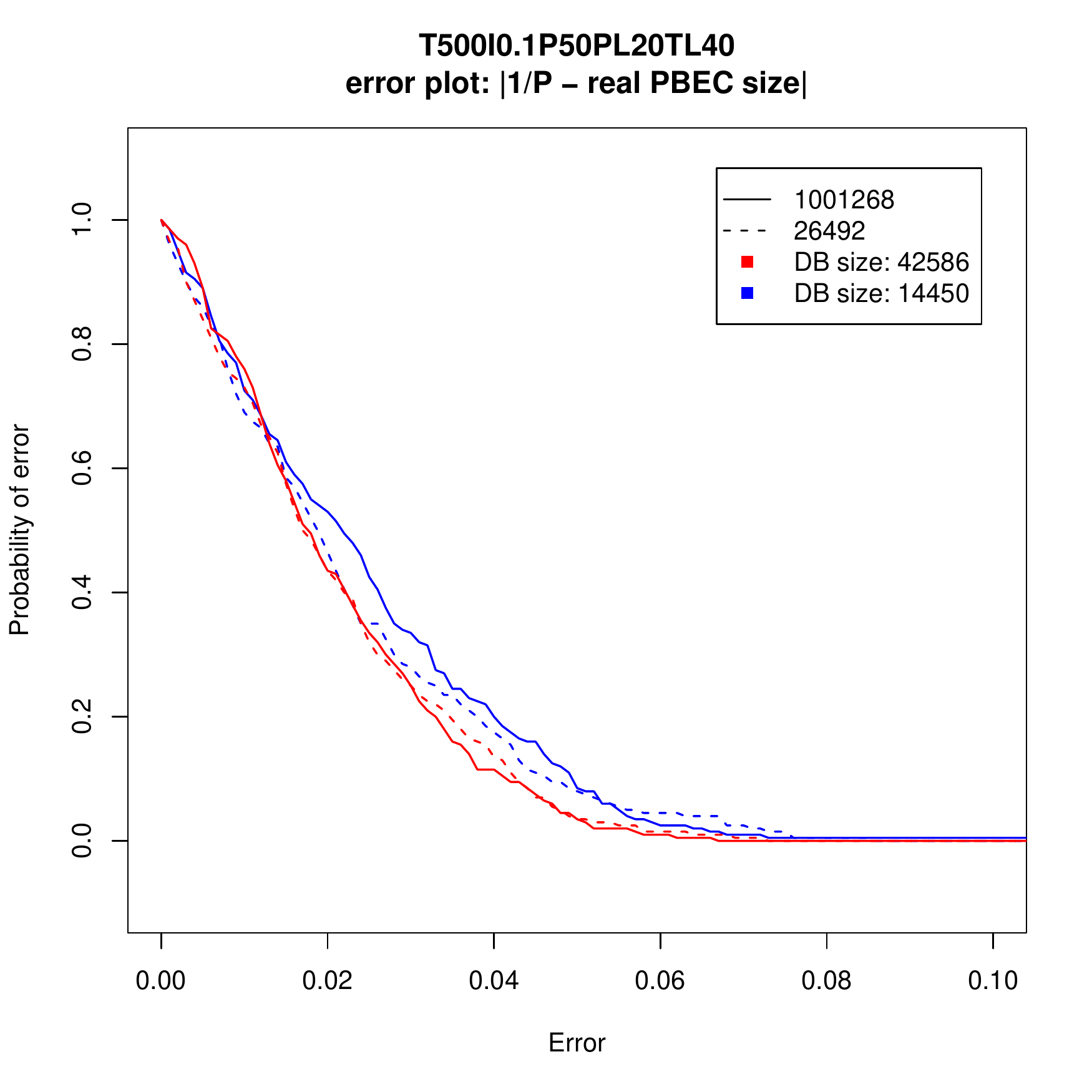}}
\scalebox{0.45}{\includegraphics{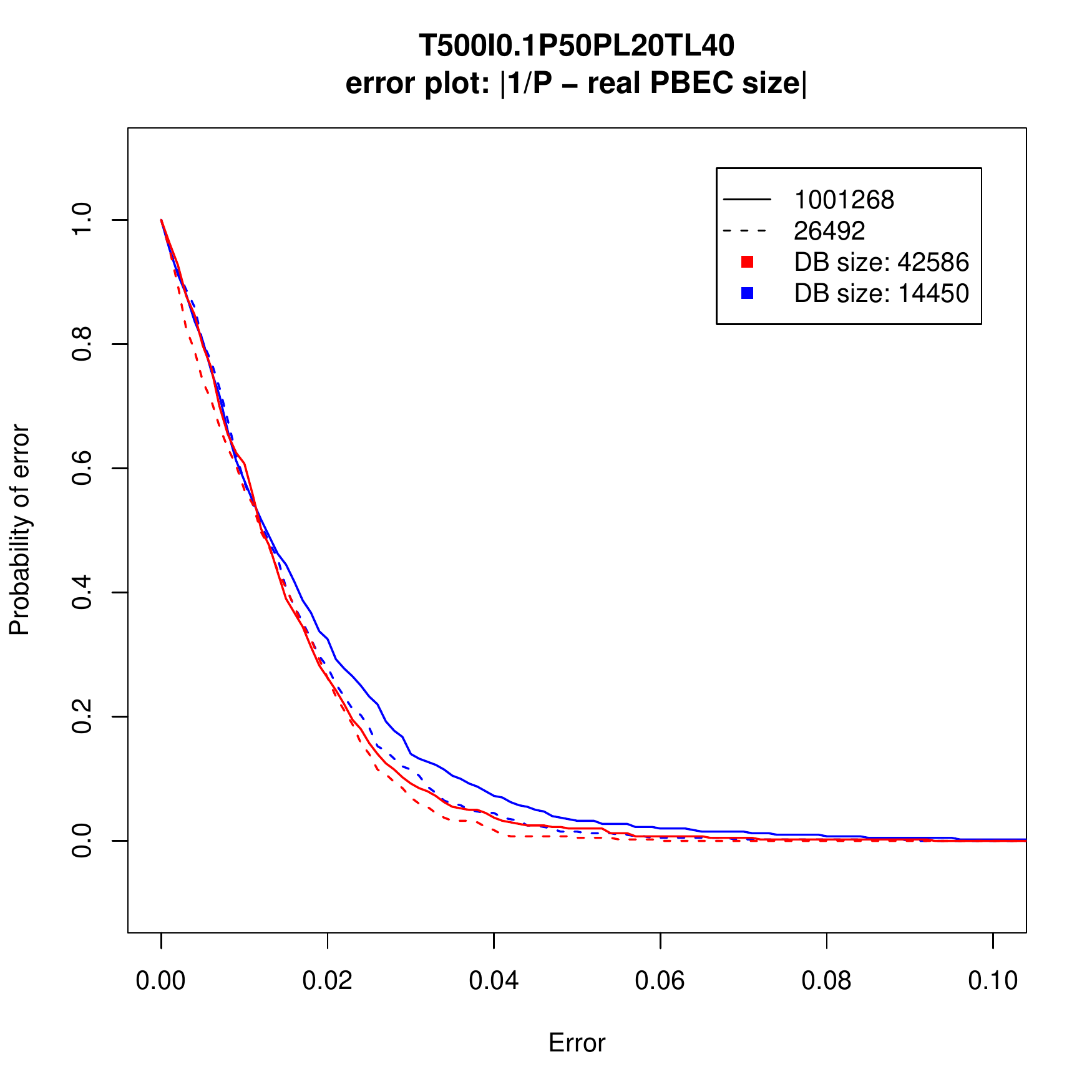}}
}}
\caption[Probability of errors made in Phases 1 and 2]{Probability of
  error of the estimation of the union of PBECs using a database
  sample created in Phase~1 and 2. Experiments made using $\procnum=5$
  processors (left) and $\procnum=10$ processors (right). The
  \texttt{T500I0.1P50PL20TL40} database.}
\label{fig:scheduler-error-4}
\end{figure*}

\begin{figure*}[!p]
\centering
\vbox{\hbox{
\scalebox{0.45}{\includegraphics{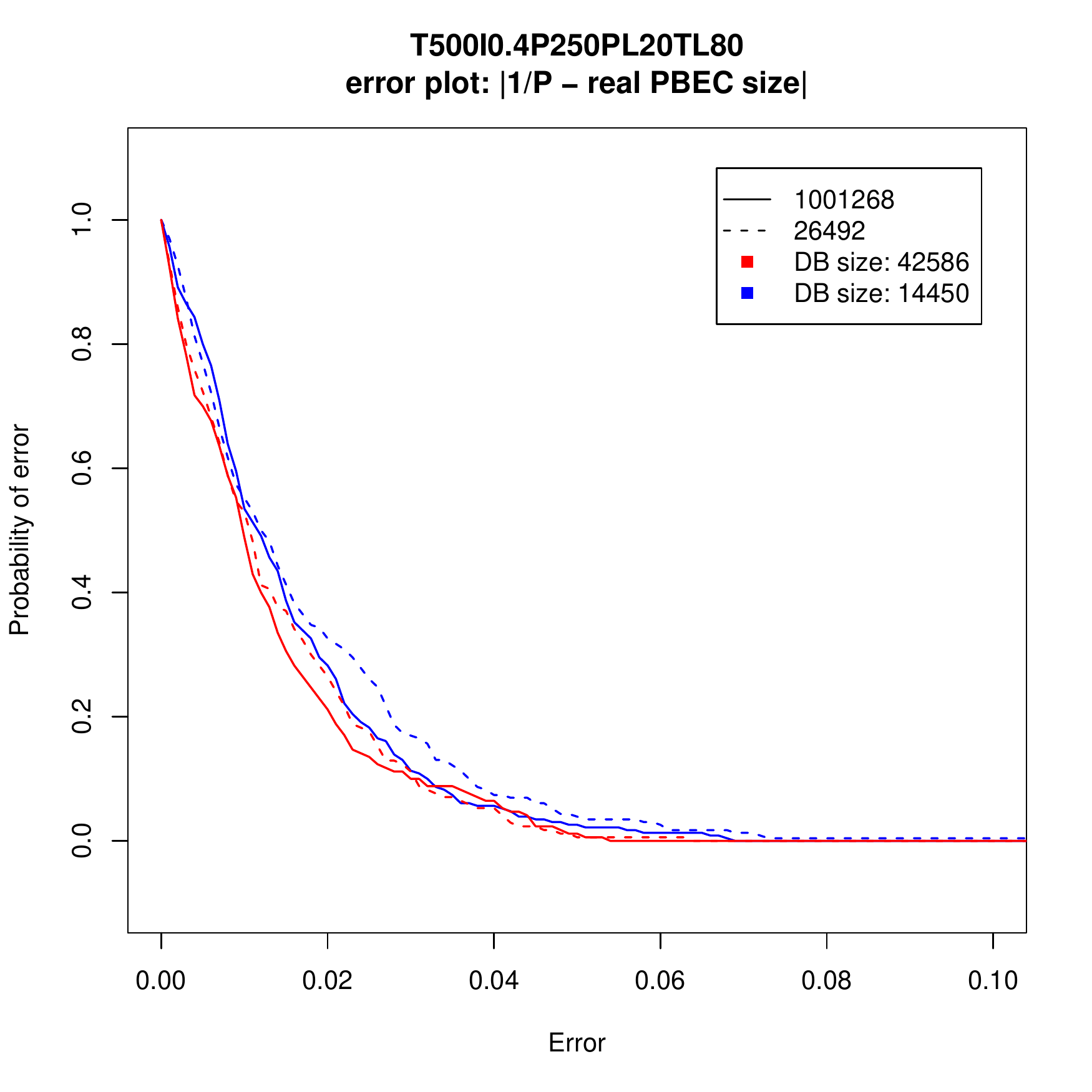}}
\scalebox{0.45}{\includegraphics{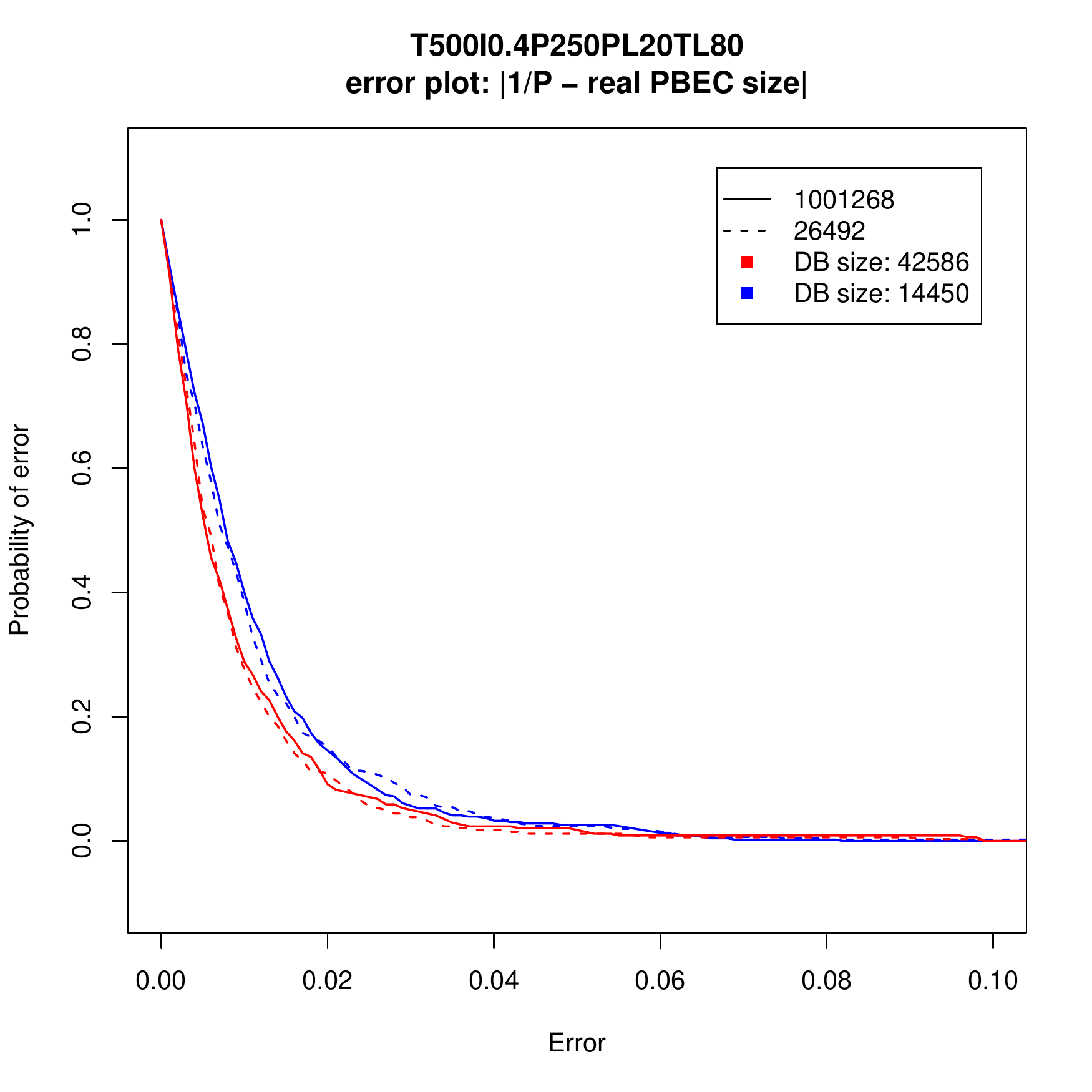}}
}}
\caption[Probability of errors made in Phases 1 and 2]{Probability of
  error of the estimation of the union of PBECs using a database
  sample created in Phase~1 and 2. Experiments made using $\procnum=5$
  processors (left) and $\procnum=10$ processors (right). The
  \texttt{T500I0.4P250PL20TL80} database.}
\label{fig:scheduler-error-5}
\end{figure*}

\begin{figure*}[!p]
\centering
\vbox{\hbox{
\scalebox{0.45}{\includegraphics{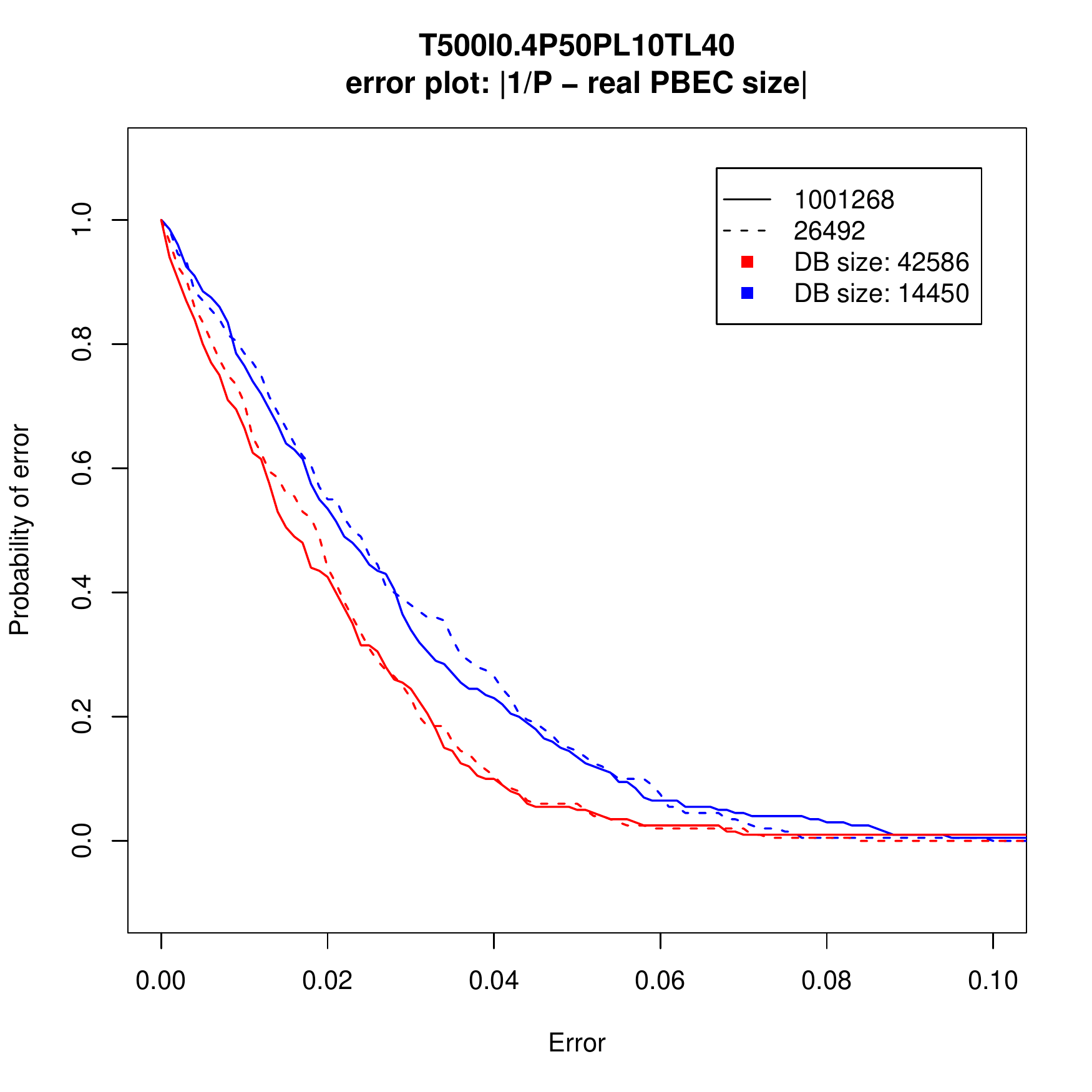}}
\scalebox{0.45}{\includegraphics{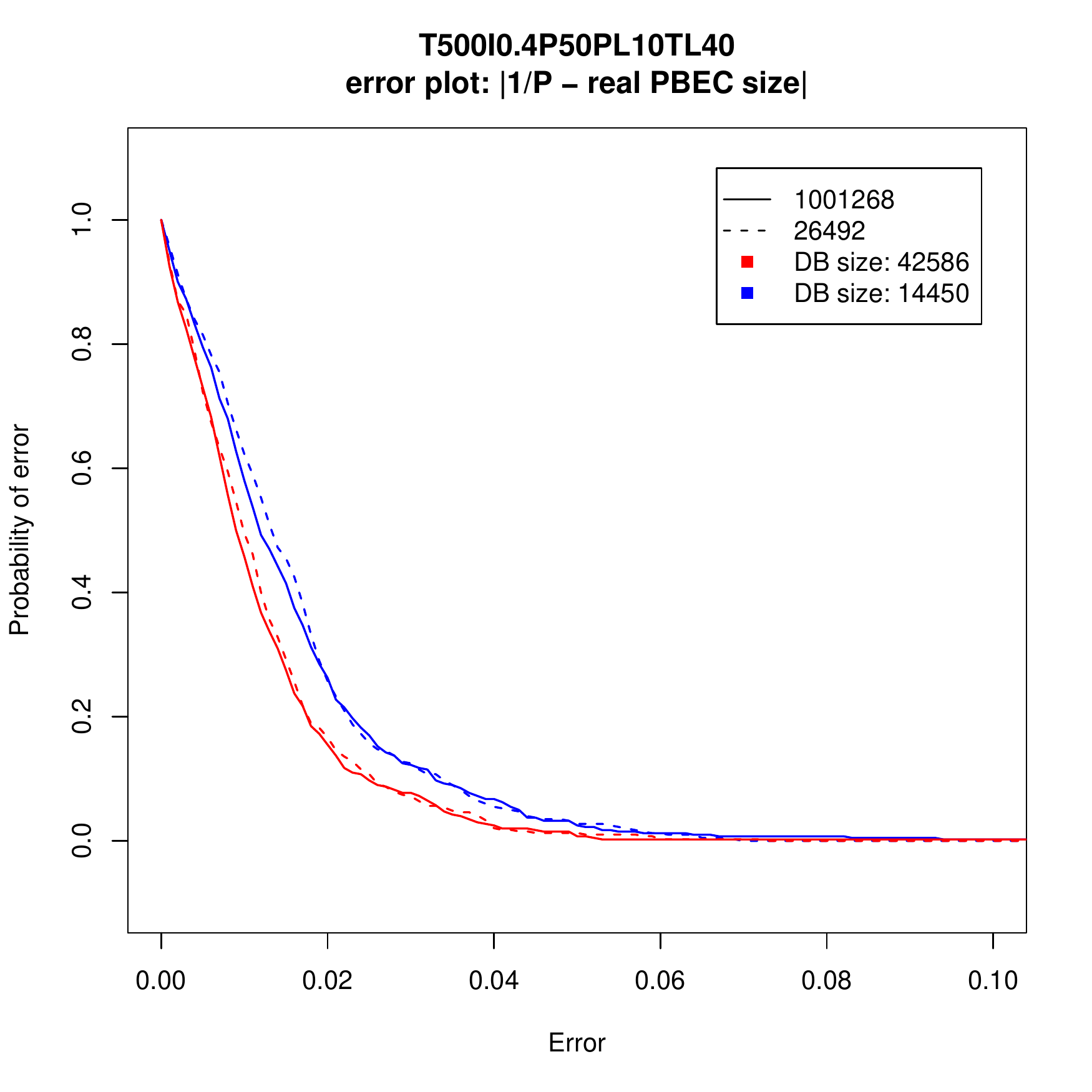}}
}}
\caption[Probability of errors made in Phases 1 and 2]{Probability of
  error of the estimation of the union of PBECs using a database
  sample created in Phase~1 and 2. Experiments made using $\procnum=5$
  processors (left) and $\procnum=10$ processors (right). The
  \texttt{T500I0.4P50PL10TL40} database.}
\label{fig:scheduler-error-6}
\end{figure*}

\begin{figure*}[!p]
\centering
\vbox{\hbox{
\scalebox{0.45}{\includegraphics{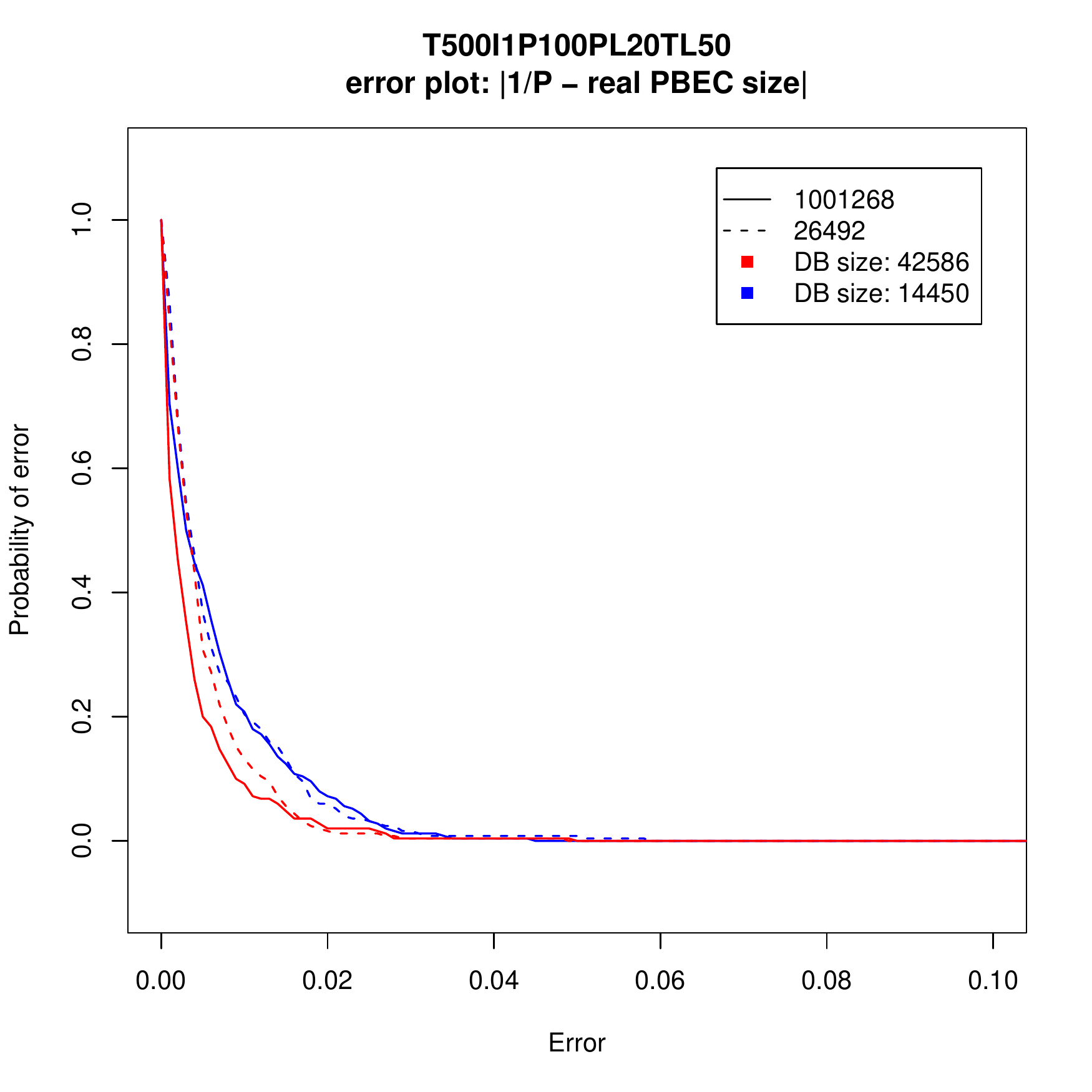}}
\scalebox{0.45}{\includegraphics{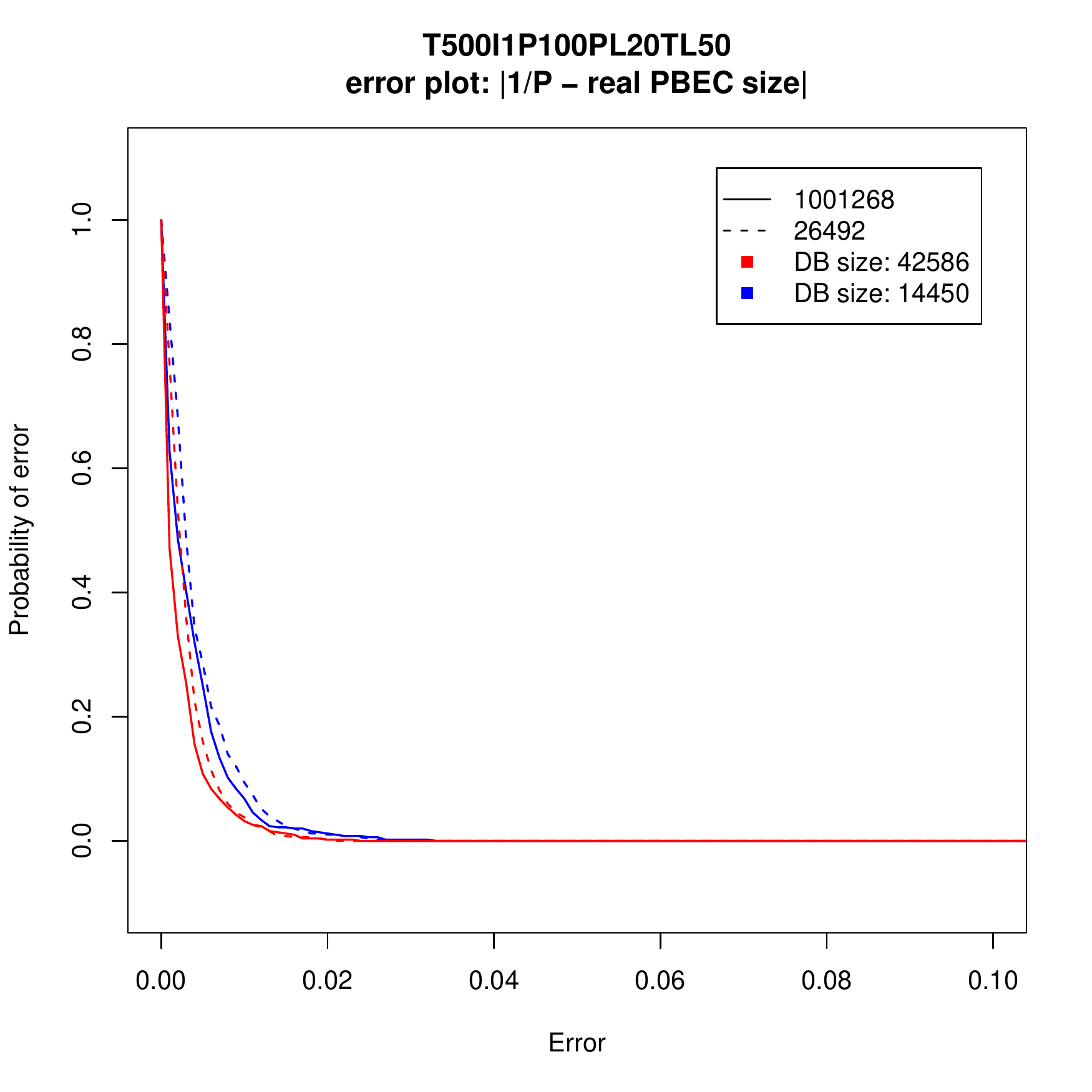}}
}}
\caption[Probability of errors made in Phases 1 and 2]{Probability of
  error of the estimation of the union of PBECs using a database
  sample created in Phase~1 and 2. Experiments made using $\procnum=5$
  processors (left) and $\procnum=10$ processors (right). The
  \texttt{T500I1P100PL20TL50} database.}
\label{fig:scheduler-error-7}
\end{figure*}

We have made two experiments on each database:

\begin{enumerate}
\item \label{meas:pbec-error} \emph{Experiment 1}. Measuring the error
  of a union of PBECs: The probability of the error of the estimation
  of the sizes of a union of PBECs: we show the probability of the
  error $\left| \frac{|\bigcup_{j\in\fipartidxset_i}[\isetU_j] \cap
    \fiapprox|}{|\fiapprox|} -
  \frac{|\bigcup_{j\in\fipartidxset_i}[\isetU_j] \cap
    \fismpl|}{|\fismpl|} \right|$. We have chosen $\procnum=5$ and $\procnum=10$.

\item \label{meas:pbec-double-sampling-error} \emph{Experiment 2.}
  Error of the estimate of the amount of work per processor: for a set
  of prefixes $\{\isetU_i\}$ and for $\procnum$ processors such that
  $\frac{|\bigcup_{j\in\fipartidxset_i}[\isetU_j] \cap
    \fismpl|}{|\fismpl|} \approx 1/\procnum$, we show a graph of
  probability of the error $\left | \frac{1}{\procnum} -
  \frac{|\bigcup_{j\in\fipartidxset_i} [\isetU_j] \cap
    \allfi|}{|\allfi|} \right|$. This graph is the most important for
  our work. See
  Figures~\ref{fig:scheduler-error-1}--\ref{fig:scheduler-error-7}
\end{enumerate}



\begin{figure*}[!pb]
\centering
\vbox{\hbox{
\scalebox{0.45}{\includegraphics{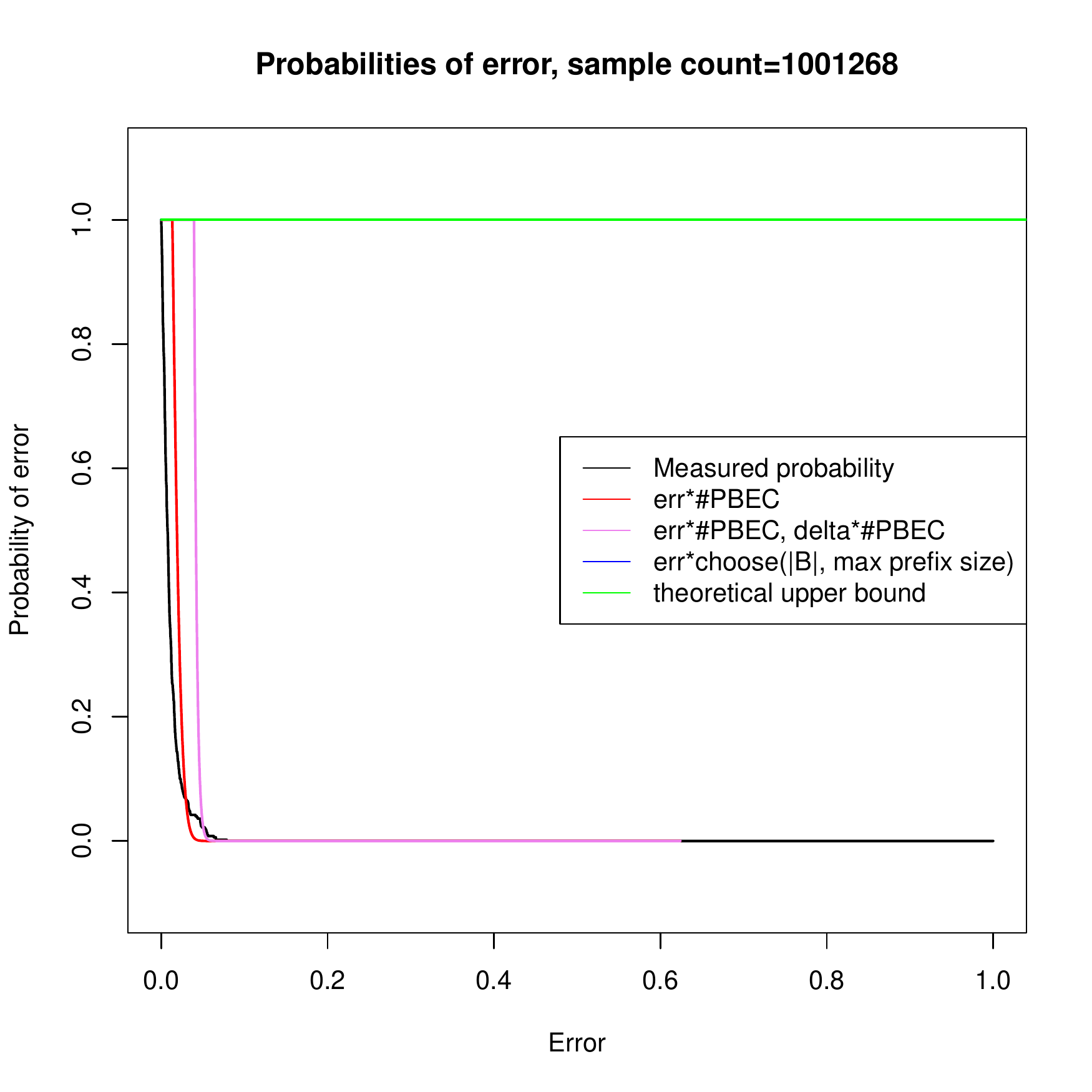}}
\scalebox{0.45}{\includegraphics{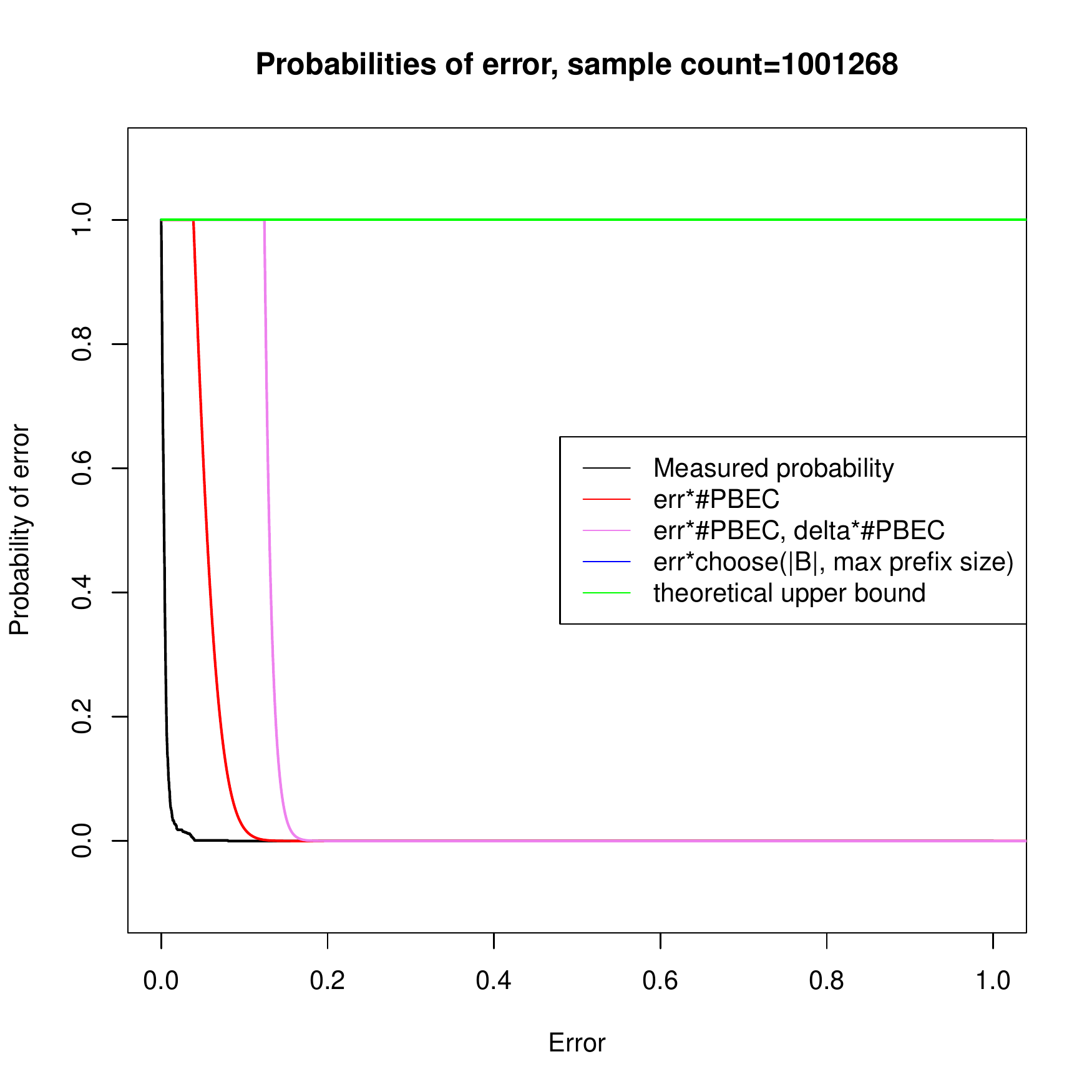}}
}}

\vbox{\hbox{
\scalebox{0.45}{\includegraphics{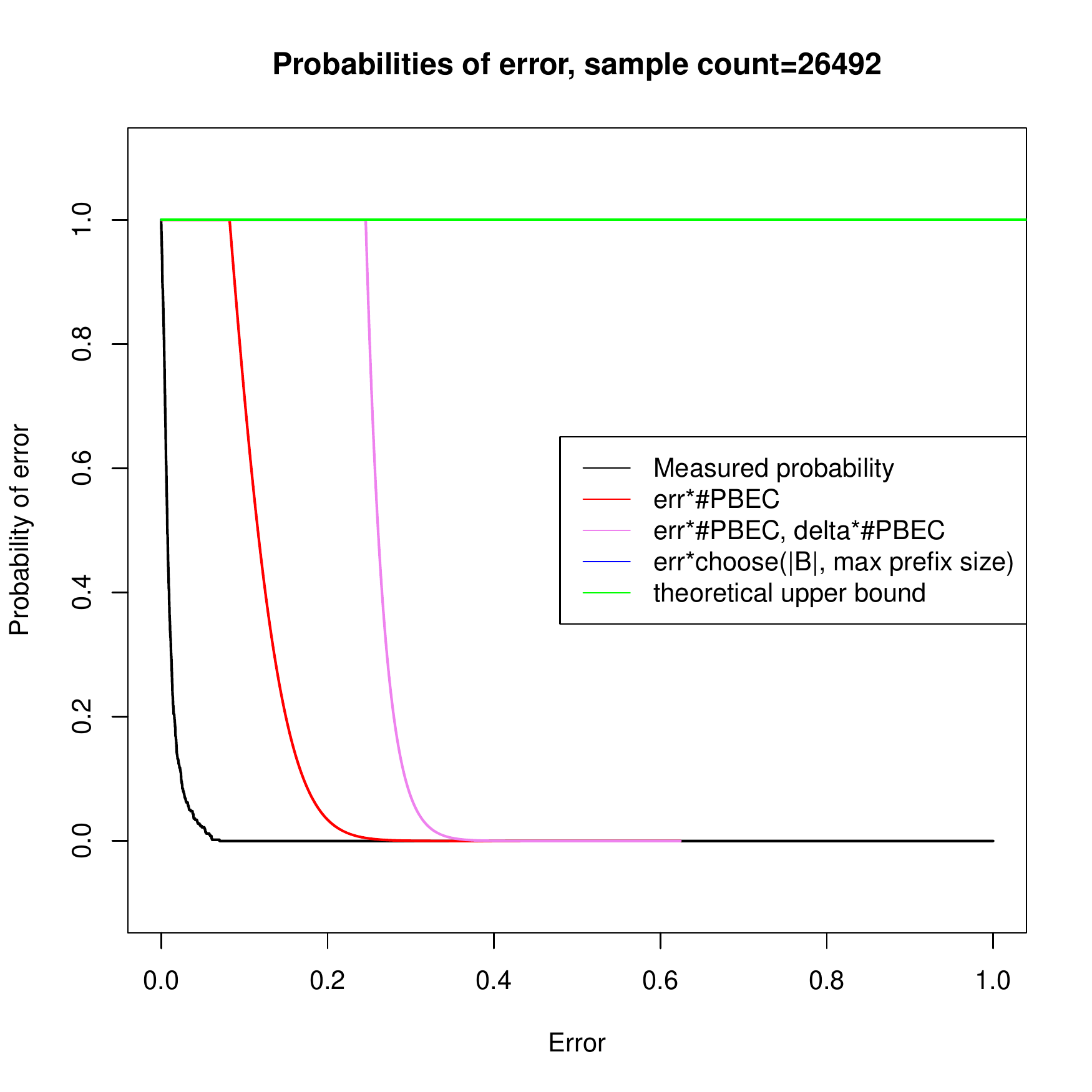}}
\scalebox{0.45}{\includegraphics{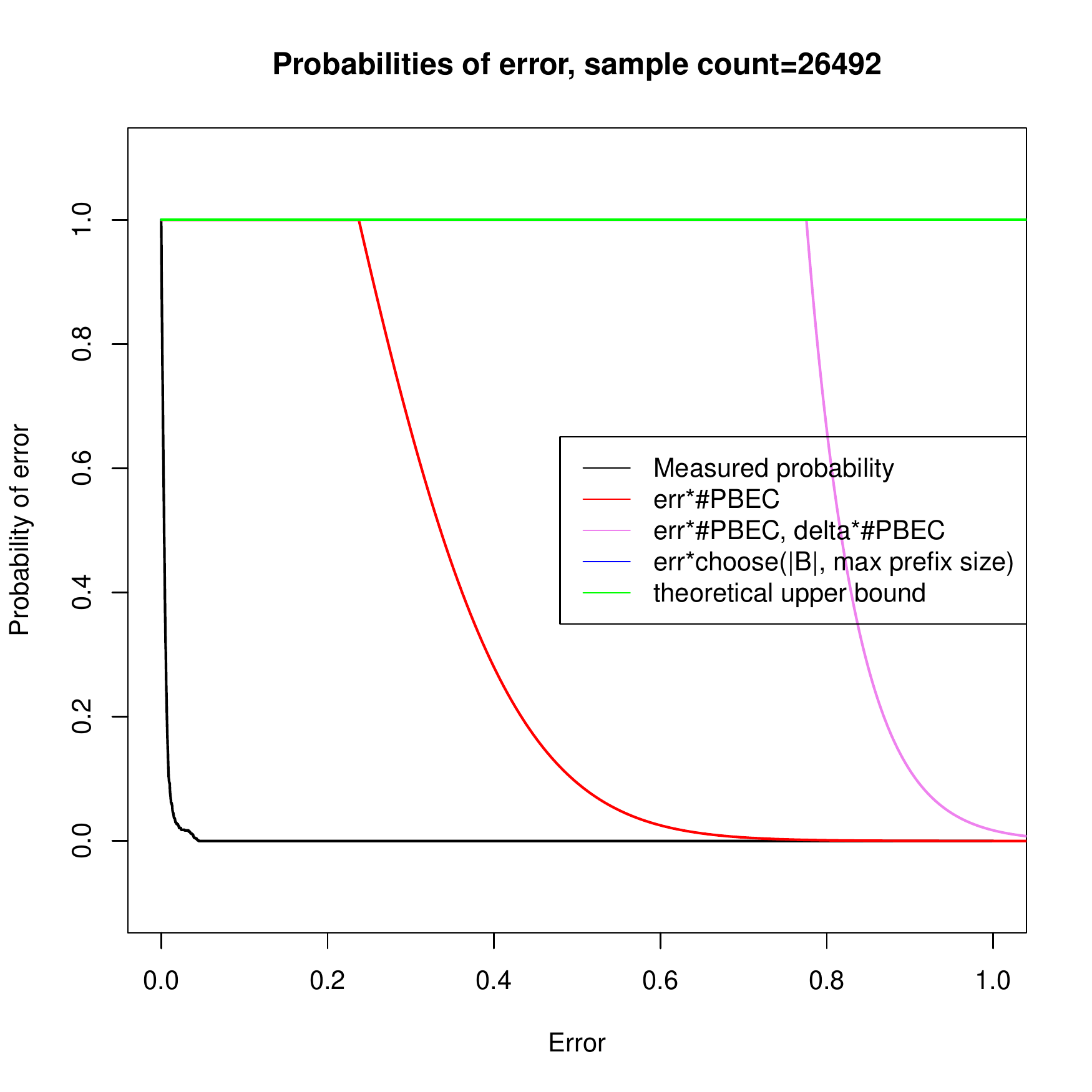}}
}}

\caption[Errors of the estimates of the sizes of union of PBECs, the
  \texttt{T500I0.1P50PL10TL40} database]{The
  \texttt{T500I0.1P50PL10TL40} database: probability of error of the
  estimation of the union of PBECs created in Phase~2 for
  $\procnum=5$ on the left hand graphs and for $\procnum=10$ on the right hand
  graphs.}
\label{fig:error-T500I0.1P50PL10TL40}\label{fig:error-first}
\end{figure*}

\begin{figure*}[!p]
\centering
\vbox{\hbox{
\scalebox{0.45}{\includegraphics{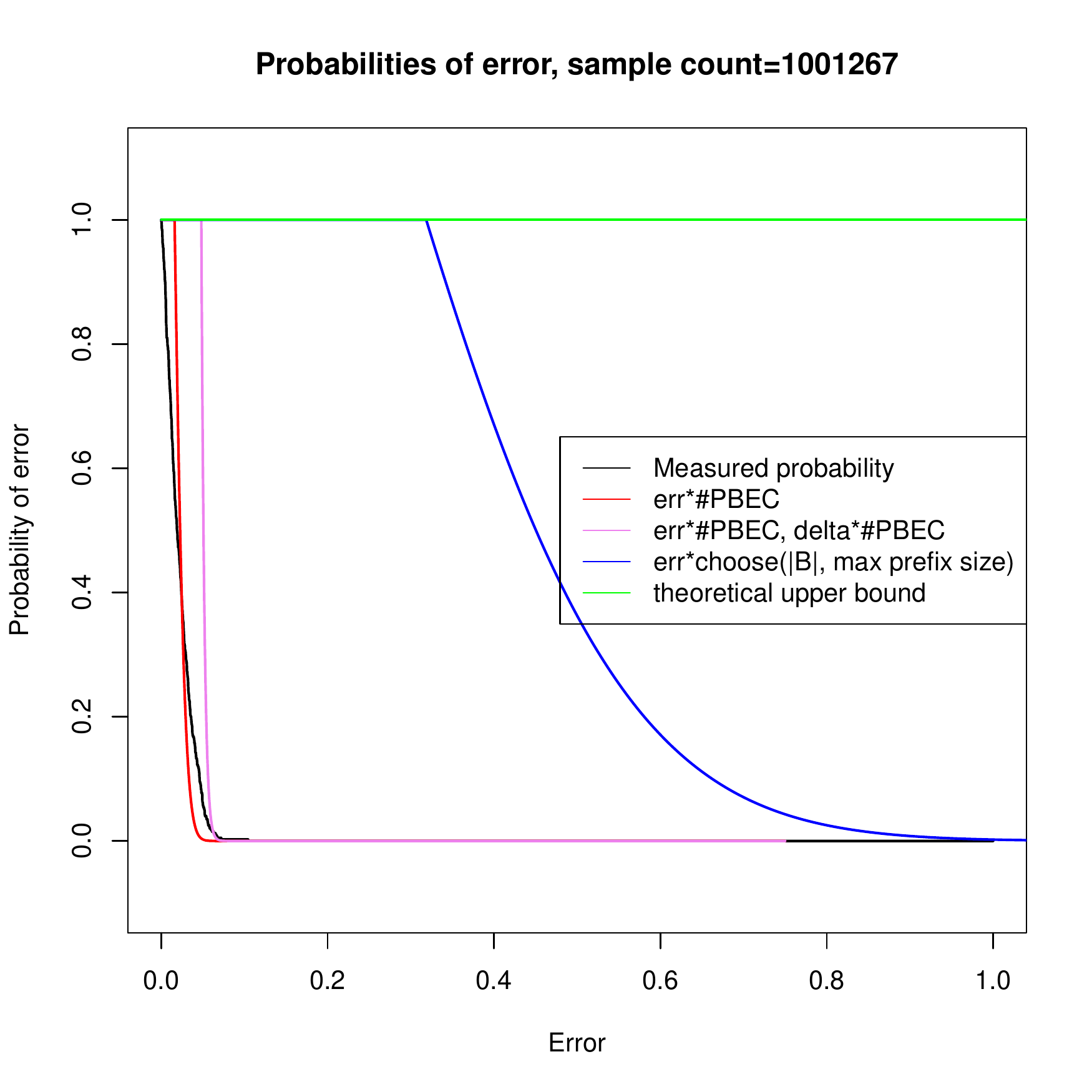}}
\scalebox{0.45}{\includegraphics{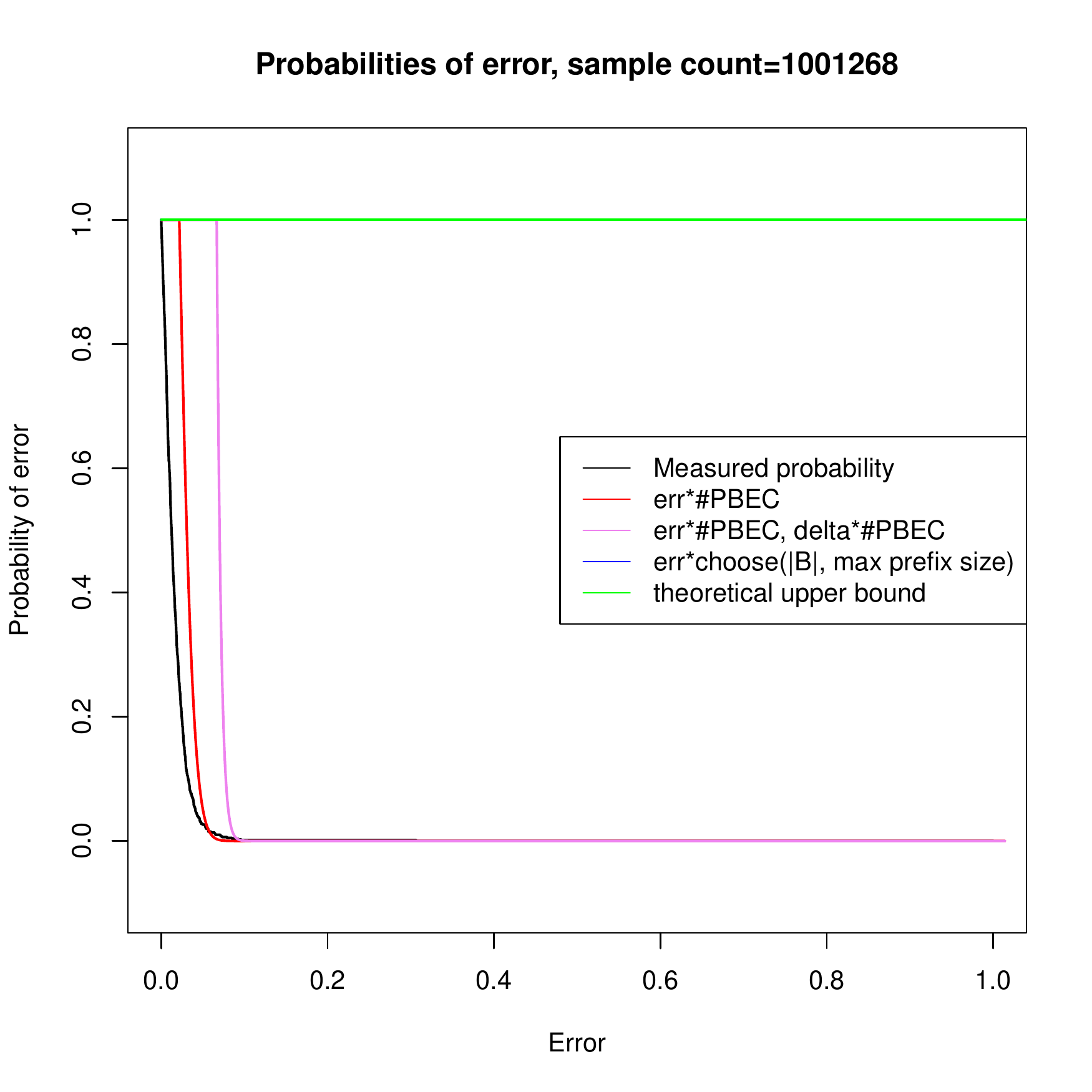}}
}}

\vbox{\hbox{
\scalebox{0.45}{\includegraphics{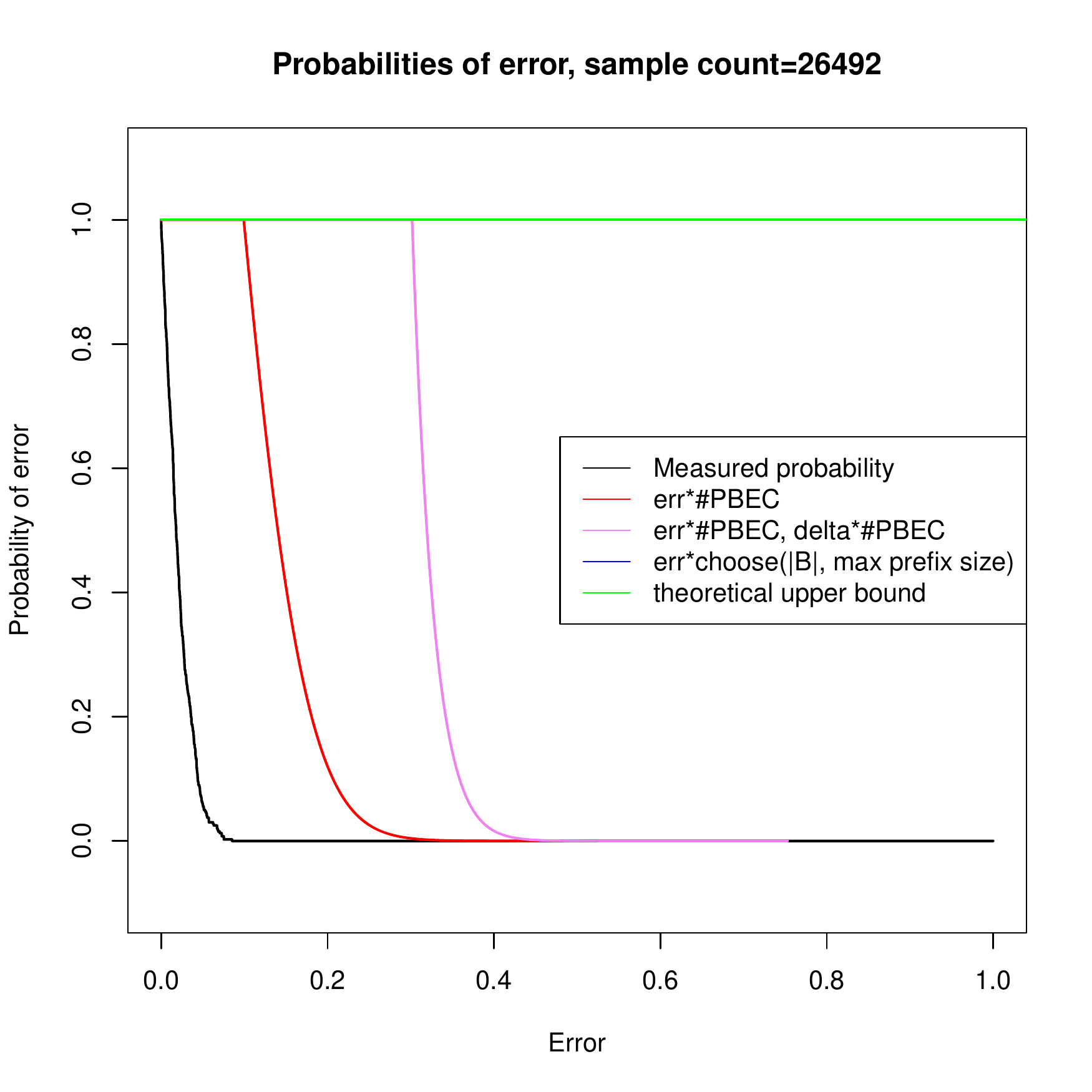}}
\scalebox{0.45}{\includegraphics{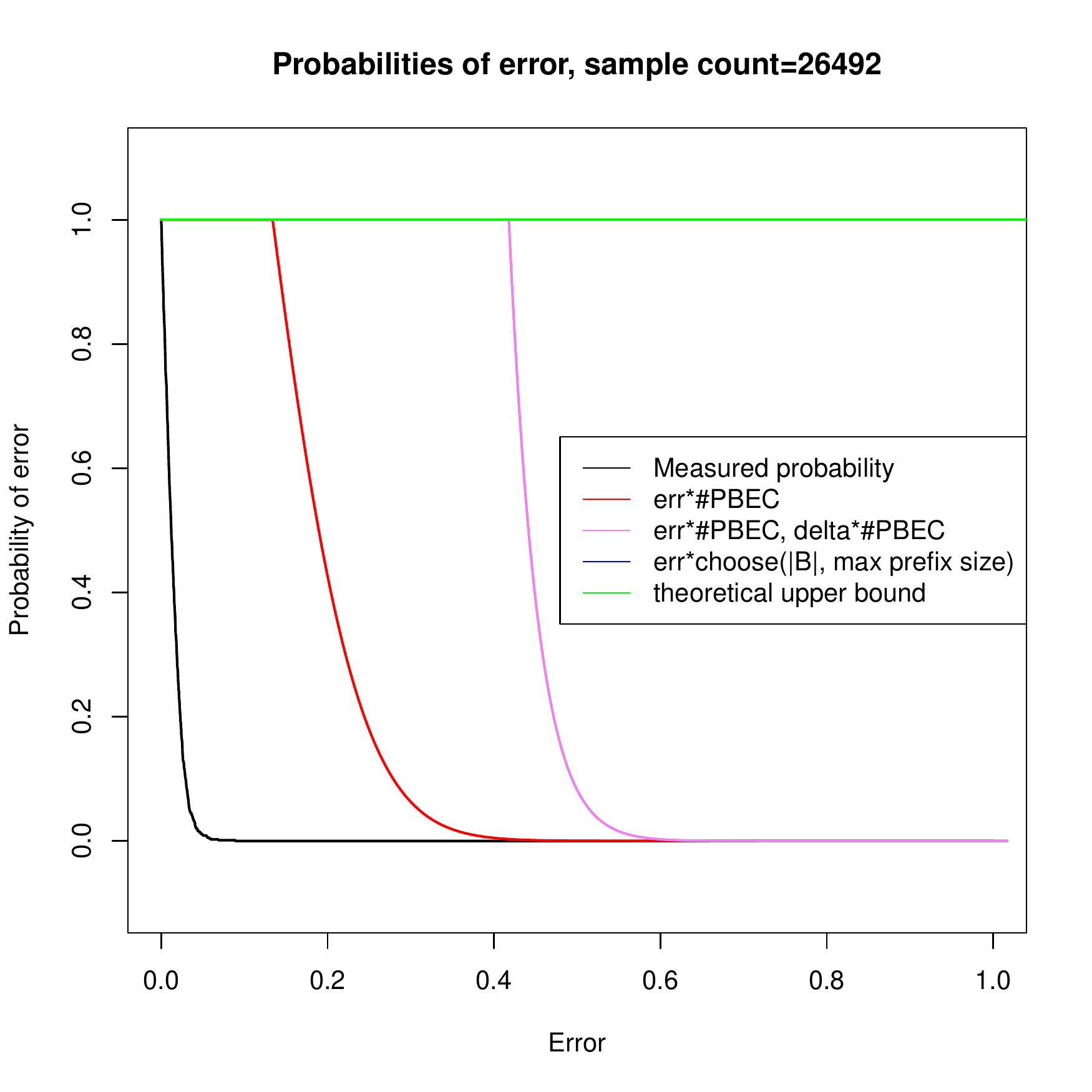}}
}}

\caption[Errors of the estimates of the sizes of union of PBECs, the
  \texttt{T500I0.1P50PL20TL40} database]{The
  \texttt{T500I0.1P50PL20TL40} database: probability of error of the
  estimation of the union of PBECs created in Phase~2 for
  $\procnum=5$ on the left hand graphs and for $\procnum=10$ on the right
  graphs.}
\label{fig:error-T500I0.1P50PL20TL40}
\end{figure*}

\begin{figure*}[!p]
\centering

\vbox{\hbox{
\scalebox{0.45}{\includegraphics{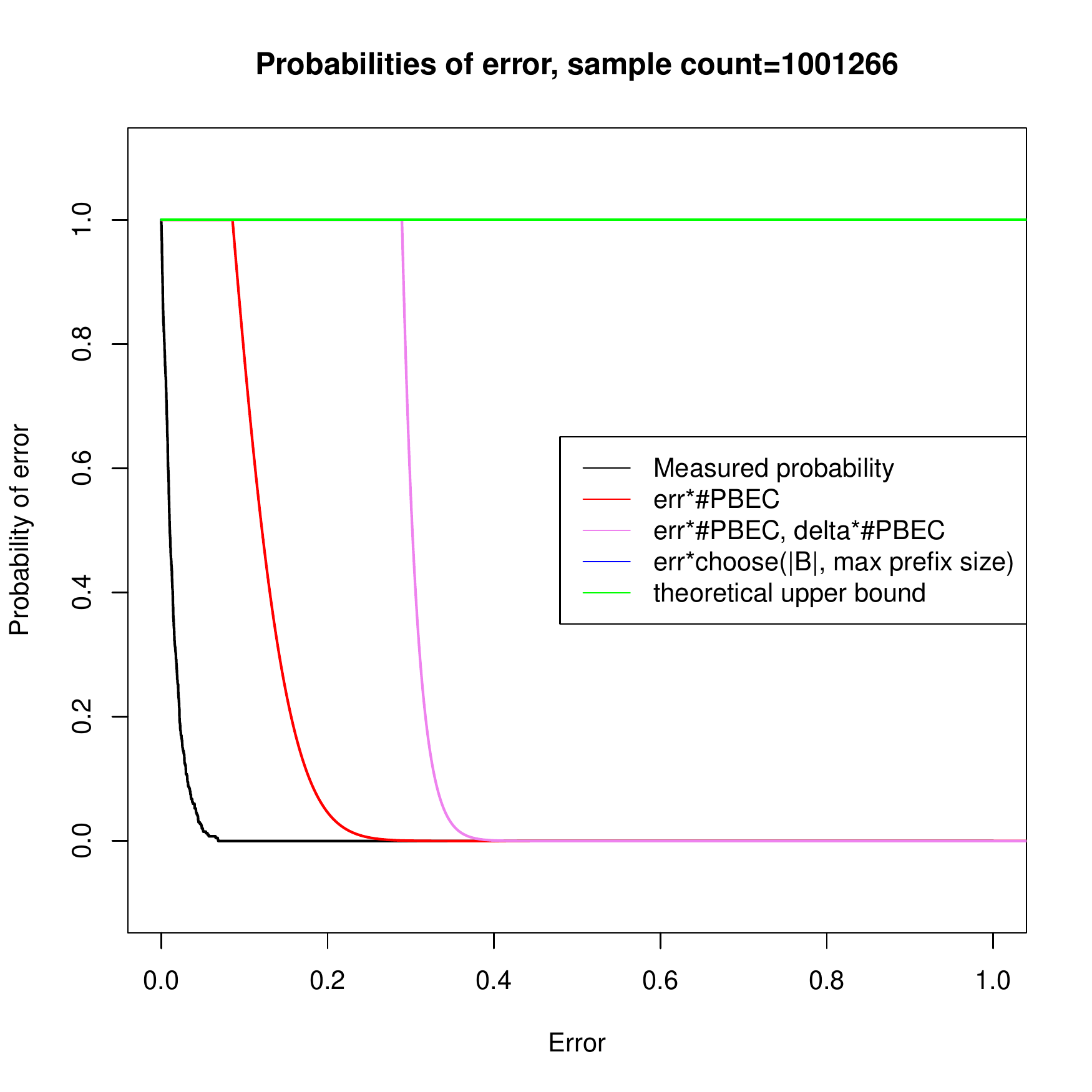}}
\scalebox{0.45}{\includegraphics{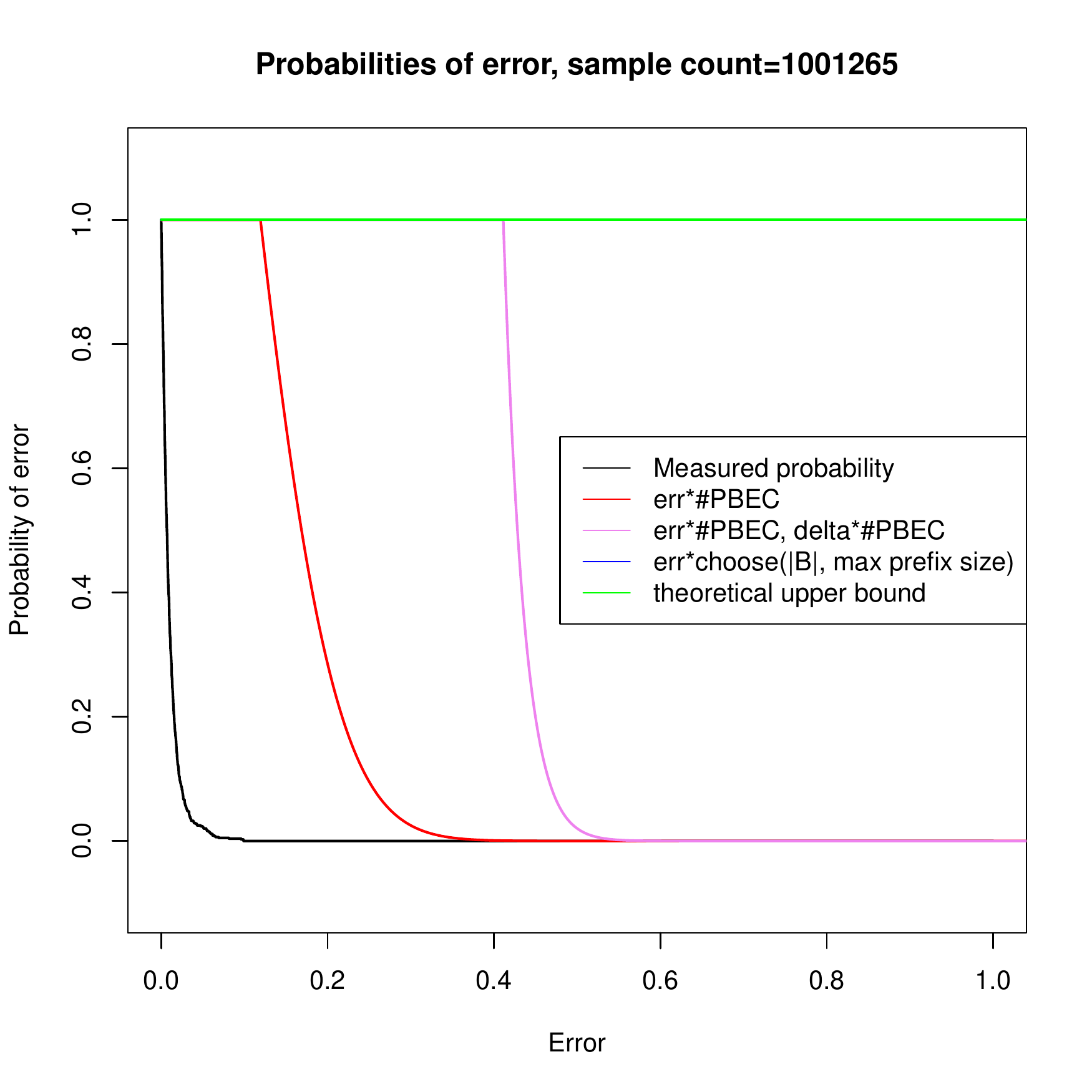}}
}}

\vbox{\hbox{
\scalebox{0.45}{\includegraphics{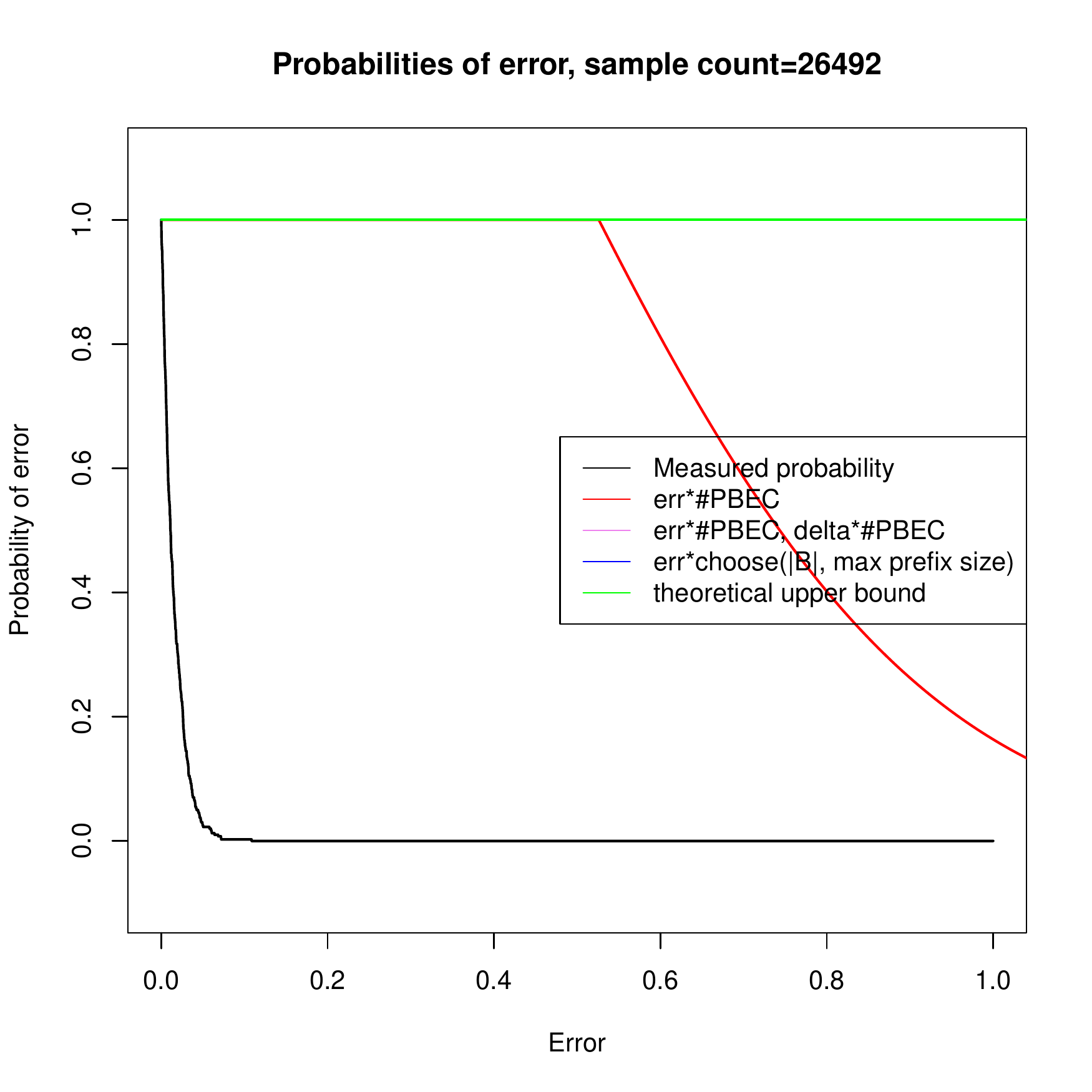}}
\scalebox{0.45}{\includegraphics{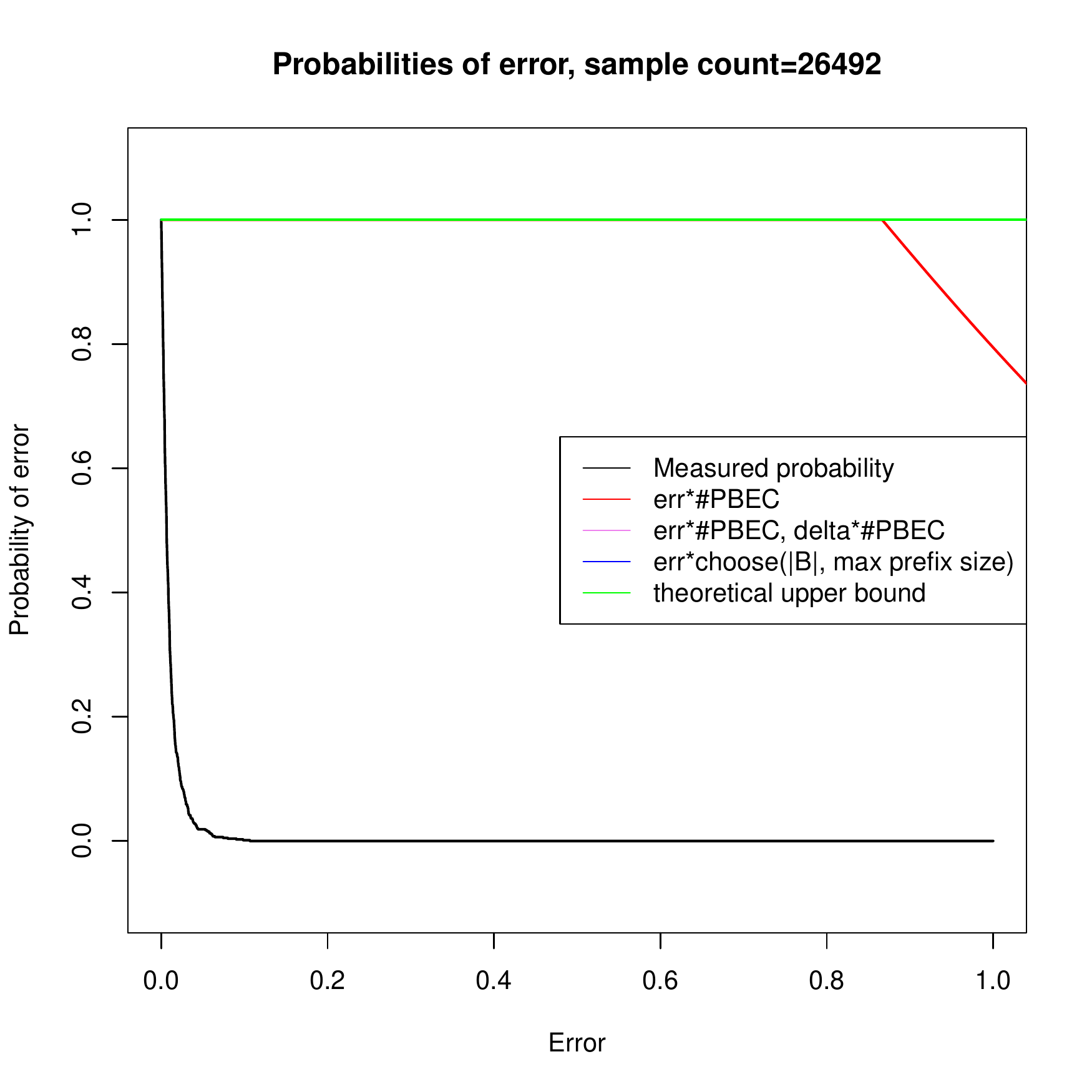}}
}}

\caption[Errors of the estimates of the sizes of union of PBECs, the
  \texttt{T500I0.4P250PL20TL80} database]{The
  \texttt{T500I0.4P250PL20TL80} database: probability of error of the
  estimation of the union of PBECs created in Phase~2 for
  $\procnum=5$ on the left hand graphs and for $\procnum=10$ on the right hand
  graphs.}
\label{fig:error-T500I0.4P250PL20TL80}
\end{figure*}

\begin{figure*}[!p]
\centering
\vbox{\hbox{
\scalebox{0.45}{\includegraphics{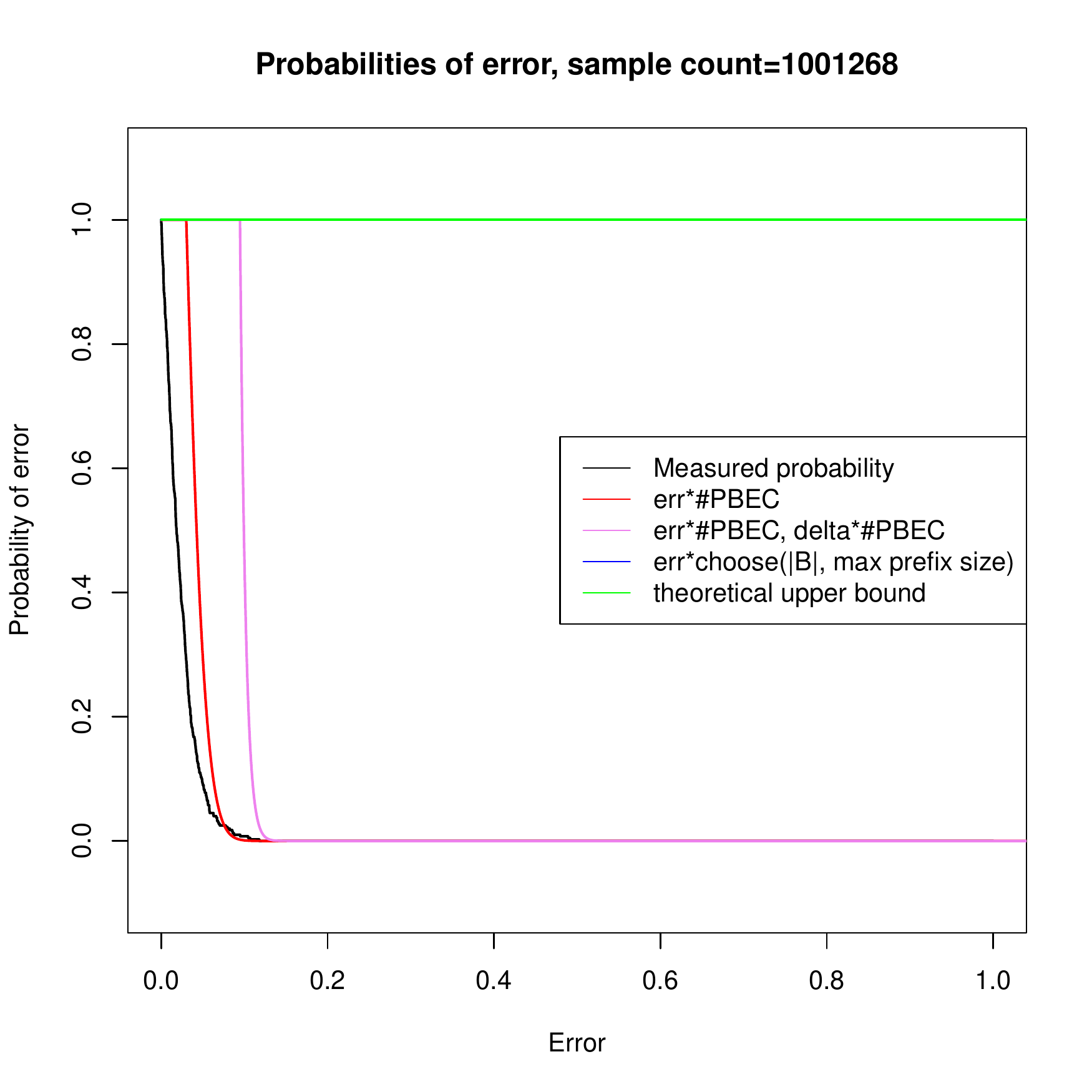}}
\scalebox{0.45}{\includegraphics{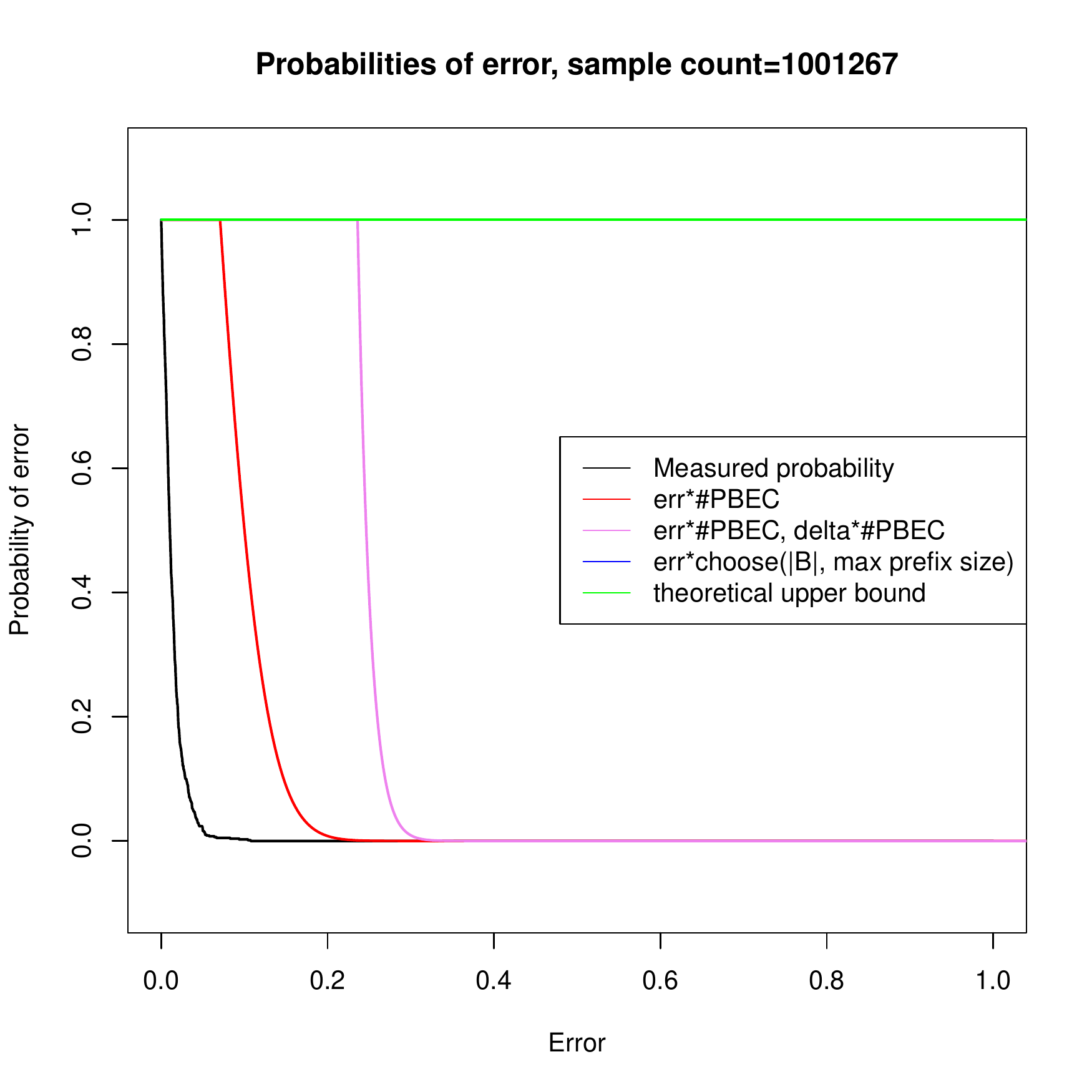}}
}}

\vbox{\hbox{
\scalebox{0.45}{\includegraphics{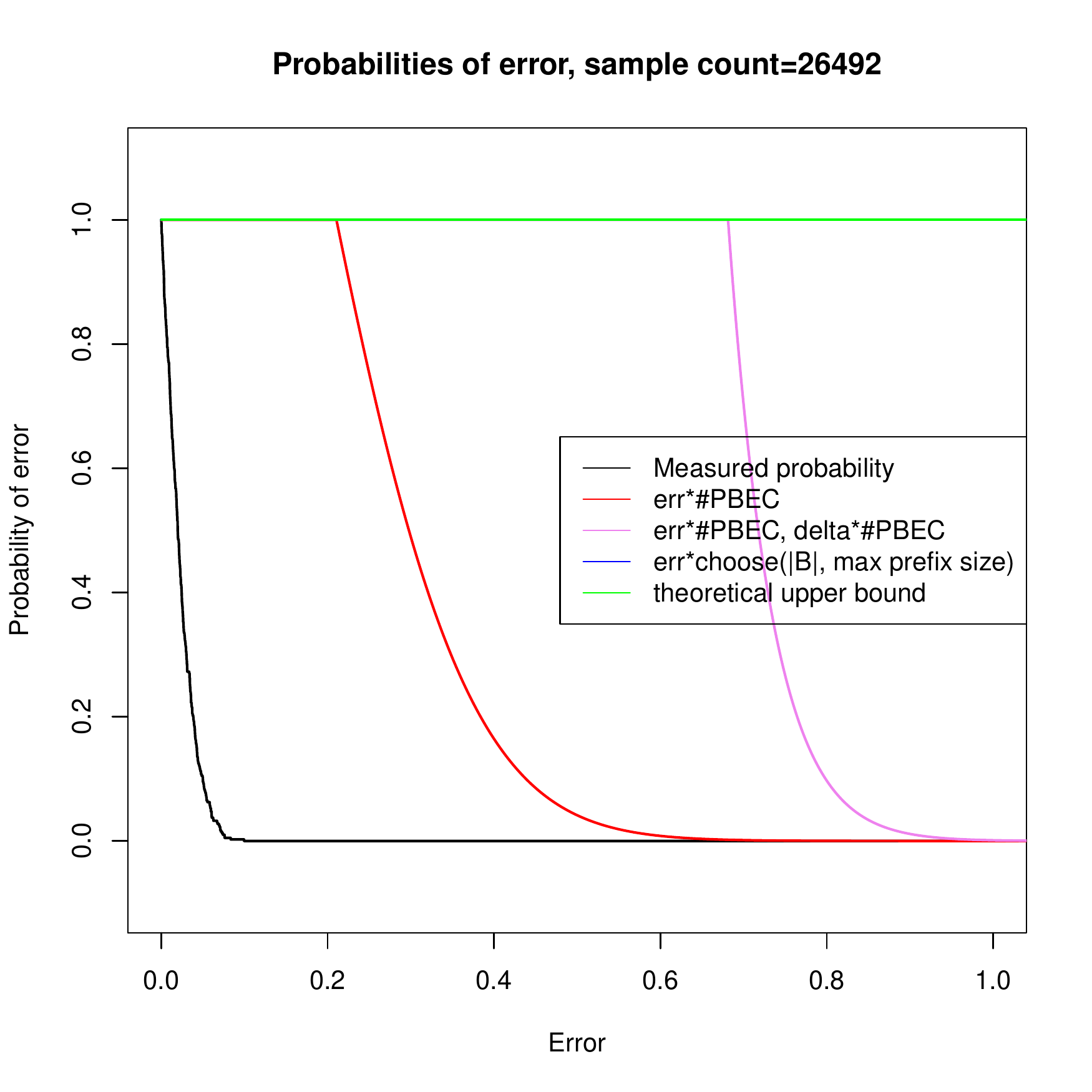}}
\scalebox{0.45}{\includegraphics{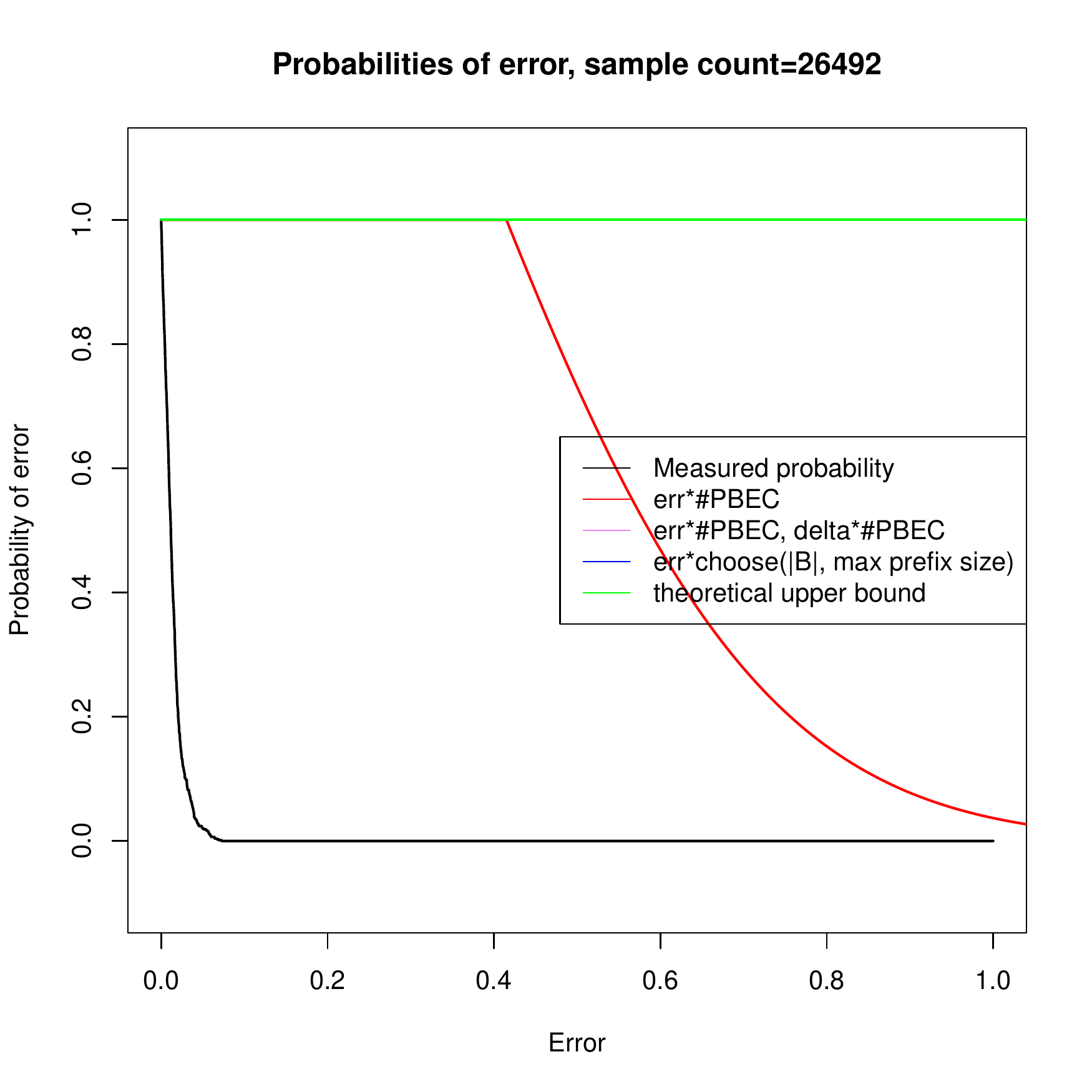}}
}}

\caption[Errors of the estimates of the sizes of union of PBECs, the
  \texttt{T500I0.4P50PL10TL40} database]{The
  \texttt{T500I0.4P50PL10TL40} database: probability of error of the
  estimation of the union of PBECs created in Phase~2 for
  $\procnum=5$ on the left hand graphs and for $\procnum=10$ on the right hand
  graphs.}
\label{fig:error-T500I0.4P50PL10TL40}
\end{figure*}

\begin{figure*}[!p]
\centering
\vbox{\hbox{
\scalebox{0.45}{\includegraphics{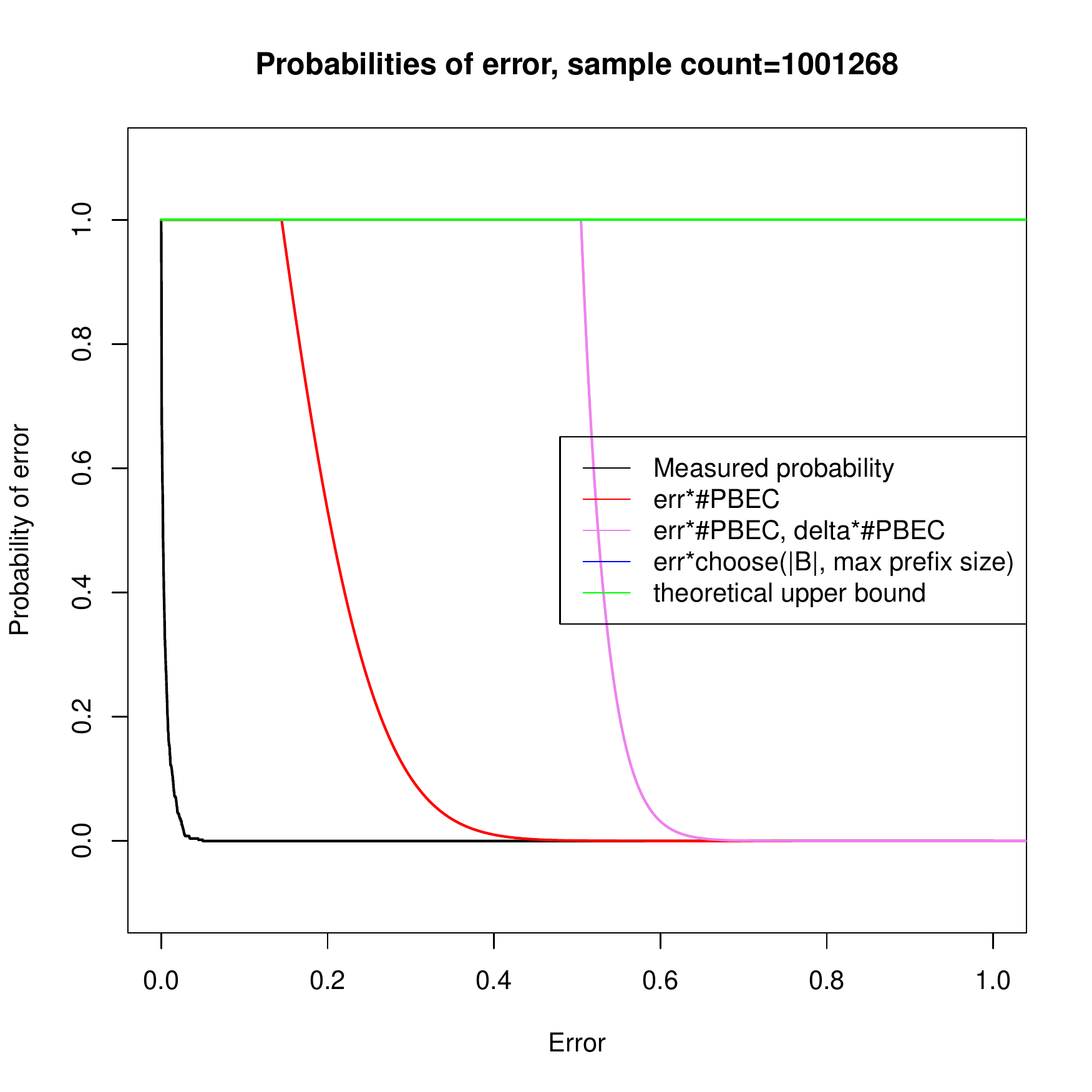}}
\scalebox{0.45}{\includegraphics{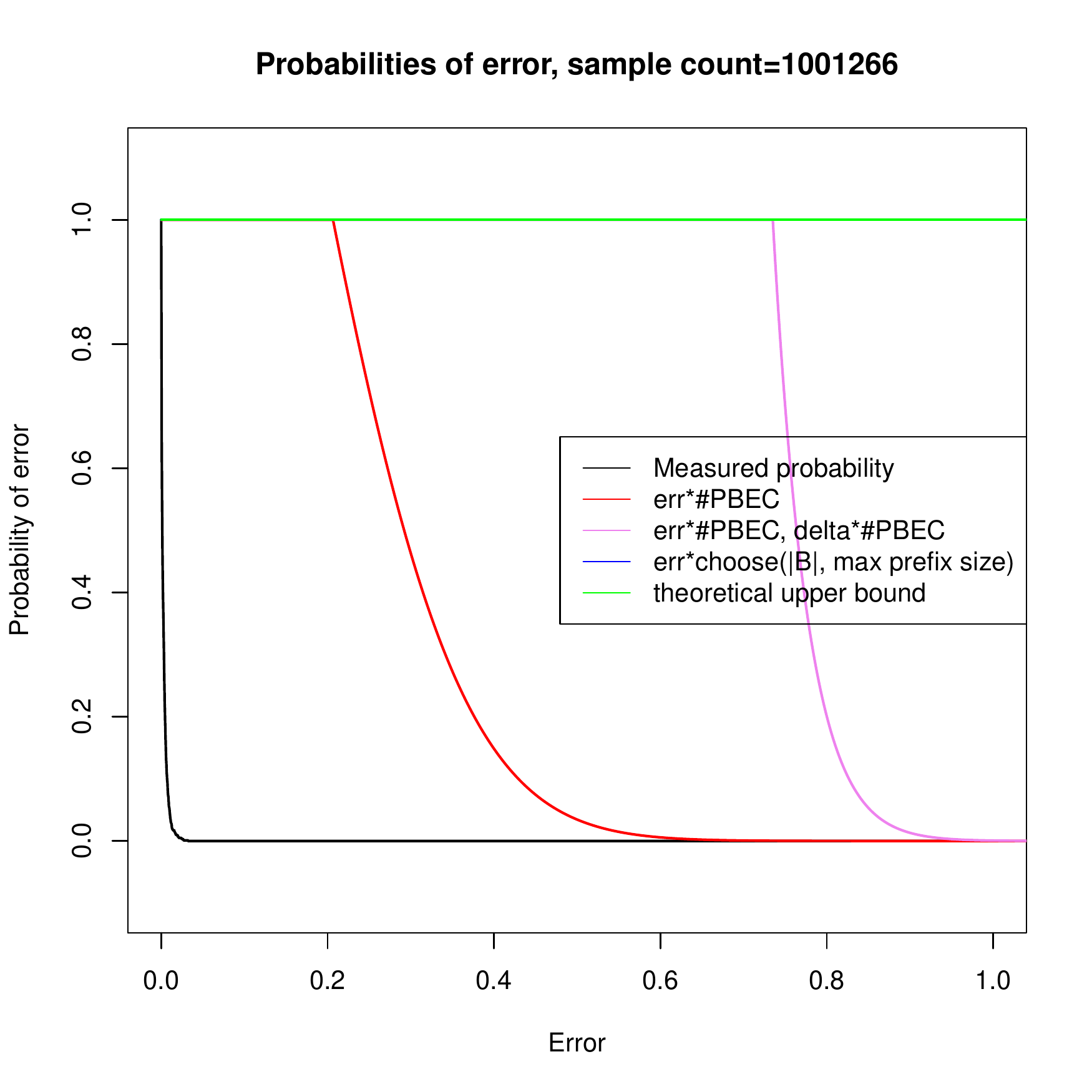}}
}}

\vbox{\hbox{
\scalebox{0.45}{\includegraphics{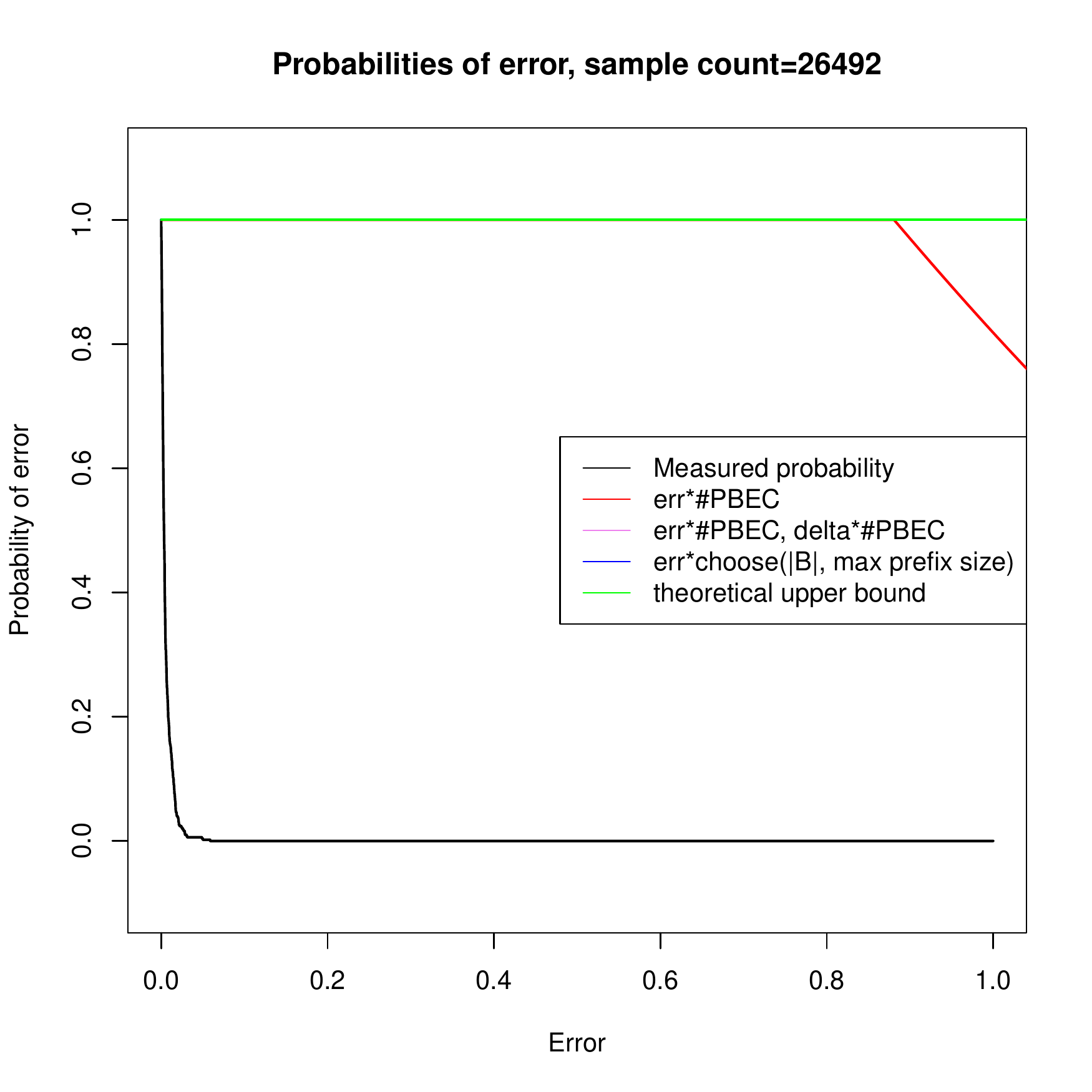}}
\scalebox{0.45}{\includegraphics{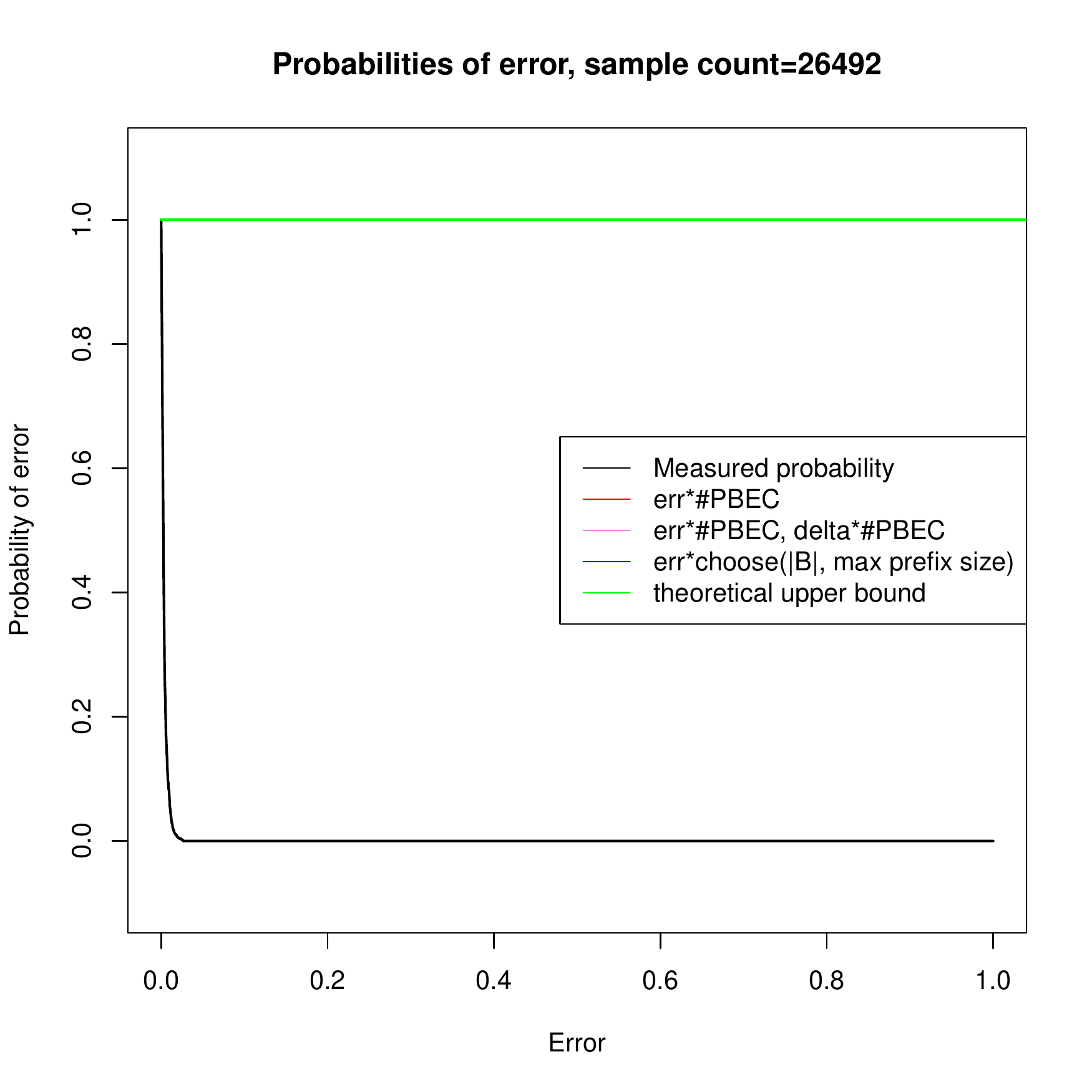}}
}}

\caption[Errors of the estimates of the sizes of union of PBECs, the
  \texttt{T500I1P100PL20TL50} database]{The
  \texttt{T500I1P100PL20TL50} database: probability of error of the
  estimation of the union of PBECs created in Phase~2 for
  $\procnum=5$ on the left hand graphs and for $\procnum=10$ on the right hand
  graphs.}
\label{fig:error-T500I1P100PL20TL50}\label{fig:error-last}
\end{figure*}

\emph{Detailed description of experiment 1}: We have performed the measurements
with the following parameters: 1) we have chosen $\procnum = 5$ processors and
$|\fismpl| = 1`001`268$ samples; 2) we have chosen $\procnum = 5$ and $|\fismpl|
= 26492$. For both measurements, we have chosen $|\dbsmpl| = 42586$ and
$|\dbsmpl| = 14450$, various values of $\rminsupp$. These parameters were mixed
resulting into four graphs: two graphs for $\procnum = 5$: 1) $|\fismpl| =
1`001`268$; 2) $|\fismpl| = 26492$ and two graphs for $\procnum = 10$: 1)
$|\fismpl| = 1`001`268$; 2) $|\fismpl| = 26492$.  We show typical results of the
measurements \ref{meas:pbec-error}: one figure for measurements~1, see
Figures~\ref{fig:error-first}--\ref{fig:error-last}.  Note that each experiment
is performed on a single database with \emph{various values of $\rminsupp$}.

The graphs in Figures~\ref{fig:error-first}--\ref{fig:error-last} show the
typical results of the measurement of the probability of the error (experiment
\ref{meas:pbec-error}).  We have measured the probability $\delta_{\fismpl}$ of
the error $\epsilon_{\fismpl}$ of the estimation of the union of PBECs that were
scheduled in Phase~2 for particular number of samples, see
Algorithm~\ref{alg:phase-2} (the \textsc{Phase-2-FI-Partitioning} Algorithm).
We denote the error of the estimation of \emph{single} PBEC by
$\epsilon_{\fismpl}$, its probability by $\delta_{\fismpl}$, and the number of
PBECs by $N$, the number of items by $|\baseset|$ and the length of the longest
prefix by $\ell$. The figures show four lines: black is the measured probability
of the error; red is the probability of the error computed as
$\epsilon_{\fismpl}\cdot N$ with probability $\delta_{\fismpl}$; violet line is
the probability of the error computed as $\epsilon_{\fismpl}\cdot N$ with
probability $\delta_{\fismpl}\cdot N$; the blue line is the probability of the
error computed as $\epsilon_{\fismpl}\cdot{|\baseset| \choose \ell}$ and the
green line is the error $\epsilon_{\fismpl}\cdot{|\baseset| \choose \ell}$ with
probability $\delta_{\fismpl}\cdot{|\baseset| \choose \ell}$. The green line is
the real theoretical upper bound on the probability. The reason is that we have
to consider \emph{independent} PBECs, however the PBECs are dependent on the
sample $\fismpl$, see the \textsc{Phase-2-FI-Partitioning} Algorithm. In the
figures, some of the lines are missing: the reason is that the lines are out of
the graph on the right. The lines should always be in the following order: (from
left to right) red, violet, blue. The green line is the correct theoretical
upper bound. We can view all PBECs with prefix size $\ell$ as independent, the
number of such prefixes is ${|\baseset| \choose \ell}$. The other lines are
shown to see how big is the influence of each factor and the dependence of PBECs
on the sample.

The result of the experiment is: the theoretical upper bound is too
loose and the probability of the error is usually reasonable for
practical purposes.

\emph{Detailed description of experiment 2}:
Figures~\ref{fig:scheduler-error-1}--\ref{fig:scheduler-error-7} show the
results of the double sampling process, i.e., the size of the union of the PBECs
created in Phase~1 and 2, processed by one processor. There are combination of
dashed and solid line with two colors: red and blue. That is: four lines per
graph. The red color indicates measurement with $|\dbsmpl|=42856$ and the blue
color indicates measurements with $|\dbsmpl|=14450$. The solid line shows the
probability of the error with $|\fismpl|=1001268$ and the dashed line shows the
probability of the error with $|\fismpl|=26492$.  The left hand graph shows the
measurements for $\procnum=5$ and the right hand graph show the measurements for
$\procnum=10$. It can be seen from the graphs that the larger database sample
the smaller the probability of the error. The probability of the error is almost
the same for different size of $|\fismpl|$.  The exception of this is
Figure~\ref{fig:scheduler-error-2}: in this figure the probability of error is
lower for larger database size and bigger for smaller database size (and the
size of the sample almost does not matter).



In addition to the measurements, we have computed for each database
the number of PBECs that make $96\%$ of the total number of FIs. We
denote the set of the prefixes of the $96\%$ of PBECs by
$S=\{\isetU\}$, i.e., $\sum_{\isetU\in S} |[\isetU]\cap\allfi| \geq
0.96\cdot |\fismpl|$. We have discovered that $96\%$ of all samples
are contained in $\approx 100$--$200$ PBECs (the number of all PBECs
varies between $\approx 300$--$3000$). Let $\isetV_{min} = \arg
\min_{\isetW\in S} | [\isetW] \cap \allfi |$ be the prefix of the
smallest PBEC created in Phase 2, we have measured the relative size
of the smallest PBEC $|[\isetV_{min}]\cap\allfi|\approx
0.0007$-$0.003$. Therefore, the value of $\rho$ can be chosen between
$0.0007$-$0.003$, depending on the database.

\clearpage

\section{Evaluation of the speedup}\label{sec:exp-eval-speedup}


Two of the proposed parallel methods, namely the \scparfimiseq{}
method (Method~\ref{method:parallel-fimi-seq}) and the
\scparfimipar{} method (Method~\ref{method:parallel-fimi-par}),
need to compute the MFIs $\mfiapprox$ from a database sample
$\dbsmpl$. In the experiments, in Phase 1, we use the \emph{fpmax*}
\cite{grahne03fpmax} algorithm that computes the MFIs. In the case of
the \scparfimiseq{} the \emph{fpmax*} algorithm is executed
sequentially on processor $\proc_1$. In the case of the
\scparfimipar{}, we execute the \emph{fpmax*} algorithm in parallel,
see Chapter~\ref{chap:parallel-mfi}.

In Phase 4 in our experiments, we use the \textsc{Eclat} algorithm for
mining of FIs. In the case of the \scparfimireserv{} method
(Method~\ref{method:parallel-fimi-reserv}), the \textsc{Eclat}
algorithm is also used in Phase 1, i.e., the reservoir sampling
samples the output of the \textsc{Eclat} algorithm.

As the parameters of our method, we use the number of samples
$|\dbsmpl|$ and $|\fismpl|$. The used parameters are summarized in
Table~\ref{tbl:experiments-number-of-samples}.

\begin{table}[h]\label{tbl:experiments-number-of-samples}
\centering

\begin{tabular}{|c|c|c|c|c|c|c|c|c|c|c|c|c|} \hline
$|\dbsmpl|$ & 10000 & 10000 & 10000 & 14450 & 14450 & 14450 & 14450 & 14450 & 20000 & 20000 & 20000 \\ \hline
$|\fismpl|$ & 19869 & 26492 & 33115 & 13246 & 19869 & 26492 & 33115 & 39738 & 19869 & 26492 & 33115 \\ \hline
\end{tabular}

\caption{Sizes of $|\dbsmpl|$ and $|\fismpl|$ used in experiments}\label{fig:datasets}
\end{table}

\begin{figure*}[!p]
\centering
\scalebox{0.35}{\includegraphics{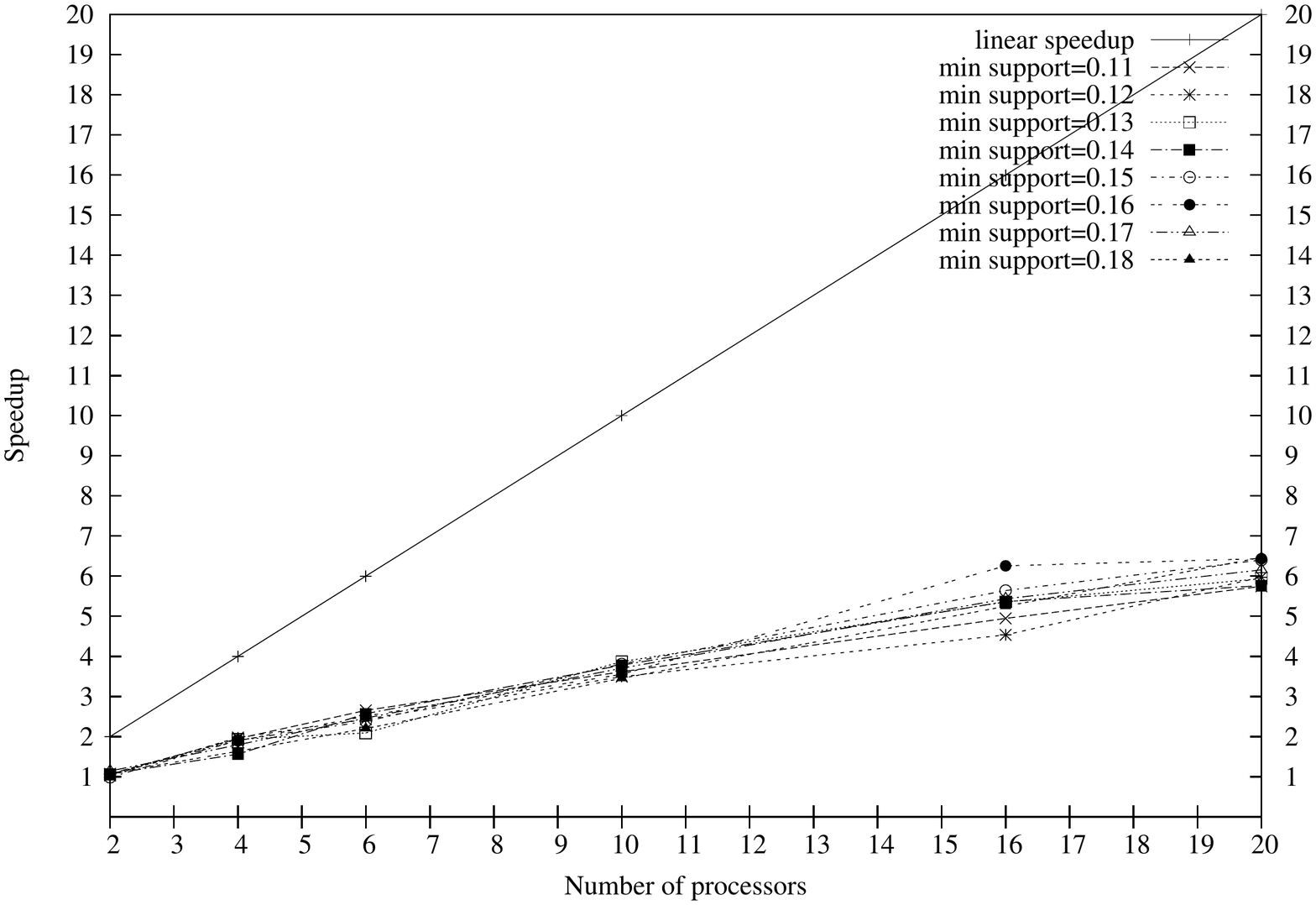}}
\scalebox{0.35}{\includegraphics{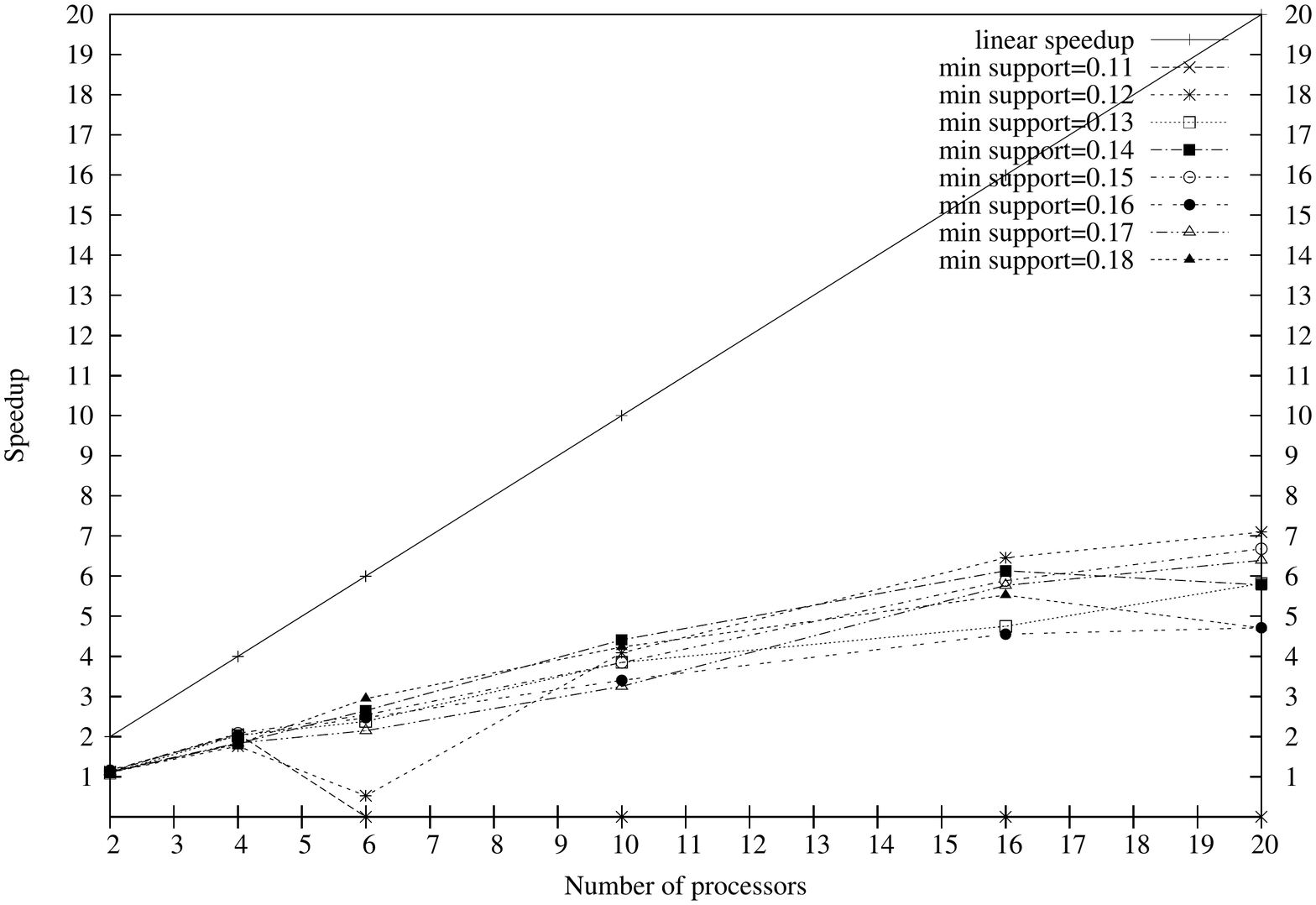}}
\scalebox{0.35}{\includegraphics{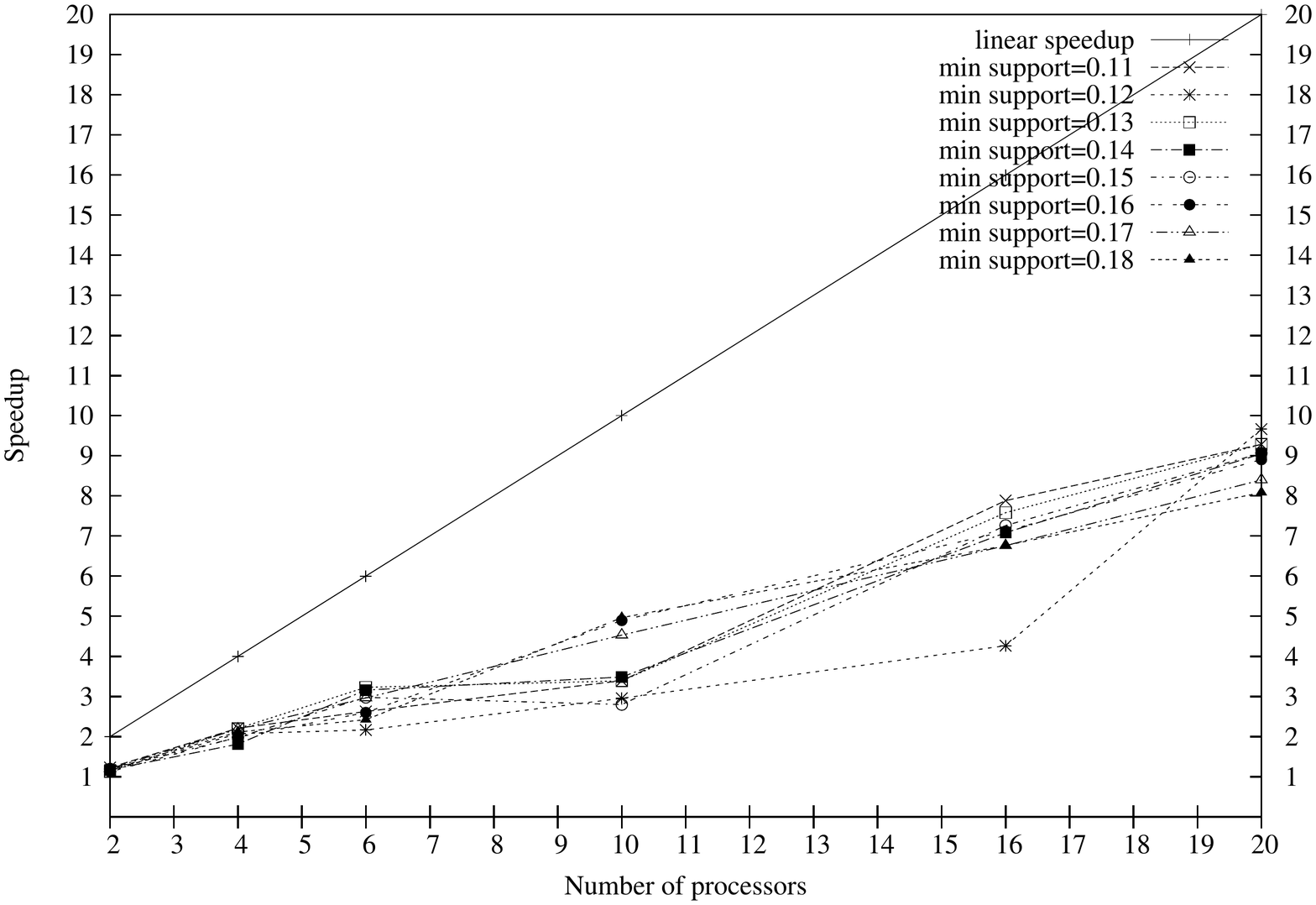}}
\caption[\hskip0.1cm Speedup measured on the {\tt T500I0.1P100PL20TL50}]{Speedups of the \scparfimiseq{}, \scparfimipar{}, and
  \scparfimireserv{} methods (from top to bottom) on the {\tt
    T500I0.1P100PL20TL50} database.}
\label{speedup-T500I0.1-1}
\end{figure*}

\begin{figure*}[!p]
\centering
\scalebox{0.35}{\includegraphics{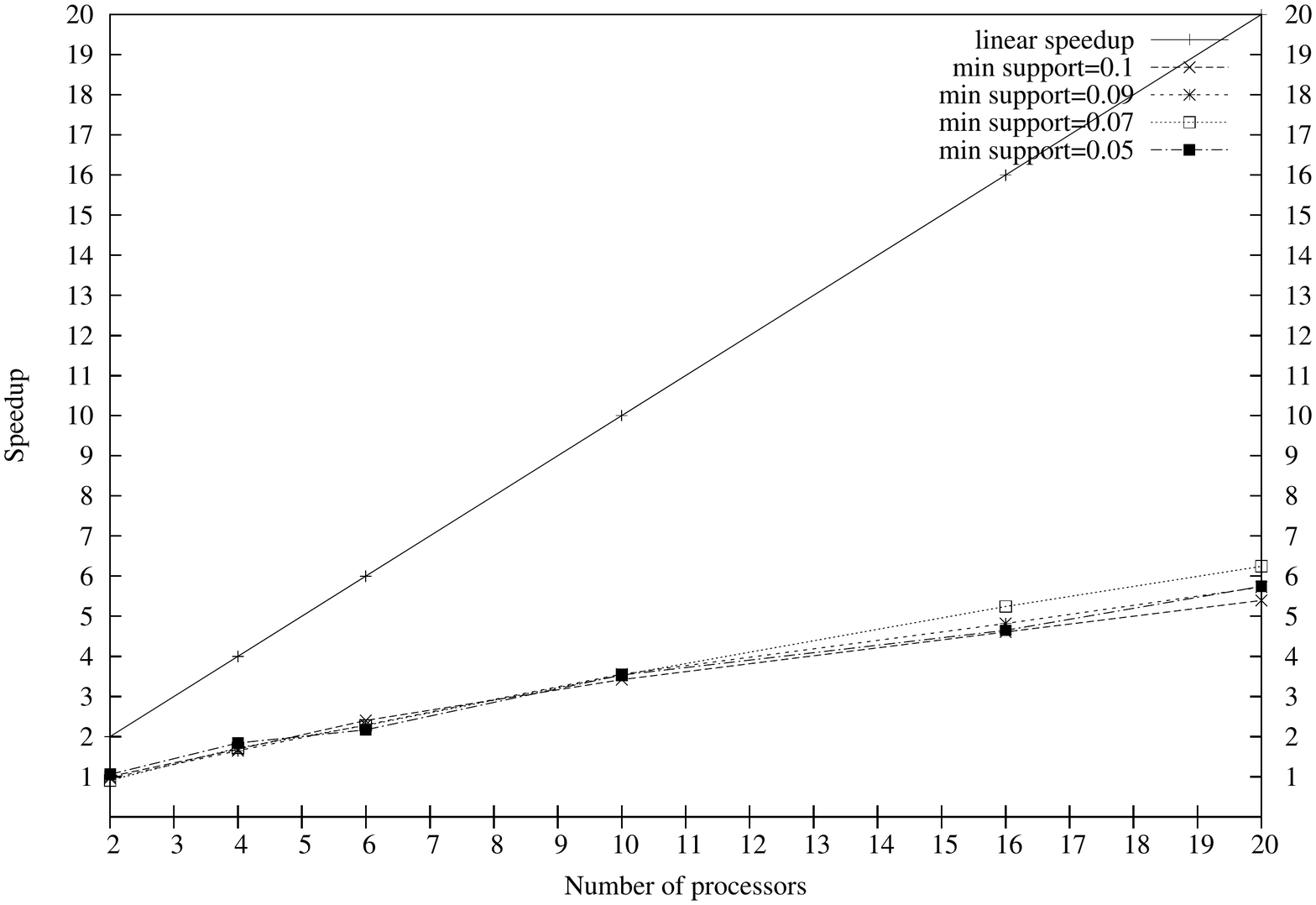}}
\scalebox{0.35}{\includegraphics{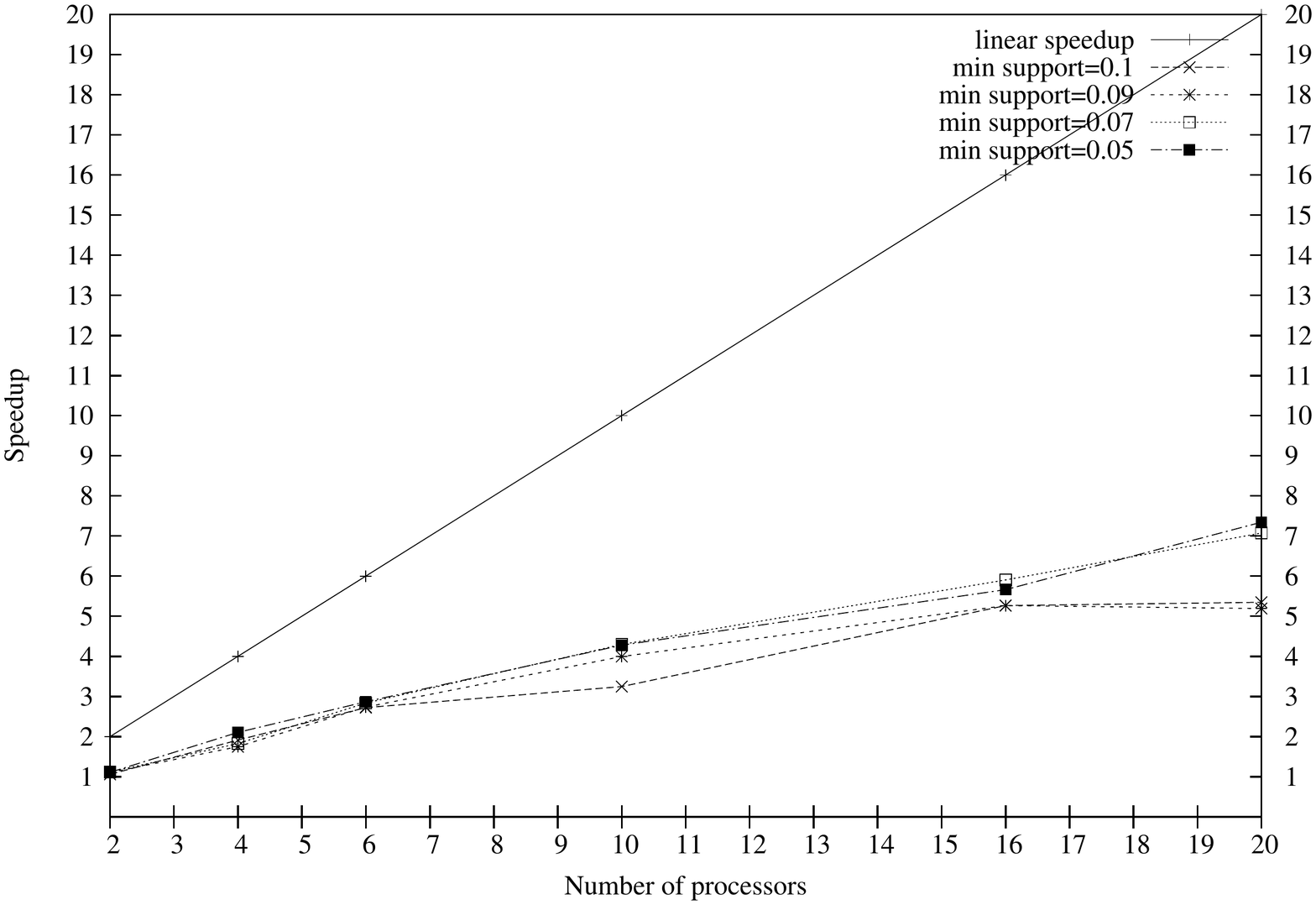}}
\scalebox{0.35}{\includegraphics{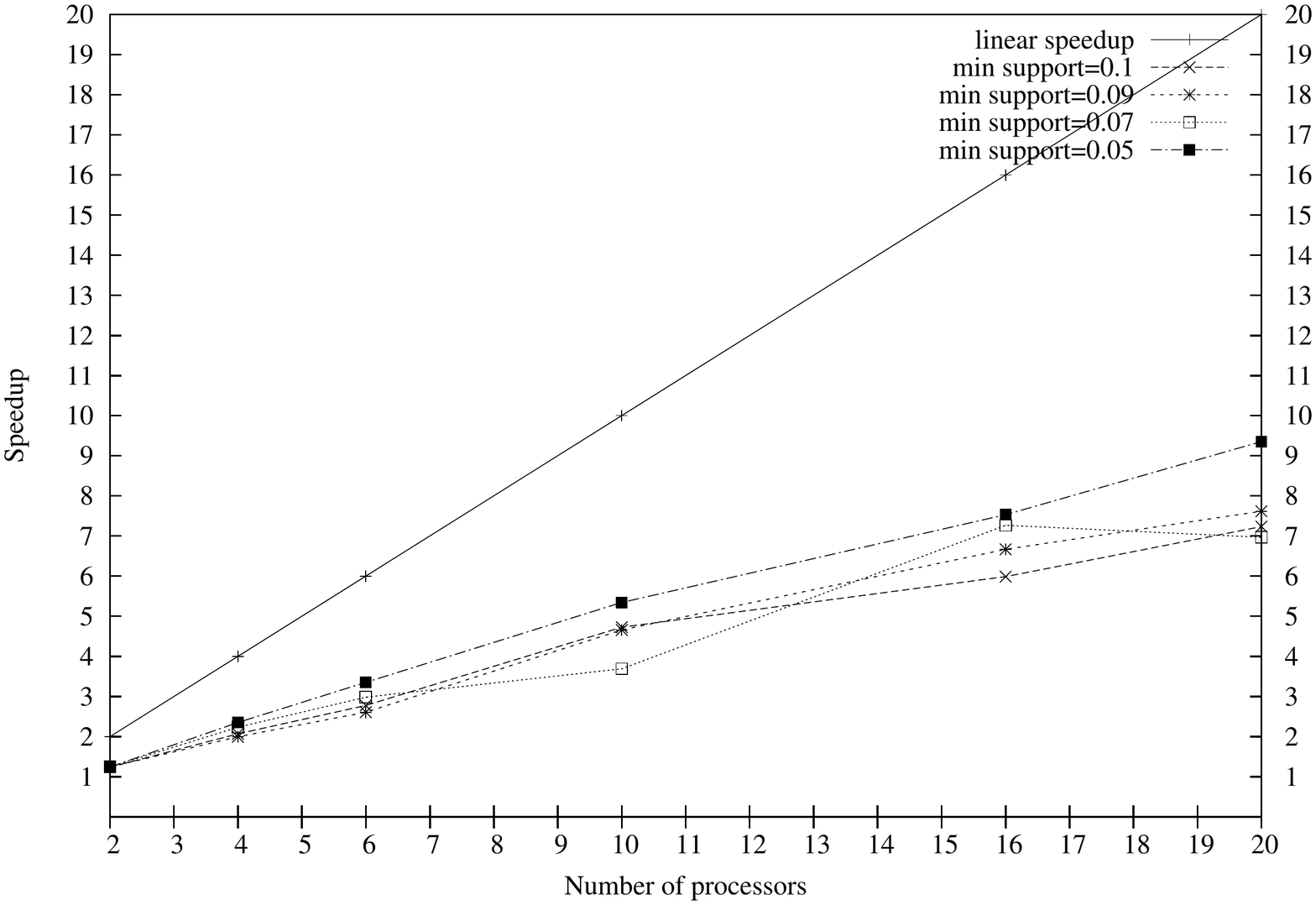}}
\caption[\hskip0.1cm Speedup measured on the {\tt T500I0.1P250PL10TL40}]{Speedups of the \scparfimiseq{}, \scparfimipar{}, and
  \scparfimireserv{} methods (from top to bottom) on the {\tt
    T500I0.1P250PL10TL40} database.}
\label{speedup-T500I0.1-2}
\end{figure*}

\begin{figure*}[!p]
\centering
\scalebox{0.35}{\includegraphics{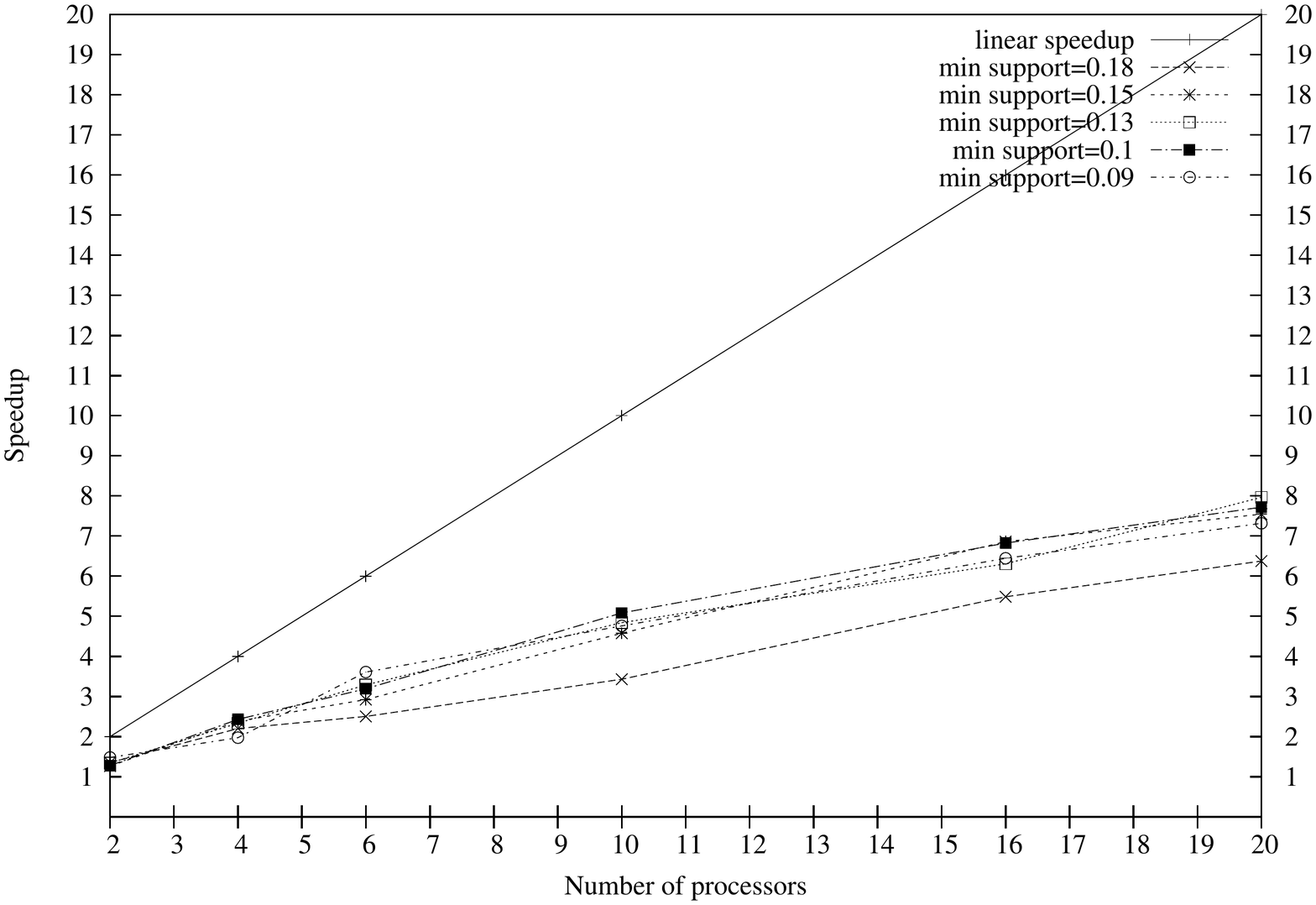}}
\scalebox{0.35}{\includegraphics{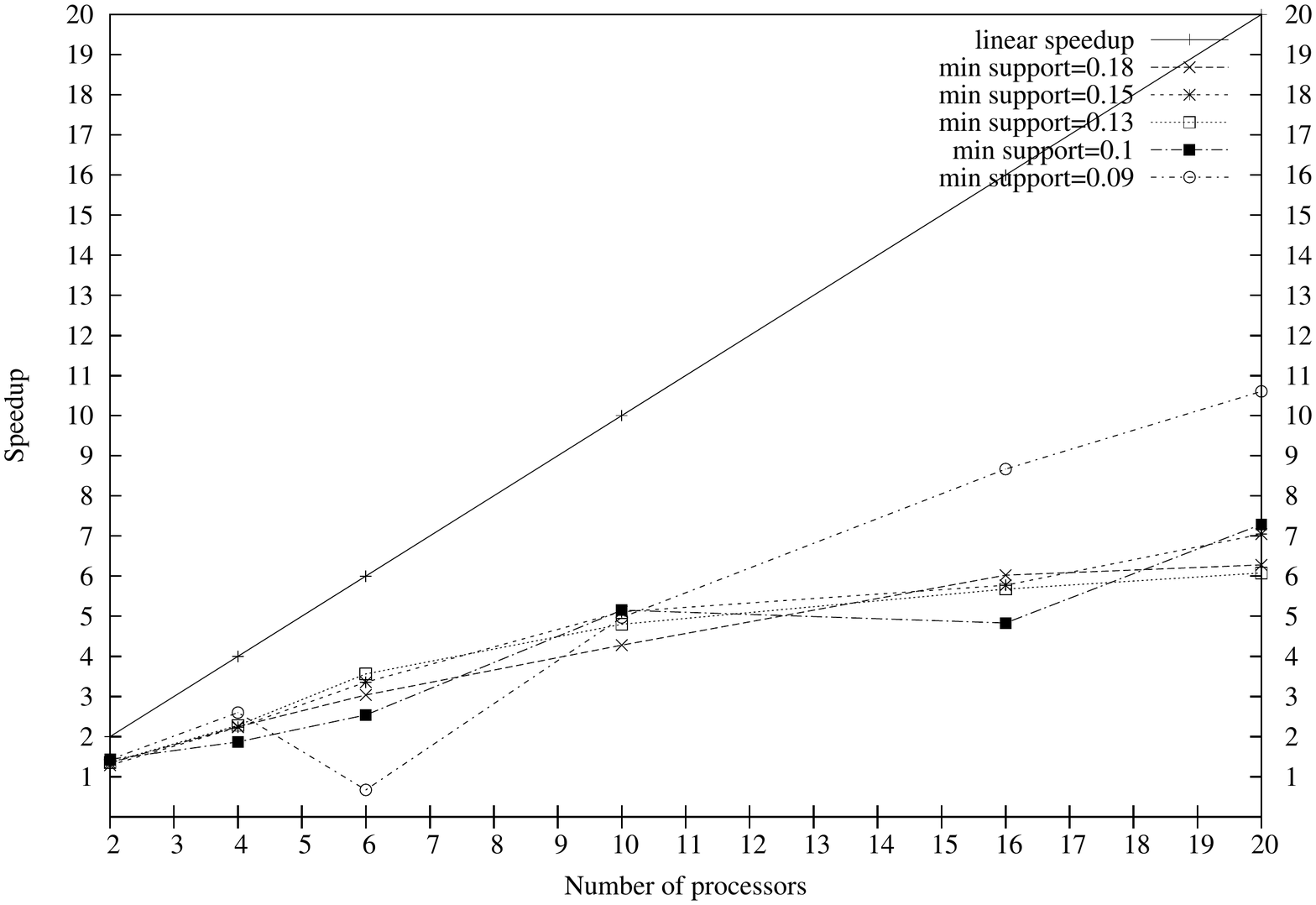}}
\scalebox{0.35}{\includegraphics{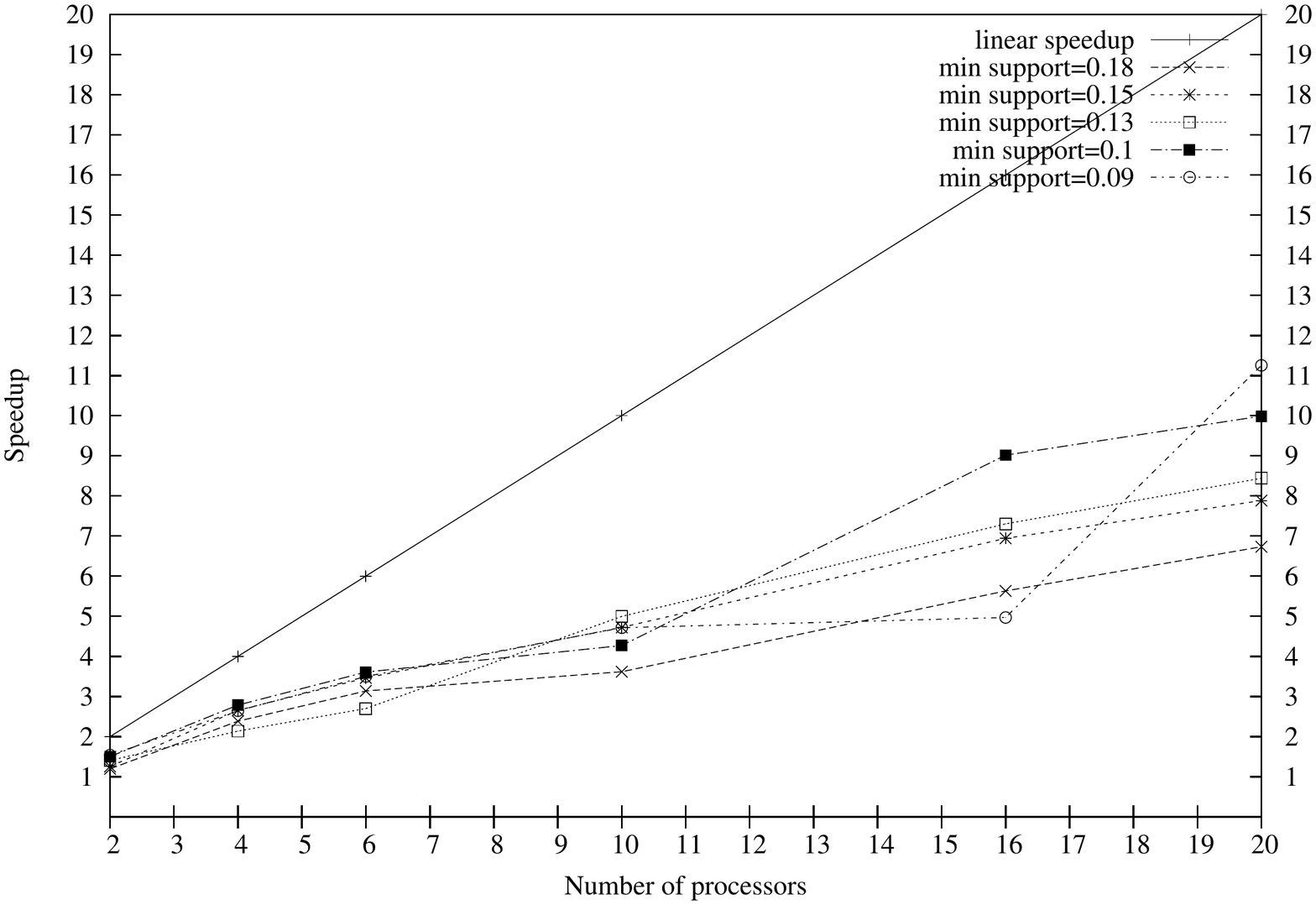}}
\caption[\hskip0.1cm Speedup measured on the {\tt T500I0.1P50PL10TL40}]{Speedups of the \scparfimiseq{}, \scparfimipar{}, and
  \scparfimireserv{} methods (from top to bottom) on the {\tt
    T500I0.1P50PL10TL40} database.}
\label{speedup-T500I0.1-3}
\end{figure*}

\begin{figure*}[!p]
\centering
\scalebox{0.35}{\includegraphics{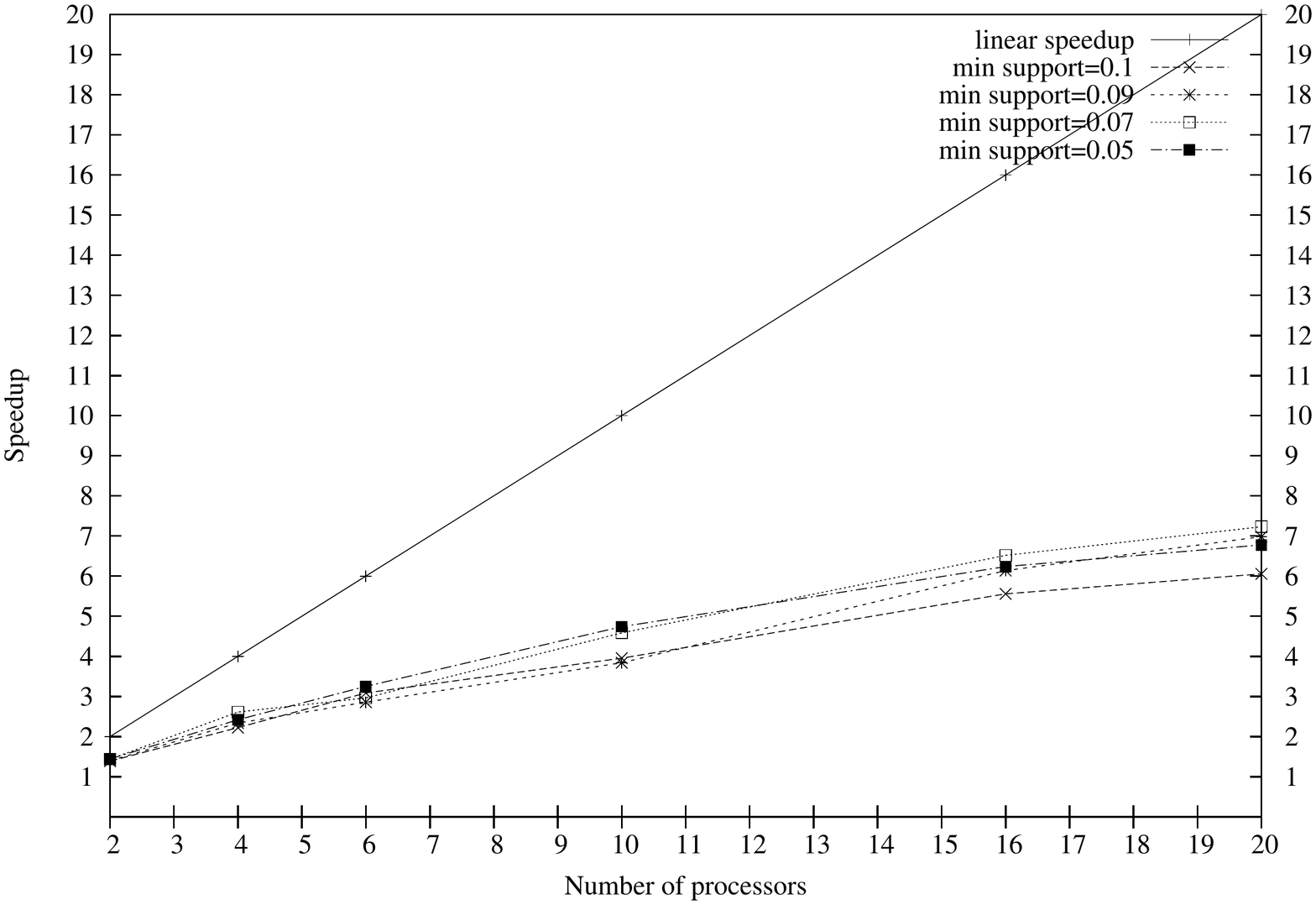}}
\scalebox{0.35}{\includegraphics{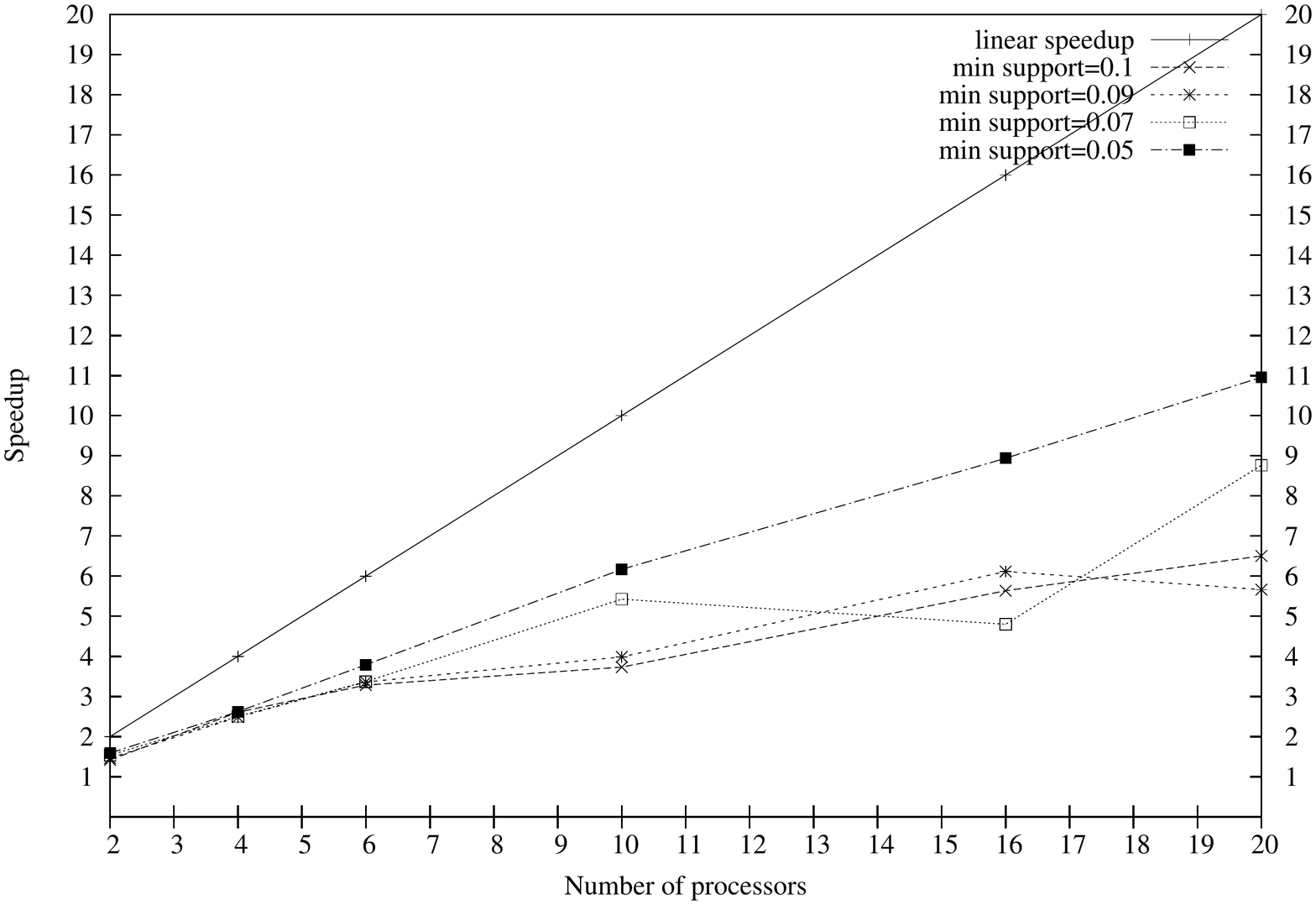}}
\scalebox{0.35}{\includegraphics{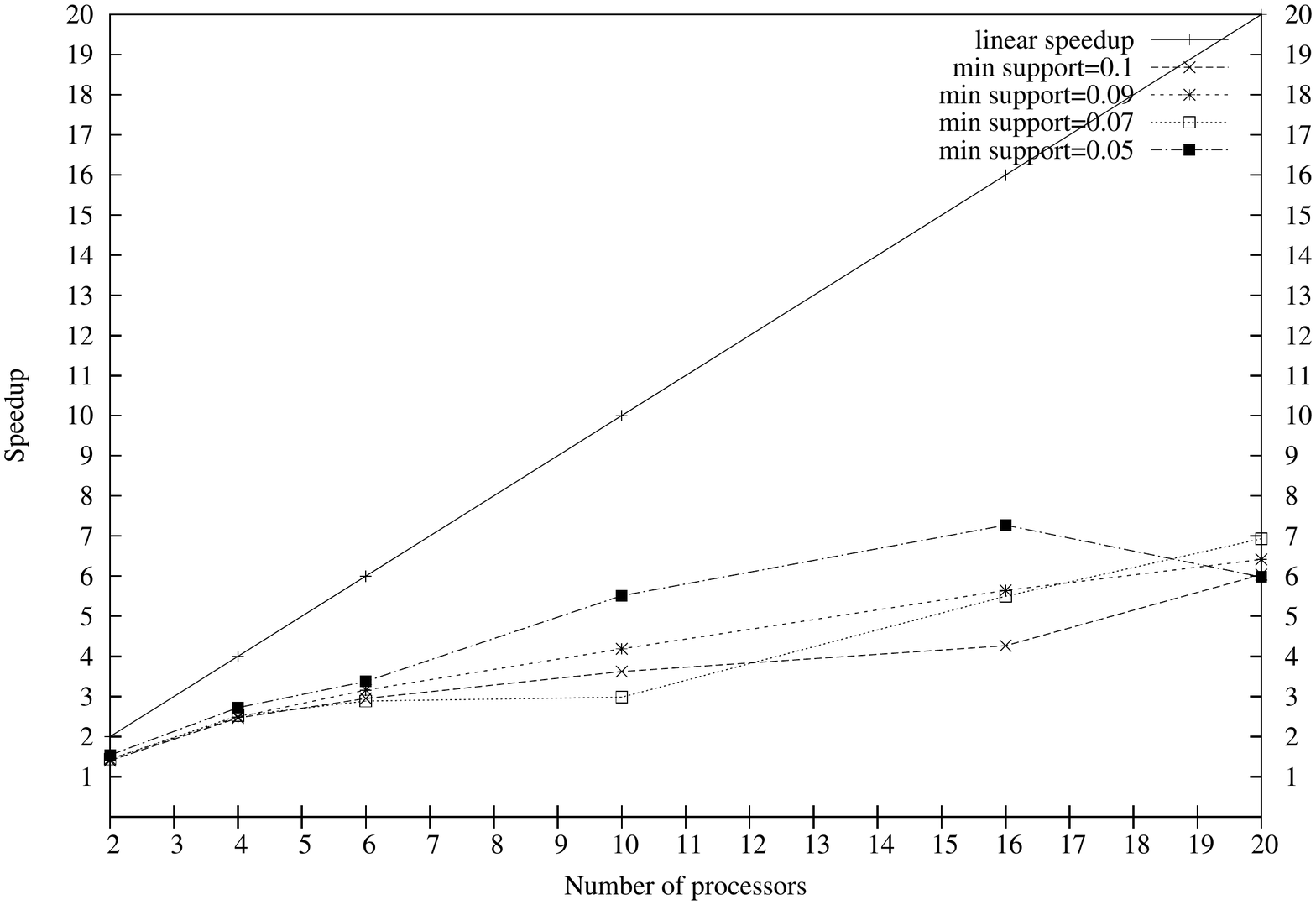}}
\caption[\hskip0.1cm Speedup measured on the {\tt T500I0.1P50PL20TL40}]{Speedups of the \scparfimiseq{}, \scparfimipar{}, and
  \scparfimireserv{} methods (from top to bottom) on the {\tt
    T500I0.1P50PL20TL40} database.}
\label{speedup-T500I0.1-4}
\end{figure*}

\begin{figure*}[!p]
\centering
\scalebox{0.35}{\includegraphics{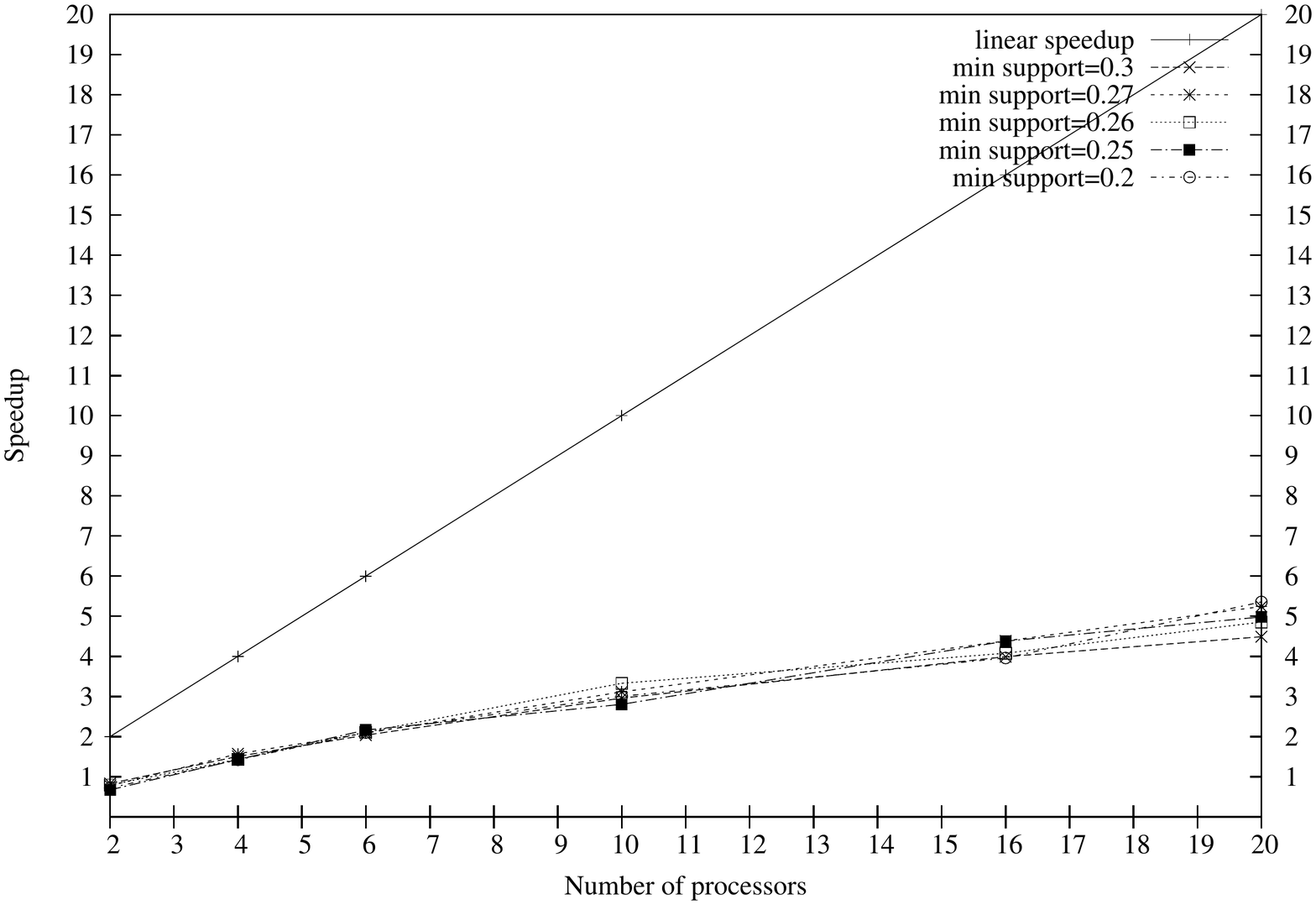}}
\scalebox{0.35}{\includegraphics{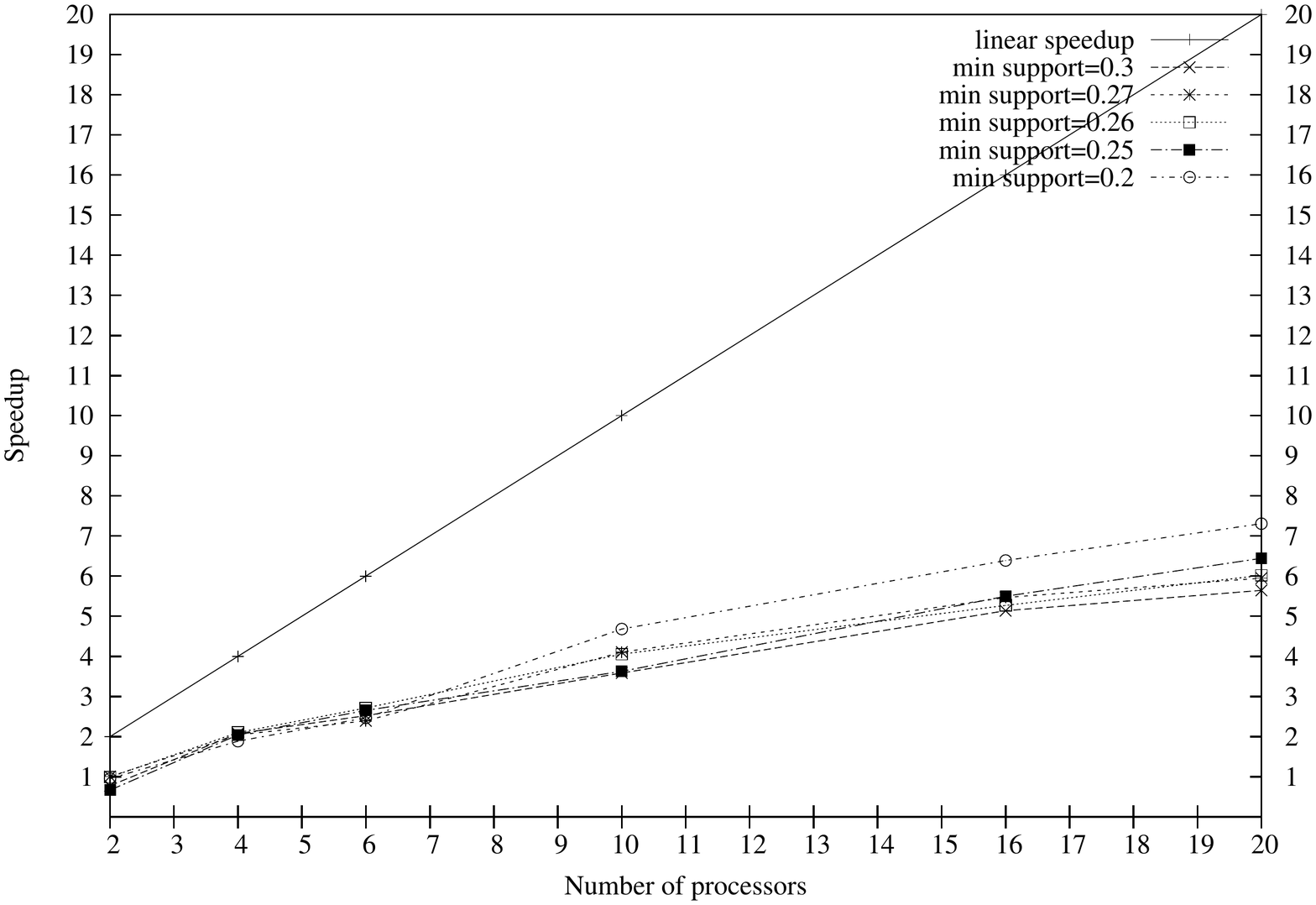}}
\scalebox{0.35}{\includegraphics{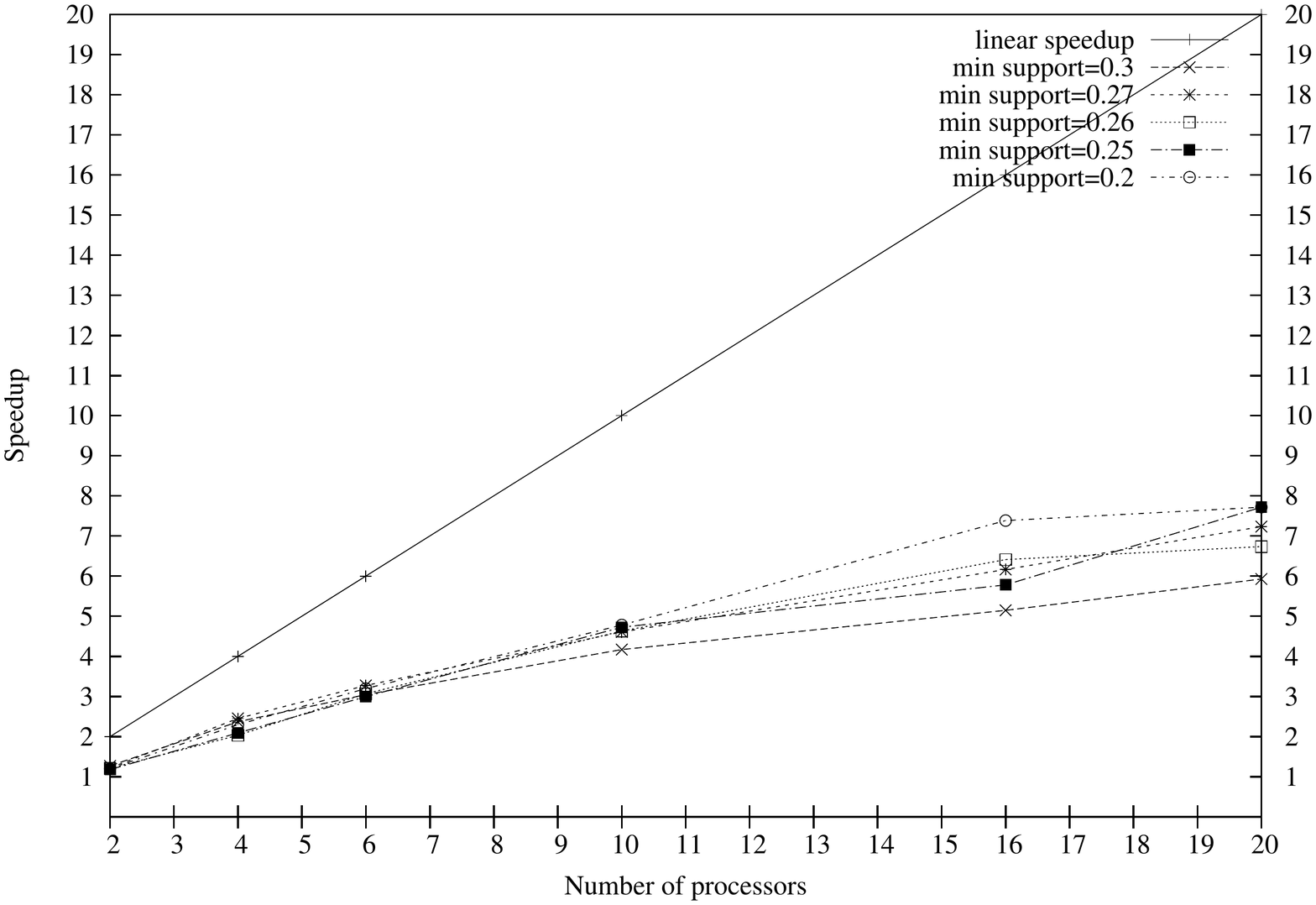}}
\caption[\hskip0.1cm Speedup measured on the {\tt T500I0.4P250PL10TL120}]{Speedups of the \scparfimiseq{}, \scparfimipar{}, and
  \scparfimireserv{} methods (from top to bottom)on the {\tt
    T500I0.4P250PL10TL120} database.}
\label{speedup-T500I0.4-2}
\end{figure*}

\begin{figure*}[!p]
\centering
\scalebox{0.35}{\includegraphics{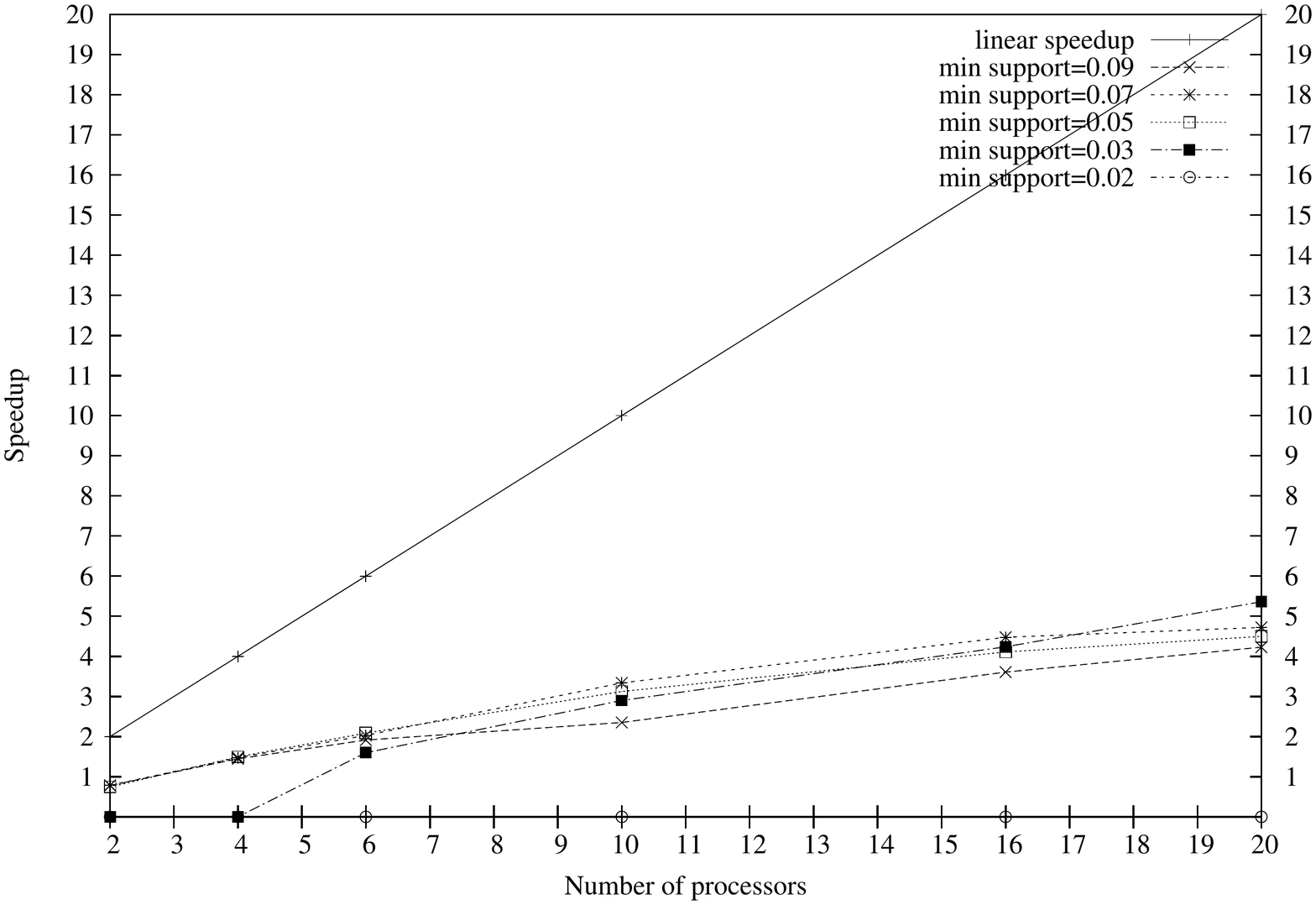}}
\scalebox{0.35}{\includegraphics{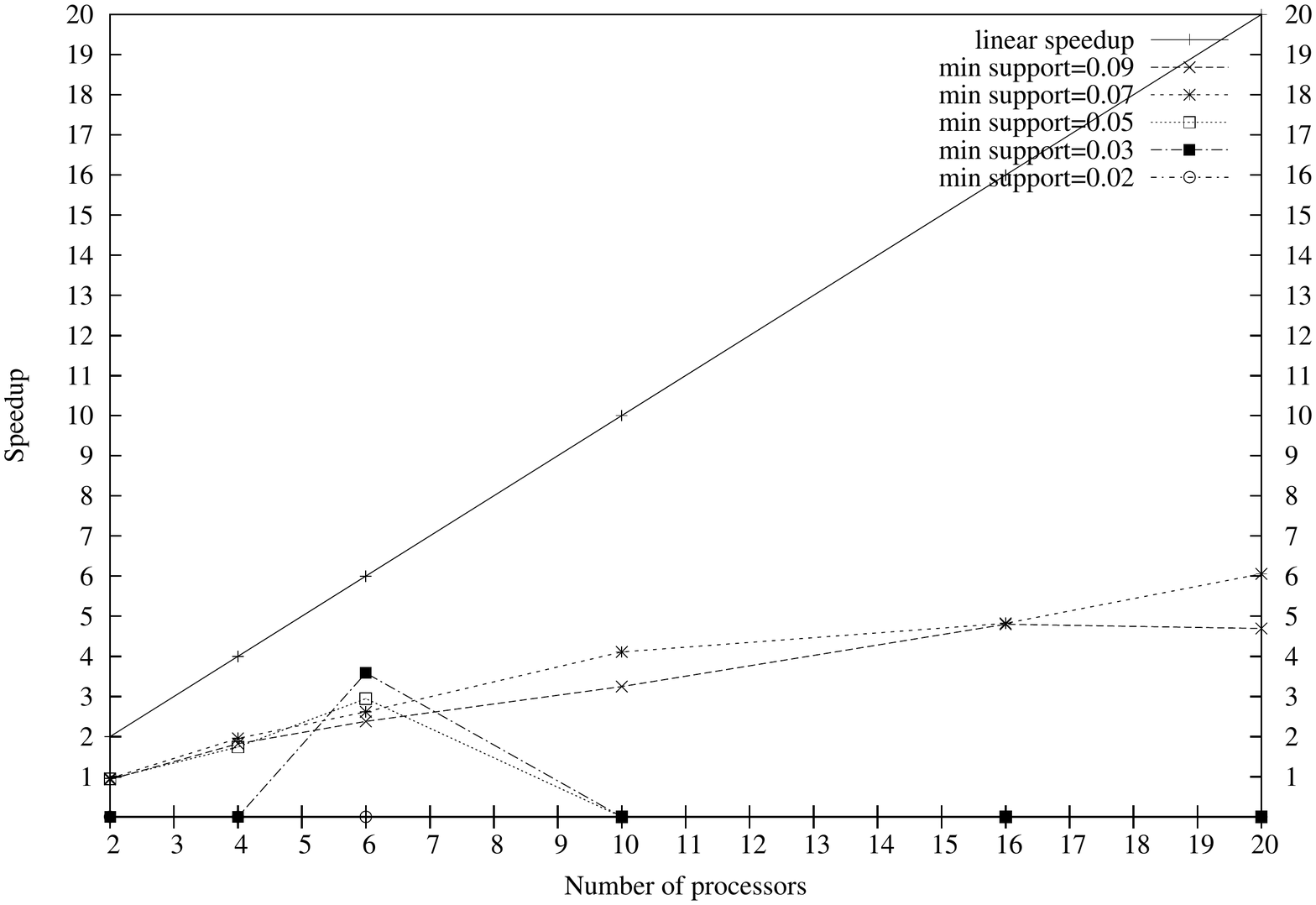}}
\scalebox{0.35}{\includegraphics{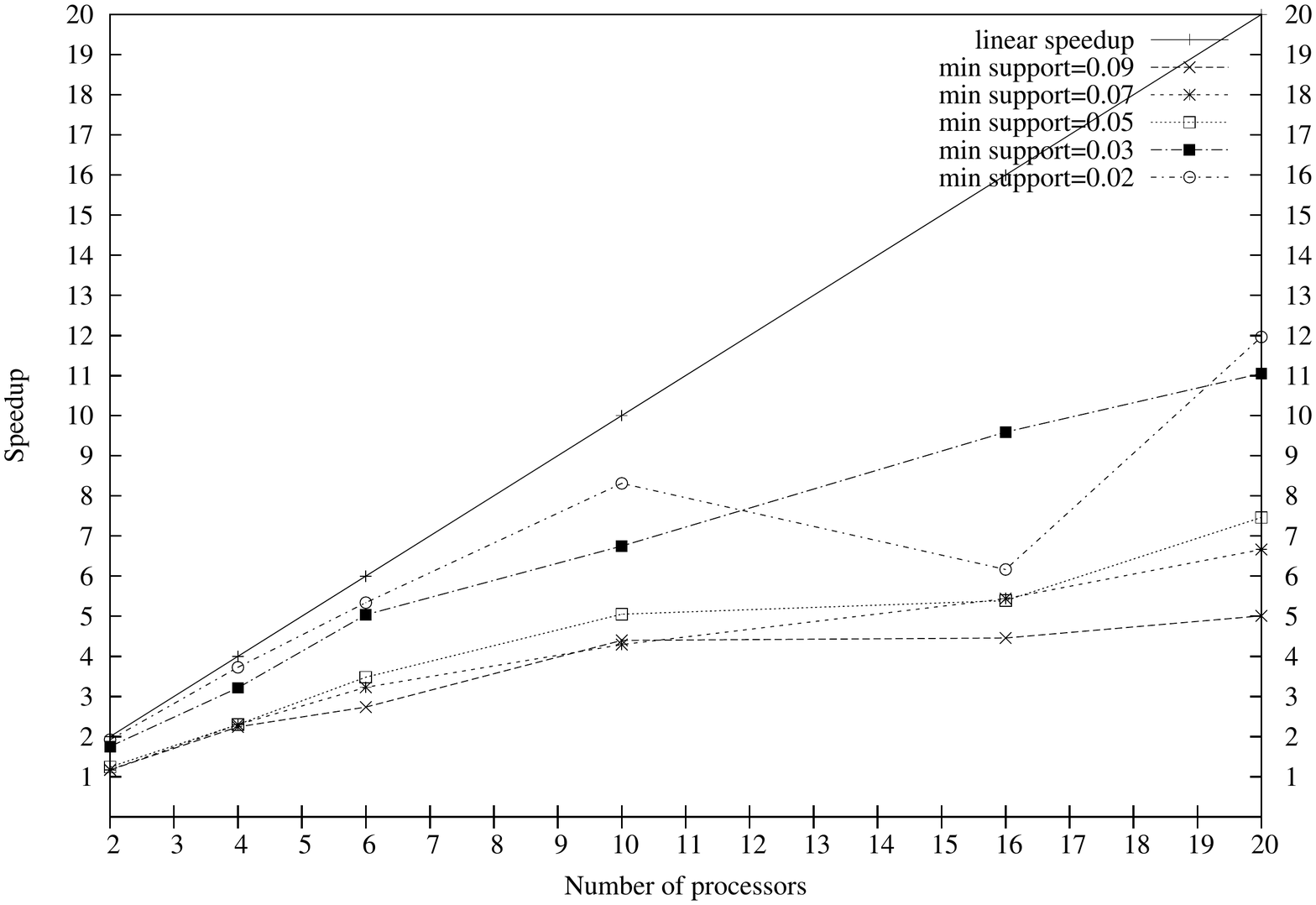}}
\caption[\hskip0.1cm Speedup measured on the {\tt T500I0.4P250PL20TL80}]{Speedups of the \scparfimiseq{}, \scparfimipar{}, and
  \scparfimireserv{} methods (from top to bottom) on the {\tt
    T500I0.4P250PL20TL80} database.}
\label{speedup-T500I0.4-3}
\end{figure*}

\begin{figure*}[!p]
\centering
\scalebox{0.35}{\includegraphics{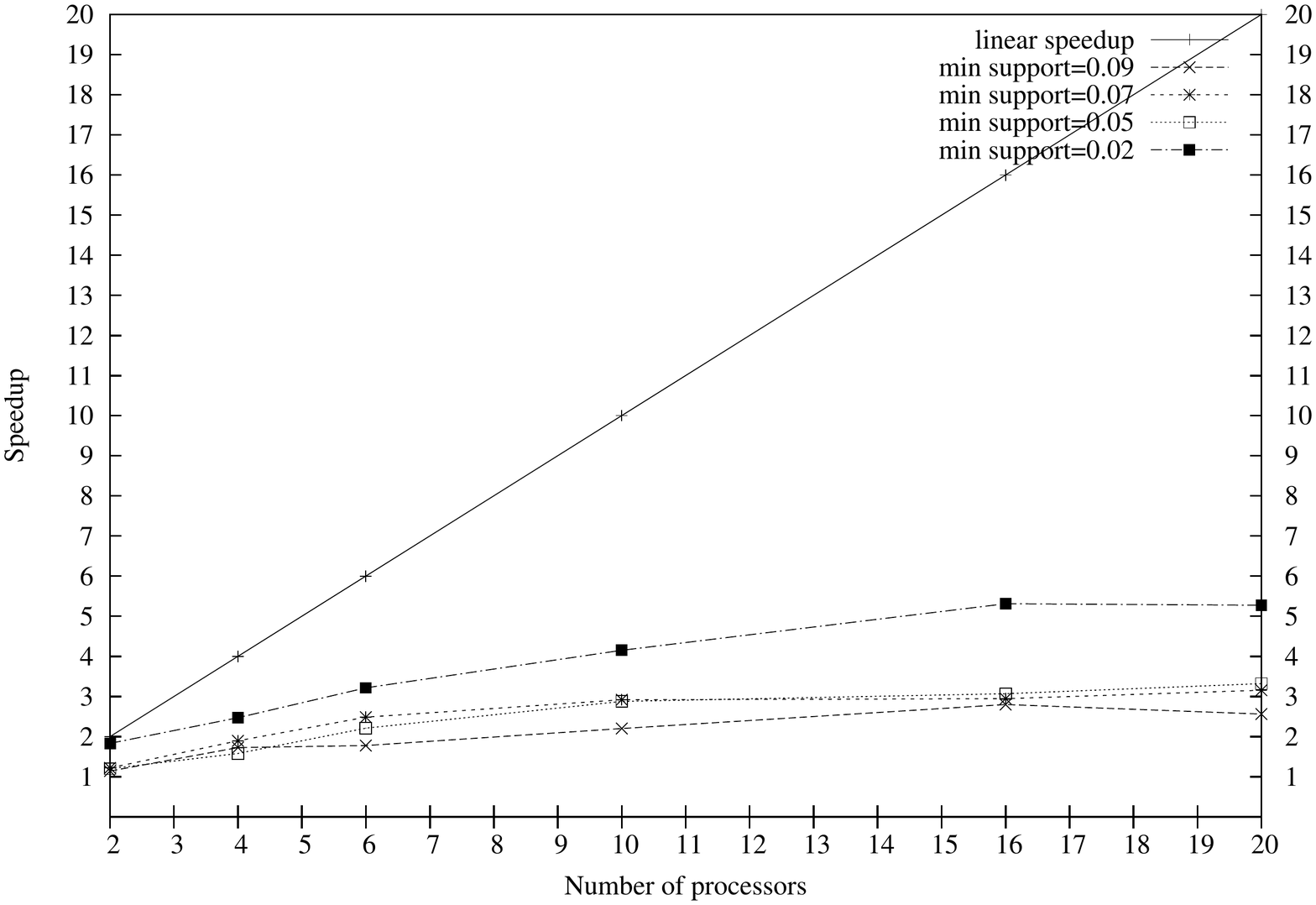}}
\scalebox{0.35}{\includegraphics{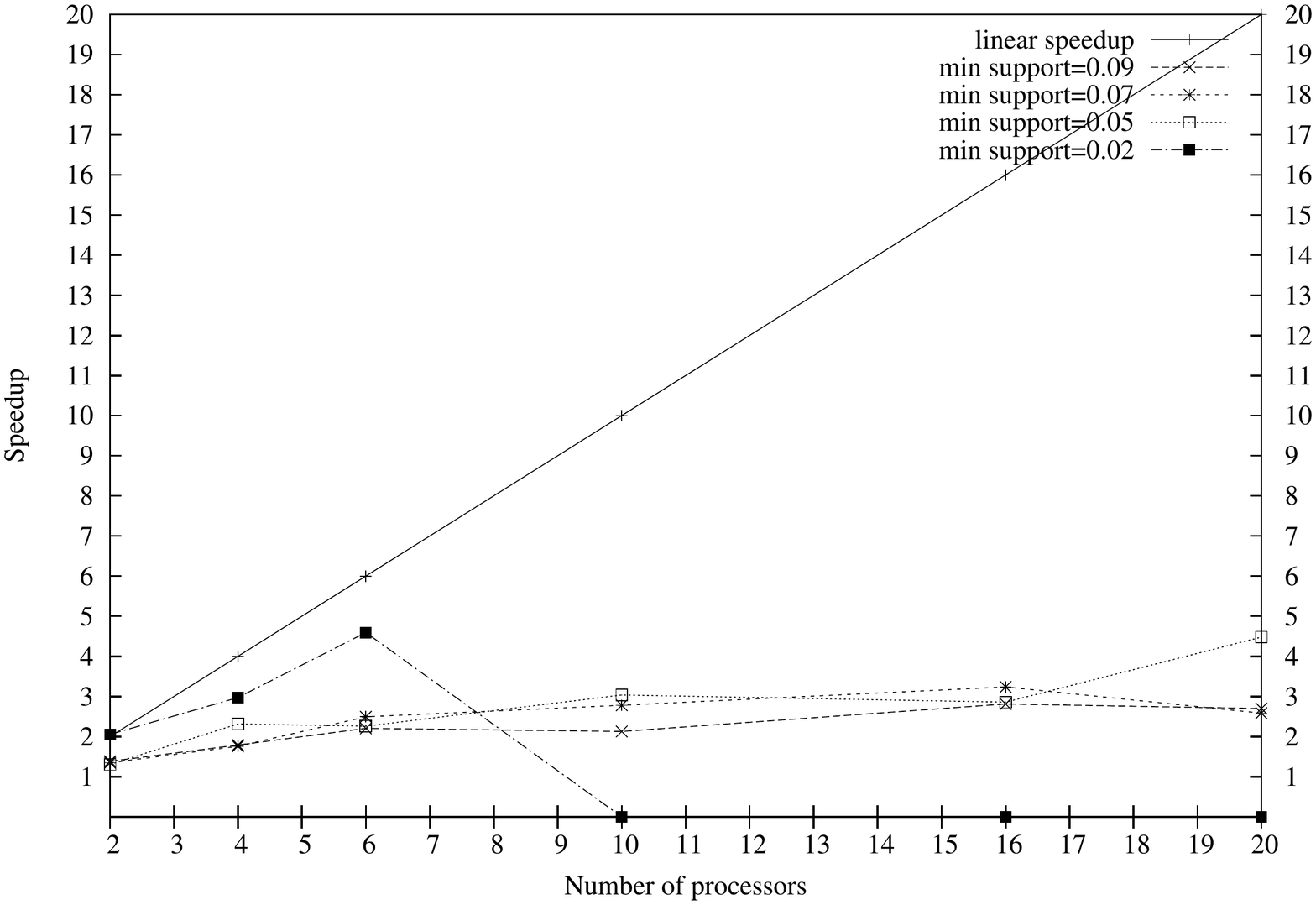}}
\scalebox{0.35}{\includegraphics{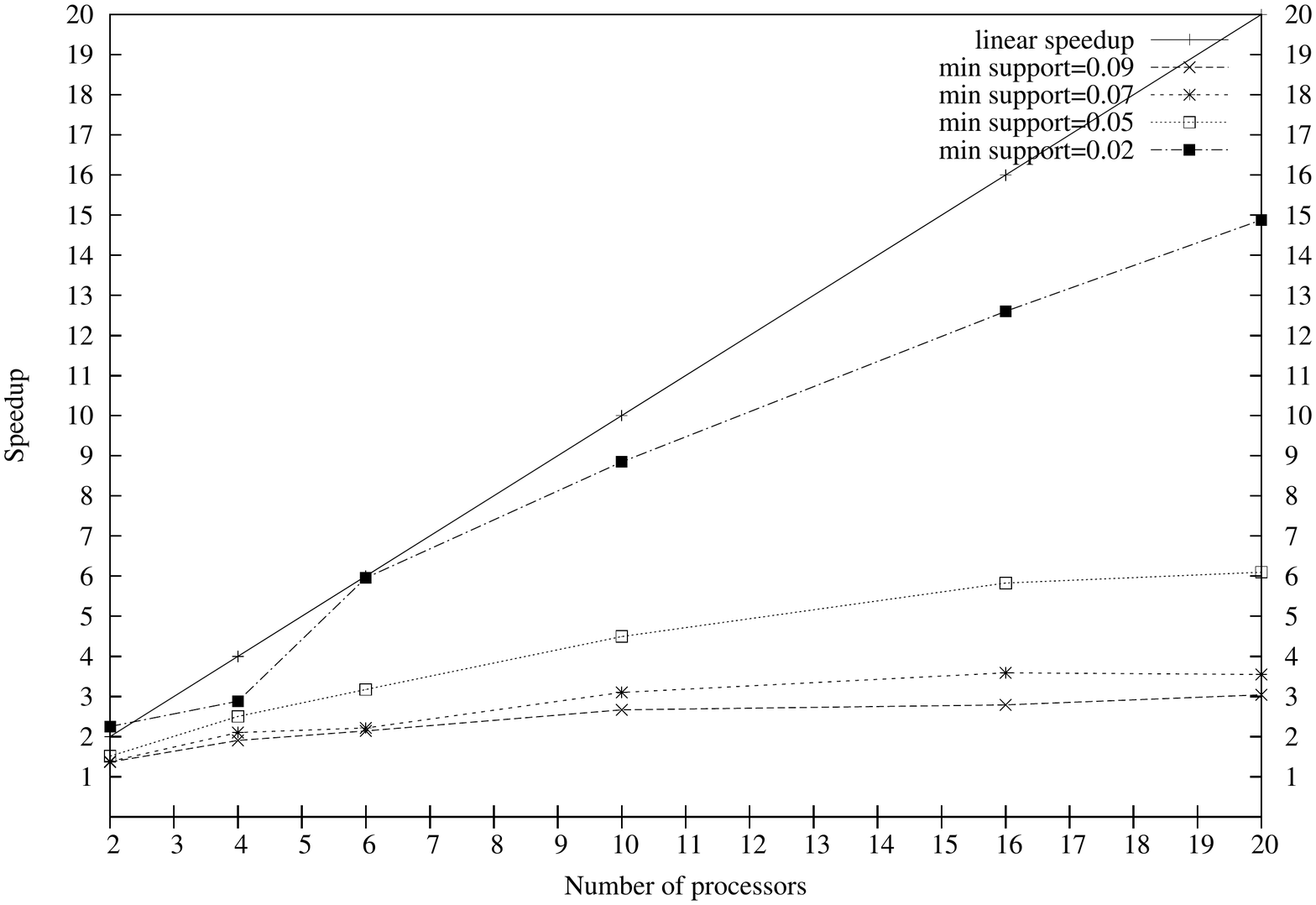}}
\caption[\hskip0.1cm Speedup measured on the {\tt T500I0.4P50PL10TL40}]{Speedups of the \scparfimiseq{}, \scparfimipar{}, and
  \scparfimireserv{} methods (from top to bottom) on the {\tt
    T500I0.4P50PL10TL40} database.}
\label{speedup-T500I0.4-4}
\end{figure*}

Because the number of graphs with speedups is prohibitive, we show the graphs
for $|\dbsmpl|=10000, |\fismpl|=19869$.  Figures
\ref{speedup-T500I0.1-1}--\ref{speedup-T500I0.4-4} clearly demonstrate that for
reasonably large and reasonably structured databases, the speedup is up to
$\approx 13$ on $20$.  The \scparfimiseq{} achieves speedup up to $\approx 8$ on
$20$ processors, the \scparfimipar{} method achieves maximal speedup up to
$\approx 11$ on 20 processors, and the \scparfimireserv{} method achieves
speedup up to $\approx 13$ on $20$ processors and in one case up to $\approx 15$
on $20$ processors, see Figure~\ref{speedup-T500I0.4-4} with measurements for database
\texttt{T500I0.4P50PL10TL40} .  The speedups $0$ indicates that the program run
out of memory. The reason of the memory exhaustion is the large amount of MFIs
and the effect of Theorem~\ref{thm:parallel-mfi-count}: due to the dynamic load
balancing each processor can found all candidates in each assigned PBEC. That
is: if the program implementing the \scparfimiseq{} method runs out of memory,
then the program implementing the \scparfimipar{} method usually also runs out
of main memory. The program implementing the \scparfimireserv{} method never
runs out of memory: the sample needs approximately the same amount of memory
independently of the value of the minimal support and the database. The
evaluation of the sampling process in Section~\ref{sec:exp-eval-pbec} shows that
the estimates are quite good. The question is, why the speedup is not almost
linear?  The answer to this question is obvious: making the sample takes some
time. Additionally, we can observe that lower values of $\rminsupp$ makes better
speedup with the two cases for the \texttt{T500I0.4P250PL20TL80} database having
a very good speedup of $\approx 13$ on $\procnum=20$ processors. The reason is
obvious: the sampling process taking the same number of sample on the database
of the same size makes better speedup, i.e., if it takes more time to compute
sequentially the FIs for given support in the given database, then the speedup
is usually better.

Tables~\ref{tbl:avg-speedup-first}--\ref{tbl:avg-speedup-last}
contain average values of the speedup for particular combination of
database and number of processors.  Some of the numbers in these
tables are typed in bold: 

\begin{enumerate}
\item First consider the tables for \scparfimipar{} and \scparfimiseq:
  the \emph{bigger} value of average speedup corresponding to the same
  database and the same number of processors is typed in bold, e.g.,
  \scparfimiseq{} has average value of speedup $1.354$ and
  \scparfimipar{} has average value $1.407$ for the database
  \texttt{T500I0.4P50PL10TL40} and $\procnum = 2$.  The value
  \textbf{$1.407$} is typed in bold because it is the maximum of the
  two values.

\item A value in the table for \scparfimireserv{} is bold if the value
  is the \emph{biggest} value of average speedup corresponding to the
  same database and the same number of processors for all three
  methods, e.g., the average speedup of \scparfimireserv{} for the
  \texttt{T500I0.4P50PL10TL40} database and $\procnum=2$, the value
  $1.543$, is typed in bold because it is the biggest of the three
  values: $1.354$, $1.407$, $1.543$.
\end{enumerate}

\begin{figure*}[!p]

\centering
\scalebox{0.35}{\includegraphics{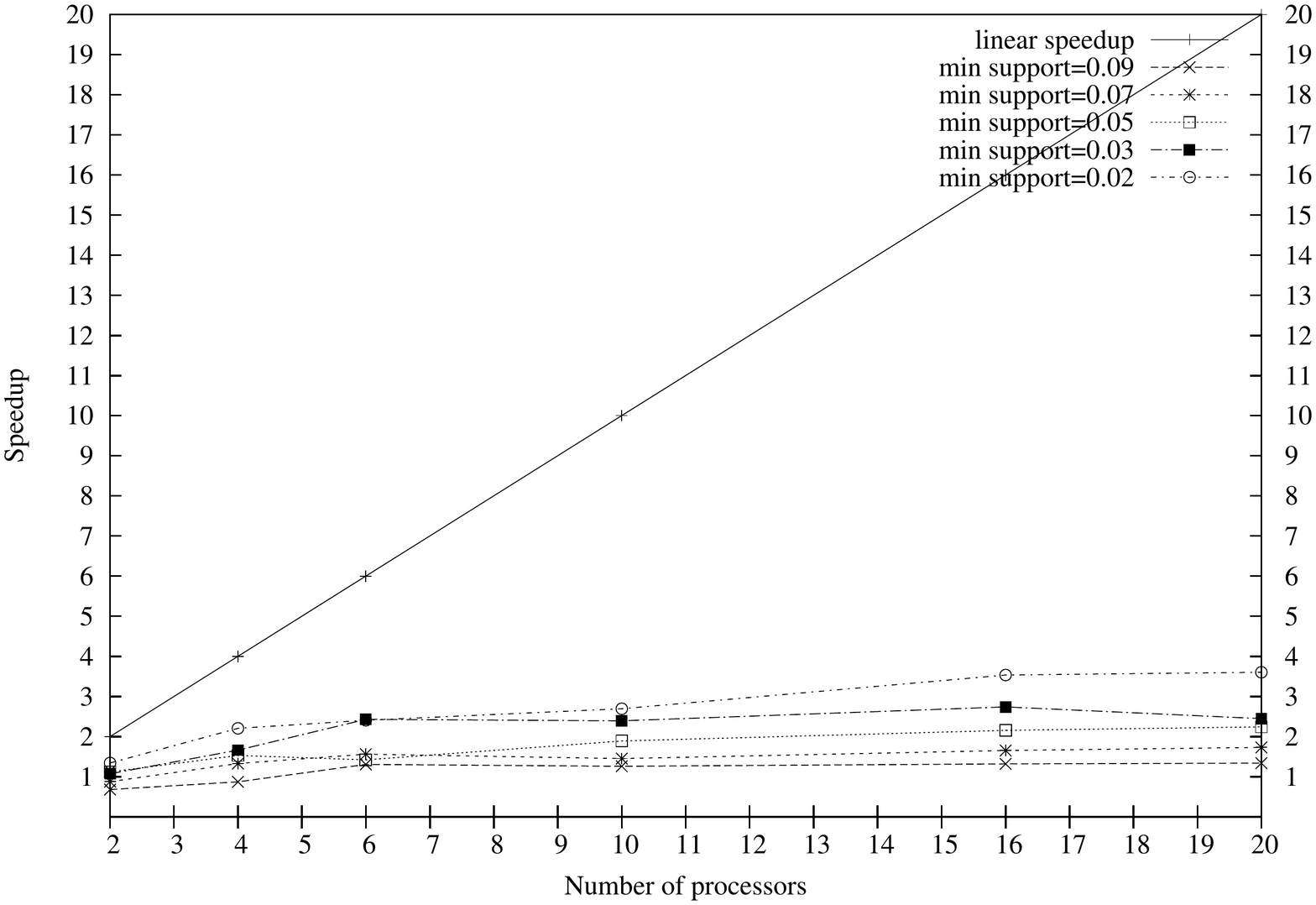}}
\scalebox{0.35}{\includegraphics{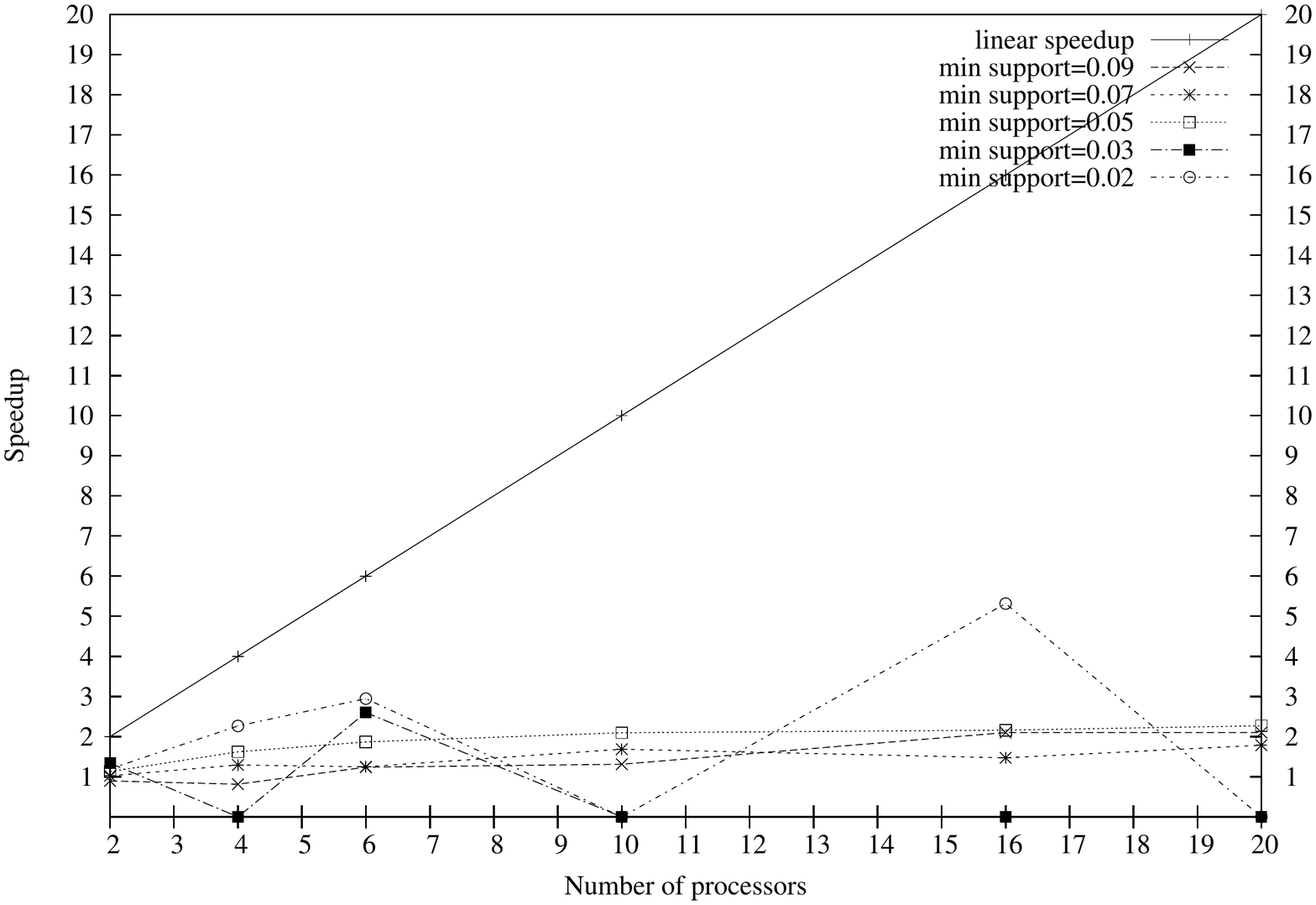}}
\scalebox{0.35}{\includegraphics{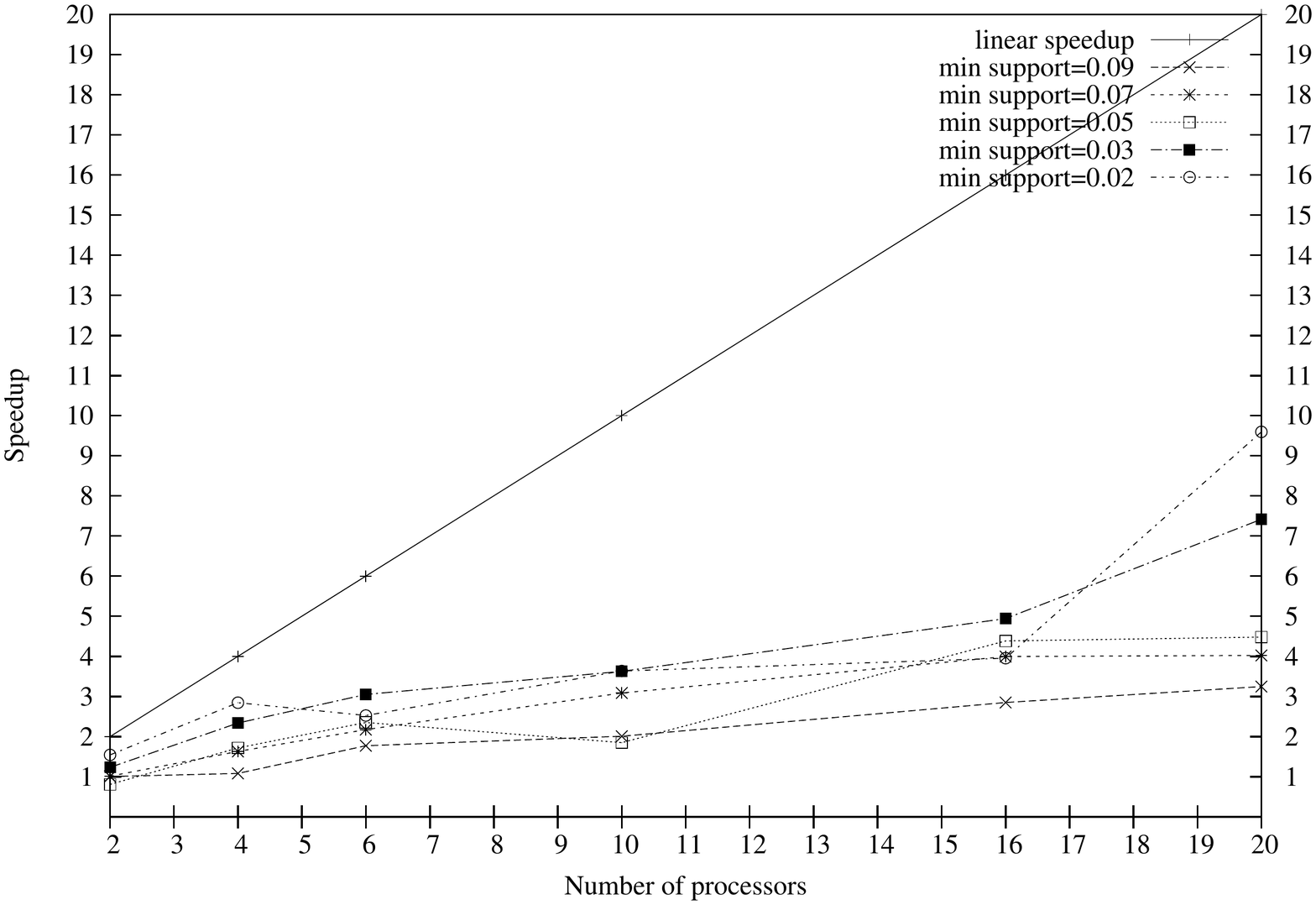}}
\caption[\hskip0.1cm Speedup measured on the {\tt T500I1P100PL20TL50}]{Speedups of the \scparfimiseq{}, \scparfimipar{}, and
  \scparfimireserv{} methods (from top to bottom) on the {\tt
    T500I1P100PL20TL50} database.}
\label{speedup-T500I1-1}
\end{figure*}

From the graphs on Figures~\ref{speedup-T500I0.1-1}--\ref{speedup-T500I1-1} and
tables on Table~\ref{tbl:avg-speedup-14450-26492} it follows that the
\scparfimipar{} is usually faster then the \scparfimiseq{} for the number of
processors $\procnum \leq 20$. The \scparfimireserv{} performs better then: a)
the \scparfimipar{} and b) the \scparfimiseq{} method. Our hypothesis is that
the \scparfimireserv{} makes better estimates of the relative size of the union
of PBECs, see Section~\ref{chap:approx-counting} for discussion of the sampling
process. Still, there is a possibility to improve the speedup of the
\scparfimireserv{} method, for discussion see Chapter \ref{chap:future-work}.
Additionally, there is an advantage of the \scparfimireserv{} over the two other
methods: it is not necessary to compute the MFIs. The number of MFIs can be very
large and the program implementing the \scparfimiseq{} method or the
\scparfimipar{} can run out of main memory. This happens for some supports of
some databases, e.g., \texttt{T500I0.4P250PL20TL80},
\texttt{T500I0.4P50PL10TL40}, and \texttt{T500I1P100PL20TL50} (indicated by the
speedup value $0$).

Very low speedups were obtained for the database with 1000 items in Figure
\ref{speedup-T500I1-1}, the {\tt T500I1P100PL20TL50} database. The reason for
such a bad speedup lies in Phase~1 and 2. There is always a processor that has
much bigger running time in Phase~4. For example, for $\rminsupp=0.02$ and for
$\procnum=10$ the execution time of Phase~4 is (in seconds): $194$, $1199$,
$319$, $245$, $536$, $357$, $477$, $212$, $332$, $212$. A sum of these times is
$4087$ seconds, the sequential algorithm runs $\approx 3800$ seconds. The
probability of error of the estimates made in Phase~2 of {\tt
  T500I1P100PL20TL50} are competitive to other databases, see
Figure~\ref{fig:error-T500I1P100PL20TL50}. The best speedup that achieved by
\scparfimireserv{} is $\approx 8$ on $20$ processors for $\rminsupp = 0.02$. In
other cases the speedup is not so good. The reason of such behaviour is unknown.

Tables~\ref{tbl:avg-speedup-first}--\ref{tbl:avg-speedup-last} show
the average speedup for the parameters shown in
Table~\ref{tbl:experiments-number-of-samples}. The best combination of
values for the \scparfimireserv{} algorithm, e.g., the best values of
$|\dbsmpl|$ and $|\fismpl|$ is the following:

\begin{table}[h]\label{tbl:experiments-number-of-samples}
\centering
\begin{tabular}{|l|c|c|} \hline
Variant of our method &$|\dbsmpl|$               & $|\fismpl|$ \\ \hline
\scparfimireserv{} & 10000 & 19869 \\ \hline
\scparfimipar{} & 10000 & 33115 \\ \hline
\scparfimiseq{} & 10000 & 19869 \\ \hline
\end{tabular}
\caption{Best combintation of $|\dbsmpl|$ and $|\fismpl|$ for $\procnum=20$}\label{fig:best-sample-sizes}
\end{table}

We have made some experiments with the parameter $\alpha$ and chosen
$\alpha=0.3$: this value of $\alpha$ assures good granularity of the
partitioning of $\allfi$ using the PBECs, i.e., the PBECs are small
enough so the \textsc{LPT-MakeSpan} algorithm makes partitions of size
$1/\procnum$. The value of $\rho$ can be chosen between
$0.0007$-$0.003$, see Section~\ref{sec:exp-eval-pbec}. Even that there
is large number of parameters, our experiments show that there is not
so big difference between the values of $|\dbsmpl|$ and
$|\fismpl|$. Our hypothesis is the value of $\alpha$ can be set to
$\alpha=0.3$ and the value of $\rho$ can be set to $\rho=0.001$. These
setting of parameters seems to be sufficient for all our experiments.




\clearpage

\begin{table}[!p]
\def\dbsmplsize{10000}
\def\pbecsmplsize{19869}
{
\centering
\input{tables/avg_speedups/avg_speedup-\dbsmplsize-\pbecsmplsize-fpmax_eclat.ltx.bold}
\medskip
\input{tables/avg_speedups/avg_speedup-\dbsmplsize-\pbecsmplsize-pfpmax_eclat.ltx.bold}
\medskip
\input{tables/avg_speedups/avg_speedup-\dbsmplsize-\pbecsmplsize-reservoir.ltx.bold}
}
\caption[Average speedup of the proposed methods for $|\dbsmpl|=\dbsmplsize $ and $|\fismpl|=\pbecsmplsize$.]{Average speedup
of the proposed methods for $|\dbsmpl|=\dbsmplsize$ and $|\fismpl|=\pbecsmplsize$.}
\label{tbl:avg-speedup-10000-19869}\label{tbl:avg-speedup-first}
\end{table}

\begin{table}[!p]
\def\dbsmplsize{10000}
\def\pbecsmplsize{26492}
{
\centering
\input{tables/avg_speedups/avg_speedup-\dbsmplsize-\pbecsmplsize-fpmax_eclat.ltx.bold}
\medskip
\input{tables/avg_speedups/avg_speedup-\dbsmplsize-\pbecsmplsize-pfpmax_eclat.ltx.bold}
\medskip
\input{tables/avg_speedups/avg_speedup-\dbsmplsize-\pbecsmplsize-reservoir.ltx.bold}
}
\caption[Average speedup of the proposed methods for $|\dbsmpl|=\dbsmplsize $ and $|\fismpl|=\pbecsmplsize$.]{Average speedup
of the proposed methods for $|\dbsmpl|=\dbsmplsize$ and $|\fismpl|=\pbecsmplsize$.}
\label{tbl:avg-speedup-10000-26492}
\end{table}

\begin{table}[!p]
\def\dbsmplsize{10000}
\def\pbecsmplsize{33115}
{
\centering
\input{tables/avg_speedups/avg_speedup-\dbsmplsize-\pbecsmplsize-fpmax_eclat.ltx.bold}
\medskip
\input{tables/avg_speedups/avg_speedup-\dbsmplsize-\pbecsmplsize-pfpmax_eclat.ltx.bold}
\medskip
\input{tables/avg_speedups/avg_speedup-\dbsmplsize-\pbecsmplsize-reservoir.ltx.bold}
}
\caption[Average speedup of the proposed methods for $|\dbsmpl|=\dbsmplsize $ and $|\fismpl|=\pbecsmplsize$.]{Average speedup
of the proposed methods for $|\dbsmpl|=\dbsmplsize$ and $|\fismpl|=\pbecsmplsize$.}
\label{tbl:avg-speedup-10000-33115}
\end{table}

\begin{table}[!p]
\def\dbsmplsize{14450}
\def\pbecsmplsize{13246}
{
\centering
\input{tables/avg_speedups/avg_speedup-\dbsmplsize-\pbecsmplsize-fpmax_eclat.ltx.bold}
\medskip
\input{tables/avg_speedups/avg_speedup-\dbsmplsize-\pbecsmplsize-pfpmax_eclat.ltx.bold}
\medskip
\input{tables/avg_speedups/avg_speedup-\dbsmplsize-\pbecsmplsize-reservoir.ltx.bold}
}
\caption[Average speedup of the proposed methods for $|\dbsmpl|=\dbsmplsize $ and $|\fismpl|=\pbecsmplsize$.]{Average speedup
of the proposed methods for $|\dbsmpl|=\dbsmplsize$ and $|\fismpl|=\pbecsmplsize$.}
\label{tbl:avg-speedup-14450-13246}
\end{table}

\begin{table}[!p]
\def\dbsmplsize{14450}
\def\pbecsmplsize{19869}
{
\centering
\input{tables/avg_speedups/avg_speedup-\dbsmplsize-\pbecsmplsize-fpmax_eclat.ltx.bold}
\medskip
\input{tables/avg_speedups/avg_speedup-\dbsmplsize-\pbecsmplsize-pfpmax_eclat.ltx.bold}
\medskip
\input{tables/avg_speedups/avg_speedup-\dbsmplsize-\pbecsmplsize-reservoir.ltx.bold}
}
\caption[Average speedup of the proposed methods for $|\dbsmpl|=\dbsmplsize $ and $|\fismpl|=\pbecsmplsize$.]{Average speedup
of the proposed methods for $|\dbsmpl|=\dbsmplsize$ and $|\fismpl|=\pbecsmplsize$.}
\label{tbl:avg-speedup-14450-19869}
\end{table}

\begin{table}[!p]
\def\dbsmplsize{14450}
\def\pbecsmplsize{26492}
{
\centering
\input{tables/avg_speedups/avg_speedup-\dbsmplsize-\pbecsmplsize-fpmax_eclat.ltx.bold}
\medskip
\input{tables/avg_speedups/avg_speedup-\dbsmplsize-\pbecsmplsize-pfpmax_eclat.ltx.bold}
\medskip
\input{tables/avg_speedups/avg_speedup-\dbsmplsize-\pbecsmplsize-reservoir.ltx.bold}
}
\caption[Average speedup of the proposed methods for $|\dbsmpl|=\dbsmplsize $ and $|\fismpl|=\pbecsmplsize$.]{Average speedup
of the proposed methods for $|\dbsmpl|=\dbsmplsize$ and $|\fismpl|=\pbecsmplsize$.}
\label{tbl:avg-speedup-14450-26492}
\end{table}

\begin{table}[!p]
\def\dbsmplsize{14450}
\def\pbecsmplsize{33115}
{
\centering
\input{tables/avg_speedups/avg_speedup-\dbsmplsize-\pbecsmplsize-fpmax_eclat.ltx.bold}
\medskip
\input{tables/avg_speedups/avg_speedup-\dbsmplsize-\pbecsmplsize-pfpmax_eclat.ltx.bold}
\medskip
\input{tables/avg_speedups/avg_speedup-\dbsmplsize-\pbecsmplsize-reservoir.ltx.bold}
}
\caption[Average speedup of the proposed methods for $|\dbsmpl|=\dbsmplsize $ and $|\fismpl|=\pbecsmplsize$.]{Average speedup
of the proposed methods for $|\dbsmpl|=\dbsmplsize$ and $|\fismpl|=\pbecsmplsize$.}
\label{tbl:avg-speedup-14450-33115}
\end{table}

\begin{table}[!p]
\def\dbsmplsize{14450}
\def\pbecsmplsize{39738}
{
\centering
\input{tables/avg_speedups/avg_speedup-\dbsmplsize-\pbecsmplsize-fpmax_eclat.ltx.bold}
\medskip
\input{tables/avg_speedups/avg_speedup-\dbsmplsize-\pbecsmplsize-pfpmax_eclat.ltx.bold}
\medskip
\input{tables/avg_speedups/avg_speedup-\dbsmplsize-\pbecsmplsize-reservoir.ltx.bold}
}
\caption[Average speedup of the proposed methods for $|\dbsmpl|=\dbsmplsize $ and $|\fismpl|=\pbecsmplsize$.]{Average speedup
of the proposed methods for $|\dbsmpl|=\dbsmplsize$ and $|\fismpl|=\pbecsmplsize$.}
\label{tbl:avg-speedup-14450-39738}
\end{table}

\begin{table}[!p]
\def\dbsmplsize{20000}
\def\pbecsmplsize{19869}
{
\centering
\input{tables/avg_speedups/avg_speedup-\dbsmplsize-\pbecsmplsize-fpmax_eclat.ltx.bold}
\medskip
\input{tables/avg_speedups/avg_speedup-\dbsmplsize-\pbecsmplsize-pfpmax_eclat.ltx.bold}
\medskip
\input{tables/avg_speedups/avg_speedup-\dbsmplsize-\pbecsmplsize-reservoir.ltx.bold}
}
\caption[Average speedup of the proposed methods for $|\dbsmpl|=\dbsmplsize $ and $|\fismpl|=\pbecsmplsize$.]{Average speedup
of the proposed methods for $|\dbsmpl|=\dbsmplsize$ and $|\fismpl|=\pbecsmplsize$.}
\label{tbl:avg-speedup-20000-19869}
\end{table}

\begin{table}[!p]
\def\dbsmplsize{20000}
\def\pbecsmplsize{26492}
{
\centering
\input{tables/avg_speedups/avg_speedup-\dbsmplsize-\pbecsmplsize-fpmax_eclat.ltx.bold}
\medskip
\input{tables/avg_speedups/avg_speedup-\dbsmplsize-\pbecsmplsize-pfpmax_eclat.ltx.bold}
\medskip
\input{tables/avg_speedups/avg_speedup-\dbsmplsize-\pbecsmplsize-reservoir.ltx.bold}
}
\caption[Average speedup of the proposed methods for $|\dbsmpl|=\dbsmplsize $ and $|\fismpl|=\pbecsmplsize$.]{Average speedup
of the proposed methods for $|\dbsmpl|=\dbsmplsize$ and $|\fismpl|=\pbecsmplsize$.}
\label{tbl:avg-speedup-20000-26492}
\end{table}

\begin{table}[!p]
\def\dbsmplsize{20000}
\def\pbecsmplsize{33115}
{
\centering
\input{tables/avg_speedups/avg_speedup-\dbsmplsize-\pbecsmplsize-fpmax_eclat.ltx.bold}
\medskip
\input{tables/avg_speedups/avg_speedup-\dbsmplsize-\pbecsmplsize-pfpmax_eclat.ltx.bold}
\medskip
\input{tables/avg_speedups/avg_speedup-\dbsmplsize-\pbecsmplsize-reservoir.ltx.bold}
}
\caption[Average speedup of the proposed methods for $|\dbsmpl|=\dbsmplsize $ and $|\fismpl|=\pbecsmplsize$.]{Average speedup
of the proposed methods for $|\dbsmpl|=\dbsmplsize$ and $|\fismpl|=\pbecsmplsize$.}
\label{tbl:avg-speedup-20000-33115}\label{tbl:avg-speedup-last}
\end{table}

\clearpage


\clearpage


\section{The evaluation of the database replication experiments}\label{sec:exp-eval-db-repl}

We have evaluated the improvement of the database replication and the database
replication itself on the real databases. We have not used the data generated by
the IBM generator, the replication factor in this data is almost $\procnum$. The
reason for the bad replication factor is the randomness of the data. We have
used the following real databases \cite{fimi03}: \texttt{kosarak},
\texttt{accidents}, \texttt{chess}, \texttt{connect}, \texttt{mushroom},
\texttt{pumsb\_star}, and \texttt{pumsb}. As the implementation of the QKP
algorithm, we have downloaded the source code from \cite{qkp-source}, an
implementation of the algorithm described in~\cite{qkp-paper}.

The results of the experiments are summarized in tables. For each
database there are three tables: improvement of QKP scheduling against
the greedy scheduling, see Algorithm \ref{alg:lpt-schedule}, the
database replication using the greedy scheduling, and the database
replication using the QKP schedule. We have chosen the number of
processors: $4$, $6$, $10$, and $14$.


The biggest improvement of the database replication ($28\%$) is on the
\texttt{mushroom} database. It can be seen that the biggest improvement
is at the relative support level $0.001$. The improvements are much
smaller, when the relative support is $>0.01$.  The \texttt{mushroom}
database is also one of the two databases where we have achieved a
replication factor after reduction $\ll\procnum-1$ (for $14$
processors). The lowest replication factor $2.7$ on $14$ processors
was measured on the \texttt{mushroom} database. In most cases the
replication factor is between $\procnum-1$ and $\procnum$. The
replication factor after reduction is also lower
($\ll\procnum-1$) on the \texttt{pumsb\_star} database.

Overall, the improvement of the replication factor mostly ranges
between $\approx 1\%$ and $\approx 13\%$. It sometimes happens that
the replication factor is worse after reduction. The worsening is
for the \texttt{pumsb} database $-0.0464\%$, \texttt{pumsb\_star}
$-2.2881\%$ and $-0.2538\%$. We consider these values as outliers.








Generally it holds that for two processors the database replication is
very high, but mostly does not reach $\procnum$ for $\procnum$
processors. However, in most cases the replication factor is between
$\procnum-1$ and $\procnum$.


The most interesting case is the \texttt{mushroom} database. From the
experiments it can be seen that the lower the support the better
results. The best database replication factor is $10$ on $14$
processors.

To conclude, from the databases we can made hypothesis that the
database replication factor is high for higher values of support and
small for lower values of supports.

\begin{table}[!p]
\centering

\noindent Improvement(in $\%$):

\begin{tabular}{|l|l|l|l|}\hline
$\procnum$/$\rminsupp$ & 0.0050 & 0.0040 & 0.0030\\ \hline
4 & 12.6096 & 11.6921 & 15.9684 \\ \hline
6 & 13.6673 & 19.4744 & 20.7549 \\ \hline
10 & 18.0931 & 18.0157 & 18.7086 \\ \hline
14 & 17.6054 & 20.2953 & 21.7529 \\ \hline
\end{tabular}

\medskip
\noindent Database replication \emph{without reduction}:

\begin{tabular}{|l|l|l|l|}\hline
$\procnum$/$\rminsupp$ & 0.0050 & 0.0040 & 0.0030\\ \hline
4 & 1.76357 & 1.9325 & 1.86456 \\ \hline
6 & 2.08358 & 2.14564 & 2.18368 \\ \hline
10 & 2.36798 & 2.4311 & 2.49938 \\ \hline
14 & 2.55512 & 2.55404 & 2.74345 \\ \hline
\end{tabular}

\medskip
\noindent Database replication \emph{after reduction (using the \textsc{DB-REPL-MIN} algorithm)}:

\begin{tabular}{|l|l|l|l|}\hline
$\procnum$/$\rminsupp$ & 0.0050 & 0.0040 & 0.0030\\ \hline
4 & 1.54119 & 1.70655 & 1.56682 \\ \hline
6 & 1.79881 & 1.72779 & 1.73046 \\ \hline
10 & 1.93954 & 1.99312 & 2.03178 \\ \hline
14 & 2.10528 & 2.03569 & 2.14667 \\ \hline
\end{tabular}

\caption[Improvement of the database replication: \texttt{kosarak}]{Improvement of the database replication of the \texttt{kosarak} database.}
\label{tbl:dbrepl-impr-kosarak}
\end{table}

\begin{table}[!p]
\centering

\noindent Improvement(in $\%$):

\begin{tabular}{|l|l|l|l|l|l|l|}\hline
$\procnum$/$\rminsupp$ & 0.06 & 0.05 & 0.04 & 0.03 & 0.02 & 0.01\\ \hline
4 & 0.0365 & 0.0150 & 0.7585 & 1.4095 & 0.0960 & 0.0833 \\ \hline
6 & 0.6032 & 0.6150 & 2.5080 & 2.8985 & 0.3852 & 4.9895 \\ \hline
10 & 3.2480 & 2.6723 & 2.5703 & 3.7139 & 4.0293 & 3.8973 \\ \hline
14 & 1.7851 & 4.0688 & 7.1765 & 6.3381 & 2.5573 & 4.3714 \\ \hline
\end{tabular}

\medskip
\noindent Database replication \emph{without reduction}:

\begin{tabular}{|l|l|l|l|l|l|l|}\hline
$\procnum$/$\rminsupp$ & 0.06 & 0.05 & 0.04 & 0.03 & 0.02 & 0.01\\ \hline
4 & 4 & 4 & 4 & 4 & 4 & 4 \\ \hline
6 & 5.99995 & 5.99996 & 5.99909 & 5.99992 & 5.99999 & 5.99837 \\ \hline
10 & 9.9964 & 9.99502 & 9.99673 & 9.99737 & 9.99766 & 9.99586 \\ \hline
14 & 13.9603 & 13.9502 & 13.9414 & 13.9648 & 13.9715 & 13.9636 \\ \hline
\end{tabular}

\medskip
\noindent Database replication \emph{after reduction (using the \textsc{DB-REPL-MIN} algorithm)}:

\begin{tabular}{|l|l|l|l|l|l|l|}\hline
$\procnum$/$\rminsupp$ & 0.06 & 0.05 & 0.04 & 0.03 & 0.02 & 0.01\\ \hline
4 & 3.99854 & 3.9994 & 3.96966 & 3.94362 & 3.99616 & 3.99667 \\ \hline
6 & 5.96376 & 5.96306 & 5.84863 & 5.82601 & 5.97688 & 5.69908 \\ \hline
10 & 9.67172 & 9.72792 & 9.73978 & 9.62608 & 9.59482 & 9.60629 \\ \hline
14 & 13.7111 & 13.3826 & 12.9409 & 13.0797 & 13.6142 & 13.3532 \\ \hline
\end{tabular}

\caption[Improvement of the database replication: \texttt{accidents}]{Improvement of the database replication of the \texttt{accidents} database.}
\label{tbl:dbrepl-impr-accidents}
\end{table}

\begin{table}[!p]
\centering
\noindent Improvement(in $\%$):

\begin{tabular}{|l|l|l|l|l|l|l|l|}\hline
$\procnum$/$\rminsupp$ & 0.7 & 0.6 & 0.5 & 0.4 & 0.3 & 0.2 & 0.1\\ \hline
4 & 0.0000 & 0.0783 & 0.0548 & 0.0235 & 0.0235 & 0.6570 & 1.0717 \\ \hline
6 & 0.0625 & 0.0313 & 0.1825 & 0.2347 & 0.4642 & 0.1512 & 1.9035 \\ \hline
10 & 1.1765 & 0.1596 & 0.0688 & 0.2472 & 0.2941 & 0.3817 & 1.0889 \\ \hline
14 & 0.2372 & 0.2257 & 0.2214 & 0.3886 & 1.0614 & 0.6143 & 4.2286 \\ \hline
\end{tabular}

\medskip
\noindent Database replication \emph{without reduction}:

\begin{tabular}{|l|l|l|l|l|l|l|l|}\hline
$\procnum$/$\rminsupp$ & 0.7 & 0.6 & 0.5 & 0.4 & 0.3 & 0.2 & 0.1\\ \hline
4 & 4 & 4 & 4 & 4 & 4 & 4 & 4 \\ \hline
6 & 6 & 6 & 6 & 6 & 6 & 6 & 6 \\ \hline
10 & 10 & 9.99875 & 10 & 10 & 10 & 10 & 10 \\ \hline
14 & 13.9994 & 13.9987 & 14 & 14 & 14 & 14 & 14 \\ \hline
\end{tabular}

\medskip
\noindent Database replication \emph{after reduction (using the \textsc{DB-REPL-MIN} algorithm)}:

\begin{tabular}{|l|l|l|l|l|l|l|l|}\hline
$\procnum$/$\rminsupp$ & 0.7 & 0.6 & 0.5 & 0.4 & 0.3 & 0.2 & 0.1\\ \hline
4 & 4 & 3.99687 & 3.99781 & 3.99906 & 3.99906 & 3.97372 & 3.95713 \\ \hline
6 & 5.99625 & 5.99812 & 5.98905 & 5.98592 & 5.97215 & 5.99093 & 5.88579 \\ \hline
10 & 9.88235 & 9.98279 & 9.99312 & 9.97528 & 9.97059 & 9.96183 & 9.89111 \\ \hline
14 & 13.9662 & 13.9671 & 13.969 & 13.9456 & 13.8514 & 13.914 & 13.408 \\ \hline
\end{tabular}

\caption[Improvement of the database replication: \texttt{chess}]{Improvement of the database replication of the \texttt{chess} database.}
\label{tbl:dbrepl-impr-chess}
\end{table}

\begin{table}[!p]
\centering

\noindent Improvement(in $\%$):

\begin{tabular}{|l|l|l|l|}\hline
$\procnum$/$\rminsupp$ & 0.3 & 0.2 & 0.1\\ \hline
4 & 0.0000 & 0.0000 & 0.0103 \\ \hline
6 & 0.2218 & 0.0005 & 2.3822 \\ \hline
10 & 0.3900 & 1.5442 & 1.2324 \\ \hline
14 & 0.7933 & 1.3633 & 1.2881 \\ \hline
\end{tabular}

\medskip
\noindent Database replication \emph{without reduction}:

\begin{tabular}{|l|l|l|l|}\hline
$\procnum$/$\rminsupp$ & 0.3 & 0.2 & 0.1\\ \hline
4 & 4 & 4 & 4 \\ \hline
6 & 6 & 6 & 6 \\ \hline
10 & 9.96607 & 9.96607 & 10 \\ \hline
14 & 13.9661 & 13.9661 & 13.9661 \\ \hline
\end{tabular}

\medskip
\noindent Database replication \emph{after reduction (using the \textsc{DB-REPL-MIN} algorithm)}:

\begin{tabular}{|l|l|l|l|}\hline
$\procnum$/$\rminsupp$ & 0.3 & 0.2 & 0.1\\ \hline
4 & 4 & 4 & 3.99959 \\ \hline
6 & 5.98669 & 5.99997 & 5.85707 \\ \hline
10 & 9.9272 & 9.81217 & 9.87676 \\ \hline
14 & 13.8553 & 13.7757 & 13.7862 \\ \hline
\end{tabular}

\caption[Improvement of the database replication: \texttt{connect}]{Improvement of the database replication of the \texttt{connect} database.}
\label{tbl:dbrepl-impr-connect}
\end{table}

\begin{table}[!p]
\centering

\noindent Improvement(in $\%$):

\begin{tabular}{|l|l|l|l|l|l|l|}\hline
$\procnum$/$\rminsupp$ & 0.1 & 0.08 & 0.06 & 0.04 & 0.02 & 0.001\\ \hline
4 & 0.6155 & 0.8617 & 0.5663 & 1.0093 & 4.7883 & 11.0753 \\ \hline
6 & 2.6015 & 3.9635 & 1.4903 & 1.8668 & 2.1500 & 14.6110 \\ \hline
10 & 3.8776 & 3.2556 & 3.5445 & 9.6659 & 8.0022 & 22.7943 \\ \hline
14 & 5.8516 & 5.9913 & 7.9623 & 7.7287 & 10.0239 & 28.9319 \\ \hline
\end{tabular}

\medskip
\noindent Database replication \emph{without reduction}:

\begin{tabular}{|l|l|l|l|l|l|l|}\hline
$\procnum$/$\rminsupp$ & 0.1 & 0.08 & 0.06 & 0.04 & 0.02 & 0.001\\ \hline
4 & 4 & 4 & 4 & 4 & 4 & 4 \\ \hline
6 & 5.99951 & 6 & 5.99606 & 6 & 6 & 6 \\ \hline
10 & 9.93599 & 9.98929 & 9.98769 & 9.99926 & 9.99852 & 9.96972 \\ \hline
14 & 13.9791 & 13.8902 & 13.9357 & 13.979 & 13.9547 & 13.8765 \\ \hline
\end{tabular}

\medskip
\noindent Database replication \emph{after reduction (using the \textsc{DB-REPL-MIN} algorithm)}:

\begin{tabular}{|l|l|l|l|l|l|l|}\hline
$\procnum$/$\rminsupp$ & 0.1 & 0.08 & 0.06 & 0.04 & 0.02 & 0.001\\ \hline
4 & 3.97538 & 3.96553 & 3.97735 & 3.95963 & 3.80847 & 3.55699 \\ \hline
6 & 5.84343 & 5.76219 & 5.9067 & 5.88799 & 5.871 & 5.12334 \\ \hline
10 & 9.55071 & 9.66408 & 9.63368 & 9.03274 & 9.19842 & 7.69719 \\ \hline
14 & 13.1611 & 13.058 & 12.8261 & 12.8986 & 12.5559 & 9.86177 \\ \hline
\end{tabular}

\caption[Improvement of the database replication: \texttt{mushroom}]{Improvement of the database replication of the \texttt{mushroom} database.}
\label{tbl:dbrepl-impr-mushroom}
\end{table}

\begin{table}[!p]
\centering
\noindent Improvement(in $\%$):

\begin{tabular}{|l|l|l|l|l|l|l|l|}\hline
$\procnum$/$\rminsupp$ & 0.25 & 0.3 & 0.32 & 0.35 & 0.42 & 0.49 & 0.56\\ \hline
4 & 7.3980 & 0.9556 & 6.3810 & 0.9308 & 1.9091 & 6.1343 & 6.5228 \\ \hline
6 & 6.9107 & 8.4605 & 2.8155 & 4.2704 & 4.4023 & 4.6576 & -2.2881 \\ \hline
10 & 7.1429 & 4.4941 & 18.9267 & 9.2538 & -0.2538 & 8.1635 & 7.4531 \\ \hline
14 & 9.0592 & 5.8842 & 5.0286 & 22.2261 & 3.5505 & 13.2112 & 14.8919 \\ \hline
\end{tabular}

\medskip
\noindent Database replication \emph{without reduction}:

\begin{tabular}{|l|l|l|l|l|l|l|l|}\hline
$\procnum$/$\rminsupp$ & 0.25 & 0.3 & 0.32 & 0.35 & 0.42 & 0.49 & 0.56\\ \hline
4 & 4 & 3.72837 & 4 & 4 & 3.72837 & 3.99865 & 3.72833 \\ \hline
6 & 5.72837 & 5.72837 & 5.72819 & 5.72801 & 5.72705 & 5.72502 & 5.51461 \\ \hline
10 & 9.72616 & 9.70714 & 9.70202 & 9.47426 & 9.59337 & 9.71667 & 9.02567 \\ \hline
14 & 13.6138 & 13.2065 & 13.3537 & 13.1593 & 13.2798 & 13.0995 & 12.124 \\ \hline
\end{tabular}

\medskip
\noindent Database replication \emph{after reduction (using the \textsc{DB-REPL-MIN} algorithm)}:

\begin{tabular}{|l|l|l|l|l|l|l|l|}\hline
$\procnum$/$\rminsupp$ & 0.25 & 0.3 & 0.32 & 0.35 & 0.42 & 0.49 & 0.56\\ \hline
4 & 3.70408 & 3.69274 & 3.74476 & 3.96277 & 3.65719 & 3.75336 & 3.48514 \\ \hline
6 & 5.3325 & 5.24372 & 5.56691 & 5.4834 & 5.47493 & 5.45837 & 5.64079 \\ \hline
10 & 9.03143 & 9.27089 & 7.86575 & 8.59753 & 9.61772 & 8.92345 & 8.35298 \\ \hline
14 & 12.3805 & 12.4294 & 12.6822 & 10.2345 & 12.8083 & 11.3689 & 10.3185 \\ \hline
\end{tabular}

\caption[Improvement of the database replication: \texttt{pumsb\_star}]{Improvement of the database replication of the \texttt{pumsb\_star} database.}
\label{tbl:dbrepl-impr-pumsbstar}
\end{table}

\begin{table}[!p]
\centering
\noindent Improvement(in $\%$):

\begin{tabular}{|l|l|l|l|}\hline
$\procnum$/$\rminsupp$ & 0.9 & 0.85 & 0.8\\ \hline
4 & 0.0040 & 0.2263 & 0.0152 \\ \hline
6 & 0.1925 & 0.2650 & 0.2111 \\ \hline
10 & -0.0464 & 0.8433 & 0.1771 \\ \hline
14 & 0.6181 & 0.0179 & 0.9075 \\ \hline
\end{tabular}

\medskip
\noindent Database replication \emph{without reduction}:

\begin{tabular}{|l|l|l|l|}\hline
$\procnum$/$\rminsupp$ & 0.9 & 0.85 & 0.8\\ \hline
4 & 4 & 4 & 4 \\ \hline
6 & 5.98371 & 5.9992 & 5.99839 \\ \hline
10 & 9.97184 & 9.98096 & 9.98304 \\ \hline
14 & 13.9612 & 13.9685 & 13.9393 \\ \hline
\end{tabular}

\medskip
\noindent Database replication \emph{after reduction (using the \textsc{DB-REPL-MIN} algorithm)}:

\begin{tabular}{|l|l|l|l|}\hline
$\procnum$/$\rminsupp$ & 0.9 & 0.85 & 0.8\\ \hline
4 & 3.99984 & 3.99095 & 3.99939 \\ \hline
6 & 5.97219 & 5.9833 & 5.98573 \\ \hline
10 & 9.97647 & 9.89679 & 9.96536 \\ \hline
14 & 13.8749 & 13.966 & 13.8128 \\ \hline
\end{tabular}

\caption[Improvement of the database replication: \texttt{pumsb}]{Improvement of the database replication of the \texttt{pumsb} database.}
\label{tbl:dbrepl-impr-pumsb}
\end{table}

\clearpage
\newpage

\chapter{Conclusion and future work}\label{chap:conclusion}

\section{Conclusion}

In our work, we have shown a method that parallelize an arbitrary
algorithm for mining of FIs. We have proposed two methods for
estimation of the size of a PBEC based on the
\textsc{Modified-Coverage-Algorithm} and explained why the sampling is
just a heuristic. In order to make better estimation results, we have
proposed estimation of the relative size of PBECs based on the
\textsc{Vitter-Reservoir-Sampling} algorithm. We have shown how big
error can be made by our ``double sampling process'' , see Theorem
\ref{theorem:bounds-pbec-size-reservoir} and Corollary
\ref{corollary:bounds-pbec-size-reservoir-probabilistic}.

We have shown how to execute an arbitrary sequential algorithm for
mining of all MFIs in parallel that mines a superset of all MFIs $M$
in order to speedup the sampling process based on the
\textsc{Modified-Coverage-Algorithm} and proved that the size of $M$
can be larger then $\mfiapprox$, see
Theorem~\ref{thm:parallel-mfi-count} and
Chapter~\ref{chap:parallel-mfi}.

Then in Chapter~\ref{chap:ppmfi-method} we have proposed our three
methods for parallel mining of MFIs, called \scparfimiseq{},
\scparfimipar{}, and \scparfimireserv{}, on a distributed memory
parallel computer. In Chapter~\ref{chap:exec-eclat} we have shown how
to efficiently execute the \textsc{Eclat} algorithm in Phase~4 of our
new method. In Chapter~\ref{chap:db-minimalization}, we have discussed
the database replication factor and the possibilities of
minimization of the database replication factor.

In Chapter~\ref{chap:experimental-evaluation} we have experimentally
evaluated the performance of our new method and the errors of the
estimates of the size of union of PBECs. Additionally, we have shown
that minimizing the database replication factor based on the solution
of the quadratic knapsack problem big improvement on all artificial
databases and makes slight improvement on some real databases.

\section{Future work}\label{chap:future-work}

We would like to improve the \scparfimireserv{} algorithm. The
inefficiency in the algorithm comes from the fact that the reservoir
sampling is embedded in a regular \textsc{Eclat} algorithm, i.e., the
support is computed for each frequent itemsets while sampling the
FIs. This inefficiency could be removed by using smarter algorithm
that would use the same optimizations as for example the \emph{fpmax*}
algorithm. The \emph{fpmax*} algorithm is an algorithm for mining of
MFIs and uses a list of MFIs to check for support of newly generated
frequent itemset.

The IBM database generator in some cases does not generate databases
similar to the real databases. We have already developed some database
characteristics, however their description is out of the scope of this
thesis. Additionally, we would like to create a database generator
that would generate more realistic and structured databases then the
IBM generator. 

\bibliography{papers}
\bibliographystyle{abbrv}

\renewcommand\bibname{Refereed publications of the author}

\renewcommand\bibname{Unrefereed publications of the author}

\newpage

\appendix

\chapter{Discrete probability distributions and tails}\label{appendix:discrete-distributions}

\section{Chernoff bounds}

The Chernoff bound is used to bound the number of sucessfull
independent Poisson experiments. Let $X_1, \ldots, X_n$ be $n$
independent random variables such that $X_i\in\{0,1\}$ and $P[X_i = 1]
= p_i \in \lcint 0, 1\rcint$. Let $X=\sum_i X_i$ and let $\mu$ be the
expectation of $X$, then the Chernoff bounds, i.e., the probability
$P[X\leq (1-\delta)\mu]$ or $P[X\geq (1+\delta)\mu]$ where
$\delta\in\lcint0,1\rcint$ is:

$$ \prob[X\leq (1-\delta)\mu] \leq e^{\frac{-\mu \delta^2}{4}}$$

$$ \prob[X\geq (1+\delta)\mu] \leq e^{\frac{-\mu \delta^2}{4}}$$

Another variant of the Chernoff bounds is provided in
\cite{book:alon-probabilistic}. Let have the following assupmtions:
$p\in\lcint 0,1 \rcint$, $X_1, \ldots, X_n$ mutually independent
random variables with $\prob [X_i = 1 - p] = p$ $ \prob [X_i = -p] = 1
- p$, and let $X = X_1 + \ldots + X_n$. Then for $a > 0$:

\begin{equation}\label{eqn:noga-chernoff}
\prob[|X| > a] < \exp^{-2a^2/n}
\end{equation}

The equation (\ref{eqn:noga-chernoff}) is the actual equation used by
Toivonen in \cite{toivonen96sampling} for proving
Theorem~\ref{theorem:chernoff-supp-estimate}.

\section{Hypergeometric distribution and tails}\label{appendix:sec:hypergeometric}

The hypergeometric distribution describes the following problem: let
us have an urn with $N$ balls of which $M$ are black and $N-M$ are
white. A sample of $n$ balls is drawn without replacement. The
distribution of $i$, the number of black balls, is:

$$ \prob[X=i] = \frac{{M \choose i}{N-M\choose n-i}}{{N \choose n}}.$$

The expectation of $i$ is $\expect[i]=n\frac{M}{N}$. For any
$\epsilon \geq 0$ the difference $\expect[i]-i$ is bound by:

%
\begin{equation}\label{eq:hypergeom-tail-upper}
\prob[i\geq \expect [i] + \epsilon\cdot n] \leq e^{-2\epsilon^2n}
\end{equation}

\noindent and

\begin{equation}\label{eq:hypergeom-tail-lower}
\prob[i\leq \expect [i] - \epsilon\cdot n] \leq e^{-2\epsilon^2n}.
\end{equation}

For more details, see \cite{hypergeometric-skala}.

A more precise bound can be computed using the Kullback-Leibler
divergence of two Bernoulli distributed random variables, denoted by
$D(\cdot||\cdot)$. Let $p=M/N$ and $\epsilon\geq 0$, then:


\begin{equation}
P[i\geq \expect[i]+\epsilon\cdot n] \leq \left(\left(\frac{p}{p+\epsilon}\right)^{p+\epsilon}  \left(\frac{1-p}{1-p-\epsilon}\right)^{1-p-\epsilon} \right)^n = e^{-nD(p+\epsilon||p)}
\end{equation}

or

\begin{equation}
P[i\leq \expect[i]-\epsilon\cdot n] \leq \left(\left(\frac{p}{p-\epsilon}\right)^{p-\epsilon}  \left(\frac{1-p}{1-p\epsilon}\right)^{1-p+\epsilon} \right)^n = e^{-nD(p-\epsilon||p)}.
\end{equation}

Therefore,

\begin{equation}\label{eqn:hypergeometric-kullback-leibler}
 P[\expect[i] - \epsilon n \leq i\leq \expect[i] + \epsilon n] \leq 1-(e^{-nD(p-\epsilon||p)} + e^{-nD(p+\epsilon||p)}).
\end{equation}


The \emph{multivariate hypergeometric distribution} is the same as the
hypergeometric distribution, except that the balls can have more
colors, defined as follows: let the number of colors be $C$ and the
number of balls colored with color $i$ is $M_i$ and the total number
of balls is $N=\sum_i M_i$. Let $X_i$, $1\leq i\leq C$, be a random
variable representing the number of balls colored by the $i$-th
color. The sample of size $n$ is drawn from balls and $X_i$ balls,
such that $n=\sum_{i=1}^{C} X_i$ are colored by the $i$th color. Then
the probability mass function is:

$$ P(X_1=k_1,\ldots, X_C=k_C) = \frac{\prod_{i=1}^{C} {M_i \choose k_i}}{{N \choose n}}.$$

where $k_i$ are integers. The expectation is
$\expect[X_i]=n\frac{M_i}{N}$. Obviously, the tail inequalities of the
multivariate hypergeometric distribution are the same as for the
hypergeometric distribution, i.e., the multivariate hypergeometric
distribution with $C=2$.



\section{Multivariate binomial distribution}\label{sec:multivariate-binomial}

The multivariate binomial distribution, or so called multinomial
distribution, is a distribution describing the outcome of $n$
independent Bernoulli trials where each trial results in $k$ possible
outcomes. The $i$th outcome of each trial has the probability $p_i,
\sum_{1\leq i\leq k} p_i = 1$. The probability mass function of the
multivariate binomial distribution is:

\begin{equation}
\begin{array}{ll} f(x_1, \ldots, x_k; n, p_1,\ldots, p_k) & = Pr(X_i = x_i, \ldots  , X_k=x_k) \\
                                                           & = \left\{\begin{array}{l} 
                                                                  \frac{n!}{x_1!\cdots x_k!}\cdot p_1^k\cdots p_k^k, \text{ when } \sum_{i=1}^k x_i = n \\ 
                                                                  0, \text{otherwise} 
                                                               \end{array} \right.
\end{array}
\end{equation}

for non-negative integers $x_1,\ldots,x_k$.

\chapter{Selected sequential algorithms}\label{appendix:seq-alg}

This appendix describes selected sequential algorithms together with
datastructures and optimizations, an edited version of our master
thesis \cite{kessl04master}. In this appendix, we describe some of the
existing algorithms for mining of FIs and some of its
optimizations. Namely: 1) the Apriori algorithm in
\secref{sec:apriori-algorithm}; 2) the FPGrowth algorithm in
\secref{sec:fpgrowth-algorithm}; and 3) the Eclat algorithm in
\secref{sec:eclat-algorithm}. In \secref{sec:dfs-optimizations} we
show optimizations of the Eclat algorithm and we finish the appendix
with the algorithm that generates the association rules from FIs, see
\secref{sec:discovering-rules}.

\section{The Apriori algorithm}\label{sec:apriori-algorithm}

The Apriori algorithm \cite{agrawal94fast} is a BFS algorithm based
solely on the monotonicity property, see Theorem~\ref{monotonicity}. The
Apriori algorithm uses the notion of \emph{candidates itemsets}, see
Definition~\ref{def:candidate-fi-itemset}. In the further text, we
denote the set of all FIs of size $k$ by $F_k$ and the set of
\emph{canidates} on frequent itemsets by $C_k$. Obviously, $F_k
\subseteq C_k$. The algorithm proceeds in steps.  In step $k>1$, it
first generates a set $C'_k$ of possibly frequent itemsets of size
$k$, such that $C_k\subseteq C'_k$, from the set of frequent itemsets
$F_{k-1}$ of size $k-1$ computed in the previous step $k-1$. The set
$C'_{k}$ is generated in the following way: from $F_{k-1}$, the set of
frequent itemsets of size $k-1$, we find all pairs of itemsets $\isetU
= (u_1, \ldots, u_{k-1}), \isetW=(w_1, \ldots, w_{k-1}) \in F_{k-1}$
that are identical in the first $k-2$ items, i.e., $u_i = w_i, i \leq
k - 2$. From each such pair $\isetU, \isetW$ a new candidate $\isetV =
\{ u_1,\ldots , u_{k-2}, u_{k-1}, w_{k-1} \}$ is constructed. The
candidates $C_k$ are generated from $C'_k$ in the following way: for
each $\isetU\in C'_k$, we apply the monotonicity principle, i.e., we
test whether each subset $\isetW\subset\isetV, k-1 = |\isetW| =
|\isetV-1|$ is present in $F_{k-1}$. The reason is that all subsets of
$\isetU$ must be frequent in order for $\isetU$ to be also frequent,
see \corref{monotonicity-subsets}. Therefore, if some subset of
$\isetU$ of size $k-1$ is not in $F_{k-1}$ then $\isetU$ is deleted
from $C_k$. The algorithm for generation of candidates follows:

\begin{algorithm}[H]
\caption{The \textsc{Generate-Candidates} function}\label{alg:generate-candidates}
\textsc{Generate-Candidates}(\inparam Itemset $F_k$)
\begin{algorithmic}[1]
  \STATE $C\leftarrow\emptyset$
  \FORALL{$\isetU=(u_1, \ldots, u_k), \isetW=(w_1, \ldots, w_k) \in F_k$}
    \IF{$u_k < w_k \wedge u_j = w_j, j < k$}
      \STATE $C\leftarrow C\cup\{(u_1, \ldots, u_{k-1}, u_k, w_k) \}$
    \ENDIF
  \ENDFOR
  \FOR{$\isetU \in C$}
    \IF{\CALL{Test-Subset}$(\fipart_k, \isetU) =$ {\tt false}}\label{alg:gener-cands:test}
      \STATE delete $\isetU$ from $C$
    \ENDIF
  \ENDFOR
  \STATE \textbf{return} $C$
\end{algorithmic}
\end{algorithm}

In the first step ($k=1$), the Apriori algorithm starts with
$C_1=\{\{b_i\}: b_i\in \baseset\}$ and counts support of each
$\isetU\in C_1$ in a single scan of the database, creating $F_1$.  In
steps $k>1$, the algorithm must compute the support of each $\isetU
\in C_k$, i.e., we create the set $F_k=\{\isetU | \isetU \in C_k,
\supp(\isetU)\geq\minsupp\}$.


The algorithm ends if: 1) all candidates are deleted; 2) all
candidates turn out not to be frequent. In both cases the resulting
$F_k$ is empty.

\emph{To make the explanation of the Apriori algorithm simple, we
  ommit the details of the \textsc{Test-Subset} and
  \textsc{Compute-Support} algorithms. The \textsc{Test-Subset} and
  \textsc{Compute-Support} algorithms are described in
  Section~\ref{sec:apriori-support-alg}. However, it is not necessary
  to understand the two algorithms in order to understand the Apriori
  algorithm.}

In the following text, we use the algorithm $\textsc{Test-Subset}(F_k,
\isetU)$ that checks whether all subsets of size $|\isetU|-1=k$ are
present in the set $F_k$. Additionally, we use the algorithm
$\textsc{Compute-Support}(\db, C_k)$ that computes the support of each
$\isetU\in C_k$.

Since the evaluation of the support for each candidate is quite a
time-consuming task, it has to be done as fast as possible on as few
candidates as possible.  Many candidates are generated uselessly,
because they turn out not to be frequent.


An example of the execution of the Apriori algorithm on a small
database is given in the Example \ref{exmpl:apriori-execution}. The
pseudocode of the Apriori algorithm can be found in
\algref{alg:apriori}.

\begin{figure}[!ht]\label{exmpl:apriori-execution}
\example[An example execution of the Apriori algorithm]{An example execution of the Apriori algorithm}\label{run-of-apriori}
\begin{center}
\renewcommand{\arraystretch}{2}
\begin{tabular*}{10cm}{ccc}
Input: &
{
  \medskip
  $D$=
  \renewcommand{\arraystretch}{1}
       \begin{tabular}{|c|c|} \hline
	 {\tt TID} & {\tt Transaction} \\ \hline
	 1    &   $\{ 1,2,5\}$    \\ \hline
	 2    &   $\{ 1,3,5\}$    \\ \hline
	 3    &   $\{ 2,4,5\}$    \\ \hline
	 4    &   $\{ 1,2,3,5\}$  \\ \hline
       \end{tabular}\hfill%
  \medskip
}
&  $\baseset=\{1,2,3,4,5\}, \minsupp=2$\\ 
$k=1$
&
{
  \medskip
  \renewcommand{\arraystretch}{1}
    \begin{tabular}{|c|c|}\hline
      $C_1$       & {\tt Support} \\ \hline
      $\{ 1\}$    &   3     \\ \hline
      $\{ 2\}$    &   3     \\ \hline
      $\{ 3\}$    &   2     \\ \hline
      $\{ 4\}$    &   1     \\ \hline
      $\{ 5\}$    &   5     \\ \hline
    \end{tabular}%
  \medskip
}
&%
{
  \renewcommand{\arraystretch}{1}
    \begin{tabular}{|c|c|} \hline
      $F_1$      \\ \hline
      $\{\{ 1\}, \{ 2\}, \{ 3\}, \{ 5\}\}$ \\ \hline
    \end{tabular}
}
\\ 
$k=2$
&
{
  \medskip
  \renewcommand{\arraystretch}{1}
    \begin{tabular}{|c|c|} \hline
      $C_2$       & {\tt Support} \\ \hline
      $\{ 1,2\}$    &   2     \\ \hline
      $\{ 1,3\}$    &   2     \\ \hline
      $\{ 1,5\}$    &   3     \\ \hline
      $\{ 2,3\}$    &   1     \\ \hline
      $\{ 2,5\}$    &   3     \\ \hline
    \end{tabular}%
  \medskip
}
&%
{
  \renewcommand{\arraystretch}{1}
    \begin{tabular}{|c|c|}  \hline
      $F_2$       \\ \hline
      $\{\{ 1,2\},\{ 1,3\},\{ 1,5\},\{ 2,5\}\}$  \\ \hline
    \end{tabular}%
}
\\ 
$k=3$
&
{
  \renewcommand{\arraystretch}{1}
   \begin{tabular}{|c|c|}\hline
            $C_3$       & {\tt Support} \\ \hline
            $\{ 1,2,5\}$    &   2     \\ \hline
   \end{tabular}%
}
&
{
  \renewcommand{\arraystretch}{1}
   \begin{tabular}{|c|} \hline
            $F_3$       \\ \hline
            $\{\{ 1,2,5\}\}$ \\ \hline
   \end{tabular}%
}
\end{tabular*}
\end{center}
\end{figure}

\begin{algorithm}[H]
\caption{The \textsc{Apriori} algorithm}\label{alg:apriori}
\hbox{\textsc{Apriori}$($\inparam Database $\db$, \inparam Integer $\minsupp$, \outparam Set  $\allfi$)}
\begin{algorithmic}[1]
  \STATE $k\leftarrow 1$
  \STATE Compute all frequent items and store them into $\baseset$
  \STATE $C_k \leftarrow \{\{\bitem\}: \bitem \in \baseset\}$
  \WHILE{$C_k$ not empty}
    \STATE\CALL{Compute-Support}$(\db, C_k)$
    \FORALL{$\isetU \in C_k$} 
      \IF{$\supp(\isetU) < \minsupp$}
      \STATE delete $\isetU$ from $C_k$ 
      \ENDIF
    \ENDFOR
    \STATE $\fipart_k\leftarrow C_k$
    \STATE $C_{k+1}\leftarrow$\CALL{Generate-Candidates}$(\fipart_k)$
    \STATE $k\leftarrow k+1$
  \ENDWHILE
  \STATE $\allfi\leftarrow \bigcup_{i=1}^{k-1} F_i$
  \STATE \textbf{return}
\end{algorithmic}
\end{algorithm}


\subsection{Prefix trie}\label{sec:prefix-trie}

\begin{definition}[Prefix trie]
Let $\{\isetU|\isetU\subseteq\baseset\}$ be a collection of
itemsets. Each $\isetU=(u_1,u_2,\ldots, u_{|\isetU|})$ is sorted
according to some order $<$, i.e. $u_{k}<u_{l}, k<l$. Let
$\nodes=\{\node_i\}$ be a set of nodes and $\edges = \{ \edge_i =
(\node_i,\node_j) | \node_i,\node_j\in \nodes \}$ be a set of edges of
an oriented accyclic graph $G=(\nodes, \edges)$. $G$ is called a
\emph{prefix trie} iff: each node $\node_j$ corresponds to a prefix
$(u_1,u_2,\ldots,u_l)$ of an itemset $\isetU$ and an edge
$(\node_j,\node_k)$ is present in $E$ iff there exist prefix
$(u_1,u_2,\ldots,u_l,u_{l+1})$ of $\isetU$. The node $\node_j$ has
associated the item $u_l$ and the node $\node_{k}$ has associated the
item $u_{l+1}$. Each node $\node_j$ is represented by a tuple $(depth,
max\_depth, support, children)$, where $depth, max\_depth$, and
$support$ are integers. The field children is a set
$\{(item,\node_j)\}$.
\end{definition}

Inserting the pair $(item,\node)$ into the field $children$, we denote
by $children[item]\leftarrow \node$. Reading the node from the field
$children$ is denoted by $\node\leftarrow children[item]$.

\noindent Note: the word \emph{trie} comes from the noun reTRIEval and
is pronounced as "tree", see \cite{nistgov}.

All operations (e.g., subset test, support increment and
\textsc{Generate-Candidates}) in the Apriori algorithm are based on
the prefix trie structure.  \figref{fig:example-prefix-trie} shows an
example of a prefix trie.  In the prefix trie, the Apriori algorithm
stores the set of candidates of size $k$, $C_k$, or the FIs of size
$k$, $F_k$. However, generally the trie can store itemsets of
arbitrary sizes.

\begin{figure}[!ht]
\example[An example of the prefix trie data structure]{An example of a
  prefix trie data structure. The set of children is represented by
  arcs with labels. The root represents the empty itemset. The content
  of each node is $(depth, max\_depth, support, children)$ (the depth
  field is counted from $0$). The prefix trie (b) for the database (a)
  is constructed by 3 calls: \textsc{Insert-PrefixTrie}$(\{1,2,3\}, root)$,
  \textsc{Insert-PrefixTrie}$(\{1,2,4\}, root)$, \textsc{Insert-PrefixTrie}$(\{1,3,4\},
  root)$. The \textsc{Insert-PrefixTrie} algorithm can be found in
  Algorithm~\ref{alg:prefix-tree-insert}}
\medskip
  \centering
  \hbox{\hsize=5cm
    \vbox to 5cm{
      Database\par\smallskip
      \begin{tabular}{|c|} \hline
	{\tt Itemset} \\ \hline
	$\{1,2,3\}$ \\ \hline
	$\{1,2,4\}$ \\ \hline
	$\{1,3,4\}$ \\ \hline
      \end{tabular} \par
      \vbox{\hsize=2.7cm\hfill (a)\hfill}\vfill
    }
    \vbox to 5cm{%
\xymatrixcolsep{1cm}    \xymatrix{
	& root=(0,3,3,\{1\})   \ar[d]^1          &           \\
	&  (1,3,3,\{2,3\})\ar[d]^2\ar[dr]^3   &           \\
	&  (2,3,2,\{3,4\})\ar[d]^4\ar[ld]^3   & (2,3,1,\{4\})\ar[d]^4 \\
	(3,3,1,\emptyset)  &  (3,3,1,\emptyset)                    & (3,3,1,\emptyset)         \\
      }
      \vbox{\hfill (b) \hfill}
    }
  }
  \label{fig:example-prefix-trie}
\end{figure}

As an example of a data operation, we describe the \textsc{Insert-PrefixTrie}
procedure that inserts an itemset into a prefix trie, see Algorithm
\ref{alg:prefix-tree-insert}. 

\bigskip
\begin{algorithm}[H]
\caption{The \textsc{Insert-PrefixTrie} procedure (prefix trie)}\label{alg:prefix-tree-insert}
\textsc{Insert-PrefixTrie}$($\inparam Itemset  $\isetU$, \inoutparam Node  N$)$
\begin{algorithmic}[1]
  \IF{$|\isetU| =$ N.depth}
    \STATE \textbf{return}
  \ENDIF
  \STATE $i\leftarrow$ N.depth
  \IF{$\isetU[i]\in$ N.children}
    \STATE\CALL{Insert-PrefixTrie}($\isetU$, N.children$[\isetU[i]]$)
  \ELSE
    \STATE N' $\leftarrow$ \textbf{new} Node
    \STATE N'.depth $\leftarrow$ N.depth
    \STATE N.children$[\isetU[i]] \leftarrow N'$
    \STATE $\CALL{Insert-PrefixTrie}(\isetU, N')$
  \ENDIF
\end{algorithmic}
\end{algorithm}

The \textsc{Insert-PrefixTrie} procedure is called: \textsc{Insert-PrefixTrie}$(\{1,2,3\}, root)$

\subsection{\textsc{Test-Subset} function and the \textsc{Compute-Support} procedure using prefix trie}\label{sec:apriori-support-alg}

In the \textsc{Generate-Candidates} function on code
line~\ref{alg:gener-cands:test} of
Algorithm~\ref{alg:generate-candidates}, we want to test if all
subsets of size $k-1$ of some itemset $\isetU$ of size $k$ are
contained in a set of itemsets of size $k-1$, e.g., $F_{k-1}$. The set
$F_{k-1}$ is represented by a prefix trie with maximal height
$max\_height = k-1$. The code line~\ref{alg:gener-cands:test} shows
that the algorithm is called by \textsc{Test-Subset}$(F_{k-1},
\isetU)$. Since we represent the set $F_{k-1}$ by a prefix trie, we
show the \textsc{Test-Subset} algorithm. The
\textsc{Test-Subset} algorithm has the first argument
replaced by a prefix trie node and has an additional helper parameter
(representing the depth of the recursion), i.e., let $R_k$ be a root
of a hash trie representing the set $F_k$ the algorithm
$\textsc{Test-Subset}(R_k, \isetU, 0)$ is then called by
$\textsc{Test-Susbet}(F_{k-1}, \isetU)$, i.e., the
\textsc{Test-Subset} shown in Algorithm~\ref{alg:generate-candidates}
could be implemented as:

\begin{algorithm}[H]
\caption{The \textsc{Test-Subset} function (using prefix trie)}\label{alg:test-subset}
\textsc{Test-Subset}$($\inparam Set $F_k$, \inparam Itemset $\isetU)$
\begin{algorithmic}[1]
  \STATE $R_k\leftarrow$ prefix trie representing the set $F_k$
  \STATE $\textsc{Test-Subset}(R_k,\isetU,0)$
\end{algorithmic}
\end{algorithm}

The algorithm \textsc{Test-Subset}, shown in
Algorithm~\ref{alg:test-subset} works as follows: in the root, we get
child for each item $b_i\in \isetU$ and recursively test
$\isetU\setminus\{b_i\}$ for all subsets of size $k-2$.  Thus, in an
interior node in which we get by following the item $b_i$, we will
recursively test all children which we get by hashing items $b_j>b_i$.
If the value returned from the recursive call is \texttt{true}, we
continue with the recursive descent, otherwise \texttt{false}
return. In a leaf node, we return \texttt{true}. If the return value
from root is \texttt{true} then all subsets of $t$ are in this prefix
trie.

The \textsc{Compute-Support} procedure works as follows: it
iterates over the database transactions $t$ incrementing the support
of some candidates itemsets using the
\textsc{Increment-Support} procedure. The
\textsc{Increment-Support} procedure increments the
support of all candidate itemsets that are subsets of the transaction
$t$. The \textsc{Increment-Support} procedure is almost
the same as the \textsc{Test-Subset} procedure except that
the support of a leaf node is incremented and nothing returned.

The pseudocode of the \textsc{Test-Subset} function and
\textsc{Compute-Support} procedure follows:

\bigskip
\begin{algorithm}[H]
\caption{The \textsc{Test-Subset} function (using prefix trie)}\label{alg:test-subset-hashtrie}
\hbox{\textsc{Test-Subset}$($\inparam Node N, \inparam Itemset $\isetU$, \inparam Integer index$)$}
\begin{algorithmic}[1]
  \FORALL{$i, index\leq i < |\isetU|$}
    \IF{$N$ is internal}
      \IF{$\isetU[index]\in N.children$ \textbf{and} $|\isetU| - i \geq  max\_depth - depth$}
	\STATE $result\leftarrow \CALL{Test-Subset}(N.children[\isetU[index]], \isetU, index+1)$
	\IF{$result=$\textbf{false}} 
        \STATE \textbf{return false}
	\ENDIF
      \ELSE 
      \STATE \textbf{return false}
      \ENDIF
    \ELSIF{$N$ is leaf} 
      \STATE\textbf{return true} 
    \ENDIF
  \ENDFOR
  \STATE \textbf{return result}
\end{algorithmic}
\end{algorithm}

\bigskip

\begin{algorithm}[H]
\caption{The \textsc{Compute-Support} procedure} \label{alg:compute-support-hashtrie}
\textsc{Compute-Support}$($\inparam Database $D$, \inoutparam Node N$)$
\begin{algorithmic}[1]
  \FORALL{$t\in \db$}
    \STATE$\CALL{Increment-Support}(N, t, 0)$
  \ENDFOR
\end{algorithmic}
\end{algorithm}

\bigskip
\begin{algorithm}[H]
\label{alg:hash-trie-increment-support}
\caption{The \textsc{Increment-Support} procedure}
\hbox{\textsc{Increment-Support}$($\inoutparam Node  $N$, \inparam Itemset $\isetU$, \inparam Integer index$)$}
\begin{algorithmic}[1]
  \FORALL{$i, index\leq i < |\isetU|$}
    \IF{$N$ is internal}
      \IF{$\isetU[index]\in N.children$ \textbf{and} $|\isetU| - i \geq  max\_depth - depth$}
	\STATE\CALL{Increment-Support}$(N.children[\isetU[index]], \isetU, index+1)$
      \ELSE 
      \STATE \textbf{return}
      \ENDIF
    \IF{$N$ is leaf} 
      \STATE $N.support\leftarrow N.support+1$
    \ENDIF
    \ENDIF
  \ENDFOR
\end{algorithmic}
\end{algorithm}
\bigskip

Since the number of subsets of size $k-1$ of some itemset of size $k$
is $k$, this algorithm needs at most $O(k^2)$ searches in the hash
trie.

\section{The FPGrowth algorithm} \label{sec:fpgrowth-algorithm}
The FPGrowth algorithm \cite{han00mining} is a DFS algorithm that does
not create candidates and thus does not count support for each
candidate. It rather creates a frequent pattern tree (or FP-Tree in
short) that represents the whole database. This algorithm needs only
two scans of the database, first to compute frequent items and second
to create an FP-Tree.

\begin{definition}[FP-Tree]
An \emph{FP-tree} is a prefix trie that has associated the tuple
\emph{(item, support, up-link, link, children)} with each node. The
\emph{support} field is the support of the prefix of the \emph{item}
field. The \emph{up-link} field is the link to the node at the
previous level. Nodes with a particular item form a list
linked by the \emph{link} field. An \emph{FP-Tree} also contains a
\emph{header table} in which pairs $(item, head)$ are stored. This
table contains heads of all linked lists.
\end{definition}

In the \emph{FP-tree}, we store $\isetU = (u_1,\ldots,u_n)$ with
$u_i\in\baseset$ and $u_i$ is a frequent item. The items are sorted
according to the support in descending order, i.e. $\supp(\{u_1\}) \leq
\ldots \leq \supp(\{u_n\})$. 


\noindent{\bf Rationale:} the following considerations explain briefly
some details of the FP-Tree: \par
\begin{enumerate}
  \item Only frequent items play a role in the mining process. Thus,
    we use only the frequent items for an FP-Tree construction.
  \item Because we sort the items in each itemset stored in the tree,
    we maximize the sharing of the prefix and therefore reduce size of
    the the tree and speed-up mining process. However, some recent
    publications states that the tree can be quite large.
  \item The FP-Tree construction is the same as that of the prefix trie
    (used in the Apriori algorithm) with one exception: we have to
    update the tail of the linked list of item $b$ when we add a new
    node with item $b$. 
  \item During the mining process, we need to find all nodes with
    a particular item. Thus each tree has a header table that has
    the form $(item, head)$ and each node has a link to another node,
    last node has {\sl null} pointer as the link value.
  \item The tree should be representation of the whole database.
\end{enumerate}

An FP-tree construction consists of two phases. First, all frequent
itemsets of size 1 are derived from the database (the first database
scan). Second, all transactions with deleted infrequent items and
items sorted by support (in descending order) are inserted into the
FP-Tree (the second database scan). This leads to the following
algorithm:

\bigskip
\begin{algorithm}[H]
\caption{Function \textsc{Construct-FP-Tree}}\label{alg:construct-fp-tree}
\textsc{Construct-FP-Tree}$($Database $\db$, Items $\baseset$, Integer $\minsupp$$)$
\begin{algorithmic}[1]
  \STATE Count support for each $b_i\in\baseset$
  \STATE create empty tree $T$
  \FORALL{transaction  $t\in\db$} 
    \STATE delete all $\bitem_i\in t, \supp(\{b_i\}, \db) < \minsupp$
    \STATE sort items in each transaction by support in descending order
    \STATE insert the transaction $t$ to the tree $T$ and update header links\label{alg:construct-fp-tree:insert}
  \ENDFOR
  \RETURN Constructed tree $T$
\end{algorithmic}
\end{algorithm}

\bigskip

The insert procedure on line \ref{alg:construct-fp-tree:insert} works
as the insert procedure for the prefix trie structure. The construction
process implies the following properties of an FP-Tree:

\medskip

\begin{proposition}\label{lemma:fp-tree-properties}
The FP-tree has the following properties:
\begin{enumerate}
\item An FP-Tree contains the complete information as the database
  from which it was constructed with the given $\minsupp$ with
  respect to the data mining process.
\item An FP-Tree size is bounded by occurrences of all frequent
  itemsets in database, the height of an FP-Tree is bounded by size of
  the longest itemset in the database.
\item All frequent itemsets containing item $b$ can be obtained by
  following an FP-Tree header links.
\end{enumerate}
\end{proposition}

To explain the FPGrowth algorithm, we need the following concepts:

\begin{definition}[Conditional pattern base of an item]
Let $b\in\baseset$ be an item and $T$ an FP-Tree. Let $N$ be the set
of nodes reachable from the links of the header list of $T$ for item
$b$. The \emph{conditional pattern base} of the item $b$ is the set of
all prefixes of the nodes $N$ (i.e. the set of all prefixes of $b$ in
$T$). Each prefix of a node $n\in N$ is assigned the support of the
node $n$.
\end{definition}

\begin{definition}[Conditional FP-Tree of an item]
  \emph{Conditional FP-Tree} of an item $b$'s is an FP-Tree that is
  constructed from $b$ conditional pattern base.
\end{definition}

\begin{definition}[Conditional FP-Tree of an itemset]
  Let $\isetU, \isetV\subseteq\baseset$ such that
  $\isetV=\isetU\setminus\{b\}$ and $T$ be an FP-tree. \emph{Conditional
    FP-Tree} of $\isetU$ is an FP-Tree $T'$ that is constructed as follows:
  \begin{itemize}
    \item[\hbox to 1cm{\hfill(i)\hfill}] Construct $b$ conditional
      FP-Tree $T_b$ from $T$.
    \item[\hbox to 1cm{\hfill(ii)\hfill}] Repeat step (i) recursively
      on $T_b$ for each itemset $b'\in\isetV$.
  \end{itemize}
\end{definition}

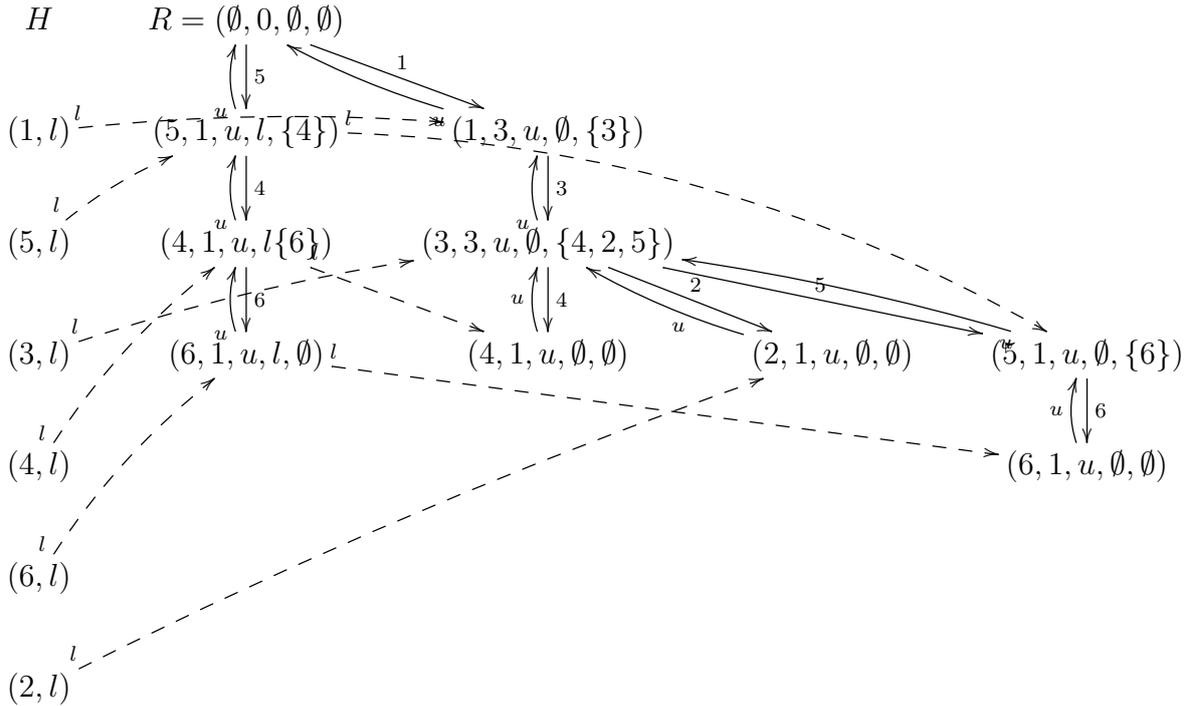
\begin{figure}[!ht]
  \centering
  \begin{tabular}{ccc}
    \hskip 2cm\begin{tabular}{|c|c|} \hline
      {\tt TID} & {\tt Transaction} \\ \hline
      1    &   $\{ 1,3,4\}$    \\ \hline
      2    &   $\{ 5,4,6\}$    \\ \hline
      3    &   $\{ 1,3,5,6\}$    \\ \hline
      4    &   $\{ 1,3,2\}$    \\ \hline
    \end{tabular} & \hskip 2cm&
    \begin{tabular}{|c|c|} \hline
      {\tt Item} & {\tt Support} \\ \hline
      1    &   3   \\ \hline
      2    &   3   \\ \hline
      3    &   2   \\ \hline
      4    &   2   \\ \hline
      5    &   1   \\ \hline
      6    &   1   \\ \hline
    \end{tabular}    \\
    \multicolumn{3}{c}{
      \xymatrixcolsep{0.8cm}\xymatrix{%
	H & R=(\emptyset,0,\emptyset,\emptyset)\ar[d]^5\ar[dr]^1 \\
        (1,l)\ar@{-->}@/^/[rr]^<l    & (5,1,u,l,\{4\})\ar[d]^4\ar@/^/[u]^<u\ar@{-->}@/^2pc/[rrrdd]^<l & (1,3,u,\emptyset,\{3\})\ar[d]^3\ar@/^/[ul]^<u & \\
        (5,l)\ar@{-->}@/^/[ru]^<l    & (4,1,u,l\{6\})\ar[d]^6\ar@/^/[u]^<u \ar@{-->}[rd]^<l& (3,3,u,\emptyset,\{4,2,5\})\ar[d]^4\ar[dr]^2\ar[drr]^5\ar@/^/[u]^<u & \\
        (3,l)\ar@{-->}@/^/[rru]^<l    & (6,1,u,l,\emptyset)\ar@/^/[u]^<u \ar@{-->}[rrrd]^<l & (4,1,u,\emptyset,\emptyset)\ar@/^/[u]^u & (2,1,u,\emptyset,\emptyset)\ar@/^/[ul]^u & (5,1,u,\emptyset,\{6\})\ar[d]^6\ar@/_/[ull]^<u \\
        (4,l)\ar@{-->}@/^/[ruu]^<l    &   &   &   & (6,1,u,\emptyset,\emptyset)\ar@/^/[u]^u \\
        (6,l)\ar@{-->}@/^/[ruu]^<l    &   &   &   & \\
	(2,l)\ar@{-->}@/^/[rrruuu]^<l    &   &   &   & \\
      }
    }\\
  \end{tabular}
  \caption[An example of an FP-Tree]{ An example of an FP-Tree. A link
    field is represented by edge with the label \emph{l} and up-link
    fields by edge with the label \emph{u}. A node contains (item,
    support, up-link, link, map). The map is represented by edge with
    a number as a label.}\label{fig:example-fp-tree}
\end{figure}

For deriving FIs, we use the property 3 of the Lemma
\ref{lemma:fp-tree-properties}. First, we choose an item and create a
conditional pattern base of this item. From the conditional pattern
base we create conditional FP-Tree and output all frequent
itemsets. This process is recursively repeated.  An example FP-Tree is
on \figref{fig:example-fp-tree}.

The path from a root of an FP-Tree will be denoted as $(b_1:s_1,
b_2:s_2, \ldots, b_n:s_n)$, where $b_j$ is an item at depth $j$ and
$s_j$ is the support of the itemset $(\bitem_1, \ldots, \bitem_j)$.
We examine data mining process by example, beginning from item 6.
First, we collect all frequent itemsets containing item 6 and derive
frequent itemset $(6)$ with support 2. And because there are two paths
$(1:3,3:3,5:1,6:1)$ and $(5:1,4:1,6:1)$, we have 6's conditional
pattern base $\{(1:1,3:1,5:1),(5:1,4:1)\}$. Each itemset from 6's
conditional pattern base occurs once in the database together with
item 6.  Construction of FP-Tree on this pattern base create 6's
conditional FP-Tree (see \figref{fig:6condtree}). Continuing in the FI
mining process only item 5 is frequent and it lead us to derive
itemset $(5,6)$ with support 2.

\begin{figure}[!ht]
  \example[6's conditional tree]{6's conditional tree constructed from the tree on \figref{fig:example-fp-tree}}\label{fig:6condtree}
  \centering
  \hskip 2cm
    \hbox{\hsize=5cm
      \vbox to 1cm {
	\renewcommand{\arraystretch}{1}
	\begin{tabular}{|c|c|} \hline
	  {\tt Transaction} & Support\\ \hline
	  $\{ 5,4\}$        &   1    \\ \hline
	  $\{ 5,3,1\}$      &   1    \\ \hline
	\end{tabular}\par
	\noindent6's conditional pattern base
      }\hskip 1cm%
      \vbox to 1cm{\xymatrixcolsep{0.8cm}\xymatrix{
	   H & R=(\emptyset,2,\emptyset,\emptyset,\{5\})\ar[d]^5 &  \\
	  (5,l)\ar@{-->}[r]^<l   & (5,2,u,\emptyset,\{1,4\})\ar[d]^1\ar@/^/[u]^u\ar[dr]^4 & \\
	  (1,l)\ar@{-->}[r]^<l   & (1,1,u,\emptyset,\{3\})\ar[d]^3\ar@/^/[u]^u & (4,1,u,\emptyset,\emptyset)\ar@/^/[ul]^u \\
	  (4,l)\ar@{-->}[rru]^<l & (3,1,u,\emptyset,\emptyset)\ar@/^/[u]^u & \\
	  (3,l)\ar@{-->}[ru]^<l  &    & \\
	}
      }
    }
\end{figure}

For item 5 the process is similar. One path is found $(1:3,3:3,5:2)$,
thus we have the conditional pattern base $\{(1:2,3:2)\}$ and we
derive frequent itemset $(5)$ with support $2$. Creating conditional
FP-Tree on this itemset creates FP-Tree with one leaf. 3's conditional
pattern base is $\{(1:2)\}$ and $(3,6)$ with support 2 is derived. 1's
conditional pattern base is empty and $(1,3,6)$ with support 2 is
derived. Looking back on $\{(1:2,3:2)\}$, and creating 1's conditional
pattern base (which is empty) lead us to derive itemset $(1,6)$. This
process leads to following observation: when the FP-Tree consists of
single path then all combinations of items in this paths derives
frequent itemset.

\clearpage

\bigskip
The pseudocode for the FPGrowth algorithm follows: \par
\begin{algorithm}[H]
\caption{The \textsc{FPGrowth} algorithm}
\hbox{\textsc{FPGrowth}$($\inparam Database $\db$, \inparam Integer $\minsupp$, \outparam Set $\allfi)$}
\begin{algorithmic}[1]
  \STATE Compute all frequent items and store them into $\baseset$
  \STATE $T\leftarrow\CALL{Construct-FP-Tree}(\db, \baseset, \minsupp)$
  \STATE $\CALL{FPGrowth-Computation}(T, \emptyset, \minsupp)$
\end{algorithmic}
\end{algorithm}

\begin{algorithm}[H]
\caption{The \textsc{FPGrowth-Computation} algorithm}
\hbox{\textsc{FPGrowth-Computation}$($\inparam FP-Tree $T$, \inparam Itemset $\isetU$, \inparam Integer $\minsupp$, \outparam $\allfi)$}
\begin{algorithmic}[1]
   \IF{$T$ contains only single path $P$} 
      \FORALL{combination $\isetW$ of nodes in the path $P$}
        \STATE $s\leftarrow min\{s: \bitem\in \isetW \wedge s=\bitem.\text{support}\}$
	\IF{$s \geq \minsupp$}
	  \STATE $\allfi\leftarrow \allfi\cup\{\isetW \cup \isetU\}$ 
	\ENDIF
      \ENDFOR
      \ELSE
      \FORALL{items $\bitem$ in the header of $T$}
	\IF{$\supp(\{\bitem\})\geq \minsupp$}
	  \STATE $\allfi\leftarrow \{\{\bitem\}\cup \isetU\}$
	\ENDIF
	\STATE Construct $\bitem$'s conditional {\bf FP-Tree} $T_\bitem$ from $T$, i.e., creating tree\par representing $\isetU\cup\{\bitem\}$
	\IF{size of the tree $T_\bitem \neq 0$}
           \STATE\CALL{FPGrowth-Computation}$(T_\bitem, \isetU\cup\{\bitem\})$
	\ENDIF
      \ENDFOR
   \ENDIF
\end{algorithmic}
\end{algorithm}
\bigskip

\clearpage
\section{The Eclat Algorithm} \label{sec:eclat-algorithm}

Papers \cite{zaki00scalable, zaki98thesis} use different approach than
the Apriori algorithm. Eclat (which stands for {\bf E}quivalence {\bf
  CL}ass {\bf T}ransformation) uses lattice-based approach that
utilizes vertical representation of a database. The Eclat algorithm is
a DFS or BFS algorithm. Whereas all of the above algorithms use
several scans of a database, this approach scans the database only
once.

\subsection{Support counting}
Let $\lattice=(\powerset(\baseset);\subseteq)$ be a lattice,
$\bitem_i\in \atoms(\lattice)$ be an atom and $\tidlist(\{\bitem_i\})$
be the tidlist of the atom $b_i$. Thus, the support of $\bitem_i$ can
be computed as $|\tidlist(\{\bitem_i\})|$. We can get set of
transaction ids containing itemset $\{\bitem_i, \bitem_j\}, i\neq j$,
as $\tidlist(\{b_i\}) \cap \tidlist(\{b_j\})$ and $\supp(\{\bitem_i,
\bitem_j\})=|\tidlist(\{\bitem_i\})\cap\tidlist(\{\bitem_j\})|$. In
general, the support of a set $S\subseteq \atoms(\lattice)$ can be
computed as $|\bigcap_{\bitem_i\in S} \tidlist(\{b_i\})|$, see Section
\ref{sec:lattice-in-algorithms}. In particular, we can use only two
subsets of $\isetV$ to compute $\supp(\isetV)$, because to create
$\isetV$ we need two $\isetU_1, \isetU_2 \subseteq \isetV,
\isetU_1\cup\isetU_2=\isetV$, i.e. $\tidlist(\isetV) =
\tidlist(\isetU_1) \cap \tidlist(\isetU_2)$.





\subsection{The depth-first search Eclat algorithm}\label{eclat-bottom-up}

In Section~\ref{sec:lattice-in-algorithms}, we have discussed the
PBECs and the hierarchy of PBECs. The hierarchy of PBECs forms a tree
that can be used in a DFS algorithm. The Eclat algorithm is an
algorithm that searches the tree of PBECs in a DFS fashion.  This
strategy utilizes the lattice decomposition of frequent itemsets
induced into smaller classes. To compute the support of any itemset,
we simply intersect list of transaction id's of any of its two subsets
in lexicographic or reverse lexicographic order.

The depth-first search tree of the join semi-lattice of all FIs is
depicted in \figref{fig:dfs-tree-eclat}. The algorithm proceeds
recursively. Example of the tidlist constructed by the algorithm are
in \exref{bottom-up-tables}. The algorithm \textsc{Eclat-DFS} is
summarized in Algorithm \ref{alg:eclat-dfs}. The algorithm is called
by $\textsc{Eclat-DFS}(\db, \minsupp, \allfi)$ and the output
stored in $\allfi$.

\begin{algorithm}[H]
\caption{The \textsc{Eclat-DFS} algorithm}\label{alg:eclat-dfs}
\textsc{Eclat-DFS}\vbox{$($\inparam Database $\db$, \inparam Support $\minsupp$, \outparam Set $\allfi)$}

\begin{algorithmic}[1]
  \STATE Create vertical representation $T$ of the database $\db$
  \STATE $\atoms\leftarrow$ all frequent items from $\db$
  \STATE $\CALL{Eclat-DFS-Computation}(\atoms, T, \emptyset, \minsupp, \allfi)$
\end{algorithmic}
\end{algorithm}

\begin{algorithm}[H]
\caption{The \textsc{Eclat-DFS-Computation} algorithm}\label{alg:eclat-dfs-computation}
\textsc{Eclat-DFS-Computation}$($\vtop{\inparam Atoms $\atoms$,\par\noindent
                                       \inparam Tidlists $T$,\par\noindent
                                       \inparam Itemset $P$,\par\noindent
                                       \inparam Support $\minsupp$,\par\noindent
                                       \outparam Set $\allfi)$}

\begin{algorithmic}[1]
  \item[\textbf{Note:}] The tidlists of itemsets $\isetU$, $\tidlist(\isetU)$, used in this algorithm are taken from $T$.
  \FORALL{\text{atom} $\atom_i \in \atoms$}
    \STATE $\atoms_i \leftarrow \emptyset$
    \FORALL{atom $\atom_j\in \atoms, \atom_i < \atom_j$}
      \IF{$|\tidlist(P\cup\{\atom_j\})| \geq \minsupp$}
	\STATE $\atoms_i \leftarrow \atoms_i\cup \{\atom_j\}$
        \STATE $f\leftarrow P\cup \{\atom_j\}$
	\STATE $\allfi \leftarrow \allfi \cup \{f\}$
      \ENDIF
    \ENDFOR 
    \STATE\CALL{Eclat-DFS-Computation}$(\atoms_i, P\cup\{a_i\}, \allfi)$
  \ENDFOR
\end{algorithmic}
\end{algorithm}

\begin{figure}[!ht] 
\centering
\includegraphics[type=mps,ext=.mps,read=.mps]{lattice_dfstree_plain}
\caption[The DFS tree of the execution of the Eclat algorithm.]{The
  DFS tree of the execution of the Eclat algorithm using the order
  $1<2<3<4<5$ of the baseset $\baseset=\{1,2,3,4,5\}$.}
\label{fig:dfs-tree-eclat}
\end{figure}

\clearpage

\begin{figure}[!hb]
\example[Example of bottom-up search strategy]{Bottom-up search strategy ($\minsupp=2$)} \label{bottom-up-tables}
\begin{center}
\noindent\begin{tabular*}{15cm}{p{7.5cm}p{7.5cm}}
\hskip 1.5cm Horizontal representation of $\db$ & \hskip 1.5cm Vertical representation of $\db$\\
{
  \hfill
  \renewcommand{\arraystretch}{1}
       \begin{tabular}{|c|c|} \hline
	 {\tt TID} & {\tt Transaction} \\ \hline
	 1    &   $\{ 1,2,3,4\}$    \\ \hline
	 2    &   $\{ 3,5\}$    \\ \hline
	 3    &   $\{ 1,3,4\}$    \\ \hline
	 4    &   $\{ 1,2\}$    \\ \hline
	 5    &   $\{ 1,3,4,5\}$    \\ \hline
	 6    &   $\{ 1,2,3,4,5\}$    \\ \hline
       \end{tabular}\hfill%
  \medskip
} &
{
  \renewcommand{\arraystretch}{1}
       \begin{tabular}{|c|c|c|c|c|c|} \hline
{\tt itemset}, $\isetU=$ & $\{1\}$ & $\{2\}$ & $\{3\}$ & $\{4\}$ & $\{5\}$ \\ \hline
        {\tt tidlist},   & $1$ & $1$ & $1$ & $1$ & $2$ \\ 
     $\tidlist(\isetU)=$ & $3$ & $4$ & $2$ & $3$ & $5$ \\ 
	                 & $4$ & $6$ & $3$ & $5$ & $6$ \\ 
	                 & $5$ &     & $5$ & $6$ &     \\ 
	                 & $6$ &     & $6$ &     &     \\ \hline
       \end{tabular}\hfill%
  \medskip
}
\\
\multicolumn{2}{c}{%
  \renewcommand{\arraystretch}{1}
       \begin{tabular}{|p{2cm}|c|c|c|c|c|c|c|c|c|c|} \hline
	 Frequent         & $\times$  & $\times$  & $\times$  & $\times$  & $\times$  & $\times$  &           & $\times$  & $\times$  & $\times$   \\ \hline
{\tt itemset}, $\isetU=$    & $\{1,2\}$ & $\{1,3\}$ & $\{1,4\}$ & $\{1,5\}$ & $\{2,3\}$ & $\{2,4\}$ & $\{2,5\}$ & $\{3,4\}$ & $\{3,5\}$ & $\{4,5\}$  \\ \hline
{\tt tidlist},            &    $1$    &    $1$    &    $1$    &    $5$    &    $1$    &    $1$    &    $6$    &    $1$    &    $2$    &    $5$     \\
$\tidlist(\isetU)=$       &    $4$    &    $3$    &    $3$    &    $6$    &    $6$    &    $6$    &           &    $3$    &    $5$    &    $6$     \\
	                  &    $6$    &    $5$    &    $5$    &           &           &           &           &    $5$    &    $6$    &            \\
	                  &           &    $6$    &    $6$    &           &           &           &           &    $6$    &           &            \\ \hline
       \end{tabular}
  \medskip%
} \\
\multicolumn{2}{c}{%
{
  \renewcommand{\arraystretch}{1}
       \begin{tabular}{|c|c|c|c|c|c|c|c|c|c|} \hline
	 Frequent         & $\times$    & $\times$    &             & $\times$    & $\times$    & $\times$    \\ \hline
{\tt itemset}, $\isetU=$  & $\{1,2,3\}$ & $\{1,2,4\}$ & $\{1,2,5\}$ & $\{1,3,4\}$ & $\{1,3,5\}$ & $\{1,4,5\}$ \\ \hline
	 {\tt tidlist}    &     $1$     &     $1$     &     $6$     &     $1$     &     $5$     &     $5$     \\ 
$\tidlist(\isetU)=$       &     $6$     &     $6$     &             &     $3$     &     $6$     &     $6$     \\ 
                          &             &             &             &     $5$     &             &             \\ 
                          &             &             &             &     $6$     &             &             \\ \hline

       \end{tabular}\hfill\medskip
}} \\
\multicolumn{2}{c}{%
{
  \renewcommand{\arraystretch}{1}
       \begin{tabular}{|c|c|c|c|} \hline
         Frequent         & $\times$    &             & $\times$    \\ \hline   
{\tt itemset}, $\isetU=$  & $\{2,3,4\}$ & $\{2,4,5\}$ & $\{3,4,5\}$ \\ \hline
{\tt tidlist},             &     $1$     &     $6$     &     $5$     \\       
$\tidlist(\isetU)=$       &     $6$     &             &     $6$     \\ \hline
       \end{tabular}
  \hfill\medskip
}} \\
\multicolumn{2}{c}{%
{
  \renewcommand{\arraystretch}{1}
       \begin{tabular}{|c|c|c|c|} \hline
         Frequent itemset &   $\times$    &   $\times$    \\ \hline   
{\tt itemset}, $\isetU=$  & $\{1,2,3,4\}$ & $\{1,3,4,5\}$ \\ \hline
         {\tt tidlist},   &     $1$       &     $5$       \\       
$\tidlist(\isetU)=$       &     $6$       &     $6$       \\ \hline
       \end{tabular}
  \hfill\medskip
}}
\end{tabular*}
\end{center}
\end{figure}

\section{Possible optimizations of the DFS sequential algorithms} \label{sec:dfs-optimizations}
\subsection{The ``closed itemsets'' optimalization}

The concept of \emph{closed itemsets}, see
Definition~\ref{def:closed-itemset} can be used for optimization of
the DFS algorithms.

Let a DFS algorithm process prefix $\isetU$ and the possible branches
(extensions) of $\isetU$ are denoted by $\prefixext$. The algorithm
can extend $\isetU$ by all items $\prefixext' = \{\bitem_i |
\bitem_i\in \prefixext, \supp(\bitem_i)=\supp(\isetU)\}$ without
computation of the intermediate tidlists $\tidlist(\isetV\cup\isetU),
\isetV\subseteq\prefixext'$, i.e., saving $O(2^{|\isetV|})$ of
intersections of tidlists.




\subsection{Ordering of items in DFS algorithms}\label{sec:dynamic-item-ordering}


Consider a baseset $\baseset$ and a database $\db$. Any DFS
algorithm should expand every prefix $\isetU$ using the extensions
$\prefixext$ sorted by the support in ascending order. This allows
for efficient computation of intermediate steps.

At a particular step of a sequential FIM algorithm, the prefix
$\itsetprefix=\{\pref_1, \ldots, \pref_k\}, \pref_i\in \baseset$, and
extensions $\prefixext=\{\ext_1, \ldots, \ext_l\}, \ext_i\in B$, and
$\rsupp(\ext_{1})\leq\rsupp(\ext_{2})\leq\ldots\leq \rsupp(\ext_{l})$.
The algorithm can choose from many possible orders. Let choose two
possible orders of $\ext_i$ for processing: 1)
$\ext_{1},\ldots,\ext_{l}$ (smallest first); 2)
$\ext_{l},\ldots,\ext_{1}$ (largest first).

\vbox{
\begin{enumerate}
\item A DFS algorithm processes every prefix $\itsetprefix$ in
  the following way: extend the prefix $\itsetprefix\cup\{\ext_1\}$
  and consider $\ext_2, \ext_3, \ldots, \ext_l$ as extensions (in that
  order). Using this order it follows that the smallest partition of the
  database gets the largest partition of the search space.

\item A DFS algorithm processes every prefix in the following
  way: extend the prefix $\itsetprefix\cup\{\ext_l\}$ and consider
  $\ext_l, \ext_{l-1}, \ldots, \ext_2$ as extensions (in that
  order). Using this order it follows that the largest partition of the
  database gets the largest partition of the search space.
\end{enumerate}
}

If we compare these two approaches, it is clear that the second case
should be much slower than the first case. The reason is that it is
more time-consuming to process an item that has large support than an
item that has a small support. Other cases are somewhere in-between of
these two cases.  The optimal solution is to compute the support of
each extension $\prefixext=\{\ext_1,\ldots,\ext_n\}$ for every prefix
$\isetU$ and reorder the items, i.e. choose the order $\ext_1 \leq
\ext_2 \leq \ldots \leq \ext_n$ such that $\supp(\isetU\cup\{\ext_1\})
\leq \supp(\isetU\cup\{\ext_2\})\leq \ldots \leq
\supp(\isetU\cup\{\ext_2\})$. Such \emph{dynamic} ordering of items is
essential for the speed of a DFS FIM algorithm.


\subsection{The concept of diffsets}

If we start to mine large database with very large lists of
transaction ids for an item (or atoms), the intersection time becomes
too large.  Furthermore, the size of a list of transaction ids of a
frequent itemset also become very large and these lists of transaction
id's cannot fit into the main memory.  These problems are solved with
so called \emph{difference sets} (or \emph{diffsets} in short)
\cite{zaki01fast}. A diffset is the difference of two list of
transaction ids.

\begin{definition}[Difference set]\label{diffset}
Let $\isetU\subset \baseset$ be an itemset and $\bitem_i\in
\baseset-\isetU$ an item. $\tidlist(\isetU)$ denotes a set of
transaction id's. \emph{Difference set} (or diffset in short) is
$\diffset(\isetU\cup\{\bitem_i\})=\tidlist(\isetU)-\tidlist(\{\bitem_i\})$
\end{definition}

First we have to note that the size of a diffset of an itemset
$\isetU\cup\{\bitem_i\}$ is no longer the support of the itemset. However,
the support of $\isetU\cup\{\bitem_i\}$ can be computed as follows:

$$Supp(\isetU\cup\{\bitem_i\})=Supp(\isetU)-|\diffset(\isetU\cup\{\bitem_i\})|$$

Now let $\isetU\subset\baseset$ be an itemset and
$\bitem_i,\bitem_j\in \baseset,\bitem_i<\bitem_j$ and
$\bitem_i,\bitem_j\notin \isetU$ be two items. We use instead of
transaction list $\tidlist (\isetU\cup\{\bitem_i\})$ ($\tidlist
(\isetU\cup\{\bitem_j\})$), diffsets $\diffset
(\isetU\cup\{\bitem_i\})$ ($\diffset (\isetU\cup\{\bitem_j\})$),
respectively. We want to compute support of
$\isetU\cup\{\bitem_i\}\cup\{\bitem_j\}$ using only diffsets. From
\defref{diffset}, we have
$Supp(\isetU\cup\{\bitem_i\}\cup\{\bitem_j\}) =
Supp(\isetU\cup\{\bitem_i\}) -
|\diffset(\isetU\cup\{\bitem_i\}\cup\{\bitem_j\})|$. But we have only
diffsets and not list of transaction id's. But it is easy to fix:

$$\diffset(\isetU\cup\{\bitem_i\}\cup\{\bitem_j\})\begin{array}[t]{l}
          \displaystyle=\tidlist(\isetU\cup\{\bitem_i\})-\tidlist(\isetU\cup\{\bitem_j\})                             \\
	  \displaystyle=\tidlist(\isetU\cup\{\bitem_i\})-\tidlist(\isetU\cup\{\bitem_j\})+\tidlist(\isetU)-\tidlist(\isetU)     \\
          \displaystyle=(\tidlist(\isetU)-\tidlist(\isetU\cup\{\bitem_j\}))-(\tidlist(\isetU)-\tidlist(\isetU\cup\{\bitem_i\})) \\
	  \displaystyle=\diffset(\isetU\cup\{\bitem_j\})-\diffset(\isetU\cup\{\bitem_i\})                             \\
\end{array} $$

The concept of diffsets is used in the Eclat algorithm. Generally it
is possible to use the diffsets in an arbitrary algorithm that uses
the \emph{vertical representation} of the database.

\section{Discovering rules}\label{sec:discovering-rules}

To complete the overview of the sequential algorithms, we show how to
create association rules from the FIs.

When we have discovered all frequent itemsets, we have to create all
rules $X\Rightarrow Y$ with given confidence. Generation of all such
rules is based on the following observation:


If we have frequent itemset $L$ and its subset $A$,
and $Conf(A,(L-A)) >\minconf$, then if $A\Rightarrow (L-A)$ does not
have enough confidence, then for all subsets $a \subseteq A$ the
rule $a\Rightarrow (L-a)$ does not have enough confidence. For
example, if the association rule $123\Rightarrow 4$ does not have
enough confidence, we need not check whether $12\Rightarrow 34$ holds.
Following algorithm uses this observation:

\bigskip
\begin{algorithm}[H]
\caption{The \textsc{Generate-All-Rules} algorithm}
\textsc{Generate-All-Rules}$($\inparam Set $\allfi)$
\begin{algorithmic}[1]
  \FORALL{frequent itemset $\isetU, |\isetU| \geq 2$}
    \STATE $\CALL{Generate-Rules}(\isetU, \isetU)$
  \ENDFOR
\end{algorithmic}
\end{algorithm}
\bigskip

\begin{algorithm}[H]
\caption{The \textsc{Generate-Rules} algorithm}
\vbox{\textsc{Generate-Rules}$($\inparam Itemset $\isetU$,\inparam Itemset $\isetW)$}
\begin{algorithmic}[1]
  \REQUIRE $|\isetU|=k, |\isetW|=m$
  \STATE $A=\{\isetV:  \isetV\subset \isetW \wedge |\isetV| = m-1 \}$
  \FORALL{$\isetV \in A$}
    \STATE compute confidence $\conf(\isetV, \isetU)$, $c \leftarrow \supp(\isetU)/\supp(\isetV)$
    \IF{$c \geq \minconf$}
      \STATE output $(\isetV\Rightarrow \isetU\setminus\isetV)$, with confidence $c$ and support $\supp(\isetU)$
      \IF{$|\isetW|-1>1$} 
        \STATE \textbf{Generate-Rules}$(\isetU, \isetW)$
      \ENDIF
    \ENDIF
  \ENDFOR
\end{algorithmic}
\end{algorithm}

\clearpage


\chapter{Lists of abbreviations}
\begin{tabular}{p{5cm}p{11cm}}
BFS            & Breadth-First Search \\
cc-NUMA        & cache-coherent Non-Uniform Memory Access \\
CI             & closed itemsets \\
DM             & Distributed Memory \\
DFS            & Depth-First Search \\
diffset        & difference set \\
Eclat          & Equivalence class transformation \\ 
FI             & Frequent itemset \\
FIMI           & Frequent Itemset MIning \\
FP-Growth      & Frequent Pattern Growth \\
FP-Tree        & Frequent Pattern Tree \\
FPM            & Fast parallel mining \\
FPGrowth       & Frequent Pattern Growth \\
LPT            & least processing time \\
MFI            & Maximal frequent itemset \\
MLFPT          & Multiple Local Frequent Pattern Tree \\
NUMA           & Non-Uniform Memory Access \\
PBEC           & Prefix-based equivalence class \\
TID            & transaction id \\
tidlist        & Transaction Id list \\
trie           & a prefix trie, from the word retrieval \\
QKP            & quadratic knapsack \\
\end{tabular}

\chapter{Used symbols}\label{chap:used-symbols}

\begin{tabular}{p{5cm}p{11cm}}
 $\baseset$                   & Base itemset $\baseset=\{\bitem_1, \bitem_2, \ldots, \bitem_{|\baseset|}\}, \bitem_1<\bitem_2<\ldots<\bitem_{|\baseset|}$ \\ 
 $\powerset(S)$               & The powerset of the set $S$, i.e., the set $\{s | s\subseteq S\}$ \\
 $\isetU,\isetV,\isetW$       & Itemsets or sets \\
 $\isetU\Rightarrow \isetV$   & Association rule \\
 $\db$                        & A database  \\
 $\dbpart_i$                  & A database partition, usually $\dbpart_i\cap\dbpart_j=\emptyset, i\neq j$ and $\db=\bigcup_i\dbpart_i$ \\
 $t=(\isetU,id)$              & A transaction from the database with unique identifier id \\
 $\tidlist(\isetU)$           & The transaction id list (or tidlist in short) of the transactions containing the itemset $\isetU$ as a subset \\
 $\dbsmpl$                    &  A sample of the database $\db$ \\
 $\procnum$                   & The number of processors \\
 $\proc_i$                    & $i$-th processor, $1\leq i\leq \procnum$ \\
 $\supp(\isetU,\db)$          & The \emph{support} of an itemset $\isetU$ in database $\db$ (or $\supp(\isetU)$ if $\db$ is clear from context) \\
 $\rsupp(\isetU,\db)$         & The \emph{relative support} of an itemset, i.e., $\rsupp(\isetU, \db) = \supp(\isetU,\db) / |\db|$ \\
 $\minsupp, \rminsupp$        & The absolute minimal support and the relative minimal support \\
 $\conf(\isetU,\isetW,\db)$   & Confidence of association rule $\isetU\Rightarrow \isetW$ (or $\conf(\isetU, \isetW)$ if $\db$ is clear from context \\
 $\minconf$                   & Minimal confidence \\
 $\bcover(\isetU,\db)$        & Cover of the itemset $\isetU$ in database $\db$, i.e., the set of transaction containing $\isetU$ as a subset \\
 $\tcover(T,\db)$             & Cover of the transactions $T\subseteq\db$, i.e., an itemsets containing all items that are contained
                                in \emph{all} transactions $T$ \\
 $\allfi$                     & The set of all frequent itemsets \\
 $\mfi$                       & The set of all maximal frequent itemsets \\
 $\mfiapprox$                 & The approximation of the MFIs in the database $\db$, computed using $\dbsmpl$ \\
 $\fipart_i, 1\leq i\leq\procnum$  & disjoint partitions of all FIs $\allfi=\bigcup_{i=1}^\procnum\fipart_i$ \\
 $\fipart_k, k > 0$           & The set of frequent itemsets of size $k$ \\ 
 $\fiapprox$                  & The approximation of the FIs in the database $\db$, i.e.,
                                $\fiapprox = \{\isetU | \exists\isetW\in\mfiapprox, \isetU \subseteq \isetW \}$ \\
\end{tabular}
\begin{tabular}{p{5cm}p{11cm}}
 $\fismpl$                    & A sample of $\fiapprox$, i.e., $\fismpl \subseteq \fiapprox$ \\
 $C_k$                        & Set of candidate itemsets on FIs of size $k$  \\
 $\top$                       & The top element of a lattice  \\
 $\bot$                       & The bottom element of a lattice  \\
 $\atoms(\lattice)$           & The set of all atoms of a lattice $\lattice$  \\
 $\tidlist(\isetU)$           & List of transaction ids of an itemset $\isetU$  \\
 $\diffset(\isetU)$           & Difference set (or diffset in short) of transaction ids of an itemset $\isetU$  \\
 $[\isetU|\prefixext]$        & Prefix based equivalence class with prefix $\isetU$, i.e., \par\noindent
                                $[\isetU|\prefixext] = \{\isetW | \isetW=\isetU\cup \isetV, \isetV\subseteq\prefixext\text{ and }
                                \forall \bitem_{\prefixext}\in\prefixext, \bitem_{\isetU}\in\isetU: \bitem_{\isetU} < \bitem_{\prefixext} \}$.
                                $\prefixext$ can be omitted if clear from context.\\

 $\epsilon_{\dbsmpl}$           & Error of an approximation of the support of an itemset $\isetU$ in database
                                $\db$ computed from a database sample $\dbsmpl$ \\
 $\delta_{\dbsmpl}$             & Probability of the error $\epsilon_{\dbsmpl}$ \\
 $\epsilon_{\fismpl}$           & Error of an approximation of the relative size of a set $\fipart\subseteq\fiapprox$,
                                i.e., an error of the size $|\fipart|/\fiapprox$ \\
 $\delta_{\fismpl}$             & Probability of the error $\epsilon_{\fismpl}$ \\
\end{tabular}

\end{document}